Collegium Sciences et Technologies
ED412 SESAMES - Synthèses, Expériences, Simulations, Applications : de la Molécule aux Édifices Supramoléculaires

Thèse présentée pour l'obtention du grade de

**Docteur de l'Université de Lorraine**

en Chimie

par **M. Bastien MUSSARD**

# Modélisations quantochimiques des forces de dispersion de London par la méthode des phases aléatoires (RPA) : développements méthodologiques

Soutenance publique le 13 Décembre 2013

Membres du jury :

| | | | |
|---|---|---|---|
| Prof. | Benoit CHAMPAGNE | Université de Namur | Rapporteur |
| Dr. | Andreas SAVIN | DR CNRS & Université Pierre et Marie Curie | Rapporteur |
| Prof. | Xavier ASSFELD | Université de Lorraine | Membre du Jury |
| Prof. | Georg JANSEN | Universität Duisburg-Essen | Membre du Jury |
| Prof. | Peter REINHARDT | Université Pierre et Marie Curie | Membre du Jury |
| Dr. | Ágnes SZABADOS | Université Eötvös de Budapest | Membre du Jury |
| Dr. | János ÁNGYÁN | DR CNRS & Université de Lorraine | Directeur de thèse |
| Dr. | Sébastien LEBEGUE | CR CNRS & Université de Lorraine | Codirecteur de thèse |

Laboratoire de Cristallographie, Résonance Magnétique et Modélisations ($CRM^2$)
Faculté des Sciences et Technologies, Université de Lorraine

*Aux Kagress, qui ont formé mon esprit*
*et détruit mon corps.*

*À Julie pour les relectures*
*et pour tout le reste.*

*À mes parents.*

« *I said, c'est qui… ?* »

# Résumé

Dans cette thèse sont montrés des développements de l'approximation de la phase aléatoire (RPA) dans le contexte de théories à séparation de portée. On présente des travaux sur le formalisme de la RPA en général, et en particulier sur le formalisme "matrice diélectrique" qui est exploré de manière systématique. On montre un résumé d'un travail sur les équations RPA dans le contexte d'orbitales localisées, notamment des développements des orbitales virtuelles localisées que sont les « orbitales oscillantes projetées » (POO). Un programme a été écrit pour calculer des fonctions telles que le trou de d'échange, la fonction de réponse, *etc.*.. sur des grilles de l'espace réel (grilles parallélépipédiques ou de type "DFT"). On montre certaines de ces visualisations. Dans l'espace réel, on expose une adaptation de l'approximation du dénominateur effectif (EED), développée originellement dans l'espace réciproque en physique du solide. Également, les gradients analytiques des énergies de corrélation RPA dans le contexte de la séparation de portée sont dérivés. Le formalisme développé ici à l'aide d'un lagrangien permet une dérivation tout-en-un des termes courte- et longue-portée qui émergent dans les expression du gradient, et qui montrent un parallèle intéressant. Des applications sont montrées, telles que des optimisations de géométries aux niveaux RSH-dRPA-I et RSH-SOSEX d'un ensemble de 16 petites molécules, ou encore le calcul et la visualisation des densités corrélées au niveau RSH-dRPA-I.

# Abstract

In this thesis are shown developments in the random phase approximation (RPA) in the context of range-separated theories. We present advances in the formalism of the RPA in general, and particularly in the "dielectric matrix" formulation of RPA, which is explored in details. We show a summary of a work on the RPA equations with localised orbitals, especially developments of the virtual localized orbitals that are the « projected oscillatory orbitals » (POO). A program has been written to calculate functions such as the exchange hole, the response function,*etc...* on real space grid (parallelepipedic or of the "DFT" type) ; some of those visualisations are shown here. In the real space, we offer an adaptation of the effective energy denominator approximation (EED), originally developped in the reciprocal space in solid physics. The analytical gradients of the RPA correlation energies in the context of range separation has been derived. The formalism developped here with a lagrangian allows an all-in-one derivation of the short- and long-range terms that emerge in the expressions of the gradient. These terms show interesting parallels. Geometry optimisations at the RSH-dRPA-I and RSH-SOSEX levels on a set of 16 molecules are shown, as well as calculations and visualisations of correlated densities at the RSH-dRPA-I level.

# Remerciements

Je remercie Claude Lecomte et Dominik Schaniel, les deux directeurs successifs du laboratoire CRM$^2$ pour m'avoir accueilli dans de bonnes conditions de travail, ainsi que pour les divers congrès et écoles auxquels j'ai pu participer pendant les trois ans de cette thèse.

Un grand merci à János, pour sa patience et pour connaître la recette du mélange idéal entre accompagnement et autonomie. Je voudrais également le remercier avec beaucoup d'émotion pour sa disponibilité en général, et pour le fait que même dans les moments où sa santé l'a éloigné de Nancy, je ne me suis jamais senti seul ou perdu.

Je remercie les membres du jury, Xavier Assfeld, Georg Jansen, Peter Reinhardt et Ágnes Szabados, ainsi que les rapporteurs, Benoit Champagne et Andreas Savin pour avoir accepté de juger ce travail. I would also like to most warmly thank all the people at the Laboratory of Theoretical Chemistry, Eötvös University in Budapest for the kind welcome I received there, especially Péter Szalay and, again, Ágnes. Je voudrais aussi mentionner Fabien Cailliez : c'est avec lui que j'ai découvert que c'est en chimie théorique que je ferai de la recherche ; et Julien Toulouse pour des échanges scientifiques toujours très intéressants et très clairs.

Merci à t.i.b.o pour le cours de "Fonctions de Green 101" au bord de l'eau dans les premiers mois de la thèse ; merci à Qbi et Benoït pour avoir relu les Annexes un peu mathématiques. Merci en fait à tout ceux qui savent qu'on se doit tout les uns les autres… Merci à Julie pour les relectures, les "cafés du petit matin" et tout le reste.

Enfin, merci à mes parents, pour tout.

Nancy, le 15 octobre 2013
B. Mussard

# Table des matières

# Table des notations

Les notations suivantes sont utilisées dans le manuscrit :

| | |
|---|---|
| $X_{a\to b}$ | désigne une suite des éléments $X_i$ : $X_a, X_{a+1}, \ldots, X_b$ |
| $\int_{X_{a\to b}}$ | de la même manière, désigne : $dX_a dX_{a+1} \ldots dX_b$ |
| $\mathcal{L}^2(\mathbb{A}, \mathbb{C})$ | est l'espace de Hilbert des fonctions de $\mathbb{A}$ dans $\mathbb{C}$ de carré intégrable. |
| $\langle f\|g\rangle$ | produit scalaire de l'espace $\mathcal{L}^2(\mathbb{A}, \mathbb{C})$ : $\int_{\mathbb{A}} f^* g$ |
| $\mathcal{L}^b$ | désigne plus librement un espace de Hilbert généré par la base $b$ |
| $\delta_{i,j}$ | produit de Kronecker (égal à 1 si $i = j$, à 0 sinon) |
| $A \doteq B$ | indique un changement de notation ou une définition |
| $A \propto B$ | est utilisé pour éviter les constantes (de normalisation par exemple) lorsqu'elles ne sont pas indispensables |

On utilisera le système d'*unités atomiques*.

Autant que possible, $|\psi_i\rangle$ et $|\phi_i\rangle$ désigneront respectivement des spin-orbitales et des orbitales spatiales ; $|\Psi\rangle$ représentera des fonctions d'onde à $N$-électrons et $|\Phi\rangle$ des fonctions d'onde à $N$-électrons mono-déterminantales ; $i, j, k, \ldots$ désigneront des orbitales moléculaires occupées, $a, b, c, \ldots$ des orbitales virtuelles, $p, q, r, \ldots$ des orbitales quelconques, et $\alpha, \beta, \ldots$ des orbitales atomiques.

On utilise la notation physique : $\langle pq|rs\rangle = \iint \phi_p^*(\mathbf{x}_1)\phi_r(\mathbf{x}_1)\hat{w}(\mathbf{x}_1, \mathbf{x}_2)\phi_q^*(\mathbf{x}_2)\phi_s(\mathbf{x}_2)$.

On utilise de manière extensive la convention de Einstein, où les indices non répétés sont implicitement sommés.

# Introduction

Le vaste domaine de recherche appelé « chimie théorique » recouvre tout ce qui vise à établir des théories fondamentales pour expliquer ou prédire les comportements qui entourent les molécules ou, plus généralement, les édifices moléculaires. En particulier la « chimie quantique » est une application de la mécanique quantique aux molécules, c'est-à-dire regroupe les travaux qui cherchent à étudier les propriétés des molécules à l'aide de théories et d'approximations directement dérivées de l'équation de Schrödinger. Dans l'ère moderne, les progrès numériques liés aux puissances de calcul disponibles ont étendu l'utilisation du calcul quantique bien au-delà des laboratoires de théoriciens. Cette démocratisation de l'accès au calcul va de paire avec un effort sans cesse renouvelé pour améliorer la compréhension de la physique des systèmes - et ainsi la précision des calculs. La quantité de domaine où il n'est pas possible de faire des calculs à la précision chimique s'est donc largement amenuisée, mais certains points précis restent problématiques. Parmi ces domaines problématiques on peut encore compter la description des interactions faibles à longue-portée, dont on parlera plus loin.

Toutes les théories qui ont été mises en place en chimie quantique ont comme point de départ une approximation du problème compliqué de la gestion de l'interaction Coulombienne instantanée entre $N$ électrons (ceci s'inscrit d'ailleurs dans le domaine beaucoup plus général du traitement des problèmes à $N$-corps, un domaine qui traverse de nombreuses disciplines). En théorie Hartree-Fock (et post-HF), on substitue l'ensemble des interactions de paires d'électrons par l'interaction des électrons avec le champ moyen généré par les $N-1$ autres électrons ; lorsque l'on applique la théorie des fonctionnelles de la densité (DFT), on approxime la fonctionnelle d'échange-corrélation par des fonctionnelles au mieux semi-locales. C'est-à-dire que comme toujours dans les raisonnements scientifiques en général, on approxime un problème d'une complexité presque impénétrable (l'interaction, *corrélée* à l'infini, de $N$ corps) par des notions humainement compréhensibles, telles que des « moyennes » et des « approximations sphériques ». Ainsi, ces modèles décrivent bien (et parfois : incroyablement bien, vue la sévérité des approximations) les comportements qui sont le résultat d'une physique du système peu éloignée d'un champ moyen et/ou par nature locale. Ceci n'est pas le cas des interactions longue-portée.

Les interactions longue portée que l'on appellera ici « interactions de Van der Waals » désignent des interactions relativement faibles qui jouent un rôle structurant dans les grands édifices moléculaires ou en l'absence des forces plus importantes que sont les interactions covalentes ou ioniques. Parmi ces interactions de Van der Waals, les forces de dispersion de London sont un type d'interaction qui émergent de la polarisation dynamique mutuelle de nuages d'électrons, surtout entre sous-systèmes neutres. Ces forces ont un caractère très longue-portée. Dans la limite de sous-systèmes qui ne se recouvrent pas, et en utilisant des développements multipolaires, on trouve que le terme dominant les forces de dispersion de London est le terme d'interaction « dipôle induit/dipôle induit », sous la forme d'un $C_6/R^6$ amorti. L'idée est que, contrairement aux forces électrostatiques entre dipôles *permanents*, les forces de dispersion de London émergent de la manière bien connue suivante : du fait de fluctuations quantiques dans une partie d'un grand système, un dipôle *éphémère* apparaît et produit un champ électrique en $1/R^3$ dans le reste du système. Ce champ électrique induit la création d'un autre dipôle *éphémère* dans une autre partie du système, et la partie du système où est apparu le premier dipôle *éphémère* ressent alors un champ électrique « en retour » dont la dépendance à la distance est : $1/R^3.1/R^3 = 1/R^6$. Une manière de voir cela (et d'expliquer pourquoi ces forces ne peuvent être recouvrées par des fonctionnelles usuelles en DFT) est de dire que les fluctuations dipolaires des parties éloignées d'un grand système sont corrélées ; c'est-à-dire que ces forces émergent à cause d'un trou de corrélation qui est de très grande portée, bien loin des approximations semi-locales.

Face à ce défi, de nombreuses solutions sont disponibles dans la littérature, notamment avec des améliorations des fonctionnelles d'échange-corrélation. Une autre solution est d'utiliser l'Approximation de la Phase Aléatoire (RPA). Il s'agit d'une approximation élaborée dans les années $50'$ pour décrire la corrélation dans les plasma. L'idée initiale est que les comportements des éléments dans un plasma sont le fruit de la collaboration/compétition entre d'une part un phénomène d'oscillations cohérentes sur l'ensemble du plasma causées par des interactions à longue-portée et d'autre part des corrections dues à des interactions à plus courte portée. Cette méthode est donc, par construction et pour des raisons physiques, particulièrement adaptée à la description des interactions longue-portée. Ceci est encore plus vrai si l'on utilise la RPA dans le cadre d'une théorie de séparation de portée des interactions inter-électronique. Dans ce cas de figure, l'interaction courte-portée (pour laquelle les approximations semi-locales des fonctionnelles sont pertinentes) est prise en charge par la DFT, et les interactions longue-portée sont laissées à la RPA. Pour différentes raisons, ce couplage rigoureux DFT+RPA élimine un certain nombre de faiblesses des deux théories, et s'est montré prometteur pour le traitement des systèmes possédant des interactions longue-portées faibles.

Les deux premiers chapitres posent les bases théoriques, les notations et les concepts qui sont utilisés tout au long du manuscrit. Le second chapitre, en particulier, décrit la RPA telle qu'elle a été développée dans les articles originaux des années $50'$ et dérive les multiples formulations dans lesquelles on peut trouver la RPA aujourd'hui. Il s'agit à la fois d'un chapitre qui résume l'état de l'art du domaine et d'un chapitre qui présente des réflexions personnelles sur le sujet et clarifie certains points de détails. J'y montre le formalisme connexion adiabatique avec théorème de fluctuation-dissipation (AC-FDT), que l'on peut considérer comme un point de départ pour développer les formulations dites de "connexion adiabatique", de "matrice diélectrique", de "plasmon", ainsi que la formulation qui utilise des équations de type "équations de Riccati " et qui est équivalente à des approximations que l'on peut faire dans le contexte de la théorie *Coupled-Cluster* (CC). Des dérivations unifiées sont montrées pour tous ces formalismes. Il existe un lien étroit entre la RPA telle qu'elle est pratiquée dans la chimie quantique moderne et des notions de théorie quantique des champs : ces liens sont rendus clairs dans l'Annexe A.

Après un court rappel des techniques et des enjeux liés à l'utilisation d'orbitales locales pour le calcul d'énergies de corrélation, je présente au chapitre 3 deux travaux assez différents. Le premier est un travail réalisé à l'occasion d'une collaboration avec le Laboratoire de Chimie Théorique (LCT) de l'Université Pierre et Marie Curie à Paris. Une procédure a été mise en place par E. Chermak et P. Reinhardt pour construire des orbitales localisées d'un dimère dans lesquelles on peut reconnaître la trace des orbitales localisées pour les monomères. Avec une telle méthode, on peut classer les di-excitations en catégories selon un critère, disons : géométrique. On a montré que, de l'ensemble des di-excitations imaginables pour le système, les couplages inter-moléculaires de mono-excitations intra-moléculaires contribuent majoritairement à l'énergie d'interaction. La contribution de cette thèse a été d'écrire un programme autonome interfacé avec `MOLPRO` qui permet le calcul d'énergies de corrélation RPA avec des orbitales locales. Dans le même chapitre, je montre des réflexions autour d'un type d'orbitales virtuelles localisées appelées POO, pour *Projected Oscillatory Orbitals*. Il s'agit d'une idée originale de Boys qui n'a pas fait l'objet de développements significatifs dans la littérature. Je dérive ici des équations de base concernant les POO, incluant des développements multipolaires des intégrales bi-électroniques dans la base des POO. Je montre que les éléments de matrice du moment dipolaire entre une orbitale occupée localisée et une orbitale virtuelle localisée de type POO se réduisent à l'expression des recouvrements entre POO. Ce résultat non trivial simplifie largement les équations RPA dans la base des POO, qui sont également dérivées ici.

Au cours de cette thèse j'ai créé des outils informatiques qui permettent de calculer sur des grilles parallélépipédiques de l'espace réel ou sur des grilles de type "DFT" des fonctions telles que $\chi(\mathbf{r}_1, \mathbf{r}_2)$, le trou d'échange $h_x(\mathbf{r}_1, \mathbf{r}_2)$, la fonction de Dirac $\delta(\mathbf{r}_1, \mathbf{r}_2)$, *etc.*.. Dans le chapitre 4, je montre de telles visualisations, notamment de la fonction de réponse $\chi$. Je souligne la structure de la fonction de réponse, et sa relation avec les fonctions de corrélation inter-électronique, notamment le trou d'échange. Ce chapitre est le fruit d'un travail pas complétement abouti, et qui vise à terme à calculer des objets que l'on rencontre dans des calculs de corrélation RPA (tels que la fonction de réponse) sur des grilles de l'espace construites dans ce but, c'est-à-dire sur des grilles où les points sont générés de telle manière à mieux échantillonner l'espace *entre* les atomes (et non, comme c'est le cas des grilles de type "DFT", l'espace où l'on s'attend à ce que la densité soit intéressante).

Dans la même optique de travail dans l'espace réel, l'objet principal du chapitre 5 est l'adaptation de la technique de l'EED (*Effective Energy Denominator*) développée par Berger *et. al.* dans l'espace réciproque. Cette technique est une généralisation de l'approximation de Unsöld dans laquelle le dénominateur effectif dépend des coordonnées et de la fréquence. Une série de relations a pu être dérivée concernant les numérateurs qui émergent lorsque l'on applique cette technique, et leur lien aux règles de somme. L'objectif de ces développements est de pouvoir calculer la fonction réponse Kohn-Sham sans faire intervenir des sommations sur les orbitales virtuelles et pouvoir exprimer à terme l'énergie de corrélation de longue portée (et les corrections de l'énergie de dispersion) comme fonctionnelle des orbitales occupées.

C'est dans le chapitre 6 qu'est présenté ce qui peut être considéré comme le travail principal de cette thèse : la dérivation d'un gradient analytique de l'énergie de corrélation RPA dans un contexte de séparation de portée. Jusqu'à présent, dans la littérature, la seule dérivation à notre connaissance de gradient d'une énergie avec séparation de portée est celle de Chabbal *et. al.*, pour RSH-(L)MP2. Il s'agit d'une adaptation *a posteriori* du gradient MP2 *sans* séparation de portée. Je montre ici, en utilisant le formalisme lagrangien, une dérivation tout-en-un du gradient de l'énergie RSH-RPA. Des notations inédites sont introduites pour gérer de manière transparente les contributions des termes courte-portée ; d'ailleurs, tout au long de la dérivation, on voit un parallèle solide entre le comportement des contributions courte- et longue-portée qui émergent. Ces nouveaux gradients sont implémentés dans le « cœur » de `MOLPRO`, c'est-à-dire d'une manière qui profite des forces et de la gestion du reste du programme. À la suite de cette programmation, les calculs de dipôle au niveau RSH-RPA ainsi que les optimisations de géométrie, par exemple, sont du coup disponibles immédiatement. Je présente ici des résultats d'optimisation sur un ensemble de 16 molécules aux niveaux RSH-dRPA-I et RSH-SOSEX, ainsi que des visualisations des densités corrélées au niveau RSH-dRPA-I.

De manière générale, un effort a été fait dans les Annexes pour décrire des notions qui sont à la limite des habitudes d'un chimiste théoricien (on trouve par exemple des notions telles que celles qui sont décrites Annexe A et qui sont plutôt connues en théorie quantique des champs, ou des notions d'intégration complexe montrées Annexe B) ainsi que pour fournir des détails de dérivations utiles à la compréhension du manuscrit. Tous ces éléments sont regroupés dans des Annexes par soucis de concision, et sont tout aussi primordiaux que ce qui est montré dans le corps du manuscrit.

# Chapitre 1

# Premières approches d'un problème à N-corps

Dans ce premier chapitre, on présente quelques notions de chimie quantique pour la plupart largement connues du lecteur averti. Le but premier est de poser des notations et des définitions qui seront utilisées tout au long du manuscrit, mais on va également préparer les notions qui se révéleront importantes dans la suite (notions générales telles que l'énergie de corrélation, les fluctuations, la connexion adiabatique, *etc.*.. ).

Tout le travail de la thèse a été pensé dans un cadre de séparation de portée électronique, c'est-à-dire dans un contexte où l'on utilise des techniques qui mélangent un traitement du système avec des méthodes dites « à fonction d'onde » et des méthodes de DFT. On présente donc rapidement, dans l'ordre, la théorie Hartree-Fock, la théorie DFT, et les techniques de séparation de portée (notamment : RSH). Toutes ces méthodes s'inscrivent dans une philosophie de traitements des systèmes à $N$-corps basés sur l'approximation du champ moyen ; ce sont des méthodes qui servent à calculer des énergies de référence qui s'écartent de l'énergie exacte d'une valeur que j'appelle "énergie de corrélation".

Le travail de cette thèse est centré sur la RPA, montrée dans le chapitre 2, qui est une méthode de calcul de l'énergie de corrélation. Dans ce chapitre on explique simplement ce qu'est cette notion d'énergie de corrélation, à la fois d'un point de vue mathématique (c'est l'énergie qui manque dans les approximations de champ moyen que l'on utilise comme référence) et d'un point de vue physique (c'est une énergie qui est due à une partie des mouvements corrélés des électrons tels qu'ils sont décrits par la fonction de corrélation de paire, extraite de la fonction de réponse linéaire). On rappel dans ce chapitre que l'on peut utiliser une connexion adiabatique pour rigoureusement faire le lien entre le système réel et le système modélisé, c'est-à-dire que l'on peut connecter l'énergie exacte à l'énergie de référence. De cette connexion émerge naturellement une expression potentiellement exacte de l'énergie de corrélation. Toutes ces notions nous amènent vers une meilleure compréhension de ce qu'est l'énergie de corrélation, et elles seront utilisées comme base pour dériver les équations de l'énergie RPA dans le chapitre 2.

## 1.1 Comprendre le problème

Une molécule peut se voir comme une collection de $M$ noyaux et $N$ électrons interagissant les uns avec les autres par force de Coulomb. Les noyaux sont modélisés par des charges ponctuelles $Z_{1\to M}$ de masses $M_{1\to M}$ et de coordonnées $\mathbf{R}_{1\to M} \in \mathbb{R}^3$ ; les électrons ont des coordonnées spatiales $\mathbf{r}_{1\to N} \in \mathbb{R}^3$ et des spins $s_{1\to N} \in \mathbb{S} = \{+\frac{1}{2}; -\frac{1}{2}\}$. On introduit la coordonnée d'espace-spin $\mathbf{x}_i \doteq (\mathbf{r}_i, s_i) \in \mathbb{X} = \mathbb{R}^3 \otimes \mathbb{S}$. On décrit un tel système par des fonctions d'onde $|\Psi\rangle \doteq |\Psi(\mathbf{x}_{1\to N}; \mathbf{R}_{1\to M})\rangle$ de l'espace de Hilbert $\mathcal{L}^2(\mathbb{X}^N, \mathbb{C})$, fonctions propres de l'équation de Schrödinger $\hat{H}\,|\Psi\rangle = E\,|\Psi\rangle$.

L'hamiltonien $\hat{H}$ est :

$$\hat{H} = \hat{T}_e + \hat{T}_n + \hat{V}_{nn} + \hat{V}_{ne} + \hat{V}_{ee} \tag{1.1.1}$$

Cet hamiltonien est la somme des énergies cinétiques des électrons et des noyaux ($\hat{T}_e$ et $\hat{T}_n$), et des interactions Coulombiennes *instantanées* de chaque paire de composants ($\hat{V}_{nn}$, $\hat{V}_{ne}$ et $\hat{V}_{ee}$).

On utilise communément l'approximation de Born-Oppenheimer (voir par exemple la référence [1]), qui consiste à négliger le mouvement des noyaux (ces derniers sont presque 2000 fois plus lourds que les électrons) c'est-à-dire à fixer les $\mathbf{R}_{1\to M}$. On considère alors que les électrons évoluent dans le champ des noyaux fixes d'énergie $E_{nn}$ et on écrit une équation de Schrödinger électronique $\hat{H}^{\text{elec}}\,|\Psi^{\text{elec}}\rangle = E^{\text{elec}}\,|\Psi^{\text{elec}}\rangle$ qui dépend paramétriquement des coordonnées $\mathbf{R}_{1\to M}$, avec :

$$\hat{H}^{\text{elec}} \doteq \hat{H} = \hat{T}_e + \hat{V}_{ne} + \hat{V}_{ee}, \tag{1.1.2}$$

où l'on peut séparer les termes mono- et bi-électroniques en deux groupes :

$$\hat{H} = \sum_i \hat{h}_i + \sum_{ij} \hat{g}_{ij} \tag{1.1.3}$$

avec :

$$\hat{T}_e + \hat{V}_{ne} = \sum_i \hat{h}_i$$

$$\hat{V}_{ee} = \sum_{ij} \hat{g}_{ij}$$

Les électrons sont des fermions et obéissent au principe d'antisymétrie : les fonctions d'onde $|\Psi^{\text{elec}}\rangle \doteq |\Psi(\mathbf{x}_{1\to N})\rangle$ sont donc des fonctions antisymétriques de $\mathcal{L}^2(\mathbb{X}^N, \mathbb{C})$. Je désignerai par $\mathcal{A}^2(\mathbb{X}^N, \mathbb{C})$ l'espace des fonctions *antisymétriques* de $\mathbb{X}^N$ dans $\mathbb{C}$ de carré intégrable.

Le problème reste ici la résolution d'un système à $N$-corps en interactions Coulombiennes *instantanées*, problème qui n'est pas soluble analytiquement : l'équation de Schrödinger ne peut être résolue en l'état, et des approximations doivent être concédées. Toutes les premières approches de chimie quantique émergent de la même idée : transformer le problème à $N$-corps en une superposition de $N$ problèmes à 1-corps.

Cela revient à utiliser une approche de type « champ moyen », où l'on remplace l'interaction de chaque paire d'électrons (le terme à deux corps $\hat{V}_{ee}$ ou $\sum \hat{g}_{ij}$) par la somme des interactions d'un électron avec le champ statique produit par les autres particules (qui est un terme à un corps).

$$\hat{H}^{\text{approx}} = \sum_i \hat{h}_i + \sum_i \hat{v}_i^{\text{moyen}} \tag{1.1.4}$$

Avec ces méthodes, on ne néglige pas complètement l'interaction entre les électrons (présence du potentiel moyenné), mais *on néglige la corrélation entre les mouvements individuels des N particules*.

**Pour toute théorie** $\mathcal{T}$ basée sur cette idée, on a donc :

$$\begin{aligned} E_{\text{totale}} &= E_{\text{elec}} + E_{nn} \\ &= E^{\mathcal{T}}_{\text{approx}} + E^{\mathcal{T}}_{c} + E_{nn}, \end{aligned} \tag{1.1.5}$$

qui définit ce que j'appelle l'énergie de corrélation (l'énergie manquante lorsque l'on néglige la corrélation) : $E^{\mathcal{T}}_{c} = E_{\text{elec}} - E^{\mathcal{T}}_{\text{approx}}$ et qui est théorie-dépendante.

Nous allons utiliser cette définition pour désigner l'énergie de corrélation : soit par rapport à la valeur moyenne de l'hamiltonien avec un déterminant Kohn-Sham, voir par exemple l'équation (83) dans la référence [2], soit par rapport à l'énergie de séparation de portée, RSH, voir l'équation (84) de la même référence (on introduit dans la suite ces développements de Kohn-Sham et de séparation de portée, RSH). La définition largement acceptée de Löwdin serait plutôt $E_{\text{exact,non-rel}} - E_{\text{RHF}}$[3] : notre définition d'une certaine manière généralisée est un raccourci, une sorte d'abus de langage.

## 1.2 Bases dans les espaces de Hilbert

Dans un souci de clarifier les notations, on introduit ici des *orbitales* (*i.e.* des fonctions d'onde d'états à une seule particule ou fonctions d'onde mono-électroniques) qui décrivent la distribution spatiale d'un électron. On considère un ensemble infini d'orbitales spatiales $\{|\phi_i\rangle\}_\infty$, qui forme une base orthonormée de $\mathcal{L}^2(\mathbb{R}^3, \mathbb{C})$ :

$$\left\langle \phi_i \middle| \phi_j \right\rangle = \delta_{ij} \tag{1.2.1}$$

Un sous-ensemble de $\{|\phi_i\rangle\}_\infty$ formera une base d'un sous-espace de $\mathcal{L}^2(\mathbb{R}^3, \mathbb{C})$.

Pour une description complète des électrons, on définit également un ensemble infini de spin-orbitales $\{|\psi_i\rangle\}_\infty$ qui décrivent chacune à la fois la distribution spatiale et le spin d'un électron. Deux fonctions orthonormées suffisent pour décrire les deux spin possibles d'un électron : $\alpha$ et $\beta$ qui représentent respectivement les spin *up* ($\frac{1}{2}$) et *down* ($-\frac{1}{2}$). On a :

$$|\psi_i(\mathbf{x})\rangle = |\psi_i(\mathbf{r}, s)\rangle = \begin{cases} |\phi_i(\mathbf{r})\alpha(s)\rangle & \text{si} \quad s = +\dfrac{1}{2} \\ \\ |\phi_i(\mathbf{r})\beta(s)\rangle & \text{si} \quad s = -\dfrac{1}{2} \end{cases} \tag{1.2.2}$$

De la même manière, ces spin-orbitales forment une base orthonormée de $\mathcal{L}^2(\mathbb{X}, \mathbb{C})$ :

$$\left\langle \psi_i \middle| \psi_j \right\rangle = \delta_{ij}, \tag{1.2.3}$$

et un sous-ensemble de $\{|\psi_i\rangle\}_\infty$ une base orthonormée d'un sous-espace de $\mathcal{L}^2(\mathbb{X}, \mathbb{C})$.

Pour les fonctions d'onde à $N$-électrons, on introduit également les déterminants de Slater, qui sont des déterminants formés à partir des orbitales spatiales ou des spin-orbitales. Dans le cas des spin-orbitales, on a :

$$|\Psi(\mathbf{x}_{1\to N})\rangle \propto \begin{vmatrix} \psi_1(\mathbf{x}_1) & \psi_2(\mathbf{x}_1) & \cdots \\ \psi_1(\mathbf{x}_2) & \ddots & \\ \vdots & & \psi_n(\mathbf{x}_n) \end{vmatrix} \doteq |\psi_{1\to N}\rangle \tag{1.2.4}$$

Les déterminants sont antisymétriques par construction ; on peut montrer que lorsque l'on dispose d'une base *complète* $\{|\psi_i\rangle\}_\infty$ de l'espace $\mathcal{L}^2(\mathbb{X}, \mathbb{C})$ des fonctions mono-électroniques, l'ensemble $\{|\Psi_A\rangle\}_\infty$ de tous les déterminants de Slater à $N$-électrons qu'il est possible de construire à partir de cette base forme une base *complète* de l'espace $\mathcal{A}^2(\mathbb{X}^N, \mathbb{C})$ des fonctions antisymétriques à $N$-électrons (voir le paragraphe 2.2.7 du livre de Szabo-Ostlund [3]). Une expression simple pour une fonction d'onde de $\mathcal{A}^2(\mathbb{X}^N, \mathbb{C})$ sera une fonction mono-déterminentale (composée d'un seul déterminant), mais des expressions plus abouties peuvent contenir une combinaison linéaire de déterminants. Lorsque l'on exprime les fonctions d'onde avec des déterminants de Slater, on dispose pour calculer des éléments de matrices des règles de Slater-Condon.

## 1.3 HF : Théorie Hartree-Fock

Dans la suite de la thèse, on sera amené à manipuler des notions et des notations qu'il est bon d'introduire dès maintenant. Je cherche ici notamment à clarifier ce que sont les énergies dites de Hartree et d'échange ainsi que les implications de l'utilisation de fonctions mono-déterminentales.

Avant d'aborder la théorie Hartree-Fock, théorie centrale en chimie théorique, il est intéressant d'introduire rapidement une théorie dite de Hartree, et ce dans le but de souligner les points forts et faibles de la théorie Hartree-Fock.

Dans la théorie dite de Hartree, on restreint $|\Psi\rangle$ à de simples produits de fonctions mono-électroniques $|\psi_i\rangle$ (ainsi on remplace sa dépendance complexe aux coordonnées $\mathbf{x}_i$ par un produit de fonctions à une variable : les coordonnées $\mathbf{x}_i$ sont *dé-corrélées* les unes des autres) :

$$|\Psi^H(\mathbf{x}_{1\to N})\rangle = |\psi_1(\mathbf{x}_1)\ldots\psi_n(\mathbf{x}_n)\rangle \tag{1.3.1}$$

Cette formulation néglige complètement toute forme de corrélation entre les électrons, comme le montre l'expression de la probabilité $P(\mathbf{r}_1, \mathbf{r}_2)d\mathbf{r}_1 d\mathbf{r}_2$ de trouver simultanément un électron dans un volume $d\mathbf{r}_1$ autour de $\mathbf{r}_1$ et un autre dans un volume $d\mathbf{r}_2$ autour de $\mathbf{r}_2$, quel que soit leur spin et les coordonnées $\mathbf{x}_i$ des autres électrons :

$$\begin{aligned} P(\mathbf{r}_1, \mathbf{r}_2)d\mathbf{r}_1 d\mathbf{r}_2 &= \int_{s_1,s_2} \int_{\mathbf{x}_{3\to N}} \Psi^H(\mathbf{x}_{1\to N})^2 d\mathbf{r}_1 d\mathbf{r}_2 \\ &= \phi_1(\mathbf{r}_1)^2 d\mathbf{r}_1 \ \phi_2(\mathbf{r}_2)^2 d\mathbf{r}_2, \end{aligned} \tag{1.3.2}$$

qui n'est que le produit des probabilités de trouver un électron en $d\mathbf{r}_1$ et de trouver *indépendemment* un autre électron en $d\mathbf{r}_2$. Pourtant, la physique du problème dicte que ces deux électrons se repoussent par l'interaction Coulombienne *instantanée* et que les mouvements des deux particules doivent être intimement liés.

**Cette *corrélation* des positions**, manquante ici, régit le trou de Coulomb : une région de l'espace qui suit un électron au cours de son mouvement et qui est interdite aux autres électrons. On peut voir cette corrélation comme une répulsion fluctuante autour de la valeur de celle provoquée par le champ moyenné. Cette corrélation est responsable d'une portion faible en magnitude mais physiquement importante de l'énergie totale d'une molécule.

La théorie de Hartree-Fock propose plutôt de restreindre $|\Psi\rangle$ aux fonctions qui sont des déterminants dits de Slater (des déterminants formés à partir des fonctions mono-électroniques $|\psi_i\rangle$) :

$$|\Psi^{\mathrm{HF}}(\mathbf{x}_{1\to N})\rangle \propto \begin{vmatrix} \psi_1(\mathbf{x}_1) & \psi_2(\mathbf{x}_1) & \cdots \\ \psi_1(\mathbf{x}_2) & \ddots & \\ \vdots & & \psi_n(\mathbf{x}_n) \end{vmatrix} \propto |\psi_{1\to N}\rangle \tag{1.3.3}$$

Comme on a vu dans la section 1.2, une telle décomposition assure l'antisymétrie de $|\Psi\rangle$ (*i.e.* le respect du caractère fermionique et indiscernable des électrons) et est de fait une approximation de type champ moyen, comme le montre la dérivation suivante : (on utilise dans (1.1.3) la séparation en composant mono-électronique $\hat{h}_i$ et bi-électronique $\hat{g}_{ij}$)

$$\begin{aligned} E^{\mathrm{HF}} &= \langle\Psi^{\mathrm{HF}}|\,\hat{H}\,|\Psi^{\mathrm{HF}}\rangle \\ &= \sum_i \langle\psi_i|\,\hat{h}_i\,|\psi_i\rangle + \frac{1}{2}\sum_{ij}\Big(\langle\psi_i\psi_j|\,\hat{g}_{ij}\,|\psi_i\psi_j\rangle - \langle\psi_i\psi_j|\,\hat{g}_{ij}\,|\psi_j\psi_i\rangle\Big), \end{aligned} \tag{1.3.4}$$

qui peut être réécrite, avec les bonnes notations :

$$
\begin{aligned}
E^{\mathrm{HF}} &= \sum_i \langle\psi_i|\, \hat{h}_i + \frac{1}{2}\sum_j \left(\hat{J}_j - \hat{K}_j\right) |\psi_i\rangle \\
&= \sum_i \langle\psi_i|\, \hat{h}_i + \hat{v}_i^{\mathrm{HF}}\, |\psi_i\rangle
\end{aligned}
\tag{1.3.5}
$$

c'est-à-dire que l'on peut écrire l'hamiltonien sous la forme (1.1.4) avec un potentiel à 1-corps $\hat{v}_i^{\mathrm{HF}} = \frac{1}{2}\sum_j \hat{J}_j - \hat{K}_j$. L'effet de la restriction $|\Psi\rangle \to |\Psi^{\mathrm{HF}}\rangle$ se traduit donc directement par une approximation $\sum_{ij} \frac{1}{\mathbf{r}_{ij}} \to \sum_i \hat{v}_i^{\mathrm{HF}}$ de type champ moyen.

On voit émerger, dans l'équation (1.3.4), l'énergie d'interaction Coulombienne classique entre deux densités :

$$
\langle\psi_i\psi_j|\, \hat{g}_{ij}\, |\psi_i\psi_j\rangle = \iint \psi_i^*(\mathbf{x}_1)\psi_i(\mathbf{x}_1)\hat{g}_{ij}\psi_j^*(\mathbf{x}_2)\psi_j(\mathbf{x}_2), \tag{1.3.6}
$$

que l'on appelle : terme de Hartree. Il est suivi d'un terme dit d'échange, qui n'a pas d'équivalent classique :

$$
\langle\psi_i\psi_j|\, \hat{g}_{ij}\, |\psi_j\psi_i\rangle = \iint \psi_i^*(\mathbf{x}_1)\psi_j(\mathbf{x}_1)\hat{g}_{ij}\psi_j^*(\mathbf{x}_2)\psi_i(\mathbf{x}_2) \tag{1.3.7}
$$

**Notons ici** qu'une approximation de la fonction d'onde en fonction mono-déterminentale produit naturellement une approximation champ moyen et une énergie d'interaction électron-électron composée uniquement de termes de type Hartree et échange. Inversement, un hamiltonien ne contenant pas de termes à deux corps aura pour vecteur propre une fonction d'onde mono-déterminentale.

La méthode Hartree-Fock est exposée dans de nombreux ouvrages de référence (par exemple : le livre de Szabo-Ostlund [3]), je me limite donc à un rappel des idées physiques du modèle. Une bonne façon de comprendre l'implication physique de la restriction sur $|\Psi\rangle$ est de regarder à nouveau la probabilité $P(\mathbf{r}_1, \mathbf{r}_2)d\mathbf{r}_1 d\mathbf{r}_2$. Deux cas de figure différents apparaissent, selon le spin des électrons 1 et 2. Considérons d'abord le cas où les électrons sont de spin opposé, on a :

$$
\begin{aligned}
P(\mathbf{r}_1, \mathbf{r}_2)d\mathbf{r}_1 d\mathbf{r}_2 &\propto \int_{s_1,s_2} [\psi_i(\mathbf{x}_1)\psi_j(\mathbf{x}_2) - \psi_j(\mathbf{x}_1)\psi_i(\mathbf{x}_2)]^2 d\mathbf{r}_1 d\mathbf{r}_2 \\
&\propto \int_{s_1,s_2} [\phi_i(\mathbf{r}_1)\alpha(s_1)\phi_j(\mathbf{r}_2)\beta(s_2) - \phi_j(\mathbf{r}_1)\beta(s_1)\phi_i(\mathbf{r}_2)\alpha(s_2)]^2 d\mathbf{r}_1 d\mathbf{r}_2 \\
&\propto [\phi_i(\mathbf{r}_1)^2\phi_j(\mathbf{r}_2)^2 + \phi_j(\mathbf{r}_1)^2\phi_i(\mathbf{r}_2)^2] d\mathbf{r}_1 d\mathbf{r}_2,
\end{aligned}
\tag{1.3.8}
$$

où seuls les termes présentant une même fonction de spin ($\alpha$ ou $\beta$) « survivent » à l'intégration sur $s_1$ et $s_2$. Cette probabilité est similaire à (1.3.2), avec simplement en plus une notion d'indiscernabilité, c'est-à-dire un témoignage du fait que les électrons 1 et 2 peuvent occuper *n'importe quelle orbitale* *i et j*. La probabilité de trouver un électron en $d\mathbf{r}_1$ et un autre en $d\mathbf{r}_2$ est donc une moyenne des produits des probabilités de trouver les électrons dans ces domaines lorsqu'ils appartiennent de manière indiscernable aux orbitales $i$ ou $j$. Ainsi, au-delà de cette nuance, la probabilité $P(\mathbf{r}_1, \mathbf{r}_2)d\mathbf{r}_1 d\mathbf{r}_2$ est exactement comme dans (1.3.2) : complètement dé-corrélée. La théorie de Hartree-Fock, comme la théorie dite de Hartree, ne reproduit pas le trou de Coulomb.

Le cas où les électrons sont de même spin, en revanche, donne :

$$\begin{aligned} P(\mathbf{r}_1, \mathbf{r}_2)d\mathbf{r}_1 d\mathbf{r}_2 &\propto \int_{s_1,s_2} [\psi_i(\mathbf{x}_1)\psi_j(\mathbf{x}_2) - \psi_j(\mathbf{x}_1)\psi_i(\mathbf{x}_2)]^2 d\mathbf{r}_1 d\mathbf{r}_2 \\ &\propto \int_{s_1,s_2} [\phi_i(\mathbf{r}_1)\alpha(s_1)\phi_j(\mathbf{r}_2)\alpha(s_2) - \phi_j(\mathbf{r}_1)\alpha(s_1)\phi_i(\mathbf{r}_2)\alpha(s_2)]^2 d\mathbf{r}_1 d\mathbf{r}_2 \\ &\propto [\phi_i(\mathbf{r}_1)^2\phi_j(\mathbf{r}_2)^2 + \phi_j(\mathbf{r}_1)^2\phi_i(\mathbf{r}_2)^2 \\ &\quad -\phi_i^*(\mathbf{r}_1)\phi_j(\mathbf{r}_1)\phi_j^*(\mathbf{r}_2)\phi_i(\mathbf{r}_2) \\ &\quad -\phi_j^*(\mathbf{r}_1)\phi_i(\mathbf{r}_1)\phi_i^*(\mathbf{r}_2)\phi_j(\mathbf{r}_2)]d\mathbf{r}_1 d\mathbf{r}_2 \end{aligned} \tag{1.3.9}$$

Les termes « survivants » ici introduisent une corrélation entre les électrons de même spin, qui émerge naturellement de la forme déterminantale de la fonction d'onde. En particulier, on voit que la probabilité de superposer ($\mathbf{r}_2 = \mathbf{r}_1$) deux électrons de même spin est nulle :

$$\begin{aligned} P(\mathbf{r}_1, \mathbf{r}_1)d\mathbf{r}_1 d\mathbf{r}_1 &\propto [\phi_i(\mathbf{r}_1)^2\phi_j(\mathbf{r}_1)^2 + \phi_j(\mathbf{r}_1)^2\phi_i(\mathbf{r}_1)^2 \\ &\quad -\phi_i^*(\mathbf{r}_1)\phi_j(\mathbf{r}_1)\phi_j^*(\mathbf{r}_1)\phi_i(\mathbf{r}_1) \\ &\quad -\phi_j^*(\mathbf{r}_1)\phi_i(\mathbf{r}_1)\phi_i^*(\mathbf{r}_1)\phi_j(\mathbf{r}_1)]d\mathbf{r}_1 d\mathbf{r}_1 = 0 \end{aligned} \tag{1.3.10}$$

**Ainsi, en théorie Hartree-Fock**, si le trou de Coulomb pour des électrons de spin opposé n'est pas reproduit, on observe bien une corrélation de position entre électrons de même spin, qui se repoussent selon ce que l'on appelle un *trou de Fermi*.

On parle souvent de l'état Hartree-Fock comme étant non corrélé, bien que les électrons de même spin le soient (trou de Fermi). C'est une référence, un « zéro », pour la corrélation. Tout le but de traitements post-HF est donc de retrouver une partie de la *corrélation*, comprise ici comme la corrélation entre deux électrons de spin opposé.

## 1.4 DFT : Théorie de la Fonctionnelle de la Densité

On cherche à résoudre l'équation de Schrödinger car on veut obtenir $|\Psi\rangle$, qui contient toutes les informations sur le système. C'est pourtant une fonction de $\mathcal{L}^2(\mathcal{X}^N, \mathbb{C})$, c'est-à-dire une fonction à $3N$ coordonnées d'espace et $N$ coordonnées de spin. La Théorie de la Fonctionnelle de la Densité (DFT pour l'anglais *Density Functional Theory*) est un effort pour mettre la densité $n(\mathbf{r})$ comme variable principale, en lieu et place de $|\Psi\rangle$. On définit $n(\mathbf{r})$ par :

$$n(\mathbf{r}) = \langle\Psi|\,\hat{n}(\mathbf{r})\,|\Psi\rangle \qquad \text{où :} \qquad \hat{n}(\mathbf{r}) = \sum_i \delta(\mathbf{r} - \mathbf{r}_i) \tag{1.4.1}$$

Il faut donc se convaincre que la quantité $n(\mathbf{r})$, qui est une fonction à 3 coordonnées d'espace seulement, contient bien toutes les informations suffisantes pour décrire à elle seule le système : c'est le rôle des théorèmes de Hohenberg et Kohn. Ces théorèmes montrent qu'une théorie de la densité est possible, c'est-à-dire qu'il est bien possible de mettre $n(\mathbf{r})$ au centre d'une théorie.

Dans l'approximation de Born-Oppenheimer, avec l'hamiltonien que l'on voit équation (1.1.2), le seul élément spécifique à un système particulier est $\hat{V}_{ne}$ (dans le formalisme DFT on parle en fait de *potentiel externe* $\hat{V}_{ext}$, qui le plus souvent se réduit au potentiel des noyaux $\hat{V}_{ne}$, mais peut également contenir un potentiel magnétique et électrique) ; les autres composants sont communs à tout système à $N$-électrons. Le premier théorème de Hohenberg et Kohn stipule justement que la densité de l'état fondamental détermine uniquement ce potentiel externe. Ainsi l'hamiltonien du système est complètement déterminé par la densité de l'état fondamental, et à sa suite toutes les observables.

$$E = \langle\Psi|\,\hat{T}_e + \hat{V}_{ee}\,|\Psi\rangle + \int n(\mathbf{r})v_{ext}(\mathbf{r}) \tag{1.4.2}$$

On peut donc voir une observable telle que l'énergie comme une fonctionnelle de la densité $n(\mathbf{r})$ (approche DFT).

Autrement dit, il existe une relation bi-univoque entre le potentiel externe et la densité d'un système. On peut écrire :

$$E[v_{ext}] = E[v_{ext}[n]] = E[n] \tag{1.4.3}$$

Un deuxième théorème de Hohenberg et Kohn démontre que l'énergie de l'état fondamental est variationnelle par rapport à la densité, c'est-à-dire que toute approximation sur la densité produit une énergie supérieure à l'énergie correspondant à la densité exacte. On trouve l'énergie correcte en minimisant l'expression (1.4.2), et la minimisation se fait en deux temps selon l'approche de Levy[4] :

$$\begin{aligned} E_0 &= \min_{n\in\mathbb{N}} \min_{\Psi\to n} \left\{ \langle\Psi|\,\hat{T}_e + \hat{V}_{ee}\,|\Psi\rangle + \int n(\mathbf{r})v_{ext}(\mathbf{r}) \right\} \\ &= \min_{n\in\mathbb{N}} \left\{ \min_{\Psi\to n} \langle\Psi|\,\hat{T}_e + \hat{V}_{ee}\,|\Psi\rangle + \int n(\mathbf{r})v_{ext}(\mathbf{r}) \right\} \\ &= \min_{n\in\mathbb{N}} \left\{ F[n] + \int n(\mathbf{r})v_{ext}(\mathbf{r}) \right\}, \end{aligned} \tag{1.4.4}$$

où $\mathbb{N}$ est l'ensemble des densités $n(\mathbf{r})$ $N$-représentables, c'est-à-dire l'ensemble des densités qui correspondent effectivement à un état lié à $N$-électrons. On utilise, pour passer de la première à la deuxième ligne, le fait que $\int n(\mathbf{r})v_{ext}(\mathbf{r})$ fournira la même valeur pour toutes fonctions d'onde qui produit la même densité $n(\mathbf{r})$. On définit la fonctionnelle universelle (à tous les systèmes à $N$ électrons) $F[n] = \min_{\Psi \to n} \langle \Psi | \hat{T}_e + \hat{V}_{ee} | \Psi \rangle \doteq T_e[n] + V_{ee}[n]$.

Les expressions de ces fonctionnelles sont inconnues : on ne connaît pas la manière dont l'énergie cinétique dépend de la densité, ni l'expression de l'interaction des électrons. Ce que l'on peut écrire de manière exacte, c'est l'énergie cinétique d'un système sans interaction $T_s[n]$, et l'énergie d'interaction Coulombienne classique (le terme de Hartree) $E_H[n]$. On a alors l'expression *exacte* de $E_0$ :

$$\begin{aligned} E_0 &= \min_{n \in \mathbb{N}} \left\{ T_s[n] + T_c[n] + E_H[n] + E_{\text{non-classique}}[n] + \int n(\mathbf{r}) v_{ext}(\mathbf{r}) \right\} \\ &= \min_{n \in \mathbb{N}} \left\{ T_s[n] + E_H[n] + E_{xc}[n] + \int n(\mathbf{r}) v_{ext}(\mathbf{r}) \right\}, \end{aligned} \tag{1.4.5}$$

où $T_c[n]$ est une partie due aux corrélations dans l'énergie cinétique et $E_{\text{non-classique}}[n]$ comporte tout ce qui dans $V_{ee}[n]$ n'est pas décrit par $E_H[n]$.

**On introduit** dans l'équation (1.4.5) la fonctionnelle $E_{xc}[n] = T_c[n] + E_{\text{non-classique}}[n]$, fonctionnelle d'échange-corrélation, qui recueille toutes les inconnues du problème. L'amélioration de la compréhension et de l'expression de cette énergie est le sujet de tous les développements actuels en DFT.

On sépare souvent l'énergie d'échange-corrélation en une composante d'échange et une composante de corrélation. La composante d'échange (comme dans le cas Hartree-Fock) est une conséquence de la nature fermionique des électrons, et la composante de corrélation provient du manquement, dans $T_s[n]$ et $E_H[n]$, du caractère lié des mouvements individuels des électrons.

Insistons ici sur le fait que la théorie DFT est exacte si $E_{xc}[n]$ est connue exactement. Il n'y a donc pas d'énergie de corrélation résiduelle au sens de l'équation (1.1.5). Pourtant, en pratique, cette fonctionnelle d'échange-corrélation doit être approximée : on parle alors de DFAs ( pour l'anglais *Density Functional Approximations*) et une énergie de « corrélation » peut donc être définie comme : $E_c^{\text{DFA}} = E_{\text{elec}} - E^{\text{DFA}}$. En DFT (en DFAs, en fait), on parle donc de la corrélation présente dans la fonctionnelle $E_{xc}[n]$ *et* de la corrélation au sens de l'équation (1.1.5).

## 1.5 Séparation de portée électronique

L'utilisation pratique de la DFT repose sur l'approximation de la fonctionnelle (inconnue) d'échange-corrélation. L'approximation originelle, l'Approximation de Densité Locale (LDA pour l'anglais *Local Density Approximation*)[5], s'est révélée étonnamment efficace et difficile à améliorer de manière systématique. Il a été montré que la LDA, étant locale par construction, est particulièrement adaptée pour décrire les corrélations à (très) courte-portée inter-électronique, mais échoue à décrire quantitativement les corrélations à longue-portée électronique[6]. Ceci reste vrai avec la plupart des améliorations post-LDA classiques, qui demeurent par *design* des approximations de nature locale.

Ces faiblesses à décrire les corrélations à longue-portée deviennent particulièrement problématiques lorsque l'on veut traiter des systèmes qui présentent des interactions de Van der Waals. Les interactions de Van der Waals sont des interactions faibles mais uniformément attractives et à longue portée d'action[7]. En l'absence d'autres interactions plus fortes (électrostatiques ou de covalence), ce sont ces contributions attractives qui gouvernent la structure du système. Parmi les interactions de Van der Waals, les forces de dispersion de London sont un type d'interaction qui émergent de la polarisation dynamique mutuelle de nuages d'électrons. Comme l'a montré en premier London[8], ces forces ont comme origine la corrélation dynamique des électrons à longue-portée, c'est-à-dire la corrélation entre des groupes électroniques distants. Ainsi, les fonctionnelles usuelles locales ou semi-locales, qui n'utilisent que des informations sur la valeur et la (ou les) dérivée(s) de la densité électronique, ne peuvent pas prendre en compte les processus physiques entrant en jeu dans l'émergence des forces de dispersion. Ce fait a été présenté dans le cadre d'un argumentaire plus-ou-moins rigoureux par plusieurs auteurs dans la littérature (voir par exemple la figure 1 de la référence [9] et la figure 2 de [10]). Le lecteur trouvera des descriptions détaillées des forces de dispersion aux références [11], [12].

Différentes solutions existent dans la littérature pour traiter ce problème. Citons par exemple, avec différents niveaux d'empirisme, les travaux autour de paramétrisations spécifiques de nouvelles fonctionnelles GGA ou méta-GGA[13–17], les travaux de type DFT+D incluant des corrections de dispersion de Grimme et d'autres[18–22], les fonctionnelles de corrélation semi-locales de Wilson-Levy[23–25] et le modèle de Becke-Johnson basé sur un couplage entre l'électron et son trou d'échange[26–31]. Parmi les solutions non-empiriques, on peut citer le modèle de Anderssson-Langreth-Lundqvist basé sur l'expression AC-FDT de l'énergie de corrélation et sur des paramètres tirés de théorie de la réponse linéaire[32–40] ou encore l'utilisation de la DFT-SAPT[41–45]. Un autre chemin possible, celui que j'emprunterai ici, est basé sur la séparation de portée d'une manière telle que les corrélations de longue portée soient prises en compte par une approche de type fonction d'onde.

Le concept d'une théorie de la fonctionnelle de la densité hybride avec la séparation de portée des interactions électron-électron (même si la terminologie est beaucoup plus récente[46, 47]) vient des travaux de Savin *et. al.* [48–50] du milieu des années '80. Savin et Stoll ont proposé la DFT *uniquement* pour décrire la partie courte-portée d'un système électronique, où elle est performante, et utiliser une méthode de théorie à $N$-corps - une méthode de type fonction d'onde - pour décrire la partie longue-portée[46, 48, 50–54]. On parle alors de « séparation de portée électronique ». En pratique, on sépare rigoureusement l'interaction Coulombienne électron-électron comme :

$$\frac{1}{r} = v_{ee}^{\mu}(\mathbf{r}) + \overline{v}_{ee}^{\mu}(\mathbf{r}) = v_{ee}^{\mathrm{LR}}(\mathbf{r}) + v_{ee}^{\mathrm{SR}}(\mathbf{r}), \tag{1.5.1}$$

où le paramètre $\mu$ contrôle la portée de l'interaction longue-portée. Au-delà d'un certain seuil, l'interaction est dominée par la contribution longue-distance, et la contribution complémentaire de courte-portée possède une singularité à l'approche de la distance inter-électronique nulle. On peut utiliser des fonctions diverses pour $v_{ee}^{\mu}$[55], notamment la fonction *erf*[50], qui présente de nombreux avantages pratiques lors de son implémentation avec des bases Gaussiennes ou des ondes planes. D'autres possibilités pour représenter les interactions inter-électroniques de courte portée sont le potentiel Yukawa[56], la fonction Gaussienne[57], ou la combinaison de *erfc* et Gaussienne (erfgau)[58].

On réécrit la fonctionnelle universelle :

$$F[n] = F^{\mathrm{LR}}[n] + E^{\mathrm{SR}}_{\mathrm{Hxc}}[n] \tag{1.5.2}$$

où :

$$F^{\mathrm{LR}}[n] = \min_{\Psi \to n} \langle\Psi| \hat{T} + \hat{V}^{\mathrm{LR}}_{ee} |\Psi\rangle \tag{1.5.3}$$

La partie longue-portée de l'interaction électron-électron est traitée par des méthodes de type fonction d'onde : $\langle\Psi| \hat{V}^{\mathrm{LR}}_{ee} |\Psi\rangle$, et la partie courte-portée est traitée en DFT : $E^{\mathrm{SR}}_{Hxc}[n]$. Lorsque $\mu = 0$, on retrouve un DFT standard : $\hat{V}^{\mathrm{LR}}_{ee}$ est nul et $E^{\mathrm{SR}}_{Hxc}[n_\Psi]$ est la fonctionnelle Hartree-échange-corrélation usuelle ; lorsque $\mu \to \infty$ on retrouve la formulation fonction d'onde habituelle : $\hat{V}^{\mathrm{LR}}_{ee}$ est l'interaction Coulombienne sans séparation de portée et $E^{\mathrm{SR}}_{Hxc}[n_\Psi]$ est nulle.

Ainsi l'équation (1.4.4) s'écrit :

$$\begin{aligned} E_0 &= \min_{n \in \mathbb{N}} \left\{ F^{\mathrm{LR}}[n] + E^{\mathrm{SR}}_{\mathrm{Hxc}}[n] + \int n(\mathbf{r}) v_{ext}(\mathbf{r}) \right\} \\ &= \min_{\Psi \to N} \left\{ \langle\Psi| \hat{T} + \hat{V}^{\mathrm{LR}}_{ee} |\Psi\rangle + E^{\mathrm{SR}}_{\mathrm{Hxc}}[n_\Psi] + \int n_\Psi(\mathbf{r}) v_{ext}(\mathbf{r}) \right\} \end{aligned} \tag{1.5.4}$$

À partir de l'équation (en principe : exacte) (1.5.4), une approximation doit être choisie pour la fonctionnelle $E^{\mathrm{SR}}_{xc}[n]$, et une procédure d'approximation doit être dérivée pour la partie longue-portée du calcul, traitée en fonction d'onde. Parmi de nombreuses procédures existant dans la littérature[47, 59], on choisit d'utiliser une procédure dite Hybride à Séparation de Portée (RSH pour l'anglais *Range Separated Hybrid*)[46], qui propose de restreindre la minimisation de l'équation (1.5.4) au sous-ensemble des fonctions d'onde mono-déterminentales :

$$E^{\mathrm{SR,LR}}_{\mathrm{RSH}} = \min_{\Phi} \left\{ \langle\Phi| \hat{T} + \hat{V}_{ext} + \hat{V}^{\mathrm{LR}}_{ee} |\Phi\rangle + E^{\mathrm{SR}}_{Hxc}[n_\Phi] \right\} \tag{1.5.5}$$

Cette procédure réduit donc le terme $\langle\Phi| \hat{V}^{\mathrm{LR}}_{ee} |\Phi\rangle$ aux seuls termes de Hartree et d'échange de type Hartree-Fock : elle n'inclue pas la corrélation longue-portée, qui est ajoutée *a posteriori* :

$$E = E^{\mathrm{SR,LR}}_{\mathrm{RSH}} + E^{\mathrm{LR}}_{c} \tag{1.5.6}$$

La fonction d'onde mono-déterminentale $\Phi_0$ qui minimise l'équation (1.5.5) est donnée par une équation d'Euler-Lagrange auto-consistante :

$$\left(\hat{T} + \hat{V}_{ext} + \hat{V}^{\mathrm{LR}}_{Hx,HF} + \hat{V}^{\mathrm{SR}}_{Hxc}\right) |\Phi_0\rangle = \mathcal{E}_0 |\Phi_0\rangle, \tag{1.5.7}$$

où $\hat{V}^{\mathrm{LR}}_{Hx,HF}$ est le potentiel longue-portée de type Hartree-Fock et $\hat{V}^{\mathrm{SR}}_{Hxc}$ est le potentiel correspondant à la fonctionnelle $E^{\mathrm{SR}}_{Hxc}[n]$.

Il reste à construire des fonctionnelles Hartree-échange-corrélation de la courte-portée. On dispose dans la littérature de fonctionnelles srLDA[52, 58, 60] construites à partir de calculs Monte-Carlo Quantique sur le gaz homogène d'électron avec une interaction longue- ou courte-portée, et divers fonctionnelles au-delà de LDA[53, 61] notamment les fonctionnelles srPBE comme HSE[62] ou construites par interpolation[63].De nombreux schémas avec différentes méthodes de corrélation longue-portée existent dans la littérature, on citera ici sans prétention d'exhaustivité les méthodes RSH-PT2[46, 64–67], RSH-CI[51, 68], RSH-CC[53, 54, 69–71], RSH-MRPT2[72, 73] et RSH-MCSCF[61, 74]. Les combinaisons RSH-RPA seront traitées plus en détail dans le chapitre 2.

Le choix optimal pour $\mu$ lors de calculs RSH a fait l'objet d'études dans la littérature, voir par exemple[47, 75], et on peut dire que les valeurs choisies sont habituellement entre $\mu = 0.4$, $\mu = 0.5$, *etc.*.. Tous les calculs montrés dans cette thèse sont d'ailleurs faits avec un paramètre de séparation de portée de $\mu = 0.5$. La signification physique du paramètre est la suivante : il contrôle la portée de l'interaction prise en compte en DFT, c'est-à-dire qu'il contrôle la portion de l'interaction électron-électron qui est traitée par la DFT (et celle qui est traitée par les méthodes de type fonction d'onde). Ainsi, les études réalisées par Fromager *et. al.* [61, 74] consistent à observer le nombre d'occupation d'orbitales naturelles et, en somme, à voir la valeur de $\mu$ pour laquelle il est pertinent d'écrire une fonction d'onde mono-déterminantale. Une toute autre manière d'aborder le problème est de choisir $\mu$ de sorte à reproduire correctement le passage HOMO-LUMO[76, 77]. Ceci ouvre en revanche la question de l'optimisation de $\mu$ pour chacun des systèmes que l'on veut traiter. Ainsi : dans le contexte d'un calcul multi-système, quel $\mu$ doit être choisi ?

## 1.6 Connexion Adiabatique

À partir de n'importe quelle théorie de type particules indépendantes, on peut utiliser une Connexion Adiabatique (AC pour l'anglais *Adiabatic Connexion*) pour se donner un cadre de compréhension des énergies de corrélation. Dans une connexion adiabatique « classique », on propose de se doter d'une série de systèmes fictifs en pondérant l'opérateur de l'interaction électron-électron d'un hamiltonien de type (1.1.2) :

$$\hat{H}_\alpha = \hat{T}_e + \hat{V}_\alpha + \alpha \hat{V}_{ee}, \tag{1.6.1}$$

de sorte à connecter linéairement le système de référence (pour $\alpha = 0$) au système réel (pour $\alpha = 1$). Je présenterai ici plutôt une connexion adiabatique généralisée[78, 79], où l'on introduit l'opérateur d'interaction généralisé $\hat{V}_{ee,\alpha}$, c'est-à-dire où l'on introduit les hamiltoniens suivants :

$$\hat{H}_\alpha = \hat{T}_e + \hat{V}_\alpha + \hat{V}_{ee,\alpha} \tag{1.6.2}$$

On choisit $\hat{V}_\alpha$ et $\hat{V}_{ee,\alpha}$ tels que, pour une valeur $\alpha_0$ de $\alpha$ on ait :

$$\begin{aligned} \hat{H}_{\alpha_0} &= \hat{T}_e + \hat{V}_{\alpha_0} \\ &= \hat{T}_e + \hat{V}_{ne} + \hat{v}^{\text{moyen}} \end{aligned} \tag{1.6.3}$$

(notons que $\hat{V}_\alpha$ est un opérateur mono-électronique et que $\hat{v}^{\text{moyen}}$ détermine quelle théorie est utilisée comme référence) ; et que pour $\alpha_1$ on retrouve l'hamiltonien réel, correspondant à l'énergie $E_{\alpha_1}$ et à la fonction d'onde $|\Psi_1\rangle$ :

$$\begin{aligned}\hat{H}_{\alpha_1} &= \hat{T}_e + \hat{V}_{\alpha_1} + \hat{V}_{ee} \\ &= \hat{T}_e + \hat{V}_{ne} + \hat{V}_{ee}\end{aligned} \tag{1.6.4}$$

Le paramètre $\alpha$ est appelé *constante de couplage* et détermine la forme de l'interaction Coulombienne du système *intermédiaire* décrit par l'hamiltonien $\hat{H}_\alpha$, de fonction d'onde $|\Psi_\alpha\rangle$ et d'énergie $E_\alpha$. On voit dans la formulation de l'équation (1.6.2) que le potentiel à 1-corps est autorisé à évoluer le long de la connexion.

**On comprend bien déjà** que l'énergie de corrélation (l'énergie manquante dans la description de l'hamiltonien de référence) est *quelque part* entre $E_{\alpha_0}$ et $E_{\alpha_1}$.

En effet, en intégrant la dérivée de l'expression de $E_\alpha = \langle\Psi_\alpha| \hat{H}_\alpha |\Psi_\alpha\rangle$, on trouve :

$$\int_{\alpha_0}^{\alpha_1} d\alpha \frac{\mathrm{d}E_\alpha}{\mathrm{d}\alpha} = \int_{\alpha_0}^{\alpha_1} d\alpha \langle\Psi_\alpha| \frac{\mathrm{d}\hat{H}_\alpha}{\mathrm{d}\alpha} |\Psi_\alpha\rangle, \tag{1.6.5}$$

où l'on a utilisé le théorème de Hellmann-Feynman. On a donc :

$$E_{\alpha_1} = E_{\alpha_0} + \int_{\alpha_0}^{\alpha_1} d\alpha \langle\Psi_\alpha| \hat{W}_\alpha |\Psi_\alpha\rangle, \tag{1.6.6}$$

où $\hat{W}_\alpha = \frac{\mathrm{d}\hat{H}_\alpha}{\mathrm{d}\alpha} = \frac{\mathrm{d}\hat{V}_{ee}}{\mathrm{d}\alpha} + \frac{\mathrm{d}\hat{V}_\alpha}{\mathrm{d}\alpha}$. Pour extraire une énergie de corrélation, comparons les expressions des énergies $E_{\alpha_0}$ et $E^{\text{ref}}$ :

$$E_{\alpha_0} = \langle\Phi_0| \hat{T}_e + \hat{V}_{\alpha_0} |\Phi_0\rangle \tag{1.6.7}$$

et :

$$\begin{aligned}E^{\text{ref}} &= \langle\Phi_0| \hat{T}_e + \hat{V}_{\alpha_1} + \hat{V}_{ee} |\Phi_0\rangle \\ &= E_{\alpha_0} + \langle\Phi_0| \hat{V}_{ee} + \hat{V}_{\alpha_1} - \hat{V}_{\alpha_0} |\Phi_0\rangle\end{aligned} \tag{1.6.8}$$

On obtient alors, via l'équation (1.6.6), une énergie de corrélation :

$$E_c^{\text{AC}} = \int_{\alpha_0}^{\alpha_1} d\alpha \left[\langle\Psi_\alpha| \hat{W}_\alpha |\Psi_\alpha\rangle - \langle\Phi_0| \hat{W}_{\alpha_0} |\Phi_0\rangle\right] \tag{1.6.9}$$

Les termes impliquant $\frac{\mathrm{d}\hat{V}_\alpha}{\mathrm{d}\alpha}$ dans $\hat{W}$ *équilibrent* si besoin le potentiel à 1-corps qui apparaît dans $E_{\alpha_0}$, et qui ne correspond pas forcément au potentiel physique de $E_{\alpha_1}$. Le choix de la connexion adiabatique (du chemin de la connexion) ne change pas le résultat de l'intégration équation (1.6.6), et n'est donc pas unique. Choisir, par exemple, une connexion linéaire $\hat{H}_\alpha = \hat{T}_e + \hat{V} + \alpha\hat{V}_{ee}$ produit un potentiel à 1-corps $\hat{V}$ constant par rapport à $\alpha$ : il n'y a pas de rééquilibrage nécessaire à partir du potentiel à 1-corps de $E_{\alpha_0}$, et $\frac{\mathrm{d}\hat{V}_\alpha}{\mathrm{d}\alpha} = 0$. Cela revient finalement à une résolution de type théorie des perturbations de Rayleigh-Schrödinger, où $\hat{H}_0 = \hat{T}_e + \hat{V}_{ne}$ et $\hat{W}_\alpha = \hat{V}_{ee}$.

Dans le cas RSH, on considère l'énergie dépendante de $\alpha$ suivante (voir, par exemple référence [80]) :

$$E_\alpha = \min_\Psi \left\{ \langle\Psi| \,\hat{T} + \hat{V}_{ext} + (1-\alpha)\hat{V}^{\mathrm{LR}}_{Hx,HF} + \alpha\hat{V}^{\mathrm{LR}}_{ee}\, |\Psi\rangle + E^{\mathrm{SR}}_{Hxc}[n_\Psi] \right\}, \tag{1.6.10}$$

c'est-à-dire les hamiltoniens des systèmes fictifs :

$$\hat{H}_\alpha = \hat{T} + \hat{V}_{ext} + (1-\alpha)\hat{V}^{\mathrm{LR}}_{Hx,HF} + \alpha\hat{V}^{\mathrm{LR}}_{ee} + \hat{V}^{\mathrm{SR}}_{Hxc}, \tag{1.6.11}$$

qui connectent bien le système RSH pour $\alpha = 0$ au système réel pour $\alpha = 1$. L'objet $\hat{W}_\alpha$ de l'équation (1.6.9) s'écrit alors : $\hat{W}_\alpha = \hat{V}^{\mathrm{LR}}_{ee} - \hat{V}^{\mathrm{LR}}_{Hx,HF}$.

Les détails mathématiques de cette connexion adiabatique généralisée, et notamment des dérivations des situations où l'on choisit la fonction $\mathrm{erfc}(\alpha\mathbf{r})/\mathbf{r}$ ou sa forme Yukawa $\exp(-\alpha\mathbf{r})/\mathbf{r}$ pour le potentiel généralisé, sont à lire aux références[78, 81].

## 1.7 Conclusion

On a présenté dans ce chapitre les notations et notions de base qui vont servir dans le reste du manuscrit (notions telles que la corrélation électronique et la connexion adiabatique) ; ainsi que les deux manières assez différentes d'aborder le problème de la résolution de l'équation de Schrödinger électronique : les méthodes Hartree-Fock et DFT. On a également rappelé les bases de la séparation de portée électronique, qui propose de traiter un système en utilisant les forces de ces deux points de vue que sont les méthodes de type "fonction d'onde" et les méthodes basées sur la densité.

# Chapitre 2

# RPA : Approximation de la Phase Aléatoire

Ce chapitre peut être considéré comme une revue de l'approximation de la phase aléatoire, la RPA : il présente un bilan de l'état de l'art ainsi que des réflexions qui sont le fruit du travail de cette thèse. La RPA est le sujet central de ce manuscrit et il est important de présenter la dérivation et les racines de l'approximation de manière extensive. On montre les raisons premières, historiques, qui ont menées à l'idée de la RPA, qui est de traiter un problème à $N$-corps en séparant un comportement organisé des particules à longue-portée (les oscillations de plasma) et une correction explicite à courte-portée. C'est dans l'origine de la RPA, notamment, que l'on trouve l'explication du nom de la méthode, « phase aléatoire ».

On combine les notions présentées dans le chapitre premier pour former un contexte dit "AC-FDT" (connexion adiabatique et théorème de fluctuation-dissipation). Au cours de cette dérivation émergent les fonctions de réponse $\chi$ qui seront finalement les objets qui seront approximés en RPA (dans notre dérivation, c'est en fait le noyau de l'équation de Bethe-Salpeter qui est l'objet approximé). Ce chapitre fonctionne en lien étroit avec les Annexes A, B et C, qui se veulent les plus détaillées possible. À partir de l'équation AC-FDT, le calcul pratique de la RPA peut se faire de deux manières complètement différentes qui se rejoignent et se complètent : la formulation dite "connexion adiabatique" et la formulation "matrice diélectrique", que l'on peut dériver chacune plus avant pour obtenir la formulation de "plasmon". Toutes ces formulations sont largement explicitées dans le chapitre.

En fin de chapitre, un résumé de toutes les méthodes décrites ici est donné sous la forme d'une hiérarchie des équations (selon le nombre d'intégrations analytiques effectuées), et on donne la syntaxe à utiliser pour lancer ces calculs en utilisant notre implémentation dans le programme `MOLPRO`.

## 2.1 Point de vue historique et physique, Revue de la littérature

Il semble intéressant de se pencher sur l'aspect historique et physique de la RPA, pour la simple raison qu'il est d'une certaine manière caché dans la plus grande partie du formalisme de la RPA en chimie quantique moderne. L'Approximation de la Phase Aléatoire (RPA pour l'anglais *Random Phase Approximation*) apparaît dans les années 50′[82–85] comme une méthode de résolution du problème à $N$-corps et naît de la volonté de mieux décrire (c'est-à-dire mieux que dans une approximation de champ moyen) la physique du gaz uniforme d'électron, où la corrélation entre les positions des électrons à longue-portée est importante.

En effet on observe dans un gaz d'électrons des oscillations collectives (dites oscillations de plasma) qui sont la conséquence directe de la corrélation longue-portée entre les électrons. Bohm et Pines, qui introduisent la RPA, proposent de placer ces oscillations collectives au centre de la résolution du problème à $N$-corps, espérant qu'une bonne description des unes apportera une bonne compréhension de l'autre. L'idée première[82] est de découpler, par une transformation canonique, les variables des particules de celles du champ. Schématiquement, on passe alors de l'hamiltonien habituel du système :

$$\hat{H} = \hat{H}_{\text{particules}} + \hat{H}_{\text{champ}} + \hat{H}_{\text{interaction particules/champ}} \tag{2.1.1}$$

à (voir figure 2.1) :

$$\hat{H}^{\text{RPA}} = \hat{H}_{\text{particules}} + \hat{H}_{\text{oscillations}} + \hat{H}^{\text{SR}}_{\text{interaction particules/particules}} \tag{2.1.2}$$

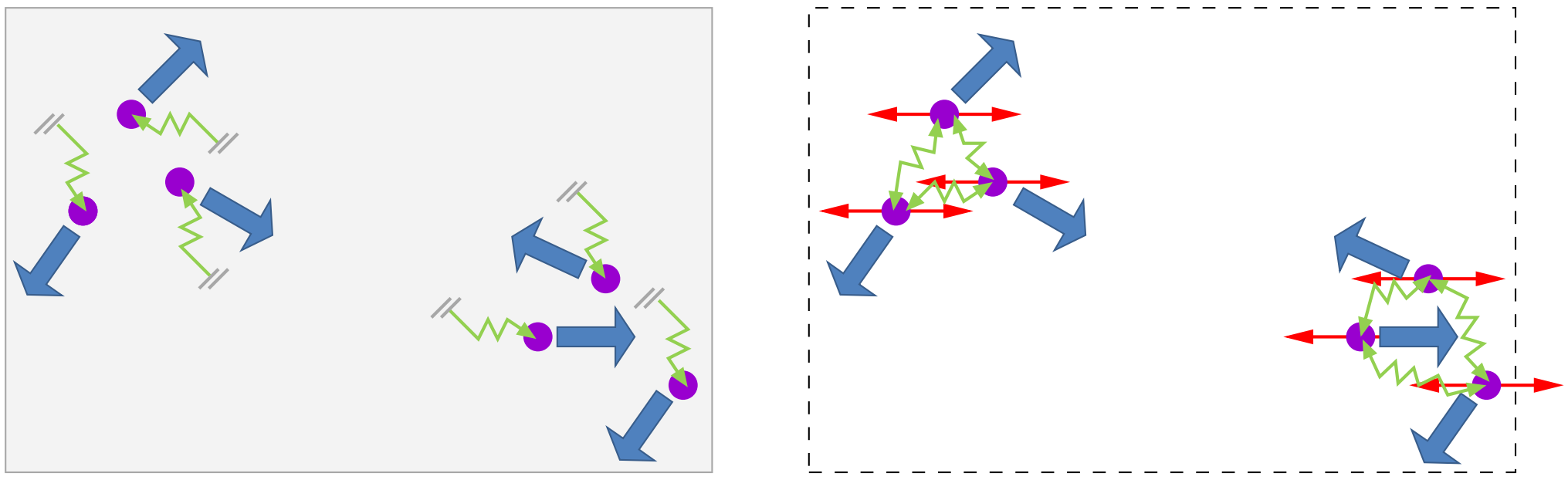

FIGURE 2.1: À gauche : un schéma représentant l'hamiltonien habituel que l'on voit équation (2.1.1) et qui est la somme d'une composante cinétique des particules (flèches bleues pleines), de l'énergie du champ moyen (arrière plan, en gris) et de l'interaction entre les particules et le champ (flèches en zig-zag vertes). À droite : l'hamiltonien tel qu'il est écrit équation (2.1.2), somme d'une composante cinétique (flèches bleues pleines), des oscillations collectives du plasma (double flèches droites rouges) et de l'interaction -à courte-portée- entre particules proches (double flèches en zig-zag vertes).

**On parvient alors** à une description avec deux comportements physiques bien distincts : d'une part un comportement collectif (des oscillations qui s'installent dues à une interaction à grande distance entre les électrons), et d'autre part une interaction explicite écrantée entre des particules proches les unes des autres et pour lesquelles la description en terme d'oscillations collectives ne suffit plus.

Cette séparation entre deux comportements (longue-portée et courte-portée) émerge naturellement de la transformation canonique et fait intervenir une distance caractéristique. Au-dessus de cette distance, la force écrantée à courte-portée est négligeable et le système est dominé par les oscillations du plasma, et au contraire au-dessous de cette distance la force écrantée décrit mieux le système. Autrement dit, en RPA, l'énergie s'exprime comme une somme d'énergies d'oscillateurs (comportement à longue-portée) et d'une correction à courte-portée (interaction résiduelle écrantée), ce qui en fait une méthode qui semble, par construction, particulièrement adaptée à un traitement RSH.

Pour conduire la transformation canonique, Bohm et Pines proposent quelques approximations, et notamment celle de ne considérer pour la description des oscillations du plasma que la réponse des particules qui sont en phase avec le champ oscillant. La réponse des autres particules, dont les phases dépendent de leur position et sont donc *aléatoires*, est négligée. C'est l'origine du nom « Approximation de la Phase Aléatoire » (la figure 2.2 donne un extrait de Bohm et Pines où l'approximation est motivée).

> (3) We distinguish between two kinds of response of the electrons to a wave. One of these is in phase with the wave, so that the phase difference between the particle response and the wave producing it is independent of the position of the particle. This is the response which contributes to the organized behavior of the system. The other response has a phase difference with the wave producing it which depends on the position of the particle. Because of the general random location of the particles, this second response tends to average out to zero when we consider a large number of electrons, and we shall neglect the contributions arising from this. This procedure we call the "random phase approximation."

FIGURE 2.2: Un extrait du premier papier de Bohm et Pines, référence [82], explicitant la motivation de l'approximation de la phase aléatoire

---

Parmi les méthodes non-empiriques qui semblent bien décrire les interactions exigeantes la RPA est devenue populaire du fait de son large champ d'application, qui va des atomes et molécules[86–92] aux solides infinis[93–95]. Il a été montré que la RPA ne donne pas de bons résultats lorsqu'elle est utilisée sur des systèmes à corrélation de courte-portée[86, 96, 97], et avec des bases Gaussiennes converge lentement avec la taille des bases[86] : la RPA est donc utilisée de nos jours dans un contexte RSH[98–102], ce qui semble palier à ces deux désavantages. La RPA a été utilisée avec un certain succès dans la littérature pour décrire des systèmes contenant des interactions de Van der Waals et en particulier impliquant des forces de dispersion[12, 37, 39, 103–105], qui sont connues pour être difficiles à traiter[12]. Les interactions faibles en général[86–89, 92, 94, 95, 106–110], ainsi que certaines dissociations de liaison[86, 88] ont également fait l'objet d'un traitement RPA.

## 2.2 AC-FDT : Théorème de Fluctuation-Dissipation avec Connexion Adiabatique

On va voir tout au long de ce manuscrit que la RPA peut être dérivée à partir de points de vue fort différents : j'utilise ici comme point de départ l'expression de l'énergie de corrélation par le théorème de fluctuation-dissipation avec connexion adiabatique[111–113]. Comme on a vu plus tôt, la connexion adiabatique est souvent utilisée dans les traitements post-HF ou post-KS, et c'est dans ce contexte que l'on rappelle ici l'équation (1.6.9) :

$$E_c^{\mathrm{AC}} = \int_0^1 d\alpha \left[ \langle \Psi_\alpha | \hat{W}_\alpha | \Psi_\alpha \rangle - \langle \Phi_0 | \hat{W}_{\alpha=0} | \Phi_0 \rangle \right], \tag{1.6.9}$$

dont on peut réécrire l'intégrande $I = \left[ \langle \Psi_\alpha | \hat{W} | \Psi_\alpha \rangle - \langle \Phi_0 | \hat{W} | \Phi_0 \rangle \right]$, dans une représentation avec des coordonnées d'espace-spin $\mathbf{x}$ et avec le potentiel $w$ correspondant à $\hat{W}$ :

$$I = \frac{1}{2} \iint w(\mathbf{x}_1, \mathbf{x}_2) \left[ \langle \Psi_\alpha | \hat{n}(\mathbf{x}_1) (\hat{n}(\mathbf{x}_2) - \delta(\mathbf{x}_1, \mathbf{x}_2)) | \Psi_\alpha \rangle - \langle \Phi_0 | \hat{n}(\mathbf{x}_1) (\hat{n}(\mathbf{x}_2) - \delta(\mathbf{x}_1, \mathbf{x}_2)) | \Phi_0 \rangle \right] \tag{2.2.1}$$

On manipule les deux termes en écrivant l'opérateur densité comme un écart (une *fluctuation*) autour de sa moyenne $\hat{n}(\mathbf{x}) = n(\mathbf{x}) + \delta\hat{n}(\mathbf{x})$, ainsi pour n'importe quelle fonction d'onde $A$ :

$$\langle A | \hat{n}(\mathbf{x}_1) (\hat{n}(\mathbf{x}_2) - \delta(\mathbf{x}_1, \mathbf{x}_2)) | A \rangle = \langle A | \delta\hat{n}(\mathbf{x}_1) \delta\hat{n}(\mathbf{x}_2) | A \rangle + n_A(\mathbf{x}_1) n_A(\mathbf{x}_2) - n_A(\mathbf{x}_1) \delta(\mathbf{x}_1, \mathbf{x}_2), \tag{2.2.2}$$

d'où :

$$I = \frac{1}{2} \iint w(\mathbf{x}_1, \mathbf{x}_2) \left[ \langle \Psi_\alpha | \delta\hat{n}(\mathbf{x}_1) \delta\hat{n}(\mathbf{x}_2) | \Psi_\alpha \rangle - \langle \Phi_0 | \delta\hat{n}(\mathbf{x}_1) \delta\hat{n}(\mathbf{x}_2) | \Phi_0 \rangle + \Delta n_\alpha \right], \tag{2.2.3}$$

avec $\Delta n_\alpha = n_\alpha(\mathbf{x}_1) n_\alpha(\mathbf{x}_2) - n_\alpha(\mathbf{x}_1) \delta(\mathbf{x}_1, \mathbf{x}_2) - n_0(\mathbf{x}_1) n_0(\mathbf{x}_2) + n_0(\mathbf{x}_1) \delta(\mathbf{x}_1, \mathbf{x}_2)$.

**On retrouve** une expression de la corrélation comme manifestation de fluctuation autour d'une moyenne (celle due au champ statique). On peut lier les fonctions de corrélation des densités (les *fluctuations*) qui apparaissent dans l'équation (2.2.3) à la partie imaginaire des fonctions de réponse du système $\chi_\alpha$ (les *dissipations*) par le théorème de fluctuation-dissipation de l'équation (2.2.4).

$$\langle \Psi_\alpha | \delta\hat{n}(\mathbf{x}_1) \delta\hat{n}(\mathbf{x}_2) | \Psi_\alpha \rangle = \int_{-\infty}^{\infty} \frac{-d\omega}{2\pi i} \chi_\alpha(\mathbf{x}_1, \mathbf{x}_2, \omega) \tag{2.2.4}$$

On trouve dans la littérature différentes façons d'écrire les expressions en relation avec la partie imaginaire de la fonction de réponse. Les quelques lignes suivantes devraient lever d'éventuels doutes sur le sujet[114] :

$$\int_{-\infty}^{\infty} d\omega \,\mathrm{Im}\,(\chi(\omega)) = \mathrm{Im}\left(\int_{-\infty}^{\infty} d\omega\, \chi(\omega)\right) = \mathrm{Im}\left(\int_{-\infty}^{\infty} d(i\omega)\, \chi(i\omega)\right) = \int_{-\infty}^{\infty} d\omega\, \chi(i\omega), \tag{2.2.5}$$

et :

$$\frac{1}{i} i \int_{-\infty}^{\infty} d\omega\, \chi(i\omega) = \frac{1}{i}\int_{-\infty}^{\infty} d(i\omega)\, \chi(i\omega) = \frac{1}{i}\int_{-\infty}^{\infty} d\omega\, \chi(\omega), \tag{2.2.6}$$

ceci explique, notamment, les versions du théorème de fluctuation-dissipation trouvées dans les références [80, 98].

Le lecteur intéressé trouvera en Annexe A.1 une discussion d'arrière plan sur le théorème de fluctuation-dissipation et diverses discussions sur la fonction de réponse $\chi$. On utilise de manière équivalente pour désigner $\chi$ les dénominations « fonction de réponse » ou « propagateur », qui font sens dans des contextes différents où $\chi$ peut apparaître. Lorsqu'une fonction de réponse est évaluée avec des opérateurs $\hat{A} \doteq \hat{r}_\alpha$ et $\hat{B} \doteq \hat{r}_\beta$ (voir A.1 et A.6.2), on parle de polarisabilité.

On trouve finalement pour l'énergie de corrélation l'expression suivante :

$$E_c^{\text{AC-FDT}} = \frac{1}{2}\int_0^1 d\alpha \iint \int_{-\infty}^{\infty} \frac{-d\omega}{2\pi i} w(\mathbf{x}_1, \mathbf{x}_2)\left[\chi_\alpha(\mathbf{x}_1, \mathbf{x}_2, \omega) - \chi_0(\mathbf{x}_1, \mathbf{x}_2, \omega) + \Delta n_\alpha\right] \tag{2.2.7}$$

## 2.3 L'approximation RPA

L'équation (2.2.7) ne contient aucune approximation, et est une expression exacte de l'énergie de corrélation dans laquelle seules les fonctions de réponse $\chi_\alpha$ des systèmes fictifs le long de la connexion adiabatique sont à approximer, par exemple en utilisant la RPA. L'ambition de ce chapitre est de donner un certain nombre de détails de la dérivation des équations RPA. Par soucis de clarté, dans cette section, la dérivation est légèrement simplifiée mais le lecteur trouvera en Annexe A la plupart des éléments laissés de côté.

Les expressions que l'on trouve équation (2.2.7) mettent en jeu des objets à deux corps ; elles peuvent être vues comme des diagonales d'expressions à 4 points[80] (on utilise la notation $1 \doteq \mathbf{x}_1$) : le potentiel $w(1,2) \doteq w(1,2;1,2)$ est la diagonale de $w(1,2;1',2')$ , et la fonction de réponse $\chi_\alpha(1,2) \doteq \chi_\alpha(1,2;1,2)$ de $\chi_\alpha(1,2;1',2')$. L'expression – exacte – de l'énergie de corrélation (2.2.7) se réécrit alors, avec les objets à 4-points :

$$E_c^{\text{AC-FDT}} = \frac{1}{2}\int_0^1 d\alpha \iint_{1,2;1',2'} \int_{-\infty}^{\infty} \frac{-d\omega}{2\pi i} w(1,2;1',2')\left[\chi_\alpha(1,2;1',2',\omega) - \chi_0(1,2;1',2',\omega) + \Delta n_\alpha\right] \tag{2.3.1}$$

L'expression de la fonction de réponse $\chi_0$ est connue exactement, sous la forme de l'expression de Lehmann suivante (Annexe A.6.1) :

$$\chi_0(1,2;1',2',\omega) = \sum_{ia} \frac{\psi_i^*(1')\psi_a(1)\psi_a^*(2')\psi_i(2)}{\omega - (\varepsilon_a - \varepsilon_i) + i\eta^+} + \frac{\psi_i^*(2')\psi_a(2)\psi_a^*(1')\psi_i(1)}{-\omega - (\varepsilon_a - \varepsilon_i) + i\eta^+}, \tag{2.3.2}$$

où $\psi_i$, $\varepsilon_i$ ($\psi_a$, $\varepsilon_a$) sont des orbitales et énergies occupées (virtuelles). La petite quantité positive $\eta^+ \to 0$ assure que $\chi_0$ est analytique, et est une conséquence directe de la nature physique du propagateur $\chi_0$ (comme visible dans l'Annexe A.6.1).

On est également capable d'écrire $\chi_\alpha$ en utilisant une relation de Bethe-Salpeter suivante :

$$\chi_\alpha(1,2;1',2',\omega)^{-1} = \chi_0(1,2;1',2',\omega)^{-1} - f_\alpha^{\mathrm{BSE}}(1,2;1',2',\omega) \tag{2.3.3}$$

Nous appliquons l'approximation adiabatique au noyau à quatre points de l'équation Bethe-Salpeter, $f_\alpha^{\mathrm{BSE}}$, c'est-à-dire que l'on travaille avec un noyau indépendant de la fréquence. Dans l'approximation dite "direct-RPA", ou dRPA (voir plus de détails section 2.4.2, ainsi que dans l'Annexe A.5) le noyau contient une composante de Hartree seule ; dans l'approximation dite "RPA-échange", ou RPAx (voir plus de détails aux mêmes endroits) il contient une composante de Fock, *i.e.* d'échange, supplémentaire. Ainsi on approxime $f_\alpha^{\mathrm{BSE}}$ par :

$$\begin{aligned} & f_\alpha^{\mathrm{BSE}}(1,2;1',2',\omega) \to \alpha f^{\mathrm{RPA}}(1,2;1',2') \\ \text{avec :}\quad & f^{\mathrm{RPA}}(1,2;1',2') = w(1,2)\left[\delta(1,1')\delta(2,2') - \zeta\delta(1,2')\delta(2,1')\right], \end{aligned} \tag{2.3.4}$$

où $\zeta$ permet de passer de dRPA ($\zeta = 0$) à RPAx ($\zeta = 1$), c'est-à-dire d'exclure/inclure une composante d'échange dans le noyau de l'équation de Bethe-Salpeter.

**C'est ici** que se fait l'approximation RPA : tout est formellement exact jusqu'à l'approximation du noyau de l'équation de Bethe-Salpeter. Une plus longue discussion sur les racines de cet aspect de l'approximation RPA peut être trouvée dans l'Annexe A, de A.2 à A.5.

L'expression finale de l'énergie de corrélation RPA sous sa forme AC-FDT[89] est donc :

$$\begin{aligned} E_c^{\text{AC-FDT}} = \frac{1}{2}\int_0^1 d\alpha \iint_{1,2;1',2'} \int_{-\infty}^{\infty} \frac{-d\omega}{2\pi i} w(1,2;1',2') \big[\chi_\alpha^{\mathrm{RPA}}(1,2;1',2',\omega) \\ -\chi_0(1,2;1',2',\omega)\big], \end{aligned} \tag{2.3.5}$$

où l'approximation RPA a comme conséquence : $\Delta n_\alpha = 0$.

À propos de ce terme correctif : lors d'une application dans un contexte de pure DFT, on a naturellement une densité égale en tout point de la connexion adiabatique à la densité du système réel, c'est-à-dire que dans le contexte de pure DFT, on a sans problème $\Delta n_\alpha = 0$. Dans n'importe quel autre contexte plus général (notamment en RSH vue comme une *Generalized Kohn-Sham*), ceci n'est plus trivialement vrai et doit être supposé. Toutefois, en RPA, l'approximation qui consiste à négliger les termes de corrélation (c'est-à-dire qui consiste à garder les termes de Coulomb et possiblement d'échange) dans la *self-energy* de l'équation de Dyson dont est dérivée l'équation de Bethe-Salpeter a comme conséquence directe que les fonctions de Green à une particule sont constantes le long de la connexion, c'est-à-dire que $\Delta n_\alpha = 0$[80].

L'obtention de l'énergie de corrélation réclame donc une double intégration (sur la fréquence $\omega$ et sur la constante de couplage $\alpha$). Plusieurs scénarios sont alors envisageables, selon les modes d'intégration choisis (analytiques ou numériques). On préférera pratiquer des intégrations analytiques sur une, ou sur les deux variables, à l'utilisation répétée de quadratures. Une intégration analytique sur (1) la fréquence $\omega$ suivie d'une intégration numérique sur (2) la constante de couplage $\alpha$ résulte en une formulation où apparaît la partie corrélation de la densité à deux particules $P_{c,\alpha}$, formulation dite "connexion adiabatique" ou "matrice densité" ; une intégration analytique sur (1) la constante de couplage $\alpha$ suivie d'une intégration numérique sur (2) la fréquence $\omega$ résulte en une formulation où émerge la matrice diélectrique, appelée formulation "matrice diélectrique" . Avec une double intégration analytique (c'est-à-dire avec une intégration analytique le long de la coordonnée $\alpha$ à partir d'une expression de type "connexion adiabatique", ou le long de la coordonnée $\omega$ à partir d'une expression de type "matrice diélectrique" ), ces deux formulations produisent la même expression de l'énergie de corrélation, dite "formule de plasmon". Dans la suite, je présente : la dérivation à partir de l'énergie (2.3.5) de la formulation "connexion adiabatique", suivie de la dérivation de la formulation de "plasmon". Finalement, en repartant de l'expression de l'énergie AC-FDT (2.3.5), je montre la dérivation de la formulation "matrice diélectrique", et prouve que l'on peut retrouver la formule de "plasmon" par cette voie.

## 2.4 Formulation "connexion adiabatique"

Dans la formulation "connexion adiabatique", l'énergie présentée équation (2.3.5) est intégrée analytiquement le long de la coordonnée de fréquence $\omega$. En observant l'équation (2.2.3), on introduit la partie corrélation de la densité à deux particules, $P_{c,\alpha}$ :

$$P_{c,\alpha}(\mathbf{x}_1, \mathbf{x}_2) = \langle \Psi_\alpha | \, \delta\hat{n}(\mathbf{x}_1)\delta\hat{n}(\mathbf{x}_2) \, | \Psi_\alpha \rangle - \langle \Phi_0 | \, \delta\hat{n}(\mathbf{x}_1)\delta\hat{n}(\mathbf{x}_2) \, | \Phi_0 \rangle + \Delta n_\alpha, \tag{2.4.1}$$

qui, avec le théorème de fluctuation-dissipation et sous l'approximation RPA, s'écrit :

$$P^{\mathrm{RPA}}_{c,\alpha}(1,2) = \int_{-\infty}^{\infty} \frac{-d\omega}{2\pi i} \left[ \chi^{\mathrm{RPA}}_\alpha(1,2,\omega) - \chi_0(1,2,\omega) \right], \tag{2.4.2}$$

de sorte que l'énergie que l'on cherche à calculer ici est :

$$E^{\mathrm{AC}}_c = \frac{1}{2} \int_0^1 d\alpha \iint_{1,2;1',2'} w(1,2;1',2') P^{\mathrm{RPA}}_{c,\alpha}(1,2;1',2'), \tag{2.4.3}$$

où $P^{\mathrm{RPA}}_{c,\alpha}(1,2) \doteq P^{\mathrm{RPA}}_{c,\alpha}(1,2;1,2)$ est la diagonale de $P^{\mathrm{RPA}}_{c,\alpha}(1,2;1',2')$ et est calculée par l'intégration analytique de l'équation (2.4.2).

### 2.4.1 Forme matricielle

Ici, de nombreuses notations vont être introduites qui serviront dans la suite du manuscrit et qui sont centrales en RPA. Le seul but de la suite de section est d'exprimer matriciellement les équations (2.3.2), (2.3.3), puis (2.4.3). Chaque élément de ces équations peut parfaitement être caractérisé par une matrice dans l'espace $\mathcal{L}^{occ} \otimes \mathcal{L}^{vir} \oplus \mathcal{L}^{vir} \otimes \mathcal{L}^{occ}$[2, 92], c'est-à-dire dans la base des produits d'orbitales occupées-virtuelles $\psi_i^*(1')\psi_a(1)$ et virtuelles-occupées $\psi_a^*(1')\psi_i(1)$. On calcule l'élément $pq, rs$ de la matrice $\mathbb{A}$ qui représente un objet $A$ comme :

$$\mathbb{A}_{pq,rs} = \int_{1,2;1',2'} \psi_p(1')\psi_q^*(1)A(1,2;1',2')\psi_r^*(2)\psi_s(2') \tag{2.4.4}$$

L'énergie de corrélation de l'équation (2.4.3) s'écrit donc sous forme matricielle :

$$E_c^{\mathrm{AC}} = \frac{1}{2}\int_0^1 d\alpha\ \mathrm{Tr}\left(\mathbb{W}.\mathbb{P}_{c,\alpha}^{\mathrm{RPA}}\right), \tag{2.4.5}$$

où « $\mathrm{Tr}(\sqcup)$ » représente la trace sur l'espace $\mathcal{L}^{occ} \otimes \mathcal{L}^{vir} \oplus \mathcal{L}^{vir} \otimes \mathcal{L}^{occ}$, $\mathbb{W}_{pq,rs} = \langle ps|\,\hat{w}\,|qr\rangle$ est la représentation matricielle de $w$ et $\mathbb{P}_{c,\alpha}^{\mathrm{RPA}}$ celle de $P_{c,\alpha}^{\mathrm{RPA}}$ :

$$\mathbb{P}_{c,\alpha}^{\mathrm{RPA}} = \int_{-\infty}^{\infty} \frac{-d\omega}{2\pi i}\left[\Pi_\alpha^{\mathrm{RPA}}(\omega) - \Pi_0(\omega)\right] \tag{2.4.6}$$

On a, pour les éléments de matrice de $\Pi_0$ et $\mathbb{F}^{\mathrm{RPA}}$ caractérisant respectivement $\chi_0$ et $f^{\mathrm{RPA}}$ :

$$\begin{aligned}
(\Pi_0)_{ia,jb} &= \frac{\delta_{ij}\delta_{ab}}{\omega - (\varepsilon_a - \varepsilon_i) + i\eta^+} \\
(\Pi_0)_{ai,bj} &= \frac{\delta_{ij}\delta_{ab}}{-\omega - (\varepsilon_a - \varepsilon_i) + i\eta^+} \\
(\Pi_0)_{ai,jb} &= (\Pi_0)_{ia,bj} = 0 \\
\mathbb{F}_{pq,rs}^{\mathrm{RPA}} &= \langle qr|\,\hat{w}\,|ps\rangle - \zeta\,\langle qr|\,\hat{w}\,|sp\rangle = \langle qr|ps\rangle - \zeta\,\langle qr|sp\rangle
\end{aligned} \tag{2.4.7}$$

Soit les représentations matricielles dans l'espace $\mathcal{L}^{occ} \otimes \mathcal{L}^{vir} \oplus \mathcal{L}^{vir} \otimes \mathcal{L}^{occ}$ de $(\chi_0)^{-1}$ et $(\chi_\alpha^{\mathrm{RPA}})^{-1}$ :

$$\begin{aligned}
(\Pi_0)^{-1} &= \omega\Delta - \Lambda_0 \\
(\Pi_\alpha^{\mathrm{RPA}})^{-1} &= \omega\Delta - \Lambda_0 - \alpha\mathbb{F}^{\mathrm{RPA}} = \omega\Delta - \Lambda_\alpha,
\end{aligned} \tag{2.4.8}$$

avec les super-matrices par blocs suivantes (chaque bloc est de dimension égale à $n_{\mathrm{occ}} \times n_{\mathrm{vir}}$) :

$$\Delta = \begin{pmatrix} \mathbf{1} & \mathbf{0} \\ \mathbf{0} & -\mathbf{1} \end{pmatrix}; \quad \Lambda_0 = \begin{pmatrix} \boldsymbol{\varepsilon} & \mathbf{0} \\ \mathbf{0} & \boldsymbol{\varepsilon} \end{pmatrix}; \quad \mathbb{F}^{\mathrm{RPA}} = \begin{pmatrix} \mathbf{A}' & \mathbf{B} \\ \mathbf{B} & \mathbf{A}' \end{pmatrix}; \quad \Lambda_\alpha = \begin{pmatrix} \mathbf{A}_\alpha & \alpha\mathbf{B} \\ \alpha\mathbf{B} & \mathbf{A}_\alpha \end{pmatrix}, \tag{2.4.9}$$

et avec :

$$\begin{aligned}
(\boldsymbol{\varepsilon})_{ia,jb} &= (\varepsilon_a - \varepsilon_i)\delta_{ij}\delta_{ab} \\
(\mathbf{A}')_{ia,jb} &= \langle ib|aj\rangle - \zeta\,\langle ib|ja\rangle = K_{ia,jb} - \zeta J_{ia,jb} \\
(\mathbf{B})_{ia,jb} &= \langle ab|ij\rangle - \zeta\,\langle ab|ji\rangle = K_{ia,jb} - \zeta K'_{ia,jb} \\
\mathbf{A}_\alpha &= \boldsymbol{\varepsilon} + \alpha\mathbf{A}'
\end{aligned} \tag{2.4.10}$$

**On se souvient que** le noyau RPA (voir équation (2.3.4)) est le lieu de l'approximation, et donne naissance à deux approximations distinctes : l'approximation dRPA et l'approximation RPAx. On ne fait pas ici la différence, c'est-à-dire que l'on utilise un formalisme décrivant les deux cas de figure ; on désigne simplement les matrices concernées pas l'exposant "RPA".

On est en mesure d'obtenir la représentation spectrale de $\Pi_\alpha^{\text{RPA}}$ en considérant l'équation aux valeurs et vecteurs propres (voir (2.4.8)) :

$$\mathbb{A}_\alpha \mathbb{C}_{\alpha,n} = \omega_{\alpha,n}\Delta\mathbb{C}_{\alpha,n}, \tag{2.4.11}$$

où les $\omega_{\alpha,n}$ sont les énergies d'excitations RPA du mode $n$. La symétrie du problème (voir (2.4.9)) impose qu'à un vecteur solution $\mathbb{C}_{\alpha,n} = \begin{pmatrix} \mathbf{x}_{\alpha,n} \\ \mathbf{y}_{\alpha,n} \end{pmatrix}$, de valeur propre $\omega_{\alpha,n}$, soit lié le vecteur solution $\mathbb{C}_{\alpha,-n} = \begin{pmatrix} \mathbf{y}_{\alpha,n} \\ \mathbf{x}_{\alpha,n} \end{pmatrix}$, de valeur propre $\omega_{\alpha,-n} = -\omega_{\alpha,n}$. La représentation spectrale de $\Pi_\alpha^{\text{RPA}}$ est (voir Annexe B.2.1) :

$$\Pi_\alpha^{\text{RPA}}(\omega) = \sum_{n>0} \left\{ \frac{\mathbb{C}_{\alpha,n}\mathbb{C}_{\alpha,n}^\dagger}{\omega - \omega_{\alpha,n} + i\eta^+} + \frac{\mathbb{C}_{\alpha,-n}\mathbb{C}_{\alpha,-n}^\dagger}{-\omega + \omega_{\alpha,-n} + i\eta^+} \right\} \tag{2.4.12}$$

Dans la formulation "connexion adiabatique", l'intégrale en fréquence de l'équation (2.4.6) est calculée analytiquement par intégration curviligne sur un contour du plan complexe supérieur (voir Annexe B.2.2). On obtient les résidus des pôles de $\Pi_\alpha^{\text{RPA}}$ et $\Pi_0$ :

$$\mathbb{P}_{c,\alpha}^{\text{RPA}} = \sum_{n>0} \left\{ \mathbb{C}_{\alpha,-n}\mathbb{C}_{\alpha,-n}^\dagger - \mathbb{C}_{0,-n}\mathbb{C}_{0,-n}^\dagger \right\} \tag{2.4.13}$$

La structure par bloc de $\mathbb{P}_{c,\alpha}^{\text{RPA}}$ est :

$$\mathbb{P}_{c,\alpha}^{\text{RPA}} = \begin{pmatrix} \mathbf{Y}_\alpha\mathbf{Y}_\alpha^\dagger & \mathbf{Y}_\alpha\mathbf{X}_\alpha^\dagger \\ \mathbf{X}_\alpha\mathbf{Y}_\alpha^\dagger & \mathbf{X}_\alpha\mathbf{X}_\alpha^\dagger \end{pmatrix} - \begin{pmatrix} \mathbf{0} & \mathbf{0} \\ \mathbf{0} & \mathbf{1} \end{pmatrix} \tag{2.4.14}$$

où les matrices $\mathbf{X}_\alpha$ et $\mathbf{Y}_\alpha$ rassemblent les vecteurs $\mathbf{x}_{\alpha,n}$ et $\mathbf{y}_{\alpha,n}$.

### 2.4.2 Différentes *flavors* de RPA

En se rappelant de l'expression matricielle *exacte* de l'énergie de corrélation dans le cadre AC-FDT :

$$E_c^{\text{AC}} = \frac{1}{2}\int_0^1 d\alpha \; \text{Tr}\left(\mathbb{W}^{\text{I}}.\mathbb{P}_{c,\alpha}\right), \tag{2.4.15}$$

on peut utiliser l'antisymétrie de $P_{c,\alpha}$ pour écrire une expression équivalente :

$$E_c^{\text{AC}} = \frac{1}{4}\int_0^1 d\alpha \; \text{Tr}\left(\mathbb{W}^{\text{II}}.\mathbb{P}_{c,\alpha}\right), \tag{2.4.16}$$

où l'on distingue la représentation matricielle d'origine : $\mathbb{W} \doteq \mathbb{W}^{\text{I}} = \langle ps|\,\hat{w}\,|qr\rangle$ (intégrales non antisymétrisées) de la représentation matricielle $\mathbb{W}^{\text{II}}_{pq,rs} = \langle ps|\,\hat{w}\,|qr\rangle - \langle pr|\,\hat{w}\,|qs\rangle$ (intégrales antisymétrisées). La notation usuelle $\langle ps|\,|qr\rangle$ pour les intégrales antisymétrisées explique la dénomination *II*, pour *double-barre*, par opposition à des intégrales $\langle pq|rs\rangle$ qui donnent lieu à des formulations *I*, pour *simple-barre*.

Comme il a été vu précédemment, $\mathbb{P}_{c,\alpha}$ est approximée dans le cadre RPA, c'est-à-dire que les fonctions de réponse $\Pi_\alpha$ sont approximées : elles sont calculées avec un noyau $\mathbb{F}_\alpha$ de l'équation de Bethe-Salpeter approximé. Utiliser un noyau contenant uniquement des termes de Hartree résulte en une approximation dite direct-RPA (dRPA) de la partie corrélation de la matrice densité à deux particules : $\mathbb{P}_{c,\alpha}^{\text{dRPA}}$, dans laquelle seul l'effet de l'écrantage de l'interaction Coulombienne est pris en compte ; utiliser un noyau contenant des termes de Hartree *et* d'échange résulte en une approximation dite RPA-échange (RPAx) : $\mathbb{P}_{c,\alpha}^{\text{RPAx}}$, pour laquelle on prend en compte l'effet de l'écrantage de type échange. Ces objets approximés ne sont pas parfaitement antisymétriques et les formulations (2.4.15) et (2.4.16) ne sont plus strictement équivalentes. Cela donne naissance à quatre expressions[2] de l'énergie de corrélation RPA dans le cadre AC-FDT :

$$E_c^{\text{dRPA-I}} = \frac{1}{2}\int_0^1 d\alpha \; \text{Tr}\left(\mathbb{W}^{\text{I}}.\mathbb{P}_{c,\alpha}^{\text{dRPA}}\right) \tag{2.4.17a}$$

$$E_c^{\text{dRPA-II}} = \frac{1}{2}\int_0^1 d\alpha \; \text{Tr}\left(\mathbb{W}^{\text{II}}.\mathbb{P}_{c,\alpha}^{\text{dRPA}}\right) \tag{2.4.17b}$$

$$E_c^{\text{RPAx-I}} = \frac{1}{2}\int_0^1 d\alpha \; \text{Tr}\left(\mathbb{W}^{\text{I}}.\mathbb{P}_{c,\alpha}^{\text{RPAx}}\right) \tag{2.4.17c}$$

$$E_c^{\text{RPAx-II}} = \frac{1}{4}\int_0^1 d\alpha \; \text{Tr}\left(\mathbb{W}^{\text{II}}.\mathbb{P}_{c,\alpha}^{\text{RPAx}}\right) \tag{2.4.17d}$$

où l'on contracte la partie corrélation de la matrice densité à deux particules $\mathbb{P}_{c,\alpha}^{\text{dRPA}}$ ou $\mathbb{P}_{c,\alpha}^{\text{RPAx}}$ avec les intégrales $\mathbb{W}^{\text{I}}$ ou $\mathbb{W}^{\text{II}}$.

Notons que d'après l'équation (2.4.16), c'est un facteur un quart qui devrait apparaître équation (2.4.17b). La question semble avoir fait débat dans la littérature, mais le facteur un quart ne doit apparaître que lorsque l'on travaille avec des objets $\mathbb{P}_{c,\alpha}$ complètement antisymétriques. Même si ce n'est pas le cas d'une manière exacte pour la version RPAx-II, $\mathbb{P}_c^{\text{RPAx}}$ est approximativement antisymétrique, tandis que ce n'est pas le cas du tout pour la version dRPA-II. On ajoute que dans ce dernier cas c'est avec un facteur un demi que la limite MP2 est recouvrée (voir section 2.4.6), tandis que pour RPAx-II la limite MP2 se retrouve avec le facteur d'un quart.

De la même manière que, dans l'équation (2.3.4), $\zeta$ permettait d'exclure/inclure l'échange dans le noyau de l'équation de Bethe-Salpeter, ici l'expression par bloc de $\mathbb{W}^{\text{I/II}}$ est :

$$\mathbb{W}^{\text{I/II}} = \begin{pmatrix} \mathbf{A}' & \mathbf{B} \\ \mathbf{B} & \mathbf{A}' \end{pmatrix}, \tag{2.4.18}$$

avec :

$$\begin{aligned} \mathbf{A}'_{ia,jb} &= \langle ib| \,\hat{w}\, |aj\rangle - \xi \langle ib| \,\hat{w}\, |ja\rangle \\ \mathbf{B}_{ia,jb} &= \langle ab| \,\hat{w}\, |ij\rangle - \xi \langle ab| \,\hat{w}\, |ji\rangle , \end{aligned} \tag{2.4.19}$$

où cette fois $\xi$ permet de passer d'une formulation *simple-barre* ($\xi = 0$) à une formulation *double-barre* ($\xi = 1$). L'introduction de cet « interrupteur » $\xi$ est une idée originale de cette thèse, et permet un traitement complètement unifié de toutes les variantes de RPA, comme rendu clair dans la suite.

**Notons que**, dans un souci d'alléger les notations, je ne distingue pas les matrices $\mathbf{A}'$ et $\mathbf{B}$ qui sont construites avec $\zeta$ dans le but de générer $\mathbb{P}_{c,\alpha}^{\text{RPA}}$ (équation (2.4.10)) de celles qui sont construites avec $\xi$ dans le but de générer $\mathbb{W}^{\text{I/II}}$ (équation (2.4.19)) : elles apparaissent dans des contextes différents mais obéissent à la même démarche d'antisymétriser des intégrales bi-électroniques (pour inclure l'échange) : les rôles des « interrupteurs » $\zeta$ et $\xi$ sont complètement similaires.

### 2.4.3 Équation unique

L'équation la plus générale possible pour exprimer les énergies de corrélation de type connexion adiabatique (AC-RPA) (2.4.17a), (2.4.17b), (2.4.17c) et (2.4.17d) est :

$$E_c = \frac{1}{2}\int_0^1 d\alpha \,\text{Tr}\left\{\begin{pmatrix} \mathbf{A}' & \mathbf{B} \\ \mathbf{B} & \mathbf{A}' \end{pmatrix} \cdot \left(\begin{pmatrix} \mathbf{Y}_\alpha \mathbf{Y}_\alpha^\dagger & \mathbf{Y}_\alpha \mathbf{X}_\alpha^\dagger \\ \mathbf{X}_\alpha \mathbf{Y}_\alpha^\dagger & \mathbf{X}_\alpha \mathbf{X}_\alpha^\dagger \end{pmatrix} - \begin{pmatrix} \mathbf{0} & \mathbf{0} \\ \mathbf{0} & \mathbf{1} \end{pmatrix}\right)\right\}, \tag{2.4.20}$$

où les éléments $\mathbf{A}'$ et $\mathbf{B}$ peuvent être définis pour correspondre à un scénario *simple-barre* ou à un scénario *double-barre* et les éléments $\mathbf{X}_\alpha$ et $\mathbf{Y}_\alpha$ peuvent être obtenus dans un cadre direct-RPA ou RPA-échange. (Le facteur un demi doit être remplacé par un quart dans le cas RPAx-II). Le développement de cette formulation donne :

$$
\begin{aligned}
E_c &= \frac{1}{2}\int_0^1 d\alpha \; \mathrm{tr}\left\{\mathbf{A}'.\left(\mathbf{Y}_\alpha\mathbf{Y}_\alpha^\dagger + \mathbf{X}_\alpha\mathbf{X}_\alpha^\dagger - \mathbf{1}\right) + \mathbf{B}.\left(\mathbf{X}_\alpha\mathbf{Y}_\alpha^\dagger + \mathbf{Y}_\alpha\mathbf{X}_\alpha^\dagger\right)\right\} \\
&= \frac{1}{2}\int_0^1 d\alpha \; \mathrm{tr}\left\{\frac{1}{2}\left(\mathbf{A}'+\mathbf{B}\right)\left(\mathbf{X}_\alpha+\mathbf{Y}_\alpha\right)\left(\mathbf{X}_\alpha+\mathbf{Y}_\alpha\right)^\dagger \right. \\
&\qquad\qquad \left. + \frac{1}{2}\left(\mathbf{A}'-\mathbf{B}\right)\left(\mathbf{X}_\alpha-\mathbf{Y}_\alpha\right)\left(\mathbf{X}_\alpha-\mathbf{Y}_\alpha\right)^\dagger - \mathbf{A}'\right\},
\end{aligned}
\tag{2.4.21}
$$

où l'on souligne la diminution de dimension par la notation « $\mathrm{tr}\{\sqcup\}$ », la trace sur l'espace $\mathcal{L}^{occ}\otimes\mathcal{L}^{vir}$. En définissant $\mathbf{Q}_\alpha = \left(\mathbf{X}_\alpha+\mathbf{Y}_\alpha\right)\left(\mathbf{X}_\alpha+\mathbf{Y}_\alpha\right)^\dagger$ on obtient :

$$
E_c = \frac{1}{2}\int_0^1 d\alpha \; \mathrm{tr}\left\{\frac{1}{2}\left(\mathbf{A}'+\mathbf{B}\right)\mathbf{Q}_\alpha + \frac{1}{2}\left(\mathbf{A}'-\mathbf{B}\right)\mathbf{Q}_\alpha^{-1} - \mathbf{A}'\right\},
\tag{2.4.22}
$$

où l'on peut aisément montrer qu'en effet $\left(\mathbf{X}_\alpha-\mathbf{Y}_\alpha\right)\left(\mathbf{X}_\alpha-\mathbf{Y}_\alpha\right)^\dagger$ est l'inverse de $\left(\mathbf{X}_\alpha+\mathbf{Y}_\alpha\right)\left(\mathbf{X}_\alpha+\mathbf{Y}_\alpha\right)^\dagger$ en utilisant la complétude de la base $\{\mathbb{C}_\alpha\}$.

### 2.4.4 Éviter le calcul des vecteurs propres

Obtenir explicitement les matrices $\mathbf{X}_\alpha$ et $\mathbf{Y}_\alpha$ requiert la résolution de l'équation aux valeurs et vecteurs propres (2.4.11) de (grande) dimension $n_{\text{occ}}\times n_{\text{vir}}+n_{\text{occ}}\times n_{\text{vir}}$. On peut réduire ces dimensions de moitié en comprenant que l'équation (2.4.11) est équivalente[86] à (somme et différence des lignes des matrices par blocs) :

$$
\begin{cases}
\left(\boldsymbol{\varepsilon}+\alpha\mathbf{A}'+\alpha\mathbf{B}\right)\left(\mathbf{x}_{\alpha,n}+\mathbf{y}_{\alpha,n}\right) = \omega_{\alpha,n}\left(\mathbf{x}_{\alpha,n}-\mathbf{y}_{\alpha,n}\right) \\
\left(\boldsymbol{\varepsilon}+\alpha\mathbf{A}'-\alpha\mathbf{B}\right)\left(\mathbf{x}_{\alpha,n}-\mathbf{y}_{\alpha,n}\right) = \omega_{\alpha,n}\left(\mathbf{x}_{\alpha,n}+\mathbf{y}_{\alpha,n}\right),
\end{cases}
\tag{2.4.23}
$$

et qu'ainsi les solutions de (2.4.11) sont accessibles exactement par la résolution de l'équation aux valeurs et vecteurs propres symétriques de dimension $n_{\text{occ}}\times n_{\text{vir}}$ :

$$
\mathbf{M}_\alpha\mathbf{z}_{\alpha,n} = \omega_{\alpha,n}^2\,\mathbf{z}_{\alpha,n}
\tag{2.4.24}
$$

où :

$$
\mathbf{M}_\alpha = \left(\boldsymbol{\varepsilon}+\alpha\mathbf{A}'-\alpha\mathbf{B}\right)^{\frac{1}{2}}\left(\boldsymbol{\varepsilon}+\alpha\mathbf{A}'+\alpha\mathbf{B}\right)\left(\boldsymbol{\varepsilon}+\alpha\mathbf{A}'-\alpha\mathbf{B}\right)^{\frac{1}{2}}
\tag{2.4.25}
$$

$$
\mathbf{z}_{\alpha,n} = \sqrt{\omega}\left(\boldsymbol{\varepsilon}+\alpha\mathbf{A}'-\alpha\mathbf{B}\right)^{-\frac{1}{2}}\left(\mathbf{x}_{\alpha,n}+\mathbf{y}_{\alpha,n}\right) = \frac{1}{\sqrt{\omega}}\left(\boldsymbol{\varepsilon}+\alpha\mathbf{A}'-\alpha\mathbf{B}\right)^{\frac{1}{2}}\left(\mathbf{x}_{\alpha,n}-\mathbf{y}_{\alpha,n}\right),
\tag{2.4.26}
$$

où on utilise la deuxième ligne de (2.4.23) pour remplacer le terme $\left(\mathbf{x}_{\alpha,n}-\mathbf{y}_{\alpha,n}\right)$ de la première ligne. Avec la matrice $\mathbf{M}_\alpha$ et les vecteurs $\mathbf{z}_{\alpha,n}$ ainsi définis, on obtient pour $\mathbf{Q}_\alpha$ :

$$
\mathbf{Q}_\alpha = \left(\boldsymbol{\varepsilon}+\alpha\mathbf{A}'-\alpha\mathbf{B}\right)^{\frac{1}{2}}\left(\mathbf{M}_\alpha\right)^{-\frac{1}{2}}\left(\boldsymbol{\varepsilon}+\alpha\mathbf{A}'-\alpha\mathbf{B}\right)^{\frac{1}{2}}
\tag{2.4.27}
$$

**En résumé**, toutes les expressions des énergies de corrélation RPA dans un cadre AC-FDT sont obtenues à partir de l'unique équation (2.4.22), en utilisant des matrices qui sont construites uniquement avec les éléments $\boldsymbol{\varepsilon}$, $\mathbf{A}'$ et $\mathbf{B}$, c'est-à-dire sans résoudre ni l'équation à grande dimension (2.4.11), ni même l'équation (2.4.24). L'étape déterminante du calcul est la construction de $(\mathbf{M}_\alpha)^{-\frac{1}{2}}$.

Pour élever $\mathbf{M}_\alpha$ à la puissance $-\frac{1}{2}$, il faut supposer qu'elle est définie positive. Dans la version dRPA, on peut vérifier qu'en effet la matrice $\mathbf{M}_\alpha$ est bien définie positive ; en revanche dans le cas RPAx, on peut rencontrer des instabilités qui font que le traitement par l'équation (2.4.27) échoue. Dans un formalisme adapté de spin (voir la section 2.6), on peut avoir affaire à la fois à des instabilités singulets (dans le cas où des liaisons sont dissociées) et à des instabilités triplets (dans des systèmes tels que le $Be_2$).

J'insiste à nouveau sur le fait que les matrices $\mathbf{A}'$ et $\mathbf{B}$ de l'équation (2.4.22) sont construites avec $\xi$, celles de l'équation (2.4.27) sont construites avec $\zeta$.

Pour être complet, on peut noter qu'une formulation alternative des objets $\mathbf{M}_\alpha$ et $\mathbf{z}_\alpha$ existe dans la littérature[115], et consiste à réécrire l'équation (2.4.23) en :

$$\begin{cases} \boldsymbol{\varepsilon}^{\frac{1}{2}}\left(\boldsymbol{\varepsilon}+\alpha\mathbf{A}'+\alpha\mathbf{B}\right)\boldsymbol{\varepsilon}^{\frac{1}{2}}\boldsymbol{\varepsilon}^{-\frac{1}{2}}\left(\mathbf{x}_{\alpha,n}+\mathbf{y}_{\alpha,n}\right) = \omega_{\alpha,n}\boldsymbol{\varepsilon}^{\frac{1}{2}}\left(\mathbf{x}_{\alpha,n}-\mathbf{y}_{\alpha,n}\right) \\ \boldsymbol{\varepsilon}^{\frac{1}{2}}\left(\mathbf{1}+\boldsymbol{\varepsilon}^{-\frac{1}{2}}\left(\alpha\mathbf{A}'-\alpha\mathbf{B}\right)\boldsymbol{\varepsilon}^{-\frac{1}{2}}\right)\boldsymbol{\varepsilon}^{\frac{1}{2}}\left(\mathbf{x}_{\alpha,n}-\mathbf{y}_{\alpha,n}\right) = \omega_{\alpha,n}\left(\mathbf{x}_{\alpha,n}+\mathbf{y}_{\alpha,n}\right) \end{cases} \tag{2.4.28}$$

Comme précédemment, on utilise la deuxième ligne pour remplacer le terme $(\mathbf{x}_{\alpha,n}-\mathbf{y}_{\alpha,n})$ de la première ligne, pour obtenir :

$$\mathbf{M}_\alpha^{\text{alt}}\mathbf{z}_{\alpha,n}^{\text{alt}} = \omega_{\alpha,n}^2\,\mathbf{z}_{\alpha,n}^{\text{alt}} \tag{2.4.29}$$

où :

$$\mathbf{M}_\alpha^{\text{alt}} = \left(\mathbf{1}+\boldsymbol{\varepsilon}^{-\frac{1}{2}}\left(\alpha\mathbf{A}'-\alpha\mathbf{B}\right)\boldsymbol{\varepsilon}^{-\frac{1}{2}}\right)^{\frac{1}{2}}\left(\boldsymbol{\varepsilon}^2+\boldsymbol{\varepsilon}^{\frac{1}{2}}\left(\alpha\mathbf{A}'+\alpha\mathbf{B}\right)\boldsymbol{\varepsilon}^{\frac{1}{2}}\right)\left(\mathbf{1}+\boldsymbol{\varepsilon}^{-\frac{1}{2}}\left(\alpha\mathbf{A}'-\alpha\mathbf{B}\right)\boldsymbol{\varepsilon}^{-\frac{1}{2}}\right)^{\frac{1}{2}} \tag{2.4.30}$$

$$\mathbf{z}_{\alpha,n}^{\text{alt}} = \sqrt{\omega}\left(\mathbf{1}+\boldsymbol{\varepsilon}^{-\frac{1}{2}}\left(\alpha\mathbf{A}'-\alpha\mathbf{B}\right)\boldsymbol{\varepsilon}^{-\frac{1}{2}}\right)^{-\frac{1}{2}}\boldsymbol{\varepsilon}^{-\frac{1}{2}}\left(\mathbf{x}_{\alpha,n}+\mathbf{y}_{\alpha,n}\right) \tag{2.4.31}$$

L'intérêt de cette formulation est l'expression sous la forme $(1+x)^{\frac{1}{2}}$ qui ouvre la voie à des approximations d'expansion de Taylor lorsque l'on est amené à calculer $\mathbf{M}_\alpha^{\frac{1}{2}}$. Notons que si l'on cherche à exprimer la matrice $\mathbf{M}_\alpha$ de la même manière, on trouve :

$$\mathbf{M}_\alpha = \boldsymbol{\varepsilon}^{\frac{1}{4}}\left(\mathbf{1}+\boldsymbol{\varepsilon}^{-\frac{1}{2}}\left(\alpha\mathbf{A}'-\alpha\mathbf{B}\right)\boldsymbol{\varepsilon}^{-\frac{1}{2}}\right)^{\frac{1}{2}}\boldsymbol{\varepsilon}^{-\frac{1}{4}}\left(\boldsymbol{\varepsilon}^2+\boldsymbol{\varepsilon}^{\frac{1}{2}}\left(\alpha\mathbf{A}'+\alpha\mathbf{B}\right)\boldsymbol{\varepsilon}^{\frac{1}{2}}\right)\boldsymbol{\varepsilon}^{-\frac{1}{4}}\left(\mathbf{1}+\boldsymbol{\varepsilon}^{-\frac{1}{2}}\left(\alpha\mathbf{A}'-\alpha\mathbf{B}\right)\boldsymbol{\varepsilon}^{-\frac{1}{2}}\right)^{\frac{1}{2}}\boldsymbol{\varepsilon}^{\frac{1}{4}}, \tag{2.4.32}$$

qui n'est pas en général égal à $\mathbf{M}_\alpha^{\text{alt}}$.

### 2.4.5 Approximations dites "-IIa"

On a conservé un formalisme le plus général possible, qui englobe toutes les combinaisons (d/x)× (I/II). Notons simplement ici que pour les énergies de corrélation *simple-barre* on a : $\mathbf{A}'^{\mathrm{I}} = \mathbf{B}^{\mathrm{I}} = \mathbf{K}$, c'est-à-dire $(\mathbf{A}' + \mathbf{B})^{\mathrm{I}} = 2\mathbf{K}$ et $(\mathbf{A}' - \mathbf{B})^{\mathrm{I}} = \mathbf{0}$, d'où l'expression particulièrement simple de l'équation (2.4.22) :

$$E_c^{\text{d/x-I}} = \frac{1}{2} \int_0^1 d\alpha \ \mathrm{tr} \left\{ \left( \mathbf{Q}_\alpha^{\text{d/x}} - \mathbf{1} \right) \mathbf{K} \right\}, \tag{2.4.33}$$

où les exposants d/x rappellent que les matrices impliquées dans la construction de $\mathbf{Q}_\alpha$ sont exprimées dans un cadre dRPA ou RPAx. On peut également approximer les équations (2.4.22) dans les cas *double-barre* pour ressembler aux équations simples (2.4.33) des cas *simple-barre*[2]. On écrit :

$$\mathbf{Q}_\alpha = \mathbf{1} + \mathbf{P}_\alpha \quad \text{et} \quad \mathbf{Q}_\alpha^{-1} = (\mathbf{1} + \mathbf{P}_\alpha)^{-1} \approx \mathbf{1} - \mathbf{P}_\alpha = \mathbf{2} - \mathbf{Q}_\alpha, \tag{2.4.34}$$

ce qui réduit les équations des énergies dRPA-II et RPAx-II aux approximations dRPA-IIa et RPAx-IIa :

$$E_c^{\text{dRPA-IIa}} = \frac{1}{2} \int_0^1 d\alpha \ \mathrm{tr} \left\{ \left( \mathbf{Q}_\alpha^{\text{dRPA}} - \mathbf{1} \right) \mathbf{B}^{\mathrm{II}} \right\} \tag{2.4.35}$$

$$E_c^{\text{RPAx-IIa}} = \frac{1}{4} \int_0^1 d\alpha \ \mathrm{tr} \left\{ \left( \mathbf{Q}_\alpha^{\text{RPAx}} - \mathbf{1} \right) \mathbf{B}^{\mathrm{II}} \right\} \tag{2.4.36}$$

### 2.4.6 Limites MP2 au second ordre de l'interaction

Avec une théorie des perturbations basée sur l'interaction électron-électron, on peut montrer que dans une approximation au second ordre les énergies de corrélation des versions dRPA-II, RPAx-I et RPAx-II correspondent à l'énergie MP2[2, 46, 64]. On cherche à déterminer les corrections de premier ordre d'une expansion en puissance de $\alpha$ :

$$\begin{aligned} \omega_{\alpha,n} &= \omega_{0,n} + \alpha\omega_n^{(1)} + \dots \\ \mathbb{C}_{\alpha,n} &= \mathbb{C}_{0,n} + \alpha\mathbb{C}_n^{(1)} + \dots, \end{aligned} \tag{2.4.37}$$

Notons que pour $\alpha = 0$, le système des équations (2.4.11) est considérablement simplifié : on a $\mathbf{y}_{0,n} = \mathbf{0}$ (ceci est rendu clair en considérant les équations (2.4.23)), et l'on est amené à résoudre $\boldsymbol{\varepsilon}\mathbf{x}_{0,n} = \omega_{0,n}\mathbf{x}_{0,n}$ : les modes $n$ correspondent à des transitions $i \to a$ du système de référence, les valeurs propres sont les éléments diagonaux de $\boldsymbol{\varepsilon}$ : $\omega_{0,n} = \varepsilon_a - \varepsilon_i$, et les vecteurs propres sont de la forme :

$$\mathbf{x}_{0,n} = \mathbf{1}_n \doteq \begin{pmatrix} 0 \\ \vdots \\ 1 \\ 0 \\ \vdots \end{pmatrix} \leftarrow \text{position } n \tag{2.4.38}$$

Avec les expressions de $\omega_{\alpha,n}$ et $\mathbb{C}_{\alpha,n}$ données (2.4.37), l'équation (2.4.11) multipliée à gauche par $\mathbb{C}^\dagger_{0,n}$, $\mathbb{C}^\dagger_{0,m}$ et $\mathbb{C}^\dagger_{0,-m}$ fournit les informations suivantes :

$$\omega_n^{(1)} = \mathbb{C}^\dagger_{0,n}\mathbb{W}\mathbb{C}_{0,n} = A'_{n,n} \tag{2.4.39}$$

$$\mathbb{C}^\dagger_{0,m}\mathbb{\Delta}\mathbb{C}_n^{(1)} = \frac{\mathbb{C}^\dagger_{0,m}\mathbb{W}\mathbb{C}_{0,n}}{-\omega_{0,m} + \omega_{0,n}} = \frac{A'_{m,n}}{-\varepsilon_m + \varepsilon_n} \tag{2.4.40}$$

$$\mathbb{C}^\dagger_{0,-m}\mathbb{\Delta}\mathbb{C}_n^{(1)} = \frac{\mathbb{C}^\dagger_{0,-m}\mathbb{W}\mathbb{C}_{0,n}}{\omega_{0,m} + \omega_{0,n}} = \frac{B'_{m,n}}{\varepsilon_m + \varepsilon_n}, \tag{2.4.41}$$

où l'on applique simplement la définition $\mathbb{C}_{0,n} = \begin{pmatrix} \mathbf{1}_n \\ \mathbf{0} \end{pmatrix}$. Ici les indices $n$ et $m$ se confondent avec les super-indices $ia$ et $jb$. Pour obtenir une expression de $\mathbb{C}_n^{(1)}$, on somme sur $m$ les équations (2.4.40) et (2.4.41) multipliées à gauche respectivement par $\mathbb{\Delta}\mathbb{C}_{0,m}$ et $\mathbb{\Delta}\mathbb{C}_{0,-m}$ pour obtenir :

$$\mathbb{\Delta}\left(\sum_{m\neq n}\mathbb{C}_{0,m}\mathbb{C}^\dagger_{0,m} + \sum_m \mathbb{C}_{0,-m}\mathbb{C}^\dagger_{0,-m}\right)\mathbb{\Delta}\mathbb{C}_n^{(1)} = \sum_{m\neq n}\frac{A'_{m,n}}{-\varepsilon_m+\varepsilon_n}\mathbb{\Delta}\mathbb{C}_{0,m} + \sum_m \frac{B'_{m,n}}{\varepsilon_m+\varepsilon_n}\mathbb{\Delta}\mathbb{C}_{0,-m} \tag{2.4.42}$$

On utilise à gauche du signe égal la résolution de l'identité $\sum_m \mathbb{C}_{0,m}\mathbb{C}^\dagger_{0,m} + \sum_m \mathbb{C}_{0,-m}\mathbb{C}^\dagger_{0,-m} = \mathbb{1}$ (où c'est l'orthogonalité $\mathbb{C}^\dagger_{0,n}\mathbb{\Delta}\mathbb{C}_n^{(1)} = 0$ qui permet d'ajouter le terme $m = n$) ainsi que l'identité $\mathbb{\Delta}^2 = \mathbb{1}$ pour obtenir :

$$\mathbb{C}_n^{(1)} \doteq \begin{pmatrix} \mathbf{x}_n^{(1)} \\ \mathbf{y}_n^{(1)} \end{pmatrix} = \sum_{m\neq n}\frac{A'_{m,n}}{-\varepsilon_m+\varepsilon_n}\begin{pmatrix} \mathbf{1}_n \\ \mathbf{0} \end{pmatrix} + \sum_m \frac{B_{m,n}}{\varepsilon_m+\varepsilon_n}\begin{pmatrix} \mathbf{0} \\ -\mathbf{1}_n \end{pmatrix} \tag{2.4.43}$$

Pour écrire l'énergie RPA générale de l'équation (2.4.22), on a besoin des matrices $\mathbf{Q}_\alpha$ et $\mathbf{Q}_\alpha^{-1}$ :

$$\begin{aligned}
\mathbf{Q}_\alpha &= (\mathbf{X}_\alpha + \mathbf{Y}_\alpha)(\mathbf{X}_\alpha + \mathbf{Y}_\alpha)^\dagger \\
&= \sum_n (\mathbf{x}_{\alpha,n} + \mathbf{y}_{\alpha,n})(\mathbf{x}_{\alpha,n} + \mathbf{y}_{\alpha,n})^\dagger \\
&= \sum_n \left(\mathbf{1}_n + \alpha\mathbf{x}_\alpha^{(1)} + \alpha\mathbf{y}_\alpha^{(1)}\right)\left(\mathbf{1}_n^\dagger + \alpha\mathbf{x}_\alpha^{(1)\dagger} + \alpha\mathbf{y}_\alpha^{(1)\dagger}\right) + O(\alpha^2) \\
&= \sum_n \mathbf{1}_n\mathbf{1}_n^\dagger + \alpha\sum_n\left(\mathbf{x}_\alpha^{(1)}\mathbf{1}_n^\dagger + \mathbf{1}_n\mathbf{x}_\alpha^{(1)\dagger}\right) + \alpha\sum_n\left(\mathbf{y}_\alpha^{(1)}\mathbf{1}_n^\dagger + \mathbf{1}_n\mathbf{y}_\alpha^{(1)\dagger}\right) + O(\alpha^2)
\end{aligned} \tag{2.4.44}$$

Le lecteur trouvera aisément que les trois sommations sur les modes $n$ sont, respectivement : la matrice identité $\mathbf{1}$, zéro et $-2\overline{\mathbf{B}}$ où on définit :

$$\overline{B}_{m,n} = \frac{B_{m,n}}{\varepsilon_m + \varepsilon_n} \qquad \text{c'est-à-dire :} \qquad \overline{B}_{ia,jb} = \frac{B_{ia,jb}}{\varepsilon_a - \varepsilon_i + \varepsilon_b - \varepsilon_j} \tag{2.4.45}$$

Ainsi les matrices $\mathbf{Q}$ et $\mathbf{Q}_\alpha^{-1}$ (que l'on obtient par un même raisonnement) s'écrivent-elles :

$$\mathbf{Q}_\alpha = \mathbf{1} - 2\alpha\overline{\mathbf{B}} + O(\alpha^2) \tag{2.4.46}$$

$$\mathbf{Q}_\alpha^{-1} = \mathbf{1} + 2\alpha\overline{\mathbf{B}} + O(\alpha^2), \tag{2.4.47}$$

ce qui mène à l'approximation au second ordre de l'énergie (2.4.22) :

$$E_c \approx \frac{1}{2}\int_0^1 d\alpha \,\mathrm{tr}\left\{\frac{1}{2}(\mathbf{A}' + \mathbf{B})\left(\mathbf{1} - 2\alpha\overline{\mathbf{B}}\right) + \frac{1}{2}(\mathbf{A}' - \mathbf{B})\left(\mathbf{1} + 2\alpha\overline{\mathbf{B}}\right) - \mathbf{A}'\right\} = \frac{1}{2}\int_0^1 d\alpha \,\mathrm{tr}\left\{-2\alpha\mathbf{B}\overline{\mathbf{B}}\right\}, \tag{2.4.48}$$

où $\mathbf{B}$ est construit avec l'« interrupteur » $\xi$ des formulations I/II et $\overline{\mathbf{B}}$ avec l'« interrupteur » $\zeta$ pour les formulations dRPA/RPAx. Ainsi les expressions des approximations au second ordre des quatre versions de RPA sont :

$$\begin{aligned}
E^{\text{dRPA-I(2)}} &= \frac{1}{2}\int_0^1 d\alpha \,\mathrm{tr}\left\{-2\alpha\mathbf{K}\overline{\mathbf{K}}\right\} = -\frac{1}{2}\mathrm{tr}\left\{\mathbf{K}\overline{\mathbf{K}}\right\} = E^{\text{dMP2}} \\
E^{\text{dRPA-II(2)}} &= \frac{1}{2}\int_0^1 d\alpha \,\mathrm{tr}\left\{-2\alpha\mathbf{B}\overline{\mathbf{K}}\right\} = -\frac{1}{2}\mathrm{tr}\left\{\mathbf{B}\overline{\mathbf{K}}\right\} = E^{\text{MP2}} \\
E^{\text{RPAx-I(2)}} &= \frac{1}{2}\int_0^1 d\alpha \,\mathrm{tr}\left\{-2\alpha\mathbf{K}\overline{\mathbf{B}}\right\} = -\frac{1}{2}\mathrm{tr}\left\{\mathbf{K}\overline{\mathbf{B}}\right\} = E^{\text{MP2}} \\
E^{\text{RPAx-II(2)}} &= \frac{1}{4}\int_0^1 d\alpha \,\mathrm{tr}\left\{-2\alpha\mathbf{B}\overline{\mathbf{B}}\right\} = -\frac{1}{4}\mathrm{tr}\left\{\mathbf{B}\overline{\mathbf{B}}\right\} = E^{\text{MP2}}
\end{aligned} \tag{2.4.49}$$

**Ainsi** toutes les versions de RPA se réduisent à une énergie MP2 au second ordre de la perturbation, sauf la version $E^{\text{dRPA-I}}$ qui se réduit à une version "directe" de l'énergie MP2[67, 86], c'est-à-dire une version de l'énergie MP2 sans terme d'échange.

## 2.5 Formulation de "plasmon"

On s'intéresse ici à une formulation, dite "de plasmon"[92, 116], des énergies RPA où les deux intégrations de l'équation (2.3.5) sont réalisées de manière analytique. Dans cette section, on intègre analytiquement sur la constante de couplage $\alpha$ des expressions du type (2.4.5), où l'intégration sur la fréquence $\omega$ a déjà été effectuée, pour obtenir une formule de plasmon. Plus tard dans le chapitre, on montrera que l'on peut obtenir la même formule en intégrant analytiquement sur la fréquence $\omega$ une expression de l'énergie RPA déjà intégrée sur la constante de couplage $\alpha$.

### 2.5.1 Équation aux valeurs et vecteurs propres

Pour des raisons qui seront rendues claires dans la suite, on va trouver utile d'exprimer de manière compacte les $n$ équations :

$$\mathbb{A}_\alpha \mathbb{C}_{\alpha,n} = \omega_{\alpha,n} \mathbb{\Delta} \mathbb{C}_{\alpha,n}, \tag{2.4.11}$$

en écrivant :

$$\mathbb{A}_\alpha \mathbb{C}_\alpha = \mathbb{\Delta} \mathbb{C}_\alpha \mathbf{\Omega}_\alpha, \tag{2.5.1}$$

où l'on définit :

$$\underset{(2n_{occ}n_{vir}.n_{occ}n_{vir})}{\underset{\uparrow}{\mathbb{C}_\alpha}} = \begin{pmatrix} \mathbf{X}_\alpha \\ \mathbf{Y}_\alpha \end{pmatrix} = (\mathbb{C}_{\alpha,1}\ \mathbb{C}_{\alpha,2}\ \ldots\ \mathbb{C}_{\alpha,n}) = \begin{pmatrix} \mathbf{X}_{\alpha,1}\mathbf{X}_{\alpha,2} & \ldots & \mathbf{X}_{\alpha,n} \\ \mathbf{Y}_{\alpha,1}\mathbf{Y}_{\alpha,2} & \ldots & \mathbf{Y}_{\alpha,n} \end{pmatrix}, \tag{2.5.2}$$

et la matrice $\mathbf{\Omega}_\alpha$ qui porte les valeurs propres $\omega_{\alpha,n}$ sur sa diagonale. Chaque colonne des matrices à gauche et à droite du signe égal de l'équation (2.5.1) correspond aux vecteurs à gauche et à droite du signe égal d'une équation (2.4.11).

### 2.5.2 Dérivation

Souvenons-nous à nouveau ici que les matrices de l'équation (2.4.11) sont exprimées dans un cadre direct-RPA ou RPA-échange (approximation du noyau $f_\alpha^{\mathrm{BSE}}$) ; ce fait est souligné dans la suite par les exposants "d/x". La normalisation de l'équation (2.4.11) :

$$\mathbb{C}_{\alpha,n}^{\mathrm{d/x},\dagger} \mathbb{\Delta} \mathbb{C}_{\alpha,n}^{\mathrm{d/x}} = (\mathbf{x}_{\alpha,n}^{\mathrm{d/x},\dagger} \quad \mathbf{y}_{\alpha,n}^{\mathrm{d/x},\dagger}) \begin{pmatrix} \mathbf{1} & \mathbf{0} \\ \mathbf{0} & -\mathbf{1} \end{pmatrix} \begin{pmatrix} \mathbf{x}_{\alpha,n}^{\mathrm{d/x}} \\ \mathbf{y}_{\alpha,n}^{\mathrm{d/x}} \end{pmatrix} = \mathbf{x}_{\alpha,n}^{\mathrm{d/x},\dagger} \mathbf{x}_{\alpha,n}^{\mathrm{d/x}} - \mathbf{y}_{\alpha,n}^{\mathrm{d/x},\dagger} \mathbf{y}_{\alpha,n}^{\mathrm{d/x}} = 1, \tag{2.5.3}$$

permet d'écrire les relations suivantes pour $\omega_{\alpha,n}$ :

$$\mathbb{C}_{\alpha,n}^{\mathrm{d/x},\dagger} \mathbb{A}_\alpha^{\mathrm{d/x}} \mathbb{C}_{\alpha,n}^{\mathrm{d/x}} = \omega_{\alpha,n}^{\mathrm{d/x}} \mathbb{C}_{\alpha,n}^{\mathrm{d/x},\dagger} \mathbb{\Delta} \mathbb{C}_{\alpha,n}^{\mathrm{d/x}} = \omega_{\alpha,n}^{\mathrm{d/x}} \tag{2.5.4}$$

$$\mathbb{C}_{\alpha,-n}^{\mathrm{d/x},\dagger} \mathbb{A}_\alpha^{\mathrm{d/x}} \mathbb{C}_{\alpha,-n}^{\mathrm{d/x}} = -\omega_{\alpha,n}^{\mathrm{d/x}} \mathbb{C}_{\alpha,-n}^{\mathrm{d/x},\dagger} \mathbb{\Delta} \mathbb{C}_{\alpha,-n}^{\mathrm{d/x}} = \omega_{\alpha,n}^{\mathrm{d/x}} \tag{2.5.5}$$

En dérivant l'équation (2.5.5), on obtient :

$$\frac{\mathrm{d}\omega^{\mathrm{d/x}}_{\alpha,n}}{\mathrm{d}\alpha} = \mathbb{C}^{\mathrm{d/x},\dagger}_{\alpha,-n}\frac{\mathrm{d}\mathbb{A}^{\mathrm{d/x}}_{\alpha}}{\mathrm{d}\alpha}\mathbb{C}^{\mathrm{d/x}}_{\alpha,-n} = \mathbb{C}^{\mathrm{d/x},\dagger}_{\alpha,-n}\mathbb{W}^{\mathrm{I/II}}\mathbb{C}^{\mathrm{d/x}}_{\alpha,-n}, \tag{2.5.6}$$

où la dérivée de la matrice $\mathbb{A}^{\mathrm{d/x}}_{\alpha}$, construite avec $\xi = 0$ ou $\xi = 1$, correspond aux matrices $\mathbb{W}^{\mathrm{I/II}}$, construites avec $\zeta = 0$ ou $\zeta = 1$. On reprend donc les expressions des énergies de corrélation pour lesquelles (2.5.6) est applicable, c'est-à-dire pour lesquelles $\xi = \zeta$ (équations (2.4.17a) et (2.4.17d)). En considérant la définition de la partie corrélation de la matrice densité à deux particules (2.4.13), l'invariance de la trace par permutation circulaire, et en intégrant sur la constante de couplage, on obtient :

$$\begin{aligned} E_c^{\mathrm{dRPA\text{-}I}} &= \frac{1}{2}\sum_n \left( \omega^{\mathrm{dRPA}}_{1,n} - \omega^{\mathrm{dRPA}}_{0,n} - \left.\frac{\mathrm{d}\omega^{\mathrm{dRPA}}_{\alpha,n}}{\mathrm{d}\alpha}\right|_{\alpha=0} \right) \\ E_c^{\mathrm{RPAx\text{-}II}} &= \frac{1}{4}\sum_n \left( \omega^{\mathrm{RPAx}}_{1,n} - \omega^{\mathrm{RPAx}}_{0,n} - \left.\frac{\mathrm{d}\omega^{\mathrm{RPAx}}_{\alpha,n}}{\mathrm{d}\alpha}\right|_{\alpha=0} \right), \end{aligned} \tag{2.5.7}$$

où les deux énergies de corrélation RPA s'écrivent comme la différence entre les énergies d'excitation RPA à pleine constante de couplage ($\omega^{\mathrm{d/x}}_{\alpha=1,n}$) et la somme des ordres zéro et un des énergies d'excitation RPA : $\omega^{\mathrm{d/x}}_{0,n} + \left.\frac{\mathrm{d}\omega^{\mathrm{d/x}}_{\alpha n}}{\mathrm{d}\alpha}\right|_{\alpha=0}$.

**On retrouve** dans cette formulation des énergies RPA la vision physique qui a motivé la dérivation de l'approximation RPA par Bohm et Pines, c'est-à-dire l'idée que l'on peut décrire le comportement des électrons (à l'origine : dans un plasma) par une somme d'hamiltoniens de type oscillateurs harmoniques, contenant la physique de l'énergie de corrélation longue-portée (voir 2.1).

On peut réécrire les termes d'ordre zéro et un comme :

$$\omega^{\mathrm{d/x}}_{0,n} + \left.\frac{\mathrm{d}\omega^{\mathrm{d/x}}_{\alpha,n}}{\mathrm{d}\alpha}\right|_{\alpha=0} = \omega^{\mathrm{d/x}}_{0,n} + \mathbb{C}^{\mathrm{d/x},\dagger}_{0,-n}\mathbb{W}^{\mathrm{I/II}}\mathbb{C}^{\mathrm{d/x}}_{0,-n} = \mathbb{C}^{\mathrm{d/x},\dagger}_{0,-n}\mathbb{A}^{0,\mathrm{d/x}}\mathbb{C}^{\mathrm{d/x}}_{0,-n} + \mathbb{C}^{\mathrm{d/x},\dagger}_{0,-n}\mathbb{W}^{\mathrm{I/II}}\mathbb{C}^{\mathrm{d/x}}_{0,-n} = \mathbb{C}^{\mathrm{d/x},\dagger}_{0,-n}\mathbb{A}^{\mathrm{d/x}}_{1}\mathbb{C}^{\mathrm{d/x}}_{0,-n}, \tag{2.5.8}$$

et, en notant qu'imposer $\mathbf{y}_{\alpha,n} = \mathbf{0}$ dans (2.4.11) revient à poser $\mathbf{B} = \mathbf{0}$ (voir les équations (2.4.23) pour s'en convaincre) :

$$\mathbb{C}^{\mathrm{d/x},\dagger}_{0,-n}\mathbb{A}^{\mathrm{d/x}}_{1}\mathbb{C}^{\mathrm{d/x}}_{0,-n} = \begin{pmatrix} \mathbf{0} & \mathbf{x}_{0,n} \end{pmatrix} \begin{pmatrix} \boldsymbol{\varepsilon} + \mathbf{A}' & \mathbf{0} \\ \mathbf{0} & \boldsymbol{\varepsilon} + \mathbf{A}' \end{pmatrix} \begin{pmatrix} \mathbf{0} \\ \mathbf{x}_{0,n} \end{pmatrix} = \varepsilon_a - \varepsilon_i + A'_{ia,ia} = \omega^{\mathrm{TDA,d/x}}_{1,n}, \tag{2.5.9}$$

où l'on voit que la somme des ordres zéro et un correspond à une approximation Tamm-Dancoff des excitations RPA[117, 118] à pleine constante de couplage. On voit également que les sommes sur les modes $n$ des équations (2.5.7) peuvent s'écrire avec des traces, comme suit :

$$
\begin{aligned}
E_c^{\text{dRPA-I}} &= \frac{1}{2}\sum_n \omega_{1,n}^{\text{dRPA}} - \omega_{1,n}^{\text{dTDA}} = \frac{1}{2}\,\text{tr}\left\{\boldsymbol{\Omega}_1^{\text{dRPA}} - \boldsymbol{\varepsilon} - \mathbf{A}'^{\text{dRPA}}\right\} \\
E_c^{\text{RPAx-II}} &= \frac{1}{4}\sum_n \omega_{1,n}^{\text{RPAx}} - \omega_{1,n}^{\text{TDAx}} = \frac{1}{4}\,\text{tr}\left\{\boldsymbol{\Omega}_1^{\text{RPAx}} - \boldsymbol{\varepsilon} - \mathbf{A}'^{\text{RPAx}}\right\}
\end{aligned}
\tag{2.5.10}
$$

### 2.5.3 Formulation avec équations de "Riccati"

On peut montrer que l'équation (2.5.1) est équivalente[106] aux équations dites de Riccati, c'est-à-dire que :

$$
\mathbb{A}_\alpha \mathbb{C}_\alpha = \mathbb{A}\mathbb{C}_\alpha \boldsymbol{\Omega}_\alpha \underset{(b)}{\overset{(a)}{\rightleftharpoons}} \mathbf{R}_\alpha[\mathbf{T}_\alpha] = \mathbf{0}, \tag{2.5.11}
$$

où $\mathbf{R}_\alpha[\mathbf{T}_\alpha]$ sera explicité dans la suite. L'implication de gauche à droite $(a)$ se démontre en multipliant l'expression par blocs de (2.5.1) à droite par $\mathbf{X}_\alpha^{-1}$ :

$$
\begin{pmatrix} \boldsymbol{\varepsilon} + \alpha\mathbf{A}' & \alpha\mathbf{B} \\ \alpha\mathbf{B} & \boldsymbol{\varepsilon} + \alpha\mathbf{A}' \end{pmatrix}\begin{pmatrix} \mathbf{1} \\ \mathbf{Y}_\alpha\mathbf{X}_\alpha^{-1} \end{pmatrix} = \begin{pmatrix} \mathbf{1} \\ -\mathbf{Y}_\alpha\mathbf{X}_\alpha^{-1} \end{pmatrix}\mathbf{X}_\alpha\boldsymbol{\Omega}_\alpha\mathbf{X}_\alpha^{-1} \tag{2.5.12}
$$

En multipliant à gauche par $\begin{pmatrix} \mathbf{Y}_\alpha\mathbf{X}_\alpha^{-1} \\ \mathbf{1} \end{pmatrix}^\dagger$ on trouve que si $\mathbb{C}_\alpha = \begin{pmatrix} \mathbf{X}_\alpha \\ \mathbf{Y}_\alpha \end{pmatrix}$ est solution de l'équation (2.5.1) , alors un objet $\mathbf{T}_\alpha$ définit comme $\mathbf{T}_\alpha = \mathbf{Y}_\alpha\mathbf{X}_\alpha^{-1}$ respecte l'équation de Riccati suivante :

$$
\mathbf{R}_\alpha[\mathbf{T}_\alpha] = \alpha\left(\mathbf{B} + \mathbf{A}'\mathbf{T}_\alpha + \mathbf{T}_\alpha\mathbf{A}' + \mathbf{T}_\alpha\mathbf{B}\mathbf{T}_\alpha\right) + \boldsymbol{\varepsilon}\mathbf{T}_\alpha + \mathbf{T}_\alpha\boldsymbol{\varepsilon} = \mathbf{0} \tag{2.5.13}
$$

À l'inverse, $(b)$ : en supposant que l'on a résolu l'équation de Riccati pour $\mathbf{T}_\alpha$, on peut « toujours » écrire une décomposition de Schur :

$$
\boldsymbol{\varepsilon} + \alpha\left(\mathbf{A}' + \mathbf{B}\mathbf{T}_\alpha\right) = \mathbf{x}_\alpha\mathbf{S}_\alpha\mathbf{x}_\alpha^\dagger \tag{2.5.14}
$$

L'équation (2.5.13) impose alors que :

$$
\boldsymbol{\varepsilon}\mathbf{T}_\alpha + \alpha\left(\mathbf{B} + \mathbf{A}\mathbf{T}_\alpha\right) = -\mathbf{T}_\alpha\left(\boldsymbol{\varepsilon} + \alpha\left(\mathbf{A}' + \mathbf{B}\mathbf{T}_\alpha\right)\right) = -\mathbf{T}_\alpha\mathbf{x}_\alpha\mathbf{S}_\alpha\mathbf{x}_\alpha^\dagger \tag{2.5.15}
$$

En multipliant les équations (2.5.14) et (2.5.15) à droite par $\mathbf{x}_\alpha$ et en définissant $\mathbf{y}_\alpha = \mathbf{T}_\alpha\mathbf{x}_\alpha$, on obtient :

$$\begin{pmatrix} \boldsymbol{\varepsilon} + \alpha\mathbf{A}' & \alpha\mathbf{B} \\ \alpha\mathbf{B} & \boldsymbol{\varepsilon} + \alpha\mathbf{A}' \end{pmatrix} \begin{pmatrix} \mathbf{x}_\alpha \\ \mathbf{y}_\alpha \end{pmatrix} = \begin{pmatrix} \mathbf{1} & \mathbf{0} \\ \mathbf{0} & -\mathbf{1} \end{pmatrix} \begin{pmatrix} \mathbf{x}_\alpha \\ \mathbf{y}_\alpha \end{pmatrix} \mathbf{S}_\alpha, \tag{2.5.16}$$

qui est l'équation aux valeurs et vecteurs propres ($\mathbf{S}_\alpha$ n'est pas diagonale mais contient sur sa diagonale les valeurs propres de la matrice RPA $\mathbb{A}_\alpha$). De plus on peut réécrire les équations de plasmon (2.5.10) en fonction de $\mathbf{T} \doteq \mathbf{T}_1$ (de nouveau, il faut se rappeler que toutes ces matrices peuvent être écrites dans un cadre dRPA ou RPAx) :

$$\begin{aligned} E_c^{\text{dRPA-I-Riccati}} &= \frac{1}{2}\,\mathrm{tr}\left\{\mathbf{B}^{\text{dRPA}}\mathbf{T}^{\text{dRPA}}\right\} \\ E_c^{\text{RPAx-II-Riccati}} &= \frac{1}{4}\,\mathrm{tr}\left\{\mathbf{B}^{\text{RPAx}}\mathbf{T}^{\text{RPAx}}\right\}, \end{aligned} \tag{2.5.17}$$

en considérant la première ligne de l'équation matricielle (2.5.12), ou l'équation (2.5.14).

On retrouve avec cette formulation (énergies RPA calculées avec des amplitudes déterminées par des équations de Riccati) des équations qui peuvent être dérivées en théorie *Coupled Cluster*. La version *Coupled Cluster Double* (CCD) met en effet en jeu des équations qui ressemblent beaucoup aux équations de Riccati, et il a été montré que la RPA est en fait complètement équivalente à une approximation "ring" de ces équations *Coupled Cluster Double* (CCD). Plus précisément, la version dRPA-I correspond à l'approximation dite drCCD (direct-ring CCD) et la version RPAx-II correspond à l'approximation rCCD (ring CCD). À ce jour, toutes les formes alternatives de RPA concevables dans le cadre rCCD n'ont pas d'analogues facilement reconnaissables dans la formulation "connexion adiabatique". D'ailleurs, une correspondance *stricte* n'existe que pour les variantes dRPA-I et RPAx-II[119–121].

Pour ce qui est de la résolution des équations de Riccati (2.5.13) pour $\alpha = 1$, on sépare les parties diagonales et hors-diagonales de la matrice $\mathbf{A}'$, pour écrire :

$$\left(\boldsymbol{\varepsilon} + \mathbf{A}'_{\text{diag}} +\right)\mathbf{T} + \mathbf{T}\left(\boldsymbol{\varepsilon} + \mathbf{A}'_{\text{diag}}\right) = -\left(\mathbf{B} + \mathbf{A}'_{\text{hors-diag}}\mathbf{T} + \mathbf{T}\mathbf{A}'_{\text{hors-diag}} + \mathbf{TBT}\right), \tag{2.5.18}$$

que l'on résout par une procédure itérative :

$$T_{ia,jb}^{(n)} = -\frac{\left(\mathbf{B} + \mathbf{A}'_{\text{hors-diag}}\mathbf{T}^{(n-1)} + \mathbf{T}^{(n-1)}\mathbf{A}'_{\text{hors-diag}} + \mathbf{T}^{(n-1)}\mathbf{B}\mathbf{T}^{(n-1)}\right)_{ia,jb}}{\left(\boldsymbol{\varepsilon} + \mathbf{A}'_{\text{diag}}\right)_{ia,ia} + \left(\boldsymbol{\varepsilon} + \mathbf{A}'_{\text{diag}}\right)_{jb,jb}} \tag{2.5.19}$$

L'expérience montre que le meilleur choix pour l'itération 0 de l'amplitude est le suivant :

$$T_{ia,jb}^{(0)} = -\frac{B_{ia,jb}}{\left(\boldsymbol{\varepsilon} + \mathbf{A}'_{\text{diag}}\right)_{ia,ia} + \left(\boldsymbol{\varepsilon} + \mathbf{A}'_{\text{diag}}\right)_{jb,jb}}, \tag{2.5.20}$$

qui fait penser à une partition Epstein-Nesbet en théorie des perturbations. On peut aussi tout à fait écrire une itération 0 qui ressemble plus à une partition de type Møller-Plesset :

$$T^{(0)}_{ia,jb} = -\frac{B_{ia,jb}}{\varepsilon_{ia,ia} + \varepsilon_{jb,jb}}, \tag{2.5.21}$$

qui nous ramène à l'analyse perturbative que l'on a réalisée section 2.4.6 et qui a montré qu'à l'ordre deux, les versions de RPA correspondent toutes à une version de MP2 (éventuellement direct-MP2).

La partition de type Epstein-Nesbet conduit en général de manière plus sûre vers une solution stabilisante des équations de Riccati.

## 2.6 Adaptation de spin

Toutes les matrices montrées jusqu'à présent sont écrites en spin-orbitales, mais les implémentations sont faites en orbitales spatiales, c'est-à-dire après une adaptation de spin[2]. Cette étape ne comporte cependant pas de difficulté particulière : toute matrice **X** utilisée dans les dérivations précédentes peut s'écrire :

$$\mathbf{X} = \begin{pmatrix} \mathbf{X}_{\uparrow\uparrow,\uparrow\uparrow} & \mathbf{X}_{\uparrow\uparrow,\downarrow\downarrow} & \mathbf{0} & \mathbf{0} \\ \mathbf{X}_{\downarrow\downarrow,\uparrow\uparrow} & \mathbf{X}_{\downarrow\downarrow,\downarrow\downarrow} & \mathbf{0} & \mathbf{0} \\ \mathbf{0} & \mathbf{0} & \mathbf{X}_{\uparrow\downarrow,\uparrow\downarrow} & \mathbf{X}_{\uparrow\downarrow,\downarrow\uparrow} \\ \mathbf{0} & \mathbf{0} & \mathbf{X}_{\downarrow\uparrow,\uparrow\downarrow} & \mathbf{X}_{\downarrow\uparrow,\downarrow\uparrow} \end{pmatrix} \tag{2.6.1}$$

(ceci est une conséquence du fait que les intégrales bi-électroniques ne peuvent être non nulle que pour des paires de spin identiques, voir Annexe C.1). Le premier bloc est appelé *no-spinflip block* (il représente des excitations qui ne changent pas le spin des électrons), le deuxième est appelé *spinflip block* (il contient des excitations qui changent le spin des électrons). La transformation :

$$\mathbf{U} = \frac{1}{\sqrt{2}} \begin{pmatrix} \mathbf{1} & \mathbf{1} & \mathbf{0} & \mathbf{0} \\ \mathbf{1} & -\mathbf{1} & \mathbf{0} & \mathbf{0} \\ \mathbf{0} & \mathbf{0} & \mathbf{1} & \mathbf{1} \\ \mathbf{0} & \mathbf{0} & \mathbf{1} & -\mathbf{1} \end{pmatrix}, \tag{2.6.2}$$

permet d'écrire des matrices adaptées de spin $\widetilde{\mathbf{X}} = \mathbf{U}^{\dagger}\mathbf{X}\mathbf{U}$. Le lecteur est vivement invité à se convaincre que toutes les matrices construites de cette manière en RPA sont bloc-diagonales (voir Annexe C.1), avec une composante singulet et trois composantes triplets :

$$\widetilde{\mathbf{X}} = \begin{pmatrix} {}^{1}\mathbf{X} & \mathbf{0} & \mathbf{0} & \mathbf{0} \\ \mathbf{0} & {}^{3,0}\mathbf{X} & \mathbf{0} & \mathbf{0} \\ \mathbf{0} & \mathbf{0} & {}^{3,1}\mathbf{X} & \mathbf{0} \\ \mathbf{0} & \mathbf{0} & \mathbf{0} & {}^{3,-1}\mathbf{X} \end{pmatrix} \tag{2.6.3}$$

On retrouve ici des matrices sans *spinflip* (singulet ${}^{1}\mathbf{X}$ et triplet ${}^{3,0}\mathbf{X}$) et les matrices *spinflip* (triplets ${}^{3,1}\mathbf{X}$ et ${}^{3,-1}\mathbf{X}$). Toutes les matrices rencontrées dans les dérivations sont construites à partir des matrices $\boldsymbol{\varepsilon}$, $\mathbf{A}'$ et $\mathbf{B}$ formulées soit dans un cadre dRPA ou RPAx, soit - avec des formules exactement similaires - dans un contexte *simple-barre* ou *double-barre*. L'adaptation de la matrice $\boldsymbol{\varepsilon}$ est triviale, et les adaptations de spin de $\mathbf{A}'$ et $\mathbf{B}$ s'écrivent :

$$\widetilde{\mathbf{A}'}^{\mathrm{d/x}}_{\mathrm{I/II}} = \begin{pmatrix} {}^1\mathbf{A}'^{\mathrm{d/x}}_{\mathrm{I/II}} & \mathbf{0} & \mathbf{0} & \mathbf{0} \\ \mathbf{0} & {}^3\mathbf{A}'^{\mathrm{d/x}}_{\mathrm{I/II}} & \mathbf{0} & \mathbf{0} \\ \mathbf{0} & \mathbf{0} & {}^3\mathbf{A}'^{\mathrm{d/x}}_{\mathrm{I/II}} & \mathbf{0} \\ \mathbf{0} & \mathbf{0} & \mathbf{0} & {}^3\mathbf{A}'^{\mathrm{d/x}}_{\mathrm{I/II}} \end{pmatrix} ; \quad \widetilde{\mathbf{B}}^{\mathrm{d/x}}_{\mathrm{I/II}} = \begin{pmatrix} {}^1\mathbf{B}^{\mathrm{d/x}}_{\mathrm{I/II}} & \mathbf{0} & \mathbf{0} & \mathbf{0} \\ \mathbf{0} & {}^3\mathbf{B}^{\mathrm{d/x}}_{\mathrm{I/II}} & \mathbf{0} & \mathbf{0} \\ \mathbf{0} & \mathbf{0} & {}^3\mathbf{B}^{\mathrm{d/x}}_{\mathrm{I/II}} & \mathbf{0} \\ \mathbf{0} & \mathbf{0} & \mathbf{0} & -\,{}^3\mathbf{B}^{\mathrm{d/x}}_{\mathrm{I/II}} \end{pmatrix} \tag{2.6.4}$$

avec :

$$\begin{aligned} {}^1\mathbf{A}'^{\mathrm{d/x}}_{\mathrm{I/II}} &= 2\mathbf{K} - {}^{\zeta}/_{\xi}\mathbf{J} \\ {}^3\mathbf{A}'^{\mathrm{d/x}}_{\mathrm{I/II}} &= \quad -{}^{\zeta}/_{\xi}\mathbf{J} \\ {}^1\mathbf{B}^{\mathrm{d/x}}_{\mathrm{I/II}} &= 2\mathbf{K} - {}^{\zeta}/_{\xi}\mathbf{K}' \\ {}^3\mathbf{B}^{\mathrm{d/x}}_{\mathrm{I/II}} &= \quad -{}^{\zeta}/_{\xi}\mathbf{K}', \end{aligned} \tag{2.6.5}$$

où les intégrales $\mathbf{K}$, $\mathbf{K}'$ et $\mathbf{J}$ sont à présent évaluées avec des orbitales spatiales. Je veux rappeler ici une nouvelle fois, avec les notations « $\sqcup^{\mathrm{d/x}}_{\mathrm{I/II}}$ », le fait que l'antisymétrie de ces matrices est « allumée » par des « interrupteurs » complètement similaires ; l'interrupteur correspondant aux situations "d/x" est $\zeta$, celui correspondant aux situations "I/II" est $\xi$. Ainsi les matrices sont les mêmes lorsqu'elles sont construites dans des contextes dRPA ($\zeta = 0$) et *simple-barre* ($\xi = 0$), et les mêmes également dans des contextes RPAx ($\zeta = 1$) et *double-barre* ($\xi = 1$). Avec les expressions des matrices $\widetilde{\mathbf{A}'}$ et $\widetilde{\mathbf{B}}$ en main, on peut montrer simplement que toutes les expressions d'énergies comportant des matrices dRPA et/ou des matrices *simple-barre* (c'est-à-dire les expressions de type dRPA-I, dRPA-II et RPAx-I) ne font pas intervenir de contribution triplet ; seules les expressions de type RPAx-II font intervenir des contributions triplets (voir Annexe C.2).

Dans la suite de la thèse, la plupart des dérivations (autour des orbitales localisées, des gradients des énergies RPA, *etc.*.. ) utilisent la formulation avec équation de "Riccati", c'est donc sur cette formulation que l'on se concentre ici. Les propriétés de symétrie des amplitudes $\mathbf{T}$[122] imposent :

$$\widetilde{\mathbf{T}} = \begin{pmatrix} {}^1\mathbf{T} & \mathbf{0} & \mathbf{0} & \mathbf{0} \\ \mathbf{0} & {}^3\mathbf{T} & \mathbf{0} & \mathbf{0} \\ \mathbf{0} & \mathbf{0} & {}^3\mathbf{T} & \mathbf{0} \\ \mathbf{0} & \mathbf{0} & \mathbf{0} & -\,{}^3\mathbf{T} \end{pmatrix} \tag{2.6.6}$$

L'adaptation de spin de l'équation de Riccati dans le cas de figure dRPA-I est donc :

$$\mathbf{0} = {}^1\mathbf{B}^{\mathrm{dRPA}} + {}^1\mathbf{A}^{\mathrm{dRPA}}\,{}^1\mathbf{T} + {}^1\mathbf{T}\,{}^1\mathbf{A}^{\mathrm{dRPA}} + {}^1\mathbf{T}\,{}^1\mathbf{B}^{\mathrm{dRPA}}\,{}^1\mathbf{T} + \boldsymbol{\varepsilon}\,{}^1\mathbf{T} + {}^1\mathbf{T}\boldsymbol{\varepsilon} \tag{2.6.7}$$

Une fois les amplitudes singulets ${}^1\mathbf{T}$ connues, l'énergie est donnée par l'adaptation de spin de la première équation (2.5.17) :

$$E_c^{\text{dRPA-I-Riccati}} = \frac{1}{2}\,\text{tr}\left\{ {}^1\mathbf{B}^{\text{dRPA}}\,{}^1\mathbf{T} \right\}, \tag{2.6.8}$$

que l'on peut compléter en introduisant une antisymétrie *a posteriori*, par :

$$E_c^{\text{dRPA-I-SOSEX}} = \frac{1}{2}\,\text{tr}\left\{ {}^1\mathbf{B}^{\text{RPAx}}\,{}^1\mathbf{T} \right\}, \tag{2.6.9}$$

dans une approximation dite *Second Order Screened EXchange* (SOSEX)[102, 123]. La dénomination provient du fait que le diagramme du terme ajouté pour passer de (2.6.8) à (2.6.9) est le diagramme d'échange au second ordre en théorie des perturbations. On peut noter à ce propos que la version dRPA-II (voir équation (2.4.17b)) est très similaire par construction au SOSEX, mais inclue des ordres supérieurs d'échange écranté. L'approximation de la version dRPA-II, que l'on a nommée dRPA-IIa, est de fait une approximation qui se rapproche du SOSEX (voir (2.4.35), où l'on tronque les ordres de $\mathbf{P}_\alpha$ ; se rappeler que $\mathbf{B}^{\text{RPAx}}$ et $\mathbf{B}^{\text{II}}$ sont les mêmes matrices). Le lien entre la formulation SOSEX dérivée ici par un formalisme avec équation de "Riccati" (c'est-à-dire par un formalisme "CC") et l'approximation dRPA-IIa dérivée dans un contexte AC-FDT a été clarifié dans les références [124] (où la version dRPA-IIa est d'ailleurs notée AC-SOSEX) et [125]. La variante dRPA-II dérivée dans le contexte "connexion adiabatique" peut être légitimement appeler *Screened Exchange* (SX) pour signifier le fait que le rôle de l'échange est plutôt passif, et il est écranté par la fonction réponse du système.

Dans ce contexte, Hesselmann a récemment proposé une version qu'il nomme RPAX2[126], et qui consiste à introduire l'échange dans les équations des Riccati en écrivant :

$$\mathbf{0} = {}^1\mathbf{B}^{\text{RPAx}} + {}^1\mathbf{B}^{\text{RPAx}}\,{}^1\mathbf{T} + {}^1\mathbf{T}\,{}^1\mathbf{B}^{\text{RPAx}} + {}^1\mathbf{T}\,{}^1\mathbf{B}^{\text{RPAx}}\,{}^1\mathbf{T} + \boldsymbol{\varepsilon}\,{}^1\mathbf{T} + {}^1\mathbf{T}\boldsymbol{\varepsilon}, \tag{2.6.10}$$

avec l'expression de l'énergie :

$$E_c^{\text{RPAX2}} = \frac{1}{2}\,\text{tr}\left\{ {}^1\mathbf{B}^{\text{dRPA}}\,{}^1\mathbf{T} \right\} \tag{2.6.11}$$

On retrouvera cette version de RPA dans les dérivations du formalisme "matrice diélectrique".

L'adaptation de spin du cas de figure RPAx-II oblige à résoudre une équation de Riccati pour ${}^1\mathbf{T}$ *et* pour ${}^3\mathbf{T}$ :

$$\begin{aligned} \mathbf{0} &= {}^1\mathbf{B}^{\text{RPAx}} + {}^1\mathbf{A}'^{\text{RPAx}}\,{}^1\mathbf{T} + {}^1\mathbf{T}\,{}^1\mathbf{A}'^{\text{RPAx}} + {}^1\mathbf{T}\,{}^1\mathbf{B}^{\text{RPAx}}\,{}^1\mathbf{T} + \boldsymbol{\varepsilon}\,{}^1\mathbf{T} + {}^1\mathbf{T}\boldsymbol{\varepsilon} \\ \mathbf{0} &= {}^3\mathbf{B}^{\text{RPAx}} + {}^3\mathbf{A}'^{\text{RPAx}}\,{}^3\mathbf{T} + {}^3\mathbf{T}\,{}^3\mathbf{A}'^{\text{RPAx}} + {}^3\mathbf{T}\,{}^3\mathbf{B}^{\text{RPAx}}\,{}^3\mathbf{T} + \boldsymbol{\varepsilon}\,{}^3\mathbf{T} + {}^3\mathbf{T}\boldsymbol{\varepsilon} \end{aligned} \tag{2.6.12}$$

L'énergie est alors donnée par l'adaptation de spin de la deuxième équation (2.5.17) :

$$E_c^{\text{RPAx-II-Riccati}} = \frac{1}{4}\,\text{tr}\left\{ {}^1\mathbf{B}^{\text{RPAx}}\,{}^1\mathbf{T} + 3\,{}^3\mathbf{B}^{\text{RPAx}}\,{}^3\mathbf{T} \right\} \tag{2.6.13}$$

L'énergie peut également être obtenue de manière théoriquement équivalente, mais en pratique différente (tout comme les équations (2.4.17c) et (2.4.17d) sont en théorie équivalentes mais sont distinctes puisque $\mathbb{P}_{c,\alpha}$ est approximée) par :

$$E_c^{\text{RPAx-II-SO2}} = \frac{1}{2}\,\text{tr}\left\{ {}^1\mathbf{B}^{\text{dRPA}}\ {}^1\mathbf{T}\right\}, \tag{2.6.14}$$

et, pour des raisons moins directes, approximée par :

$$E_c^{\text{RPAx-II-SO1}} = \frac{1}{2}\,\text{tr}\left\{ {}^1\mathbf{B}^{\text{RPAx}}\left( {}^1\mathbf{T} - {}^3\mathbf{T}\right)\right\} \tag{2.6.15}$$

Ces deux dernières versions des énergies RPA sont labellisées "SO1" et "SO2" car elles ont été originellement proposées par Szabo-Ostlund [127, 128] comme approximations d'ordre zéro d'une procédure RPA auto-cohérente[129, 130].

## 2.7 Formulation "matrice diélectrique"

[NOTE: **Les dérivations qui suivent on depuis été conduites plus élégamment dans un travail en cours de publication, où il est fait usage de fonctions de matrices.**]

Dans la formulation "matrice diélectrique", l'énergie AC-FDT (2.3.5) est intégrée analytiquement le long de la coordonnée de constante de couplage $\alpha$. On reprend les équations (2.4.5) et (2.4.6) pour écrire l'énergie de corrélation RPA :

$$E_c^{\text{AC-FDT}} = -\frac{1}{2}\int_0^1 d\alpha \int_{-\infty}^{\infty} \frac{d\omega}{2\pi i}\,\text{Tr}\left\{\mathbb{\Pi}_\alpha(\omega)^{\text{d/x}}\mathbb{W}^{\text{I/II}} - \mathbb{\Pi}_0(\omega)\mathbb{W}^{\text{I/II}}\right\}, \tag{2.7.1}$$

où il est pratique ici d'exprimer $\mathbb{\Pi}_0$ comme :

$$\mathbb{\Pi}_0(z) = \begin{pmatrix} -(\boldsymbol{\varepsilon} - z\mathbf{1})^{-1} & \mathbf{0} \\ \mathbf{0} & -(\boldsymbol{\varepsilon} + z\mathbf{1})^{-1} \end{pmatrix} = \begin{pmatrix} \mathbf{\Pi}_0^+(z) & \mathbf{0} \\ \mathbf{0} & \mathbf{\Pi}_0^-(z) \end{pmatrix}, \tag{2.7.2}$$

et où $\mathbb{\Pi}_\alpha^{\text{d/x}}(z)$ obéit à l'équation de Bethe-Salpeter :

$$\mathbb{\Pi}_\alpha^{\text{d/x}}(z)^{-1} = \mathbb{\Pi}_0^{-1}(z) - \alpha\mathbb{W}^{\text{d/x}} \quad\Leftrightarrow\quad \mathbb{\Pi}_\alpha^{\text{d/x}}(z) = \left(\mathbb{1} - \alpha\mathbb{\Pi}_0(z)\mathbb{W}^{\text{d/x}}\right)^{-1}\mathbb{\Pi}_0(z) \tag{2.7.3}$$

L'expansion en série de Taylor de la deuxième formulation de l'équation de Bethe-Salpeter donne une expression de l'énergie :

$$E_c^{\text{AC-FDT}} = -\frac{1}{2}\int_0^1 d\alpha \int_{-\infty}^{\infty} \frac{d\omega}{2\pi}\sum_{n=2}^{\infty}\alpha^{n-1}\,\text{Tr}\left\{\left(\mathbb{\Pi}_0(i\omega)\mathbb{W}^{\text{d/x}}\right)^{n-1}\mathbb{\Pi}_0(i\omega)\mathbb{W}^{\text{I/II}}\right\} \tag{2.7.4}$$

Comme dans les sections précédentes, cette formule est générale et peut en théorie correspondre à tous les scénarios (d/x) × (I/II). On cherche, comme précédemment, à réduire les dimensions du problème, c'est-à-dire à passer d'une trace « $\text{Tr}\{\sqcup\}$ » sur l'espace $\mathcal{L}^{occ}\otimes\mathcal{L}^{vir}\oplus\mathcal{L}^{vir}\otimes\mathcal{L}^{occ}$ à une trace « $\text{tr}\{\sqcup\}$ » sur l'espace $\mathcal{L}^{occ}\otimes\mathcal{L}^{vir}$. On ne peut pas ici continuer la dérivation générale de manière très avancée. Il nous faut considérer au cas par cas les versions (d/x) × (I/II).

### 2.7.1 dRPA-I

Dans le cas dRPA-I, on a : $\mathbb{W}^{\mathrm{d}} = \mathbb{W}^{\mathrm{I}} = \begin{pmatrix} \mathbf{K} & \mathbf{K} \\ \mathbf{K} & \mathbf{K} \end{pmatrix}$, ce qui simplifie considérablement la dérivation. On remarque notamment que :

$$\mathrm{Tr}\left\{(\mathbb{\Pi}_0(i\omega)\mathbb{W}^{\mathrm{d}})^n\right\} = \mathrm{Tr}\left\{\begin{pmatrix} \mathbf{\Pi}_0^+(i\omega)\mathbf{K} & \mathbf{\Pi}_0^+(i\omega)\mathbf{K} \\ \mathbf{\Pi}_0^-(i\omega)\mathbf{K} & \mathbf{\Pi}_0^-(i\omega)\mathbf{K} \end{pmatrix}^n\right\} = \mathrm{tr}\left\{(\mathbf{\Pi}_0(i\omega)\mathbf{K})^n\right\}, \tag{2.7.5}$$

où $\mathbf{\Pi}_0(i\omega) = \mathbf{\Pi}_0^+(i\omega) + \mathbf{\Pi}_0^-(i\omega)$. Ainsi l'équation (2.7.4) est-elle intégrée analytiquement le long de la coordonnée $\alpha$, pour donner (après avoir reconstitué l'expansion de Taylor d'un logarithme) :

$$E_c^{\mathrm{dRPA\text{-}I}} = \frac{1}{2}\int_{-\infty}^{\infty}\frac{d\omega}{2\pi}\,\mathrm{tr}\left\{\log(\mathbf{1} - \mathbf{\Pi}_0(i\omega)\mathbf{K}) + \mathbf{\Pi}_0(i\omega)\mathbf{K}\right\}, \tag{2.7.6}$$

où l'on reconnaît en effet la représentation matricielle de la matrice diélectrique $\boldsymbol{\varepsilon}(i\omega) = \mathbf{1} - \mathbf{\Pi}_0(i\omega)\mathbf{K}$, justifiant le nom donné à cette formulation.

### 2.7.2 dRPA-II

Le cas dRPA-II est déjà bien plus compliqué : la structure de bloc de $\mathbb{W}^{\mathrm{II}}$ ne permet pas une réduction de la dimension aussi simple que précédemment. On considère l'expression suivante pour $\mathbb{W}^{\mathrm{II}}$, séparée en une matrice principale et une correction :

$$\mathbb{W}^{\mathrm{II}} = \begin{pmatrix} \mathbf{A}' & \mathbf{B} \\ \mathbf{B} & \mathbf{A}' \end{pmatrix} = \begin{pmatrix} \mathbf{B} & \mathbf{B} \\ \mathbf{B} & \mathbf{B} \end{pmatrix} + \begin{pmatrix} \mathbf{A}' - \mathbf{B} & \mathbf{0} \\ \mathbf{0} & \mathbf{A}' - \mathbf{B} \end{pmatrix}, \tag{2.7.7}$$

de sorte que la trace de l'équation (2.7.4) s'écrit :

$$\mathrm{Tr}\left\{\begin{pmatrix} \mathbf{\Pi}_0^+\mathbf{K} & \mathbf{\Pi}_0^+\mathbf{K} \\ \mathbf{\Pi}_0^-\mathbf{K} & \mathbf{\Pi}_0^-\mathbf{K} \end{pmatrix}^{n-1}\begin{pmatrix} \mathbf{\Pi}_0^+\mathbf{B} & \mathbf{\Pi}_0^+\mathbf{B} \\ \mathbf{\Pi}_0^-\mathbf{B} & \mathbf{\Pi}_0^-\mathbf{B} \end{pmatrix} + \begin{pmatrix} \mathbf{\Pi}_0^+\mathbf{K} & \mathbf{\Pi}_0^+\mathbf{K} \\ \mathbf{\Pi}_0^-\mathbf{K} & \mathbf{\Pi}_0^-\mathbf{K} \end{pmatrix}^{n-1}\begin{pmatrix} \mathbf{\Pi}_0^+(\mathbf{A}' - \mathbf{B}) & \mathbf{0} \\ \mathbf{0} & \mathbf{\Pi}_0^-(\mathbf{A}' - \mathbf{B}) \end{pmatrix}\right\} \tag{2.7.8}$$

Cette expression se réduit à un terme dominant $\mathrm{tr}\left\{(\mathbf{\Pi}_0\mathbf{K})^{n-1}\mathbf{\Pi}_0\mathbf{B}\right\}$ suivi d'une correction de la forme :

$$\mathrm{tr}\left\{\left(\mathbf{\Pi}_0^+\mathbf{K}(\mathbf{\Pi}_0\mathbf{K})^{n-2}\mathbf{\Pi}_0^+ + \mathbf{\Pi}_0^-\mathbf{K}(\mathbf{\Pi}_0\mathbf{K})^{n-2}\mathbf{\Pi}_0^-\right)(\mathbf{A}' - \mathbf{B})\right\} \tag{2.7.9}$$

On peut montrer que le terme d'ordre 2, $n = 2$, de la correction ne contribue pas à l'énergie de corrélation. En effet, les deux éléments de l'intégrande, impliquant l'un $\mathbf{\Pi}_0^-$ et l'autre $\mathbf{\Pi}_0^+$, produisent la même intégrale - on le voit par le simple changement de variable $\omega \leftarrow -\omega$. Cette intégrale est nulle par vertu du théorème de Cauchy : les pôles, par exemple de $\mathbf{\Pi}_0^+\mathbf{K}\mathbf{\Pi}_0^+$, sont tous situés dans le même plan complexe, permettant ainsi de fermer un contour dans l'autre (voir l'Annexe B pour plus de détails concernant l'intégration complexe). On suppose de plus que les corrections des termes d'ordre supérieur, $n > 2$, sont négligeables. On trouve donc l'équation :

$$E_c^{\text{dRPA-IIa}} = -\frac{1}{2}\int_0^1 d\alpha \int_{-\infty}^{\infty} \frac{d\omega}{2\pi} \sum_{n=2}^{\infty} \alpha^{n-1} \text{ tr}\left\{(\mathbf{\Pi}_0(i\omega)\mathbf{K})^{n-1}\,\mathbf{\Pi}_0(i\omega)\mathbf{B}\right\} \tag{2.7.10}$$

On note qu'une intégration sur $\omega$ nous permet de retrouver l'équation (2.4.35) de l'approximation dRPA-IIa, dérivée dans le cadre du formalisme "connexion adiabatique". Ici, une intégration sur $\alpha$ donne (après reconstitution de l'expansion de Taylor) :

$$E_c^{\text{dRPA-IIa}} = \frac{1}{2}\int_{-\infty}^{\infty} \frac{d\omega}{2\pi} \text{ tr}\left\{\log\left(\mathbf{1} - \mathbf{\Pi}_0(i\omega)\mathbf{K}\right)\mathbf{K}^{-1}\mathbf{B} + \mathbf{\Pi}_0(i\omega)\mathbf{B}\right\} \tag{2.7.11}$$

Comme on a vu dans la section 2.6, l'approximation dRPA-IIa est très proche de l'approximation SOSEX dérivée à partir de la formulation direct-ring-CCD. Le caractère « second ordre » de SOSEX est rendu très clair dans cette dérivation, où l'on néglige les termes correctifs d'ordres supérieurs à 2 de l'expansion (2.7.8).

### 2.7.3 RPAx-Ia

De manière très similaire au cas dRPA-II, on exprime la matrice $\mathbb{W}^{\text{x}}$ qui apparaît dans la formule de l'énergie RPAx-I comme :

$$\mathbb{W}^{\text{x}} = \begin{pmatrix} \mathbf{A}' & \mathbf{B} \\ \mathbf{B} & \mathbf{A}' \end{pmatrix} = \begin{pmatrix} \mathbf{B} & \mathbf{B} \\ \mathbf{B} & \mathbf{B} \end{pmatrix} + \begin{pmatrix} \mathbf{A}' - \mathbf{B} & \mathbf{0} \\ \mathbf{0} & \mathbf{A}' - \mathbf{B} \end{pmatrix}, \tag{2.7.12}$$

de sorte que, dans l'équation (2.7.4), la matrice élevée à la puissance $n-1$, $(\mathbb{\Pi}_0(i\omega)\mathbb{W}^{\text{x}})^{n-1}$, s'écrit à présent, au premier ordre de la matrice de correction :

$$\begin{aligned}(\mathbb{\Pi}_0(i\omega)\mathbb{W}^{\text{x}})^{n-1} \approx & \begin{pmatrix} \mathbf{\Pi}_0^+\mathbf{B} & \mathbf{\Pi}_0^+\mathbf{B} \\ \mathbf{\Pi}_0^-\mathbf{B} & \mathbf{\Pi}_0^-\mathbf{B} \end{pmatrix}^{n-1} \\ & + \sum_{p=1}^{n-1} \begin{pmatrix} \mathbf{\Pi}_0^+\mathbf{B} & \mathbf{\Pi}_0^+\mathbf{B} \\ \mathbf{\Pi}_0^-\mathbf{B} & \mathbf{\Pi}_0^-\mathbf{B} \end{pmatrix}^{n-1-p} \begin{pmatrix} \mathbf{\Pi}_0^+(\mathbf{A}'-\mathbf{B}) & \mathbf{0} \\ \mathbf{0} & \mathbf{\Pi}_0^-(\mathbf{A}'-\mathbf{B}) \end{pmatrix} \begin{pmatrix} \mathbf{\Pi}_0^+\mathbf{B} & \mathbf{\Pi}_0^+\mathbf{B} \\ \mathbf{\Pi}_0^-\mathbf{B} & \mathbf{\Pi}_0^-\mathbf{B} \end{pmatrix}^{p-1},\end{aligned} \tag{2.7.13}$$

où l'approximation la plus simple de l'expression RPAx-I (négliger toute contribution provenant de la matrice corrective) donne :

$$E_c^{\text{RPAx-Ia}} = -\frac{1}{2}\int_0^1 d\alpha \int_{-\infty}^{\infty} \frac{d\omega}{2\pi} \sum_{n=2}^{\infty} \alpha^{n-1} \operatorname{tr}\left\{(\mathbf{\Pi}_0(i\omega)\mathbf{B})^{n-1}\,\mathbf{\Pi}_0(i\omega)\mathbf{K}\right\}, \qquad (2.7.14)$$

dont l'intégration le long de la coordonnée $\alpha$ fournit (après reconstitution d'un logarithme) :

$$E_c^{\text{RPAx-Ia}} = \frac{1}{2}\int_{-\infty}^{\infty} \frac{d\omega}{2\pi} \operatorname{tr}\left\{\log\left(\mathbf{1} - \mathbf{\Pi}_0(i\omega)\mathbf{B}\right)\mathbf{B}^{-1}\mathbf{K} + \mathbf{\Pi}_0(i\omega)\mathbf{K}\right\} \qquad (2.7.15)$$

Notons que cette approximation peut être considérée comme un analogue dans un formalisme AC (connexion adiabatique) de la méthode RPAX2 de Hesselmann[126], qui est basée sur des considérations plutôt d'ordre pratiques, dans le cadre rCCD. Le bien-fondé de cette méthode reste tout de même une question ouverte, car il est difficile d'associer à cette approche une fonction de réponse bien définie.

En travaillant l'expression dans le formalisme "connexion adiabatique" de cette variante, il devient clair que la réponse est traitée à un niveau "RPAx", c'est-à-dire que l'on peut écrire $\mathbf{P}_{c,\alpha}^{\text{RPAX}} = \boldsymbol{\varepsilon}^{1/2}(\mathbf{M}_\alpha^{\text{RPAX}})^{-1/2}\boldsymbol{\varepsilon}^{1/2} - \mathbf{I}$, avec $\mathbf{M}_\alpha^{\text{RPAX}} = \boldsymbol{\varepsilon}^{1/2}(\boldsymbol{\varepsilon} + 2\alpha\mathbf{B})\boldsymbol{\varepsilon}^{1/2}$. Toutefois, la signification physique de la matrice $\mathbf{M}_\alpha^{\text{RPAX}}$ reste à comprendre dans un contexte réponse linéaire.

### 2.7.4 Retrouver la formulation de plasmon

À partir de l'expression dRPA-I, on peut retrouver la formule de plasmon correspondante, originellement dérivée (voir équation (2.5.10)) par une intégration analytique (1) sur la fréquence $\omega$ suivie d'une intégration analytique sur (2) la constante de couplage $\alpha$, en faisant cette fois tout d'abord une intégration analytique (1) le long de la coordonnée $\alpha$ puis une intégration analytique sur (2) la fréquence $\omega$ (c'est-à-dire une intégration analytique de l'équation (2.7.6) sur la fréquence $\omega$). Reprenons l'équation (2.7.4) avec les expressions correctes pour le cas dRPA-I :

$$E_c^{\text{dRPA-I}} = -\frac{1}{2}\int_0^1 d\alpha \int_{-\infty}^{\infty} \frac{d\omega}{2\pi} \sum_{n=2}^{\infty} \alpha^{n-1} \operatorname{tr}\left\{(\mathbf{\Pi}_0(i\omega)\mathbf{K})^n\right\} \qquad (2.7.16)$$

L'expansion en série de Taylor peut être reconstituée pour obtenir :

$$E_c^{\text{dRPA-I}} = -\frac{1}{2}\int_0^1 d\alpha \int_{-\infty}^{\infty} \frac{d\omega}{2\pi} \operatorname{tr}\left\{(\mathbf{1} - \alpha\mathbf{\Pi}_0(i\omega)\mathbf{K})^{-1}\mathbf{\Pi}_0(i\omega)\mathbf{K} - \mathbf{\Pi}_0(i\omega)\mathbf{K}\right\}, \qquad (2.7.17)$$

où l'on peut définir $\mathbf{\Pi}_\alpha(z)$, qui obéit à l'équation de Bethe-Salpeter de plus petite dimension :

$$\mathbf{\Pi}_\alpha(z)^{-1} = \mathbf{\Pi}_0^{-1}(z) - \alpha\mathbf{K} \quad \Leftrightarrow \quad \mathbf{\Pi}_\alpha(z) = (\mathbf{1} - \alpha\mathbf{\Pi}_0(z)\mathbf{K})^{-1}\,\mathbf{\Pi}_0(z) \qquad (2.7.18)$$

Muni de cette équation de Bethe-Salpeter, on voit que l'intégration sur la constante de couplage $\alpha$ de l'équation (2.7.16) peut s'écrire (après reconstitution de l'expansion de Taylor) :

$$E_c^{\text{dRPA-I}} = \frac{1}{2}\int_{-\infty}^{\infty}\frac{d\omega}{2\pi}\,\text{tr}\left\{\log\left(\mathbf{\Pi}_0(i\omega)\mathbf{\Pi}_1^{-1}(i\omega)\right) + \mathbf{\Pi}_0(i\omega)\mathbf{K}\right\} \tag{2.7.19}$$

Cette équation réclame que l'on travaille sur les termes $\text{tr}\left\{\log\left(\mathbf{\Pi}_0(i\omega)\mathbf{\Pi}_1^{-1}(i\omega)\right)\right\}$ et $\text{tr}\left\{\mathbf{\Pi}_0(i\omega)\mathbf{K}\right\}$. On donne les expressions explicites de $\mathbf{\Pi}_0(i\omega)$ et $\mathbf{\Pi}_\alpha(i\omega)^{-1}$ suivantes :

$$\mathbf{\Pi}_0(i\omega) = \mathbf{\Pi}_0^+(i\omega) + \mathbf{\Pi}_0^-(i\omega) = \frac{-1}{\boldsymbol{\varepsilon} - i\omega\mathbf{1}} + \frac{-1}{\boldsymbol{\varepsilon} + i\omega\mathbf{1}} = \frac{-2\boldsymbol{\varepsilon}}{\boldsymbol{\varepsilon}^2 + \omega^2\mathbf{1}} = -2\boldsymbol{\varepsilon}^{\frac{1}{2}}\left(\boldsymbol{\varepsilon}^2 + \omega^2\mathbf{1}\right)^{-1}\boldsymbol{\varepsilon}^{\frac{1}{2}} \tag{2.7.20}$$

$$\mathbf{\Pi}_\alpha(i\omega)^{-1} = -\frac{1}{2}\boldsymbol{\varepsilon}^{-\frac{1}{2}}\left(\boldsymbol{\varepsilon}^2 + \omega^2\mathbf{1} - 2\alpha\boldsymbol{\varepsilon}^{\frac{1}{2}}\mathbf{K}\boldsymbol{\varepsilon}^{\frac{1}{2}}\right)\boldsymbol{\varepsilon}^{-\frac{1}{2}} = -\frac{1}{2}\boldsymbol{\varepsilon}^{-\frac{1}{2}}\left(\mathbf{M}_\alpha + \omega^2\mathbf{1}\right)\boldsymbol{\varepsilon}^{-\frac{1}{2}} \tag{2.7.21}$$

avec :

$$\mathbf{M}_\alpha = \boldsymbol{\varepsilon}^{\frac{1}{2}}\left(\boldsymbol{\varepsilon} + 2\alpha\mathbf{K}\right)\boldsymbol{\varepsilon}^{\frac{1}{2}}, \tag{2.7.22}$$

où l'on retrouve la matrice $\mathbf{M}_\alpha$ rencontrée dans la partie concernant le formalisme "connexion adiabatique" (voir l'équation (2.4.25) adaptée au cas dRPA) dont l'expression en terme de valeurs et vecteurs propres est : $\mathbf{M}_\alpha\mathbf{Z}_\alpha = \mathbf{Z}_\alpha\mathbf{\Omega}_\alpha^2$. Le deuxième terme que l'on doit clarifier donne aisément :

$$\text{tr}\left\{\mathbf{\Pi}_0(i\omega)\mathbf{K}\right\} = \text{tr}\left\{\frac{-2\boldsymbol{\varepsilon}^{\frac{1}{2}}\mathbf{K}\boldsymbol{\varepsilon}^{\frac{1}{2}}}{\boldsymbol{\varepsilon}^2 + \omega^2\mathbf{1}}\right\} = -\,\text{tr}\left\{\left(\boldsymbol{\varepsilon}^2 + \omega^2\mathbf{1}\right)^{-1}\left(\mathbf{M}_1 - \boldsymbol{\varepsilon}^2\right)\right\} \tag{2.7.23}$$

Le premier terme est plus compliqué à mettre en place. On a (voir équations (2.7.20) et (2.7.21)) :

$$\begin{aligned}\text{tr}\left\{\log\left(\mathbf{\Pi}_0(i\omega)\mathbf{\Pi}_1^{-1}(i\omega)\right)\right\} &= \text{tr}\left\{\log\left(\boldsymbol{\varepsilon}^{\frac{1}{2}}\left(\boldsymbol{\varepsilon}^2 + \omega^2\right)^{-1}\left(\mathbf{M}_1 + \omega^2\right)\boldsymbol{\varepsilon}^{-\frac{1}{2}}\right)\right\} \\ &= \text{tr}\left\{\log\left(1 + \boldsymbol{\varepsilon}^{\frac{1}{2}}\left(\boldsymbol{\varepsilon}^2 + \omega^2\right)^{-1}\left(\mathbf{M}_1 - \boldsymbol{\varepsilon}^2\right)\boldsymbol{\varepsilon}^{-\frac{1}{2}}\right)\right\},\end{aligned} \tag{2.7.24}$$

où la réécriture sous la forme $\log(1+z)$ ne sert qu'à permettre la procédure suivante : (1) expansion en série de Taylor du logarithme, (2) permutation circulaire des traces pour éliminer les $\boldsymbol{\varepsilon}^{\pm\frac{1}{2}}$, (3) recomposition de la série de Taylor pour obtenir :

$$\begin{aligned}\text{tr}\left\{\log\left(\mathbf{\Pi}_0(i\omega)\mathbf{\Pi}_1^{-1}(i\omega)\right)\right\} &= \text{tr}\left\{\log\left(1 + \left(\boldsymbol{\varepsilon}^2 + \omega^2\right)^{-1}\left(\mathbf{M}_1 - \boldsymbol{\varepsilon}^2\right)\right)\right\} \\ &= \text{tr}\left\{\log\left(\left(\boldsymbol{\varepsilon}^2 + \omega^2\right)^{-1}\left(\mathbf{M}_1 + \omega^2\right)\right)\right\}\end{aligned} \tag{2.7.25}$$

On se sert de l'expression de $\mathbf{M}_1$ en terme de valeurs et vecteurs propres : $\mathbf{M}_1\mathbf{Z} = \mathbf{Z}\mathbf{\Omega}^2$, avec $\mathbf{Z}\mathbf{Z}^\dagger = \mathbf{1}$, pour écrire :

$$\text{tr}\left\{\log\left(\mathbf{\Pi}_0(i\omega)\mathbf{\Pi}_1^{-1}(i\omega)\right)\right\} = \text{tr}\left\{\log\left(\left(\boldsymbol{\varepsilon}^2 + \omega^2\right)^{-1}\mathbf{Z}\left(\mathbf{\Omega}^2 + \omega^2\right)\mathbf{Z}^\dagger\right)\right\} \tag{2.7.26}$$

On utilise ensuite la propriété : $\text{tr}\{\log(\mathbf{X})\} = \log(\det(\mathbf{X}))$ pour éliminer les $\mathbf{Z}$ et finalement obtenir :

$$\begin{aligned}
\text{tr}\left\{\log\left(\mathbf{\Pi}_0(i\omega)\mathbf{\Pi}_1^{-1}(i\omega)\right)\right\} &= \log\left(\det\left(\left(\boldsymbol{\varepsilon}^2+\omega^2\right)^{-1}\mathbf{Z}\left(\mathbf{\Omega}^2+\omega^2\right)\mathbf{Z}^\dagger\right)\right) \\
&= \log\left(\det\left(\left(\boldsymbol{\varepsilon}^2+\omega^2\right)^{-1}\left(\mathbf{\Omega}^2+\omega^2\right)\right)\right) \\
&= \text{tr}\left\{\log\left(\left(\boldsymbol{\varepsilon}^2+\omega^2\right)^{-1}\left(\mathbf{\Omega}^2+\omega^2\right)\right)\right\}
\end{aligned} \tag{2.7.27}$$

En définitive, cela nous permet d'écrire l'équation (2.7.19) :

$$E_c^{\text{dRPA-I}} = \frac{1}{2}\int_{-\infty}^{\infty}\frac{d\omega}{2\pi}\sum_{ia}\left\{\log\left(\frac{\Omega_{ia}^2+\omega^2}{\varepsilon_{ia}^2+\omega^2}\right)-\frac{M_{1,ia,ia}-\varepsilon_{ia}^2}{\varepsilon_{ia}^2+\omega^2}\right\}, \tag{2.7.28}$$

que l'on peut intégrer analytiquement en utilisant le principe de l'argument (voir la section B.1.5 du chapitre sur l'intégration complexe en Annexe, où ce théorème est démontré, ainsi que la section B.3 où cette intégration analytique est expliquée) :

$$E_c^{\text{dRPA-I}} = \frac{1}{2}\sum_{ia}\Omega_{ia}-(\varepsilon_{ia}+K_{ia,ia}) \tag{2.7.29}$$

On retrouve donc bien l'exacte même expression de la formulation de plasmon que celle dérivée à partir d'une formulation "connexion adiabatique".

## 2.8 Hiérarchie des formulations RPA

Il sera sûrement bon ici de résumer les différentes formulations que l'on a dérivées dans ce chapitre. On peut parler d'une hiérarchie des formulations si l'on considère les intégrations analytiques et/ou numériques qui sont effectuées à partir de la formule AC-FDT montrée équation (2.3.5), et que l'on peut prendre ici comme notre point de départ. Ainsi les formulations "connexion adiabatique" et "matrice diélectrique" sont un premier niveau (une intégration est analytique et une quadrature doit être faite), et les formulations avec équation de "Riccati" et de "plasmon" sont un autre niveau (toutes les intégrations sont analytiques). Je montre ceci dans la figure 2.3.

J'ai écrit une routine dans la suite de programme `MOLPRO` [131], dont on a une licence développeur au laboratoire et sur lequel de nombreux développements on été implémentés durant cette thèse, qui rassemble et unifie les différents calculs RPA possibles, qui étaient pour la plupart précédemment implémentés. On récapitule ici les mots-clés, dans cette approche, pour lancer des calculs RPA. La base de la syntaxe est la suivante :

`{rpatddft;ecorr,DRPAI-AC},` (2.8.1)

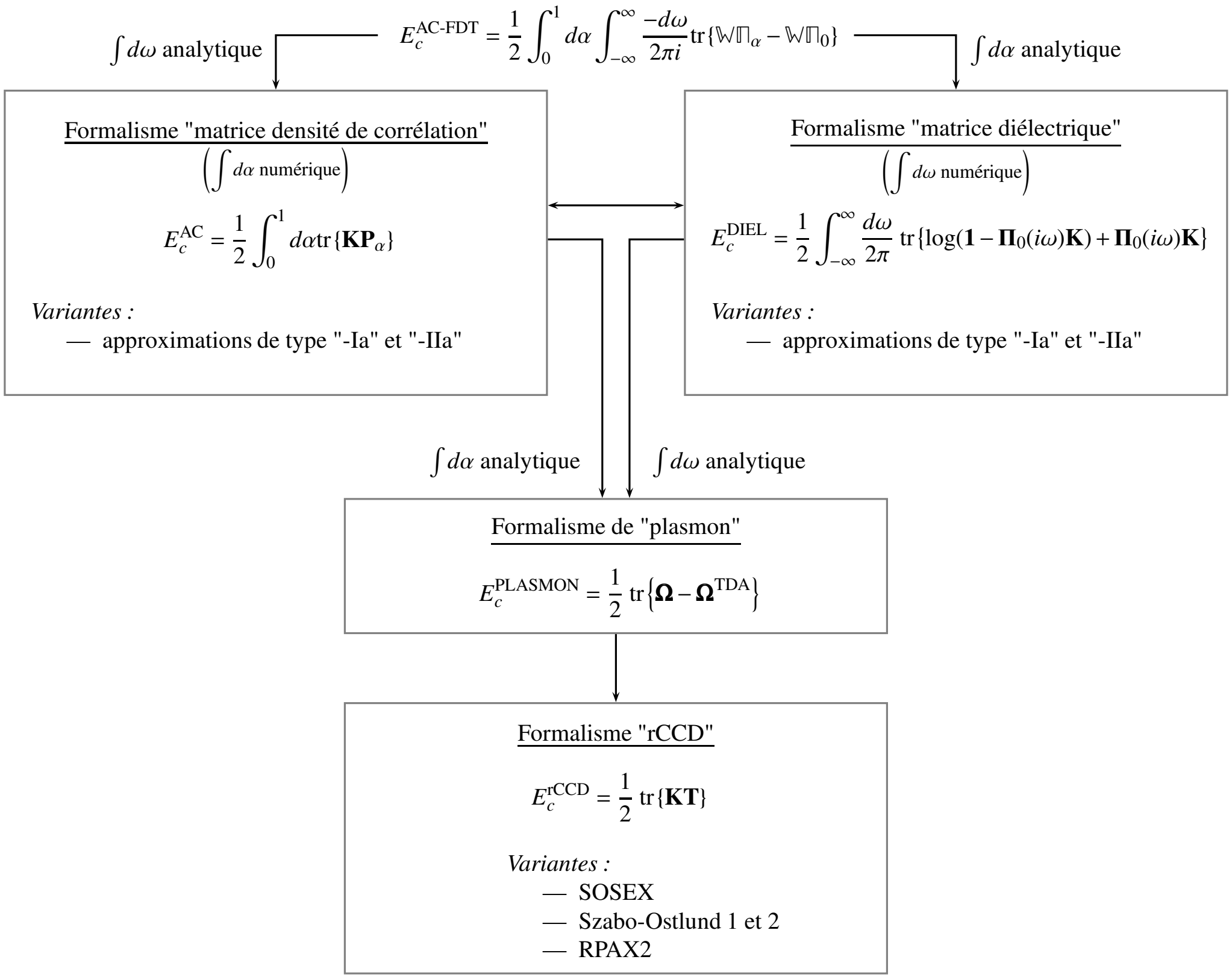

FIGURE 2.3: À partir de l'expression de l'énergie de corrélation AC-FDT que l'on voit équation (2.3.5), on peut dériver en parallèle deux formalismes. En intégrant analytiquement le long de la coordonnée de fréquence $\omega$, on donne naissance à une expression mettant en jeu une intégrale sur la coordonnée de la connexion adiabatique (à gauche). Ce formalisme est dérivé section 2.4, et aboutit aux équations (2.4.15) ou encore (2.4.20). Dans cette section on voit que l'on peut faire émerger de petites variantes dans le traitement des équations, avec notamment différentes façon de calculer la matrice $\mathbf{M}_\alpha$ ; on introduit également les approximations de type "-IIa". Si l'on intègre analytiquement sur la constante de couplage $\alpha$, on se retrouve à écrire une équation (à droite) mettant en jeu la matrice diélectrique $\boldsymbol{\varepsilon} = \mathbf{1} - \mathbf{\Pi}_0\mathbf{V}$, et qui est dérivée section 2.7. On écrit dans ce schéma l'équation (2.7.6) du cas dRPA-I : ce n'est pas l'équation la plus générale mais elle est représentative de la formulation (on voit explicitement la matrice diélectrique, notamment). Dans cette formulation, les approximations de type "-a" sont également naturellement dérivées. À partir de chacune de ces formulations, dans les cas dRPA-I et RPAx-II, on peut intégrer analytiquement sur la coordonnée restante ($\alpha$ ou $\omega$) pour obtenir de manière équivalente une équation de "plasmon" (voir équation (2.5.10)), et une formulation avec équation de "Riccati" (voir (2.5.17)). Ceci est montré dans la section 2.5. Dans la section traitant de l'adaptation de spin, section 2.6, on introduit les approximations SOSEX et Szabo-Ostlund 1 et 2.

où `rpatddft` est le mot-clé qui ouvre le module RPA dans `MOLPRO` ; `ecorr` est le mot-clé qui permet de spécifier quel calcul RPA doit être lancé (d'autres mots-clés permettent de préciser des données telles que les orbitales à utiliser pour le calcul, *etc*… ). Par défaut, un calcul de type "équation de Riccati " SO2 est lancé. Les « interrupteurs » $\zeta$ et $\xi$ peuvent être donnés par l'utilisateur au travers des mots-clés `xfac` et `antifac`, mais toutes les *flavors* de RPA ont des mot-clés et les calculs sont lancés de la manière suivante :

$$\begin{aligned} \text{dRPA-I}\ &\texttt{\{rpatddft;ecorr,SO2-RCCD\}}\text{[par défaut]} \\ \text{dRPA-II}\ &\texttt{\{rpatddft;ecorr,DRPAI-AC\}} \\ \text{RPAx-I}\ &\texttt{\{rpatddft;ecorr,RPAXI-AC\}} \\ \text{RPAx-I}\ &\texttt{\{rpatddft;ecorr,RPAXII-PLASMON\}} \\ &\ldots \end{aligned} \tag{2.8.2}$$

Dans ces calculs de type "connexion adiabatique", l'intégration sur la constante de couplage est réalisée par une quadrature de Gauss-Legendre[86] qui est paramétrable. Notons que dans le cas d'un calcul dRPA-II, la variante dRPA-IIa est également fournie. Pour de plus amples détails sur l'implémentation, le lecteur est invité à consulter le manuel utilisateur de `MOLPRO`, disponible en ligne.

On a présenté ici avec un certain nombre de détails les différents formulations de ce que l'on appelle « RPA ». Toutes les expressions dérivées peuvent être retracées vers ce qui est utilisé ici comme point d'origine : l'expression de l'énergie de corrélation AC-FDT. Ainsi, la RPA peut-être dérivée dans des contextes forts différents dans la littérature, et on s'attache ici à unifier et à discuter les relations (éventuelles) entre les différentes formulations. L'introduction de l'« interrupteur » $\xi$, le pendant de $\zeta$, participe d'un tel effort d'unification, tout comme les explorations du formalisme "matrice diélectrique". Un fait intéressant, dans le contexte de ces dérivations "matrice diélectrique", est le fait que l'on retrouve de manière indépendante l'expression RPAX2 récemment proposée par Hesselmann. Une prochaine étude permettra de voir si ses bons résultats numériques se confirment dans un contexte avec séparation de portée.

# Chapitre 3

# Orbitales localisées

On présente ici brièvement les techniques de localisation des orbitales, *a priori* et *a posteriori*, c'est-à-dire sans et avec passage par un calcul conventionnel qui produit des orbitales canoniques délocalisées. On rappelle quelques méthodes de localisation des orbitales occupées, et la technique des orbitales atomiques projetées (PAO) pour la construction d'orbitales virtuelles à caractère local.

Dans le contexte des calculs inter-moléculaires, l'utilisation des orbitales localisées peut conduire à des méthodologies particulièrement efficaces. Les méthodes de corrélation locale, rapidement rappelées ici, permettent d'atteindre une croissance linéaire du coût de calcul avec la taille du système pour la plupart des théories de corrélation. On présente dans la continuité de ces idées une procédure développée en collaboration avec le Laboratoire de Chimie Théorique de l'Université Pierre et Marie Curie de Paris qui permet de construire des orbitales d'un dimère dans lesquelles on peut reconnaître la trace des orbitales des monomères. On montre que l'on peut ainsi réduire considérablement le nombre d'excitations qui contribuent effectivement à l'énergie d'interaction.

Au centre de ce chapitre est le développement d'une méthode de construction d'orbitales virtuelles non-orthogonales appelées orbitales oscillantes projetées (POO). On dérive des relations importantes concernant la pratique de la RPA dans la base de ces POO, avec notamment une approximation d'excitations locales similaire aux approximations impliquées dans la sélection de domaines dans les méthodes de corrélation locale. Cette partie du chapitre est à lire en couple avec l'Annexe D.

## 3.1 À propos des orbitales occupées localisées

Comme il sera expliqué d'une manière plus détaillée dans la suite, la corrélation est un effet majoritairement courte-portée. Ainsi, pourvu que l'on dispose d'outils le permettant, l'expérience montre que, dans certains circonstances, on peut négliger les contributions à la corrélation de paires d'électrons qui sont éloignés l'un de l'autre, tandis que dans d'autres situations, ce sont ces corréla-tions conduisant à un gain d'énergie relativement faible, qui nous préoccupent. Une telle séparation ne peut se concevoir raisonnablement lorsque l'on travaille avec des orbitales (c'est-à-dire des électrons) délocalisées sur tout le système : les « orbitales localisées » sont les outils qui permettent de telles approximations.

On définit une orbitale localisée comme une orbitale qui est spatialement confinée dans un petit volume, montrant ainsi clairement les atomes qui sont impliqués dans une liaison. On peut donc penser que cette notion est intimement liée à la notion de paires d'électrons, c'est-à-dire qu'une orbitale localisée est censée bien représenter une liaison, une paire libre, ou même des paires d'électrons de cœur. D'ailleurs, historiquement, les orbitales localisées sont développées pour donner des informations sur la chimie du système, pour correspondre aux liaisons chimiques conceptuelles telles qu'on les idéalise : avec une séparation entre des centres (les liaisons idéalisées) et des queues (qui correspondraient alors à des interactions faibles délocalisées). Aujourd'hui, les orbitales localisées sont plutôt utilisées dans des schémas de réduction des coûts de calculs, comme l'illustre un travail mené au cours de cette thèse et présenté dans la section 3.4.

Je présente dans un premier temps un rappel des techniques de localisations des orbitales occupées bien connues, que l'on sépare en méthodes dites *a priori* et *a posteriori*. Le lecteur trouvera dans les références [132], [133] et [134] ainsi que [135–137] les écrits qui rassemblent la plupart des informations sur le sujet.

### 3.1.1 Localisation *a priori*

Habituellement les orbitales Hartree-Fock/Kohn-Sham sont obtenues en diagonalisant la matrice de Fock (de Kohn-Sham). Ces orbitales, dites *canoniques*, respectent la symétrie du système (c'est-à-dire appartiennent au groupe de symétrie de la molécule, *i.e.* sont invariantes par transformation selon des représentations irréductibles du groupe de symmétrie de la molécule), par conséquent elles sont souvent complètement délocalisées sur l'ensemble du système, et ce évidemment indépendamment des orbitales de départ qui ont été données à la procédure SCF.

**On peut chercher** à résoudre les équations générales de Hartree-Fock/Kohn-Sham, $\hat{f}\psi_i = \sum_j \varepsilon_{ij}\psi_j$, sans imposer la diagonalité de la matrice $\varepsilon$.

On trouve dans la littérature des idées autour de la résolution d'équations dérivées des équations de Hartree-Fock/Kohn-Sham mais contenant un « potentiel localisant » [138] . Les solutions trouvées de cette manière conduisent à des orbitales qui correspondent au même sous-espace des orbitales occupées que les orbitales canoniques et sont donc également des solutions des équations de Hartree-Fock/Kohn-Sham.

Une alternative possible est d'utiliser des techniques qui ne déforment pas les orbitales de départ autant qu'une procédure de diagonalisation, de sorte que si l'on fournit comme orbitales de départ des orbitales localisées (par exemple des orbitales intuitives et très approximatives, construites par combinaison linéaire d'orbitales atomiques, en lien avec la chimie du système), les orbitales optimisées et solutions des équations Hartree-Fock/Kohn-Sham en sortie de ces procédures seront assez ressemblantes et, donc, assez localisées. On peut citer les techniques de perturbation au premier ordre[139–142], où à chaque itération les orbitales sont modifiées en mélangeant les orbitales occu-pées et virtuelles de la manière suivante :

$$\begin{aligned} |i\rangle &\leftarrow |i\rangle + \sum_a c_i^a |a\rangle \\ |a\rangle &\leftarrow |a\rangle - \sum_i c_i^a |i\rangle, \end{aligned} \tag{3.1.1}$$

où les coefficients $c_i^a$ sont déterminés en cherchant à respecter le théorème de Brillouin, démarche équivalente à chercher à résoudre les équations de Hartree-Fock/Kohn-Sham[143, 144]. Ainsi, on considère que l'on atteint une convergence lorsque l'énergie n'est plus modifiée, ou, de manière équivalente, lorsque le théorème de Brillouin est satisfait. Par construction, l'orthogonalité entre les occupées et les virtuelles est conservée. En revanche on perd l'orthogonalité des orbitales occupées entre elles et des virtuelles entre elles, qui peut être restaurée à chaque itération.

### 3.1.2 Localisation *a posteriori*

On rappelle que les déterminants, solutions des équations générales d'une méthode SCF comme Hartree-Fock ou Kohn-Sham (ou RSH), sont invariants par rotations des orbitales à l'intérieur du sous-espace des orbitales occupées ou du sous-espace des orbitales virtuelles. La localisation d'orbitales occupées peut donc être réalisée *a posteriori*[145–147], à partir d'orbitales issues des calculs standards, par transformation unitaire dans le sous-espace des orbitales occupées. Notons que ces transformations ne mélangent pas les espaces occupé et virtuel, et préservent l'orthogonalité au sein des orbitales occupées.

**Ainsi**, si l'on veut localiser des orbitales canoniques obtenues lors d'un calcul conventionnel Hartree-Fock/Kohn-Sham, on peut utiliser une méthode dite de localisation *a posteriori*, où l'on utilise une transformation unitaire qui obéit à un critère de localisation, c'est-à-dire qui maximise/minimise un critère de localisation/d'étendue des orbitales.

Les méthodes de localisation *a posteriori* sont souvent classifiées en deux grandes catégories : les critères dits "externes" et les critères dits "internes". Les critères externes correspondent aux méthodes de localisation où l'on décide au préalable par un critère spatial ou par une partition de l'espace des orbitales, quels seront les fragments du système qui supporteront les orbitales localisées. Les critères internes sont formulés en termes généraux, sans faire référence à une conception au préalable concernant la forme des orbitales localisées.

Le prototype des méthodes de localisation à critères externes est celui de Magnasco et Perico[148] qui vise à maximiser la somme des populations de fragments. La population peut être définie dans le cadre de l'analyse de population de Mulliken, comme dans la version originale (voir par exemple [149]), ou les populations des atomes de Bader (voir par exemple [150]).

Parmi les critères internes celui Edmiston-Ruedenberg[145] cherche à minimiser une grandeur énergétique : la répulsion Coulombienne entre les orbitales, ou à maximiser la répulsion intra-orbitalaire, ce qui est équivalent mathématiquement. Malgré un concept physiquement transparent, c'est une méthode qui n'est plus tellement utilisée, car elle fait intervenir des intégrales bi-électroniques. Une approche différente a été proposée par Foster et Boys[146, 151] (l'approche est connue sous la dénomination de "critère de Boys") : les orbitales localisées sont obtenues en maximisant la distance entre leurs centroïdes, c'est-à-dire :

$$\max \left\{ \sum_{i<j} |\langle i|\hat{\mathbf{r}}|i\rangle - \langle j|\hat{\mathbf{r}}|j\rangle| \right\}, \tag{3.1.2}$$

ou, de manière équivalente, en minimisant leur second moment, c'est-à-dire leur variance, leur étendue :

$$\min \left\{ \sum_{i} \left\langle i \middle| (\hat{\mathbf{r}} - \langle i|\hat{\mathbf{r}}|i\rangle)^2 \middle| i \right\rangle \right\} = \min \left\{ \sum_{i} \left\langle i \middle| \hat{\mathbf{r}}^2 \middle| i \right\rangle - \langle i|\hat{\mathbf{r}}|i\rangle^2 \right\} \tag{3.1.3}$$

Le second moment d'une orbitale caractérise la déviation de sa densité autour de la position moyenne, c'est-à-dire mesure la position de l'essentiel de la densité de l'orbitale. Pour des raisons qui seront rendues claires dans la suite, on s'intéressera particulièrement aux orbitales localisées par le critère de Boys.

Il est intéressant de noter que l'autre critère de localisation souvent utilisé dans les méthodes de corrélation locale, celui de Pipek et Mezey[152], exprime une vision analogue pour quantifier la localisation sous forme d'une variance[153].

## 3.2 Orbitales virtuelles localisées

Le problème général de ces méthodes est qu'elles sont performantes pour localiser les orbitales occupées mais échouent souvent à proprement localiser simultanément les orbitales occupées et les orbitales virtuelles[154]. Dans la littérature, dans le contexte de méthodes de corrélation locale[155–160], on trouve des solutions diverses pour régler ce problème.

Une des solutions pratiques qui a rencontré beaucoup de succès, évite la localisation des virtuelles et exploite la localité de la base des orbitales atomiques. Cette méthode des *Projected Atomic Orbitals* (PAO) est due à Pulay[156]. Je montrerai dans la section 3.5 notre propre proposition sur le sujet. Une autre alternative, basée sur les orbitales localisées *a priori*, élaborée à Toulouse par Daudey et ses collaborateurs sera développée en quelques détails à propos du travail effectué en collaboration avec Peter Reinhardt et son étudiant, Edrisse Chermak, à Paris[161].

Récemment, le groupe de Poul Jørgensen à Aarhus a suggéré qu'une généralisation de la minimisation du *spread* à la puissance 4[162, 163] pourrait conduire à des orbitales virtuelles suffisamment bien localisées pour être utilisées dans des calculs de corrélation locale. La procédure s'applique aussi bien pour le critère de Boys que pour celui de Pipek-Mezey et elle a été utilisée dans les approches DEC (*divide-expand-consolidate*) MP2[159] et CCSD.

### 3.2.1 Orbitales atomiques projetées : PAO

Ici, je présente succinctement la technique de génération d'orbitales virtuelles appelée PAO, mais il existe des alternatives dans la littérature. On citera par exemple l'utilisation de paire d'orbitales naturelles (PNO)[164], d'orbitales virtuelles spécifiques[165] qui permettent de construire des bases virtuelles plus petites que les PAO, des orbitales naturelles gelées (FNO)[166, 167], la méthode de l'espace d'orbitales virtuel optimisé (OVOS)[168, 169].

**La procédure** est construite autour de l'idée que les orbitales atomiques constituent un point de départ optimal en ce qui concerne la localité des orbitales.

Comme ce sont les mêmes orbitales atomiques qui servent à construire les orbitales de l'espace occupé, les orbitales virtuelles correspondent au sous-espace « non-utilisé » de la totalité des orbitales atomiques. Cet espace complémentaire est obtenu par projection, d'où la dénomination des orbitales atomiques projetées (PAO).

Les PAO sont localisées par construction (puisqu'à l'origine, ce sont des orbitales atomiques), même si la projection détériore la localité. Tandis qu'elles sont orthogonales aux orbitales occupées, elles ne le sont pas entre elles. Ce manque d'orthogonalité entre les PAO permet de préserver un maximum de localité ; en contrepartie, c'est une source de complication quant aux équations à résoudre.

## 3.3 Méthodes de corrélation locale

Une fois que l'on dispose d'espaces occupé et virtuel qui sont construits par des orbitales localisées, on peut approximer la fonction d'onde en obéissant à la physique du système, selon des méthodes que l'on nomme méthodes de corrélation locale[156, 170–172]. Notons que le simple fait de faire des transformations linéaires entre des orbitales occupées d'une part, et entre les orbitales virtuelles de l'autre correspond à un changement de base qui n'affectera pas les observables telles que l'énergie de corrélation, à condition que leur expression soit formulée d'une manière invariante. Cette invariance est un test important de la cohérence de la méthode. L'intérêt de ces méthodes réside justement dans le fait que des approximations dont les conséquences numériques sont négligeables peuvent être introduites et conduisent à des gains importants aussi bien du point de vue du temps de calcul, du besoin de stockage ou de la loi de croissance avec la taille du système à traiter. Nous allons souligner deux types d'approximations qui permettent de réaliser ces gains importants.

Considérons d'abord le choix des déterminants excités pris en compte dans le calcul. On fait correspondre à chaque orbitale $i$ un domaine $[i]$[173] qui contient toutes les orbitales atomiques qui permettent d'approximer l'orbitale $i$ avec une précision donnée (les orbitales atomiques composant le domaine $[i]$ peuvent appartenir à différents atomes, selon la qualité de la localisation de l'orbitale $i$). Le domaine correspondant dans l'espace virtuel est construit par des PAO générées à partir des orbitales atomiques de $[i]$. Avec cette procédure, les PAO du domaine $[i]$ sont toutes spatialement proches de l'orbitale $i$.

**La première approximation** que l'on met en place pour la fonction d'onde consiste à considérer que les excitations sont purement locales.

Ainsi les simples excitations émergeant de l'orbitale occupée localisée $i$ sont restreintes aux PAO

du domaine $[i]$, et les doubles excitations d'une paire entre $i$ et $j$ aux PAO de $[ij] \doteq [i] \cup [j]$. Ceci réduit le nombre de doubles excitations à considérer, c'est-à-dire que pour une paire donnée $ij$ d'orbitales occupées localisées, le nombre d'éléments $ij \rightarrow rs$ à considérer dans une théorie n'augmente pas quadratiquement avec le nombre d'électron, mais est en fait indépendant du nombre d'électron (on passe, pour chaque paire $ij$, d'une dépendance $O(N^2)$ à $O(1)$). Notons que cette présentation de l'approximation est une simplification : on peut introduire un contrôle plus souple dans la sélection des domaines, c'est-à-dire que l'on peut aller au-delà des domaines purement locaux. La force est que des approximations moins sévères dans le choix des domaines résulteront tout autant en une méthode qui aura la même (in-)dépendance au nombre d'électrons.

Un autre aspect concerne la classification des excitations par leur nature physique, rendu possible par l'utilisation des orbitales localisées.

> **Ainsi, une seconde approximation** peut être faite si l'on se rappelle que la corrélation est un phénomène essentiellement de courte-portée. Ainsi on peut discriminer les paires entre des orbitales occupées localisées $i$ et $j$ selon leur éloignement.

Les paires entre orbitales qui sont "très distantes" sont tout simplement négligées (l'expérience montre qu'elles contribuent à l'énergie de corrélation à la hauteur de quelques micro Hartree). Les paires restantes peuvent être traitées d'une manière hiérarchique, c'est-à-dire que l'on peut à nouveau distinguer les paires "fortes" (d'orbitales très proches), traitées à un haut niveau de théorie ; les paires "faibles" (d'orbitales relativement éloignées), qui sont traitées à un plus faible niveau de théorie ; et les paires "distantes" (d'orbitales plus éloignées encore) qui sont traitées avec des intégrales bi-électroniques approximées par expansion multipolaire[174]. Le point important à comprendre est que le nombre de paires "fortes", "faibles" et "distantes" augmente linéairement avec la taille du système : seules le nombre de paires "très distantes" augmente quadratiquement avec la taille du système, et ces paires sont négligées (cette fois on passe, en ce qui concerne le nombre de paire $ij$, d'une dépendance $O(N^2)$ à $O(N)$, c'est-à-dire à une croissance linéaire du coût de calcul avec la taille du système).

Le schéma général esquissé ici doit être appliqué avec précaution, surtout lorsqu'il s'agit de l'étude des interactions faibles, comme les forces de van der Waals. Une approximation souvent exploitée pour accélérer les calculs de type "local-CCSD(T)" consiste à traiter les paires distantes au niveau MP2, ce qui peut détériorer d'une manière significative l'énergie de dispersion par rapport à la méthode Coupled-Cluster. Dans ces cas-là, soit il faut renoncer à cette approximation ou la remplacer par une autre dont les conséquences sont moins néfastes (voir ci-dessous).

Notons pour finir qu'il existe des alternatives au schéma de sélection des domaines $[i]$ que l'on décrit ici : on peut citer notamment les travaux des références [175, 176] qui cherchent à répondre à la problématique des domaines $[i]$ qui changent d'un point à l'autre lorsque l'on explore une surface d'énergie potentielle.

## 3.4 Orbitales localisées pour le calcul d'énergies d'interaction

Comme décrit juste en amont, d'un point du vue physique, la corrélation dynamique dans des systèmes non-métalliques est un effet à courte-portée, qui diminue en $1/R^6$ où $R$ est la distance inter-électronique. Ainsi il ne doit pas être nécessaire de corréler tous les électrons d'un système moléculaire étendu, et le coût (élevé) des calculs corrélés en terme de nombre d'électrons impliqués n'est qu'un *artefact* de l'utilisation des bases orthogonales, délocalisées, qui sont habituellement utilisées. Ceci n'est plus vrai lorsque l'on utilise des orbitales locales pour construire les espaces occupé *et* virtuel : avec ces orbitales, il est possible de restreindre les calculs de corrélation aux électrons proches qui contribuent effectivement à la corrélation physique du système (voir section 3.2.1).

À ce sujet, dans le contexte de calculs d'énergies d'interaction, nous avons développé à l'occasion d'une collaboration[161] avec le Laboratoire de Chimie Théorique de l'Université Pierre et Marie Curie, à Paris, une procédure de construction des orbitales d'un dimère localisées sur chaque monomères, c'est-à-dire une procédure avec laquelle on est capable d'identifier dans un dimère les orbitales provenant de chacun de ses monomères. De cette manière on est en mesure de classer des couplages de mono-excitations en différentes catégories (voir figure 3.1).

**On a pu montrer** qu'en général, dans des calculs avec séparation de portée, les couplages de type "dispersion" dans le dimère suffisent à eux seuls à décrire correctement la contribution $\Delta E_{\mathrm{RPA}}^{\mathrm{LR}}$ de l'énergie de corrélation RPA longue-portée à l'énergie d'interaction. Le coût de ces calculs croit linéairement avec la taille des molécules considérées.

Ainsi, pour le calcul de cette contribution $\Delta E_{\mathrm{RPA}}^{\mathrm{LR}}$, seul le calcul sur le dimère est nécessaire : on considère les monomères pour construire des orbitales du dimère localisées sur les monomères, mais ils ne font pas l'objet d'un calcul de corrélation. De plus, dans le calcul de corrélation du dimère, le nombre d'éléments de matrices RPA, de structure *ia*, *jb*, est drastiquement réduit.

La contribution de cette thèse dans la collaboration a principalement concerné la partie liée au calcul de l'énergie de corrélation RPA longue portée. Des procédures écrites dans le cadre d'une version de développement du programme `MOLPRO` [131], à Nancy, ont été adaptées pour être compilées dans un programme autonome. Ce programme est prévu pour être interfacé avec les scripts écrits au LCT : il lit en *input* une liste de mono-excitations et de couplages et construit les matrices des équations d'un calcul de RPA dans un formalisme de type "connexion adiabatique" adapté au contexte des orbitales localisées.

## 3.5 Orbitales oscillantes projetées : POO

[NOTE: **ce travail a depuis fait l'objet d'une publication : *Local Random Phase Approximation with Projected Oscillator Orbitals.* B. Mussard, J.G. Ángyán, Theor. Chem. Acc. 134 (2015)**]

On propose ici de revisiter une idée originellement proposée par Boys[151, 177], finalement très peu utilisée par la suite (par exemple [178]). On suggère de construire des orbitales virtuelles localisées directement à partir des orbitales occupées localisées par le critère de Boys, en les multipliant par une harmonique sphérique centrée sur le barycentre de l'orbitale. Nous allons appeler ces orbitales des orbitales oscillantes (OO). Il faut bien sûr, comme précédemment avec les PAO, les projeter sur l'espace virtuel pour supprimer les composantes servant déjà à construire l'espace occupé. On parle alors d'orbitales oscillantes projetées, POO (*Projected Oscillator Orbitals*). Ces POO partagent

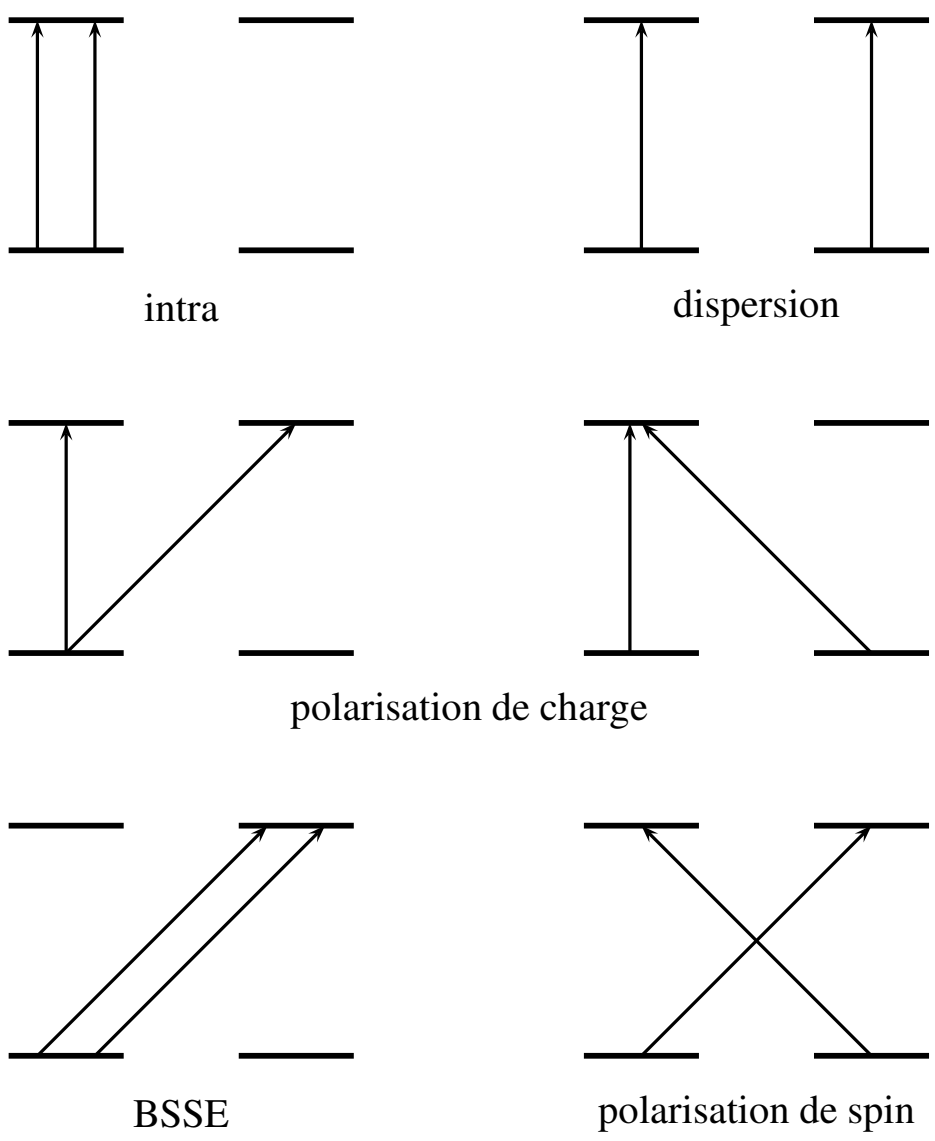

FIGURE 3.1: Couplage de deux mono-excitations (flèches) entre des orbitales occupées et virtuelles de deux monomères. Les combinaisons possibles sont : couplage de deux mono-excitations intra-moléculaires, couplage d'une mono-excitation intra-moléculaire et d'une mono-excitation inter-moléculaire, couplage de deux mono-excitations inter-moléculaires (à chaque fois : sur le même monomère ou sur deux monomères différents). Vue la localisation des orbitales, on considère que l'on peut négliger les mono-excitations inter-moléculaires : ainsi, seuls les couplages de type "intra" et "dispersion" contribuent à l'énergie. Les couplages de type "intra" sont similaires dans le dimère et ses monomères : seul les couplages de type "dispersion" contribuent à l'énergie d'interaction.

bon nombre de caractéristiques avec les PAO, elles ne sont notamment pas orthogonales entre elles.

### 3.5.1 Construction

On suppose que l'on dispose d'un ensemble d'orbitales occupées localisées $\{i\}$ (appelées LMO pour *Localized Molecular Orbitals*) et d'un sous-espace virtuel construit par des orbitales virtuelles orthogonales $\{a\}$ (appelées VMO pour *Virtual Molecular Orbitals*). Pour rendre les notions plus transparentes au lecteur, nous allons nous restreindre à un traitement d'ordre bas, sans « s'engouffrer » dans une écriture plus compliquée d'un formalisme d'ordre quelconque. On introduit donc un polynôme du premier ordre d'harmonique sphérique, $\hat{r}_\alpha - D^i_\alpha$, où $\hat{r}_\alpha$ est un composant de l'opérateur de position et $D^i_\alpha = \langle i|\hat{r}_\alpha|i\rangle$ est le vecteur position orienté vers le centroïde de l'orbitale $i$. La POO correspondante est :

$$|i_\alpha\rangle = \left(1 - \sum_j |j\rangle\langle j|\right)\left(\hat{r}_\alpha - D^i_\alpha\right)|i\rangle = \hat{r}_\alpha |i\rangle - \sum_j |j\rangle\langle j|\hat{r}_\alpha|i\rangle = \left(1 - \hat{P}\right)\hat{r}_\alpha |i\rangle, \tag{3.5.1}$$

qui définit le projecteur $\left(1 - \hat{P}\right)$. La dénomination $i_\alpha$ sert à rappeler le fait que la POO a été construite à partir de l'orbitale $i$ en utilisant $\hat{r}_\alpha$. Il est clair que l'orbitale oscillante pure est $\hat{r}_\alpha |i\rangle - |i\rangle\langle i|\hat{r}_\alpha|i\rangle$ et que contributions $-\sum_{j\neq i} |j\rangle\langle j|\hat{r}_\alpha|i\rangle$ sont des projections orthogonalisantes. Ainsi la localisation des orbitales oscillantes est détériorée par la projection, puisque nous avons des contributions de toutes les LMO $j$ qui donnent lieu à des « queues » d'orbitale. Pourtant, si ces orbitales ont été localisées avec un critère de Boys, c'est-à-dire si les étendues quadratiques ont été minimisées, on s'attend à ce que ces éléments hors-diagonaux des opérateurs $x$, $y$ et $z$ entre orbitales occupées, $\langle j|\hat{r}_\alpha|i\rangle$, ne viennent pas détériorer énormément la localisation des orbitales virtuelles[179]. En ce sens, la méthode de localisation des orbitales occupées selon Boys et cette méthode POO de localisation des orbitales virtuelles semblent naturellement aller de paire.

On peut exprimer les POO par rapport aux VMO :

$$|i_\alpha\rangle = \sum_a |a\rangle\langle a| \left(1 - \hat{P}\right)\hat{r}_\alpha |i\rangle = \sum_a |a\rangle\langle a|\,\hat{r}_\alpha |i\rangle \doteq \sum_a |a\rangle\, V_{ai_\alpha} \tag{3.5.2}$$

Les POO sont orthogonales aux orbitales occupées par construction ; elles ne sont pas orthogonales entre elles et leur recouvrement s'écrit :

$$S_{i_\alpha j_\beta} = \left\langle i_\alpha \middle| j_\beta \right\rangle = \sum_{ab} V^\dagger_{i_\alpha a}\langle a|b\rangle V_{bj_\beta} = \sum_a V^\dagger_{i_\alpha a} V_{aj_\beta} = \left(\mathbf{V}^\dagger\mathbf{V}\right)_{i_\alpha j_\beta} \tag{3.5.3}$$

Notons que notre objectif dans la suite sera de ne pas garder de référence explicite aux orbitales virtuelles : à terme, tout sera exprimé avec les orbitales occupées LMO. Le lecteur peut voir une certaine analogie avec les méthodes de corrélation dites « *dual-basis* »[180, 181], où, pour les calculs de corrélation, les bases d'orbitales sont augmentées soit par des orbitales construites soit pour correspondre à une base cible. Les bases du calcul corrélé ne correspondent donc plus aux bases sur lesquelles les calculs SCF ont été faits : il y a donc des complications liées au non-respect du théorème de Brillouin et à la structure non bloc-diagonale occupé/virtuel de la matrice de Fock dans la base augmentée. J'attire l'attention du lecteur sur le fait que ce n'est pas le cas avec les POO, qui continuent à satisfaire la condition de Brillouin, $\left\langle i\middle|\hat{f}\middle|j_\alpha\right\rangle = 0$ : il n'y a pas ici de telles complications.

Je montre figure 3.2 de simples visualisations d'une LMO correspondant à une paire libre de l'oxygène du formaldéhyde et des trois POO que l'on construit à partir de cette LMO. Je montrerai dans le chapitre 4 des visualisations plus systématiques.

### 3.5.2 Équations RPA dans la base des POO

Cet exposé a été inspiré par le traitement des méthodes de corrélation locale par Knowles et Werner[170]. On va trouver plus pratique ici de réécrire toutes les matrices $(\mathbf{X})_{ia,jb}$ trouvées dans

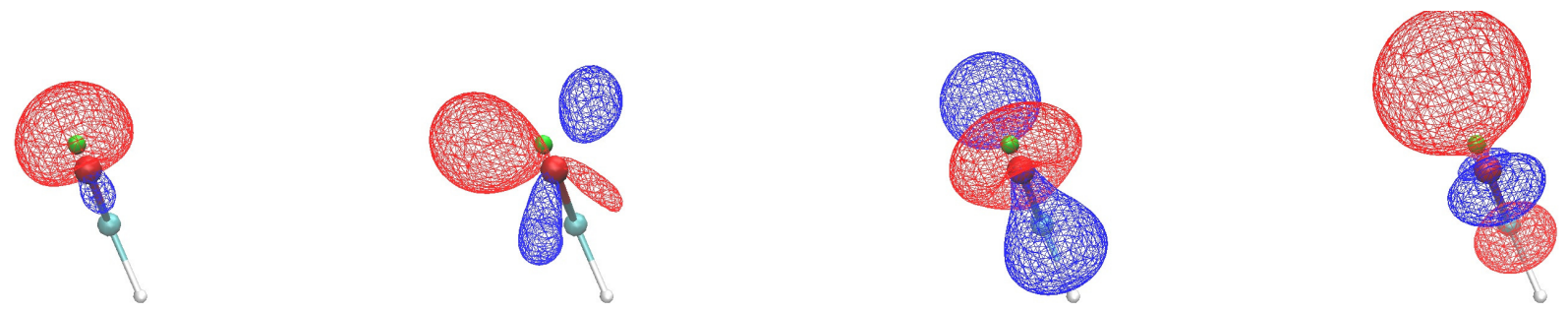

Table 3.2: De gauche à droite : isosurfaces d'une LMO correspondant à une paire libre de l'oxygène du formaldéhyde les trois POO construites à partir de cette LMO. On voit l'effet de la multiplication par les polynômes d'ordre un, selon les trois axes du système référentiel. Je présenterai au chapitre 4 des visualisations plus systématiques.

---

la dérivation des équations RPA dans sa formulation avec les équations de "Riccati ", section 2.5.3, comme : $\left(\mathbf{X}^{ij}\right)_{ab}$. Ainsi un produit $(\mathbf{XY})_{ia,jb}$ s'écrit-il $\left(\mathbf{X}^{ik}\right)_{ac}\left(\mathbf{Y}^{kj}\right)_{cb}$, et on écrit les équations de Riccati :

$$\mathbf{R}^{ij} = \mathbf{B}^{ij} + \mathbf{A}^{ik}\mathbf{T}^{kj} + \mathbf{T}^{ik}\mathbf{A}^{kj} + \mathbf{T}^{ik}\mathbf{B}^{kl}\mathbf{T}^{lj} = \mathbf{0}, \tag{3.5.4}$$

où les matrices sont de dimensions $n_{\text{VMO}} \times n_{\text{VMO}}$. On établit une relation de passage entre les amplitudes $\mathbf{T}^{ij}$ exprimées dans la base des VMO et des POO de la manière suivante :

$$\begin{aligned}
\Psi = \sum_{p_\alpha q_\beta} T^{ij}_{p_\alpha q_\beta} |^{p_\alpha q_\beta}_{ij}\rangle &= \sum_{p_\alpha q_\beta} T^{ij}_{p_\alpha q_\beta} |\ldots \phi_{p_\alpha} \ldots \phi_{q_\beta} \ldots\rangle \\
&= \sum_{p_\alpha q_\beta} \sum_{ab} T^{ij}_{p_\alpha q_\beta} |\ldots \phi_a V_{ap_\alpha} \ldots \phi_b V_{bq_\beta} \ldots\rangle \\
&= \sum_{ab} \sum_{p_\alpha q_\beta} V_{ap_\alpha} T^{ij}_{p_\alpha q_\beta} V^\dagger_{q_\beta b} |\ldots \phi_a \ldots \phi_b \ldots\rangle = \sum_{ab} T^{ij}_{ab} |^{ab}_{ij}\rangle,
\end{aligned} \tag{3.5.5}$$

c'est-à-dire que l'on dispose de la relation suivante :

$$\mathbf{T}^{ij}_{\text{VMO}} = \mathbf{V}\mathbf{T}^{ij}_{\text{POO}}\mathbf{V}^\dagger \tag{3.5.6}$$

On peut donc écrire, en multipliant les équations de Riccati à gauche par $\mathbf{V}^\dagger$ et à droite par $\mathbf{V}$ :

$$\begin{aligned}
\mathbf{V}^\dagger\mathbf{R}^{ij}\mathbf{V} = \mathbf{V}^\dagger\mathbf{B}^{ij}\mathbf{V} &+ \mathbf{V}^\dagger\mathbf{A}^{ik}\left(\mathbf{V}\mathbf{T}^{kj}_{\text{POO}}\mathbf{V}^\dagger\right)\mathbf{V} + \mathbf{V}^\dagger\left(\mathbf{V}\mathbf{T}^{ik}_{\text{POO}}\mathbf{V}^\dagger\right)\mathbf{A}^{kj}\mathbf{V} \\
&+ \mathbf{V}^\dagger\left(\mathbf{V}\mathbf{T}^{ik}_{\text{POO}}\mathbf{V}^\dagger\right)\mathbf{B}^{kl}\left(\mathbf{V}\mathbf{T}^{lj}_{\text{POO}}\mathbf{V}^\dagger\right)\mathbf{V},
\end{aligned} \tag{3.5.7}$$

c'est-à-dire :

$$\mathbf{R}^{ij}_{\text{POO}} = \mathbf{B}^{ij}_{\text{POO}} + \mathbf{A}^{ik}_{\text{POO}}\mathbf{T}^{kj}_{\text{POO}}\mathbf{S} + \mathbf{S}\mathbf{T}^{ik}_{\text{POO}}\mathbf{A}^{kj}_{\text{POO}} + \mathbf{S}\mathbf{T}^{ik}_{\text{POO}}\mathbf{B}^{kl}_{\text{POO}}\mathbf{T}^{lj}_{\text{POO}}\mathbf{S}, \tag{3.5.8}$$

avec des définitions évidentes pour $\mathbf{R}^{ij}_{\mathrm{POO}}$, $\mathbf{B}^{ij}_{\mathrm{POO}}$ et $\mathbf{A}^{ik}_{\mathrm{POO}}$. Ces matrices sont de dimensions $n_{\mathrm{POO}} \times n_{\mathrm{POO}} = 3.n_{\mathrm{LMO}} \times 3.n_{\mathrm{LMO}}$. Dans les équations précédentes, j'ai distingué les matrices exprimées dans la base des VMO de celles exprimées dans la base des POO par la dénomination "$\mathbf{X}^{ij}_{\mathrm{POO}}$", mais dans la suite ce sont les indices de dépendance qui indiqueront au lecteur la base considérée : $\left(\mathbf{X}^{ij}\right)_{ab}$ ou $\left(\mathbf{X}^{ij}\right)_{m_\alpha n_\beta}$.

Notons que si l'on veut écrire séparément les contributions de la fockienne des contributions des intégrales bi-électroniques dans les termes $\mathbf{A}^{ik}\mathbf{T}^{kj}\mathbf{S}$ (et $\mathbf{S}\mathbf{T}^{ik}\mathbf{A}^{kj}$) on se retrouve à écrire, par exemple :

$$\begin{aligned} A^{ik}_{m_\alpha l_\beta} T^{kj}_{l_\beta n_\gamma} &= V^\dagger_{m_\alpha a}\left(f_{ab}\delta_{ik} - f_{ik}\delta_{ab} + A'^{ik}_{ab}\right) V_{bl_\beta} T^{kj}_{l_\beta n_\gamma} \\ &= V^\dagger_{m_\alpha a} f_{ab} V_{bl_\beta} T^{kj}_{l_\beta n_\gamma}\delta_{ik} - f_{ik} V^\dagger_{m_\alpha a}\delta_{ab} V_{bl_\beta} T^{kj}_{l_\beta n_\gamma} - V^\dagger_{m_\alpha a} A'^{ik} V_{bl_\beta} T^{kj}_{l_\beta n_\gamma} \\ &= \left(\mathbf{f}\mathbf{T}^{ij}\right)_{m_\alpha n_\gamma} - f_{ik}\left(\mathbf{S}\mathbf{T}^{kj}\right)_{m_\alpha n_\gamma} + (\mathbf{A}'\mathbf{T})^{ij}_{m_\alpha n_\gamma}, \end{aligned} \tag{3.5.9}$$

de sorte que l'équation (3.5.8) s'écrit :

$$\mathbf{R}^{ij} = \mathbf{B}^{ij} + \left(\mathbf{f}\mathbf{T}^{ij}\mathbf{S} - f_{ik}\mathbf{S}\mathbf{T}^{kj}\mathbf{S} + \mathbf{A}'^{ik}\mathbf{T}^{kj}\mathbf{S}\right) + \left(\mathbf{S}\mathbf{T}^{ij}\mathbf{f} - \mathbf{S}\mathbf{T}^{ik}\mathbf{S}f_{kj} + \mathbf{S}\mathbf{T}^{ik}\mathbf{A}'^{kj}\right) + \mathbf{S}\mathbf{T}^{ik}\mathbf{B}^{kl}\mathbf{T}^{lj}\mathbf{S} \tag{3.5.10}$$

Les éléments de la matrice de fock dans la base des POO sont donnés dans l'Annexe D.3. Du fait de la non orthogonalité des POO et de la structure non diagonale de la matrice de Fock, le schéma de résolution habituel des équations de Riccati (montré équation (2.5.19)) doit être revu : un nouveau schéma comportant une transformation vers des orbitales virtuelles pseudo-canoniques qui diagonalisent la matrice de Fock est dérivé dans l'Annexe D.1.1.

### 3.5.3 Approximation des excitations locales

On peut considérer que les excitations sont limitées à des domaines de paires, et ainsi les dimensions effectives des équations correspondantes à une paire sont plus ou moins indépendantes de la taille du système.

**Dans les cas** les plus simples, on peut considérer que les POO à prendre en compte pour une paire $[ij]$ donnée sont les orbitales $i_\alpha$ et $j_\beta$. Ainsi, pour chaque paire $[ij]$ d'orbitales occupées, de tous les éléments de matrice $\left(\mathbf{X}^{ij}\right)_{m_\alpha n_\beta}$, seuls ceux du bloc $3\times3$ correspondant aux orbitales virtuelles localisées construites à partir des orbitales occupées $ij$ sont non nuls (voir 3.3).

Pour comprendre l'impact d'une telle simplification du modèle, reprenons les équations de Riccati vues équation (3.5.8) en explicitant les indices des matrices impliquées :

$$R^{ij}_{m_\alpha n_\beta} = B^{ij}_{m_\alpha n_\beta} + A^{ik}_{m_\alpha p_\gamma} T^{kj}_{p_\gamma q_\delta} S_{q_\delta n_\beta} + S_{m_\alpha p_\gamma} T^{ik}_{p_\gamma q_\delta} A^{kj}_{q_\delta n_\beta} + S_{m_\alpha p_\gamma} T^{ik}_{p_\gamma q_\delta} B^{kl}_{q_\delta r_\epsilon} T^{lj}_{r_\epsilon s_\xi} S_{s_\xi n_\beta}, \tag{3.5.11}$$

qui se simplifie en :

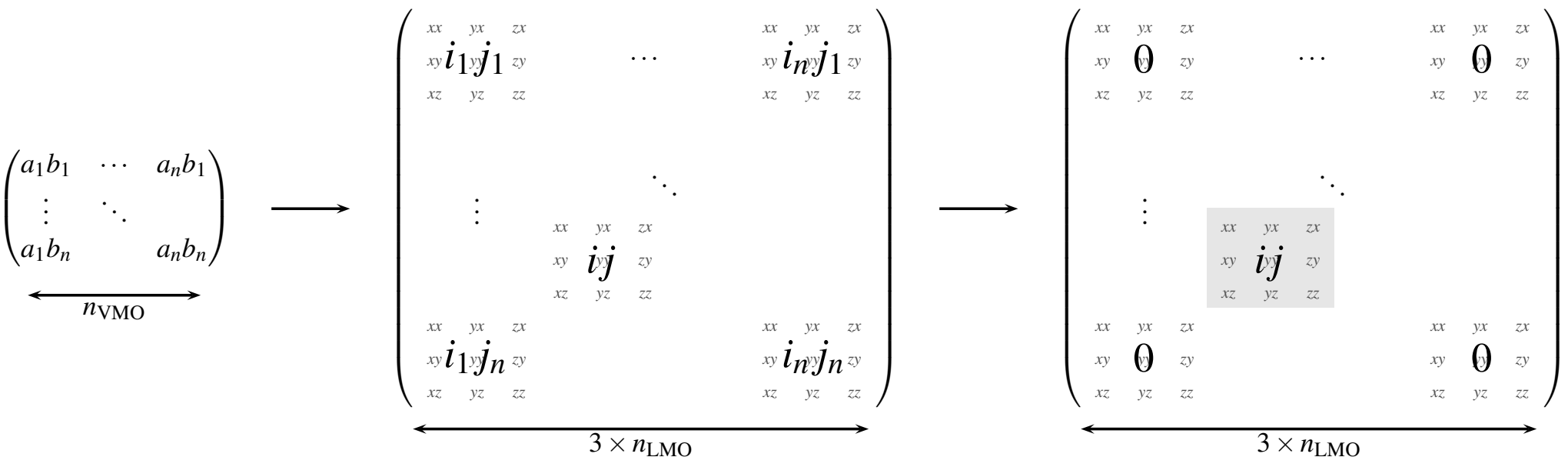

FIGURE 3.3: Dans la formulation canonique des équations de Riccati, pour chaque paire $ij$ d'orbitales occupées, on doit résoudre un problème matriciel de dimension $n_{\text{VMO}} \times n_{\text{VMO}}$ (schématiquement, à gauche). Avec un formalisme d'orbitales virtuelles localisées de type POO, chaque orbitale occupée fournit 3 orbitales oscillantes, ainsi pour chaque paire $ij$ d'orbitales occupées, on doit résoudre un problème matriciel de dimension $3.n_{\text{LMO}} \times 3.n_{\text{LMO}}$ (schéma du milieu), qui est *a priori* de plus faible dimension que $n_{\text{VMO}} \times n_{\text{VMO}}$. Avec l'approximation des excitations locales, pour chaque paire $ij$ d'orbitales occupées, seules les POO formées à partir de cette paire sont à considérer, et on ne doit plus résoudre qu'un problème matriciel de dimension $3 \times 3$ (voir le schéma de droite).

---

$$R^{ij}_{i_\alpha j_\beta} = B^{ij}_{i_\alpha j_\beta} + A^{ik}_{i_\alpha k_\gamma} T^{kj}_{k_\gamma j_\delta} S_{j_\delta j_\beta} + S_{i_\alpha i_\gamma} T^{ik}_{i_\gamma k_\delta} A^{kj}_{k_\delta j_\beta} + S_{i_\alpha i_\gamma} T^{ik}_{i_\gamma k_\delta} B^{kl}_{k_\delta l_\epsilon} T^{lj}_{l_\epsilon j_\xi} S_{j_\xi j_\beta} \tag{3.5.12}$$

Dans ce modèle d'excitations locales, on peut utiliser les notations adaptées suivantes : $\left(\mathbf{X}^{ij}\right)_{\alpha\beta} \doteq \left(\mathbf{X}^{ij}\right)_{i_\alpha j_\beta}$, c'est-à-dire que les exposants $ij$ indiquent à la fois la paire d'orbitales occupées considérée *et* les orbitales occupées qui servent de bases aux POO $\alpha$ et $\beta$. On désigne également par $\left(\mathbf{s}^{ii}\right)_{\alpha\beta} \doteq S_{i_\alpha i_\beta}$ la sous-matrice de $\mathbf{S}$ correspondant au bloc $3 \times 3$ lié à la paire $[ij]$ (voir figure 3.3). On obtient, sous forme matricielle :

$$\mathbf{R}^{ij} = \mathbf{B}^{ij} + \mathbf{A}^{ik}\mathbf{T}^{kj}\mathbf{s}^{jj} + \mathbf{s}^{ii}\mathbf{T}^{ik}\mathbf{A}^{kj} + \mathbf{s}^{ii}\mathbf{T}^{ik}\mathbf{B}^{kl}\mathbf{T}^{lj}\mathbf{s}^{jj}, \tag{3.5.13}$$

où les matrices sont de dimensions $3 \times 3$, peu importe la taille du système.

La formulation qui sépare les contributions de la fockienne des intégrales bi-électroniques dans $\mathbf{A}$ dérivée équation (3.5.10) fait émerger ici des termes $\mathbf{s}^{ik}$ et $\mathbf{s}^{kj}$, selon :

$$\begin{aligned}\mathbf{R}^{ij} = \mathbf{B}^{ij} &+ \left(\mathbf{f}^{ii}\mathbf{T}^{ij}\mathbf{s}^{jj} - f_{ik}\mathbf{s}^{ik}\mathbf{T}^{kj}\mathbf{s}^{jj} + \mathbf{A}'^{ik}\mathbf{T}^{kj}\mathbf{s}^{jj}\right)\\ &+ \left(\mathbf{s}^{ii}\mathbf{T}^{ij}\mathbf{f}^{jj} - \mathbf{s}^{ii}\mathbf{T}^{ik}\mathbf{s}^{kj} f_{kj} + \mathbf{s}^{ii}\mathbf{T}^{ik}\mathbf{A}'^{kj}\right) + \mathbf{s}^{ii}\mathbf{T}^{ik}\mathbf{B}^{kl}\mathbf{T}^{lj}\mathbf{s}^{jj},\end{aligned} \tag{3.5.14}$$

où on désigne par $\mathbf{f}^{ii}$ la sous-matrice de $\mathbf{f}$ correspondant aux orbitales virtuelles POO construites à partir des orbitales occupées $[ij]$. On décide de négliger les termes $\mathbf{s}^{ik}$ pour $i \neq k$ : il s'agit de termes $S_{i_\alpha k_\beta}$ de recouvrement d'orbitales virtuelles construites à partir d'orbitales occupées localisées potentiellement éloignées les unes des autres. On écrira alors plutôt :

$$\begin{aligned}\mathbf{R}^{ij} = \mathbf{B}^{ij} &+ \left(\mathbf{f}^{ii}\mathbf{T}^{ij}\mathbf{s}^{jj} - f_{ii}\mathbf{s}^{ii}\mathbf{T}^{ij}\mathbf{s}^{jj} + \mathbf{A}'^{ik}\mathbf{T}^{kj}\mathbf{s}^{jj}\right) \\ &+ \left(\mathbf{s}^{ii}\mathbf{T}^{ij}\mathbf{f}^{jj} - \mathbf{s}^{ii}\mathbf{T}^{ij}\mathbf{s}^{jj}f_{jj} + \mathbf{s}^{ii}\mathbf{T}^{ik}\mathbf{A}'^{kj}\right) + \mathbf{s}^{ii}\mathbf{T}^{ik}\mathbf{B}^{kl}\mathbf{T}^{lj}\mathbf{s}^{jj} \end{aligned} \tag{3.5.15}$$

Cette équation peut être résolue de manière itérative de la même manière que précédemment, comme montré dans l'Annexe D.1.2.

### 3.5.4 Intégrales bi-électroniques

Les matrices $\mathbf{A}'^{ij}$ et $\mathbf{B}^{ij}$ sont composées des éléments $\mathbf{K}^{ij}$, $\mathbf{K}'^{ij}$ et $\mathbf{J}^{ij}$ que je m'efforce ici d'expliciter. On se concentre sur l'expression de $\mathbf{K}^{ij}$, avant de discuter les expressions de $\mathbf{K}'^{ij}$ et $\mathbf{J}^{ij}$.

Les éléments de matrice des intégrales bi-électroniques peuvent être exprimés simplement par la transformation entre les VMO et les POO, en écrivant :

$$\begin{aligned}\left\langle m_\alpha j \middle| i n_\beta \right\rangle = \sum_{a,b}^{\text{VMO}} V^\dagger_{m_\alpha a} \langle aj|ib\rangle V_{bn_\beta} &= \sum_{a,b}^{\text{VMO}} \sum_{\mu\nu\rho\sigma} V^\dagger_{m_\alpha a} C_{\mu a} C_{\nu j} \langle \mu\nu|\rho\sigma\rangle C^\dagger_{i\rho} C^\dagger_{b\sigma} V_{bn_\beta} \\ &= \sum_{\mu\nu\rho\sigma} \widetilde{C}_{\mu m_\alpha} C_{\nu j} \langle \mu\nu|\rho\sigma\rangle C^\dagger_{i\rho} \widetilde{C}^\dagger_{n_\beta\sigma}\end{aligned} \tag{3.5.16}$$

Cette manière d'écrire les intégrales bi-électroniques dans la base des POO n'est pas développée plus dans la suite du manuscrit, mais pourrait se révéler intéressant à poursuivre dans le futur, comme mesure de contrôle de tous ces développements autour des POO. On peut notamment se poser des questions au sujet du comportement des POO dans le cas de deux extrêmes : les POO construites à partir de bases minimales seront-elles « pauvres », d'une certaine manière, c'est-à-dire présenteraient-elles peu de flexibilité pour décrire correctement un système ; au contraire les POO construites à partir de grandes bases possédant une grande flexibilité produiraient-elles une sorte de sous-espace optimale de virtuelles (optimale dans le sens où elle reproduirait correctement l'énergie de corrélation longue-portée). Ces questions feront l'objet de développements futurs.

Notons également que les définitions que l'on donne section 3.5.1 des POO s'arrêtent à l'ordre le plus bas des polynômes. On peut en principe générer des POO d'ordres plus élevés, et définir des matrices $\mathbf{V}$ correspondant à ces POO. Sélectionner un ordre de polynôme pour générer les POO revient à sélectionner certaines formes et propriétés de ces POO, c'est-à-dire revient en fait à sélectionner une sous-espace des orbitales virtuelles, par exemple dans le but de bien décrire telle ou telle propriété.

Dans l'esprit de mettre en place une théorie *approximée* qui profite du caractère local des orbitales, on peut également écrire une expansion multipolaire[174, 182] (ici dipolaire), par exemple pour les éléments de la matrice $\mathbf{K}$ :

$$K^{ij}_{m_\alpha n_\beta} = \left\langle m_\alpha j \middle| i n_\beta \right\rangle = \left( m_\alpha i \middle| j n_\beta \right) = \left\langle m_\alpha \middle| \hat{r}_\gamma \middle| i \right\rangle L_{i_\gamma j_\delta} \left\langle j \middle| \hat{r}_\delta \middle| n_\beta \right\rangle, \tag{3.5.17}$$

où $\left\langle m_\alpha \middle| \hat{r}_\gamma \middle| i \right\rangle$ sont des moments dipolaires et $\mathbf{L}$ est le tenseur d'interaction longue-portée de second ordre (voir Annexe D.2.1).

Un résultat remarquable dans ce développement d'orbitales POO est que les moments dipolaires que l'on trouve dans l'équation (3.5.17) s'écrivent :

$$\left\langle m_\alpha \middle| \hat{r}_\gamma \middle| i \right\rangle = \left\langle m \middle| \hat{r}_\alpha \left(1 - \hat{P}\right) \hat{r}_\gamma \middle| i \right\rangle = S_{m_\alpha i_\gamma}, \tag{3.5.18}$$

et sont en fait les recouvrements entre orbitales virtuelles POO. Ceci permet d'écrire les matrices de type **K** :

$$K^{ij}_{m_\alpha n_\beta} = S_{m_\alpha i_\gamma} L_{i_\gamma j_\delta} S_{j_\delta n_\beta}, \tag{3.5.19}$$

et, avec le modèle des excitations locales :

$$\mathbf{K}^{ij} = \mathbf{s}^{ii} \mathbf{L}^{ij} \mathbf{s}^{jj} \tag{3.5.20}$$

Les matrices $\mathbf{K}'^{ij}$ dans l'approximation des excitations locales s'écrivent :

$$\mathbf{K}'^{ij} = \mathbf{s}^{ij} \mathbf{L}^{ji} \mathbf{s}^{ij}, \tag{3.5.21}$$

et impliquent des recouvrements de densité de charge appartenant à des domaines différents, que l'on peut négliger en accord avec l'approximation des excitations locales : de toutes les matrices $\mathbf{K}'^{ij}$, seules les matrices $\mathbf{K}'^{ii} = \mathbf{s}^{ii}\mathbf{L}^{ii}\mathbf{s}^{ii}$ sont non nulles. On écrit :

$$\mathbf{B}^{ij} = \mathbf{K}^{ij} - \zeta \mathbf{K}^{ji} = \mathbf{K}^{ij}(1 - \zeta\delta_{ij}) = \mathbf{s}^{ii}\mathbf{L}^{ij}(1 - \zeta\delta_{ij})\mathbf{s}^{jj} = \mathbf{s}^{ii}\mathbf{L}^{ij}_{\mathrm{B}}\mathbf{s}^{jj}, \tag{3.5.22}$$

qui définit $\mathbf{L}^{ij}_{\mathrm{B}}$. Je choisis l'indice « B » pour rappeler que c'est un objet qui émerge dans l'expansion multipolaire de la matrice **B**. Les intégrales de type $\mathbf{J}^{ij}$ dans la formulation sans approximation s'écrivent :

$$J^{ij}_{m_\alpha n_\beta} = \left\langle i m_\alpha \middle| j n_\beta \right\rangle = \left( ij \middle| m_\alpha n_\beta \right) \tag{3.5.23}$$

Il s'agit de l'interaction de densités de charge formées par des orbitales localisées qui appartiennent à des domaines différents : on peut les négliger. On doit néanmoins garder les termes $\left(ii \middle| m_\alpha m_\beta\right)$, qui sont dérivés Annexe D.2.2. On écrit, dans l'approximation des excitations locales :

$$\mathbf{A}'^{ij} = \mathbf{K}^{ij} - \delta_{ij}\mathbf{L}^{ii}_{\mathrm{A}}, \tag{3.5.24}$$

où $\mathbf{L}_\mathrm{A}$ contient l'« interrupteur » $\zeta$ qui permet de passer d'une formulation dRPA à une formulation RPAx, et est défini dans la même Annexe. Il s'agit d'un terme long mais pas compliqué ; son intérêt est ici assez relatif. Le nom que je lui donne, $\mathbf{L}_\mathrm{A}^{ii}$, est censé rappeler le terme $\mathbf{L}_\mathrm{B}^{ij}$ que l'on a vu émerger plus haut, et rappeler qu'ici il l'objet émerge dans la dérivation de la matrice $\mathbf{A}'$.

La méthode présentée ci-dessus, déjà amplement simplifiée par l'ensemble des approximations, prend en couple les corrélations dipolaires anisotropes entre les électrons associés par les paires d'orbitales localisées. Par la résolution des équations Riccati simplifiées, on note que les effets à $N$-corps sont pris en compte à l'ordre infini. Dans la suite j'introduis quelques simplifications supplémentaires dans le but de rendre ce modèle le plus proche possible des modèles qui sont postulés, souvent d'une manière *ad hoc*, en terme de coefficients de dispersion $\mathrm{C}_6$ et d'interaction $1/R^6$ modulée par une fonction d'atténuation.

### 3.5.5 Approximation par moyenne sphérique et coefficients C6

On peut encore simplifier les matrices $3 \times 3$ discutées plus haut par des objets scalaires, en considérant une moyenne sphérique des recouvrements entre POO et des éléments de la fockienne dans la base POO :

$$S^{ii}_{\alpha\beta} \approx \tfrac{1}{3} s^i \delta_{\alpha\beta} \qquad \text{où :} \quad s^i = \left\langle i \middle| \mathbf{r}^2 \middle| i \right\rangle - \sum_k |\langle i|\mathbf{r}|k\rangle|^2 \tag{3.5.25}$$

$$f^{ii}_{\alpha\beta} \approx \tfrac{1}{3} f^i \delta_{\alpha\beta} \qquad \text{où :} \quad f^i = \tfrac{3}{4} + \sum_k f_{ik} \left\langle k \middle| \mathbf{r}^2 \middle| i \right\rangle - \sum_{kl} \langle i|\mathbf{r}|k\rangle \, f_{kl} \, \langle l|\mathbf{r}|i\rangle \tag{3.5.26}$$

(l'expression de $s^i$ se retrouve facilement à partir de la définition du recouvrement $\mathbf{S}^{ii}$ ; les dérivations qui mènent à $f^i$ sont montrées dans l'Annexe D.3). Avec cette approximation, les équations (3.5.15) s'écrivent :

$$\begin{aligned} R^{ij}_{\alpha\beta} = {} & \frac{1}{3^2} s^i {L_\mathrm{B}}^{ij}_{\alpha\beta} s^j + \left( \frac{1}{3^2} f^i \mathbf{T}^{ij} s^j - \frac{1}{3^2} f_{ii} s^i \mathbf{T}^{ij} s^j + \frac{1}{3^3} s^i L^{ik}_{\alpha\delta} s^k T^{kj}_{\delta\beta} s^j - \frac{1}{3} {L_\mathrm{A}}^{ii}_{\alpha\delta} T^{ij}_{\delta\beta} s^j \right) \\ & + \left( \frac{1}{3^2} s^i \mathbf{T}^{ij} f^j - \frac{1}{3^2} s^i \mathbf{T}^{ij} s^j f_{jj} + \frac{1}{3^3} s^i T^{ik}_{\alpha\tau} s^k L^{kj}_{\tau\beta} s^j - \frac{1}{3} s^i T^{ij}_{\alpha\tau} {L_\mathrm{A}}^{jj}_{\tau\beta} \right) + \frac{1}{3^4} s^i T^{ik}_{\alpha\tau} s^k {L_\mathrm{B}}^{kl}_{\tau\delta} s^l T^{lj}_{\delta\beta} s^j , \end{aligned} \tag{3.5.27}$$

c'est-à-dire, en multipliant par $3^2$ :

$$\begin{aligned} \mathbf{R}^{ij} = {} & s^i s^j \mathbf{L}^{ij}_\mathrm{B} + \left( f^i s^j \mathbf{I} - f_{ii} s^i s^j \mathbf{I} - 3 s^j \mathbf{L}^{ii}_\mathrm{A} \right) \mathbf{T}^{ij} + \mathbf{T}^{ij} \left( s^i f^j \mathbf{I} - f_{jj} s^i s^j \mathbf{I} - 3 s^i \mathbf{L}^{jj}_\mathrm{A} \right) \\ & + \tfrac{1}{3} s^i s^j s^k \mathbf{L}^{ik} \mathbf{T}^{kj} + \tfrac{1}{3} s^i s^j s^k \mathbf{T}^{ik} \mathbf{L}^{kj} + \tfrac{1}{3^2} s^i s^j s^k s^l \mathbf{T}^{ik} \mathbf{L}^{kl}_\mathrm{B} \mathbf{T}^{lj} \end{aligned} \tag{3.5.28}$$

On montre dans l'Annexe D.1.3 que cette équation peut être résolue sans effectuer de transformation pseudo-canonique, puisque l'on n'a plus à gérer le problème de la matrice de Fock non diagonale.

On va pouvoir dériver ici une expression de coefficients $\mathrm{C}_6$. Pour rappel rapide, les coefficients $\mathrm{C}_6$ sont les coefficients que l'on utilise dans la modélisation de la force de dispersion de London par la formule $C_6/R^6$. On peut les dériver à partir de l'équation de Casimir-Polder[183, 184] (qui décrit les forces de dispersion comme résultant de fluctuations corrélées des densités de charges entre deux couples de points situés dans deux sous-systèmes) avec une succession d'approximations qui ressemblent aux approximations qui nous ont menés à ce point (expansion multipolaire, moyenne sphérique, *etc...* ). On trouvera dans les sections II et IV.A de la référence [31] une description sommaire du procédé.

On rappelle que, notamment, la version dRPA-I de la formulation avec équations de "Riccati " se réduit au deuxième ordre de l'interaction (voir la section 2.4.6) à l'énergie MP2 directe. Dans cette formulation, et avec une adaptation de spin, l'énergie dRPA-I s'écrit (voir l'équation (2.6.8), et la section 2.6 en général) :

$$E_c^{\text{dRPA-I-Riccati}} = \frac{1}{2}\text{tr}\left\{ {}^{1}\mathbf{B}^{\text{dRPA}}\ {}^{1}\mathbf{T}^{\text{dRPA}} \right\} = \text{tr}\left\{\mathbf{KT}\right\} = \sum_{ij}\sum_{\alpha\beta} s^i L_{\alpha\beta}^{ij} s^j T_{\alpha\beta}^{ij}, \tag{3.5.29}$$

où $\mathbf{T}$ est l'amplitude obtenue avec des équations de Riccati où, dans notre cas présent, $\mathbf{L}_{\text{B}} = \mathbf{L}$ et $\mathbf{L}_{\text{A}} = \mathbf{0}$ (c'est-à-dire où $\zeta = 0$). Une approximation de la première itération de la procédure montrée équation (D.1.15) est :

$$T_{\alpha\beta}^{ij} \approx \frac{s^i s^j L_{\alpha\beta}^{ij}}{s^i f^j - s^i s^j f_{jj} + f^i s^j - s^i s^j f_{ii}}, \tag{3.5.30}$$

qui, inséré dans l'équation (3.5.29) de l'énergie, fournit une expression proche d'une sorte de MP2 directe simplifié.

Faisant usage du fait que la trace des tenseurs d'interaction longue-portée est $\sum_{\alpha\beta} L_{\alpha\beta}^{ij} L_{\alpha\beta}^{ij} = \frac{6}{R^{ij6}} F^{\mu}\left(R^{ij}\right)$ (voir l'Annexe D.2.1 où ceci est montré), on se retrouve à écrire l'énergie de corrélation de l'équation (3.5.29) sous la forme :

$$E_c^{\text{dRPA-I(2)}} \approx \sum_{ij} \frac{2}{3} \frac{s^{i2} s^{j2}}{s^i f^j - s^i s^j f_{jj} + f^i s^j - s^i s^j f_{ii}} \frac{1}{R^{ij6}} F^{\mu}(R^{ij}) = \sum_{ij} \frac{C_6^{ij}}{R^{ij6}} F^{\mu}\left(R^{ij}\right) \tag{3.5.31}$$

Cette équation définit un coefficient $C_6^{ij}$ approximé entre entre les LMO $i$ et $j$. En utilisant la forme complète de l'énergie de corrélation dans cette approximation (avec les amplitudes convergées) nous avons un modèle de type « dispersion à $N$-corps » ("*many-body dispersion*") dipolaire entre les orbitales localisées. Sur des bases assez différentes, avec des ingrédients empiriques, un modèle du même esprit a été proposé très récemment par Silvestrelli [185].

Dans une dérivation des coefficients $C_6$ à partir de la formule de Casimir-Polder, on trouve en fait ces coefficients par l'intégration :

$$C_6^{ij} = \frac{3}{\pi}\int_0^{\infty} d\omega\ \overline{\alpha}_0^i(i\omega)\overline{\alpha}_0^j(i\omega), \tag{3.5.32}$$

c'est-à-dire que, en « remontant » ce raisonnement à partir des coefficients $C_6$ que l'on a trouvé par un autre moyen, on peut établir qu'ils correspondent à une approximation terme unique de la polarisabilité dynamique dipolaire non-interagissante moyenne suivante :

$$\overline{\alpha}_0^i(i\omega) \approx \frac{2}{3}\frac{\frac{f^i s^j - s^i s^j f_{ii}}{s^i s^j}}{\left(\frac{f^i s^j - s^i s^j f_{ii}}{s^i s^j}\right)^2 + \omega^2} s^i \tag{3.5.33}$$

Le lecteur peut vérifier que cette expression de $\overline{\alpha}_0^i(i\omega)$, injectée dans l'équation (3.5.32), fournit bien les coefficients $C_6$ vus équation (3.5.31). On observe que dans l'équation (3.5.33), l'énergie d'excitation effective qui permet d'écrire la polarisabilité moyenne est dans notre cas $\overline{\omega}_i = \frac{f^i s^j - s^i s^j f_{ii}}{s^i s^j}$. On reverra cette façon d'exprimer les objets tels que les fonctions de réponse (et, ici, les polarisabilités) avec des énergies effectives moyennes dans le chapitre 5, où on adapte la technique EED à l'espace réel. Dans l'expression (3.5.33), le fait que l'on peut exprimer $\overline{\alpha}_0^i(i\omega)$ uniquement en fonction des objets $f^{\sqcup}$ et $s^{\sqcup}$ est une conséquence de la propriété remarquable que le second moment entre une orbitale occupée et une POO correspond au recouvrement entre POO (voir équation (3.5.18)).

L'énergie de corrélation RPA non approximée peut être obtenue par itération, c'est-à-dire par la formule logarithmique (voir la section 2.7 sur le sujet) :

$$E_c^{\text{dRPA-I}} \approx -\frac{1}{2}\int \frac{d\omega}{2\pi} \text{tr}\left\{\text{Log}\left(\mathbb{I} - \mathbb{\Pi}_0(i\omega)\mathbb{K}\right) + \mathbb{\Pi}_0(i\omega)\mathbb{K}\right\}, \tag{3.5.34}$$

où $\mathbb{\Pi}_0(i\omega)$ est la fonction de réponse non-interagissante dans un exemple à deux centres :

$$\mathbb{\Pi}_0(i\omega) = \begin{pmatrix} \overline{\alpha}_0^1(i\omega) & \mathbf{0} \\ \mathbf{0} & \overline{\alpha}_0^2(i\omega) \end{pmatrix} \qquad \text{où :} \quad \overline{\alpha}_0^i(i\omega) = \begin{pmatrix} \overline{\alpha}_0^i(i\omega) & 0 & 0 \\ 0 & \overline{\alpha}_0^i(i\omega) & 0 \\ 0 & 0 & \overline{\alpha}_0^i(i\omega) \end{pmatrix}, \tag{3.5.35}$$

et :

$$\mathbb{K} = \begin{pmatrix} \mathbf{K}^{11} & \mathbf{K}^{12} \\ \mathbf{K}^{21} & \mathbf{K}^{22} \end{pmatrix} \qquad \text{où :} \quad \mathbf{K}^{ij} = \begin{pmatrix} s^i L_{xx}^{ij} s^j & s^i L_{xy}^{ij} s^j & s^i L_{xz}^{ij} s^j \\ s^i L_{yx}^{ij} s^j & s^i L_{yy}^{ij} s^j & s^i L_{yz}^{ij} s^j \\ s^i L_{zx}^{ij} s^j & s^i L_{zy}^{ij} s^j & s^i L_{zz}^{ij} s^j \end{pmatrix} \tag{3.5.36}$$

On retrouve avec cette formulation alternative un modèle qui lui aussi contient des sommations d'interaction à $N$-corps à l'ordre infini, comme on a mentionné plus haut avec l'utilisation des amplitudes convergées.

## 3.6 Conclusion et perspectives

Dans ce chapitre j'ai esquissé quelques pistes possibles pour exploiter les avantages de formuler le problème RPA dans une base localisée, notamment dans le contexte de calculs inter-moléculaires (collaboration avec le LCT). Les dérivations des équations avec les orbitales oscillantes projetées montrent que l'on peut, sous certaines approximations que l'on considère adaptées, exprimer la totalité des équations impliquées dans les calculs RPA *sans* mention explicite des orbitales virtuelles. Cet axe de travail est d'un intérêt certain du point de vue pratique du temps de calcul. Également, ceci est un message encourageant quant à la perspective de construire des fonctionnelles de corrélation de qualité RPA mais qui ne nécessitent pas l'utilisation des orbitales virtuelles. Après un chapitre concernant un travail sur les visualisations dans l'espace réel, on va retrouver cette idée d'éviter la connaissance du spectre complet de l'hamiltonien dans l'approximation EED (chapitre 5).

# Chapitre 4

# Visualisations dans l'espace réel

On propose dans ce chapitre des visualisations sur des grilles de l'espace réel. Un programme a été écrit, capable de calculer des fonctions de l'espace réel telles que le trou d'échange, la fonction de réponse, ou encore une approximation Gaussienne de la fonction de Dirac, *etc.*.. Ce programme utilise en *input* des orbitales d'un calcul `MOLPRO` quelconque calculées sur des grilles parallélépipédiques ou sur des grilles dites de type "DFT", et écrites dans des fichiers formatés ou non.

Ce travail est une sorte de préambule à un travail plus important qui consisterait à repenser la génération des points des grilles de type "DFT" pour mieux échantillonner l'espace *entre* les atomes et mieux correspondre aux besoins de calculs de corrélation longue-portée.

On montre des visualisations des orbitales POO décrites au chapitre 3, ainsi que des visualisations du trou d'échange et de la fonction de réponse.

## 4.1 À propos des grilles

Représenter une fonction $f(\mathbf{r})$ dans l'espace réel suppose de mettre en place une grille de l'espace, avec un ensemble de points discrets $\{A\}$, de sorte que l'on puisse représenter la fonction $f(A)$ sur les points de grille. Les grilles les plus communément utilisées lorsque l'on veut atteindre une représentation visuelle des objets sont des grilles parallélépipédiques, ou grilles "régulières", dans lesquelles les points sont régulièrement disposés dans les directions $x$, $y$ et $z$, voir figure 4.1. Dans le cas de ces grilles "régulières", les coordonnées des points sont entièrement connues lorsque l'on connaît un point d'origine et l'espacement entre les points dans chaque direction. On peut également utiliser des grilles développées pour la DFT, qui sont des grilles dans lesquelles les points sont judicieusement placés pour correctement représenter l'espace là où la densité est supposée être « intéressante », et sont épars voire inexistants là où la densité est attendue nulle. En pratique, on place des points munis de poids autour des atomes avec des paramètres dépendant des propriétés des atomes. Ceci permet d'obtenir des grilles avec beaucoup moins de points, et où les points sont tous importants (voir par exemple figure 4.1). Avec ces grilles, on doit explicitement donner toutes les coordonnées, ainsi que les poids qui sont associés à chaque point.

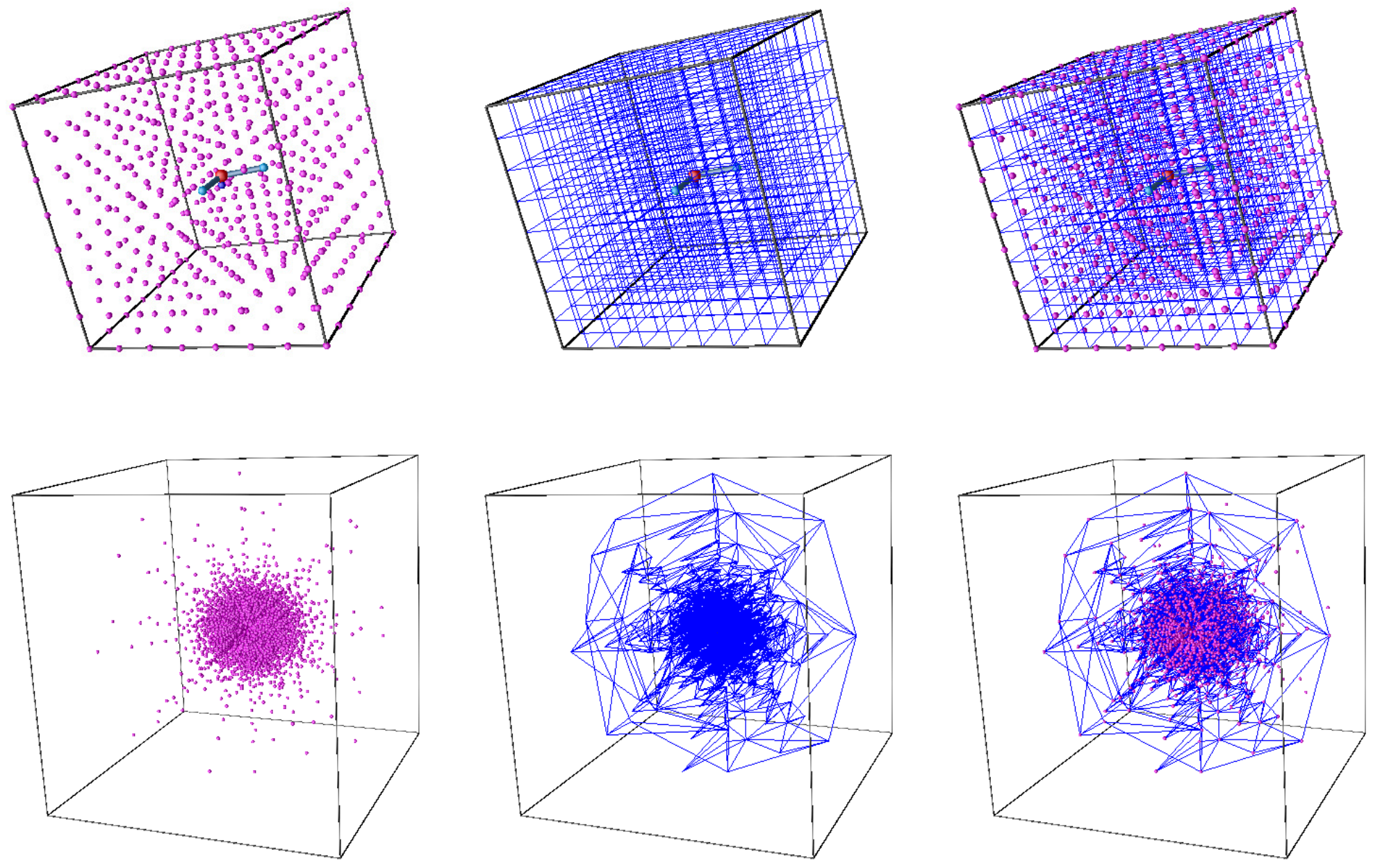

TABLE 4.1: Grilles "régulières" (haut) et de type "DFT" (bas). On montre les points de grilles à gauche, les connexions qui permettent de tracer des isosurfaces *etc.*.. au milieu, et les points avec leurs connexions à droite. (Dans le cas de la grille "régulière" seulement un point sur dix est montré, pour des raisons de clarté). On voit nettement les différences de densité et de distribution dans les deux méthodes. Notons que dans le cas des grilles de type "DFT", quelques points sont placés loin de la molécule, ce qui oblige ici à visualiser une plus grande boîte. Dans l'exemple que l'on montre ici, la grille "régulière" contient $80 \times 80 \times 80 = 512000$ points ; la grille DFT en contient 5835.

**Nous pensons** qu'un développement intéressant autour de l'utilisation de ces grilles de type "DFT" pour calculer et représenter des objets ayant trait à la corrélation longue-portée serait une reprogrammation de la génération des points de manière à échantillonner plus précisément l'espace *entre* les atomes.

J'ai écrit un programme interfacé avec `MOLPRO` [131] capable de calculer sur des grilles (de type "régulières" ou de type "DFT") des fonctions telles que la fonction de réponse $\chi(\mathbf{r}_1, \mathbf{r}_2) \doteq \chi(A, B)$, le trou d'échange $h_X(\mathbf{r}_1, \mathbf{r}_2) \doteq h_X(A, B)$, *etc.*.. À partir d'un calcul `MOLPRO` au choix, les orbitales sont récupérées dans des fichiers `CUBE` classiques dans le cas des grilles "régulières", et, dans le cas de grilles de type "DFT", dans des fichiers `CUBE` modifiés qui semblent décrire des grilles $1 \times 1 \times N_{\text{tot}}$ où $N_{\text{tot}}$ est le nombre total de points de grille. Une liste des coordonnées et des poids est alors lue dans le fichier de sortie du calcul `MOLPRO`. Les fichiers peuvent être sous forme formatée ou non.

De nombreux outils de visualisation existent pour représenter des fonctions de l'espace à l'aide de grilles réelles (citons `VMD`, ou `MoProViewer`). J'ai mis en place une procédure pour représenter des fonctions de l'espace calculées sur des grilles de type "DFT", en utilisant `OpenDX DataExplorer` (par brièveté : `DX`). Ainsi, les fonctions calculées sur des grilles par notre programme sont données en sortie, soit à nouveau sous forme de fichiers `CUBE` dans le cas des grilles "régulières", soit sous une forme lisible par `DX` dans le cas des grilles de type "DFT".

## 4.2 Orbitales localisées POO

Je propose la procédure suivante de visualisation systématique des orbitales POO : pour un ensemble LMO/POO $\left\{i, i_x, i_y, i_z\right\}$ donné, on montre les isocontours sur les trois plans de coupe ($xy$, $yz$, $zx$) centrés sur le centroïde de la LMO $i$.

Dans la figure 4.2 est présenté le résultat d'une telle procédure pour une orbitale LMO correspondant à une paire libre de l'oxygène d'une molécule d'eau et figure 4.3 pour une orbitale correspondant à une liaison C-C de l'éthylène. À chaque fois, pour chaque plan de coupe, je montre, de gauche à droite : le plan de coupe, l'isocontour de la LMO, et les isocontours des 3 POO construites à partir de la LMO. Le centroïde de la LMO, centre des plans de coupes, est représenté par un *point* noir labelisé "Q".

On voit bien l'effet de la multiplication par un élément $\hat{r}_\alpha$ dans chaque direction, c'est-à-dire la création de plans nodaux et la succession de zones de contours positifs et négatifs autour du centroïde. Pour chaque direction (c'est-à-dire pour chaque POO) un des plans de coupe est « inutile » (par exemple un plan de coupe $yz$ pour une POO créée par multiplication par $\hat{r}_x$). On voit également la détérioration relative de la localité par rapport à l'orbitale LMO.

## 4.3 Trou d'échange

[NOTE: **ce travail a depuis fait l'objet d'une publication : *Relationships between charge density response functions, exchange holes and localized orbitals.* B. Mussard, J.G. Ángyán, Comp. Theor. Chem. 1053 44-52 (2015)**]

Une idée intéressante pour évaluer la qualité des orbitales localisées est d'utiliser une relation bien connue[186–189] concernant l'expression du trou d'échange dans le contexte des orbitales localisées. On considère l'expression du trou d'échange avec des fonctions d'onde monodéterminantales :

$$h_x(\mathbf{r}_1, \mathbf{r}_2) = \frac{-1}{n(\mathbf{r}_1)} \sum_{ij} \phi_i^*(\mathbf{r}_1)\phi_j(\mathbf{r}_1)\phi_i^*(\mathbf{r}_2)\phi_j(\mathbf{r}_2), \tag{4.3.1}$$

qui, dans le cas d'orbitales parfaitement localisées, s'écrit :

$$h_x(\mathbf{r}_1, \mathbf{r}_2) = -\sum_i \frac{|\phi_i(\mathbf{r}_1)|^2}{n(\mathbf{r}_1)} |\phi_i(\mathbf{r}_2)|^2 \tag{4.3.2}$$

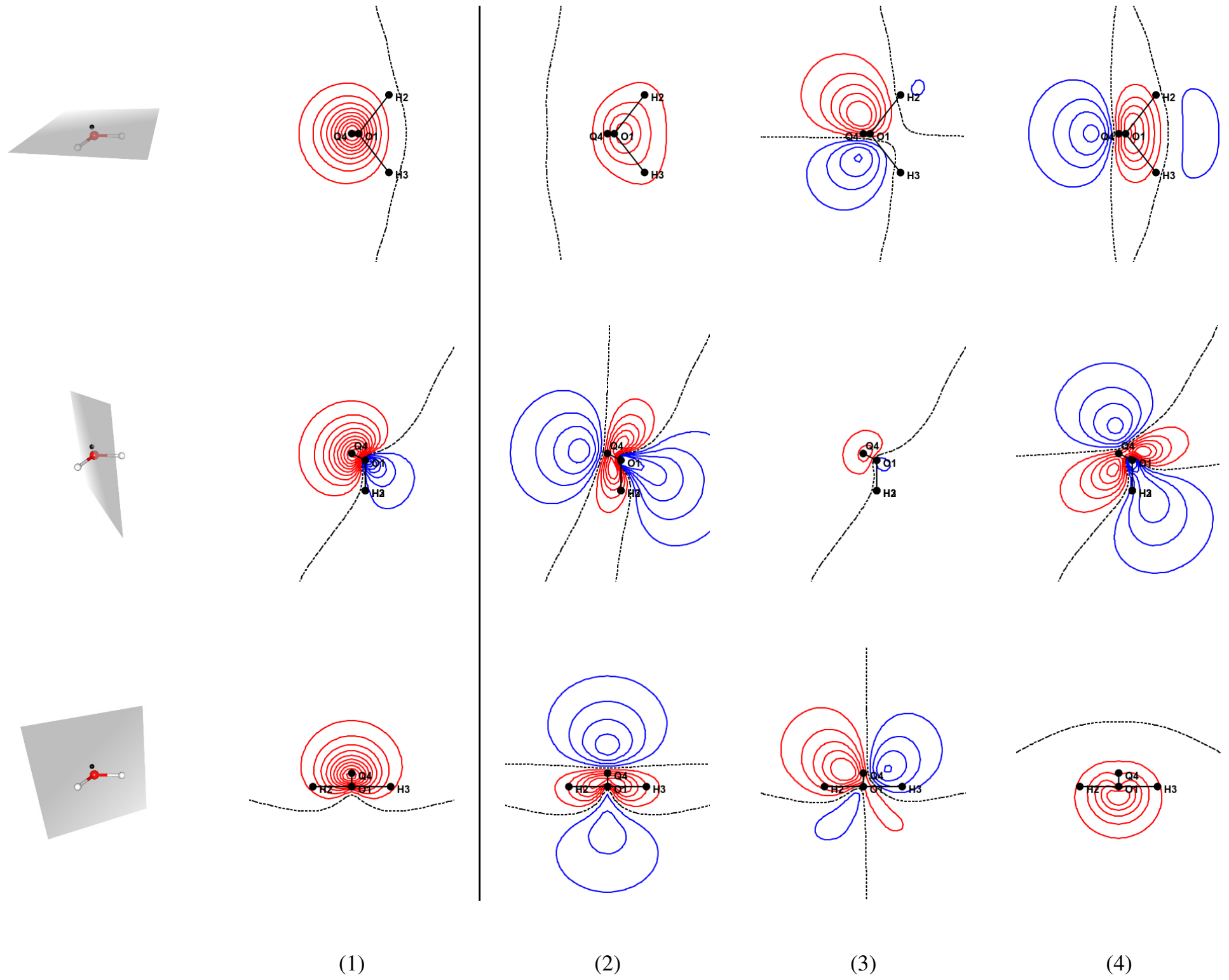

Figure 4.2: Isocontours des orbitales oscillantes projetées (POO, colonnes (2), (3) et (4)) construites à partir d'une orbitale localisée occupée (LMO, colonne (1)) correspondant à un doublet libre de l'oxygène de la molécule d'eau. On montre (tout à gauche) un schéma indiquant les plans de coupe sur lesquels sont calculés les isocontours ; on représente par un *point* noir labelisé "Q" le centroïde de la LMO.

---

On peut en effet considérer que l'espace de la molécule est divisé en domaines[31] $\Omega_i$ dans lesquels $\frac{|\phi_i(\mathbf{r}_1)|^2}{n(\mathbf{r}_1)} \approx 1$. Ainsi, avec la fonction fenêtre $\Theta_i(\mathbf{r}_1)$ (qui est égale à 1 si $\mathbf{r}_1$ est dans $\Omega_i$ et est nulle autrement), on écrit :

$$h_x(\mathbf{r}_1, \mathbf{r}_2) = -\sum_i \Theta_i(\mathbf{r}_1)\, |\phi_i(\mathbf{r}_2)|^2 \,, \tag{4.3.3}$$

c'est-à-dire que l'on peut approximer le trou d'échange lié à un point de référence $\mathbf{r}_A$ par le carré de l'orbitale qui est localisée dans le domaine de l'espace de ce point : $h_x(\mathbf{r}_A, \mathbf{r}_2) = |\phi_A(\mathbf{r}_2)|^2$, où $A$ désigne l'orbitale pour laquelle $\Theta_A(\mathbf{r}_A) = 1$.

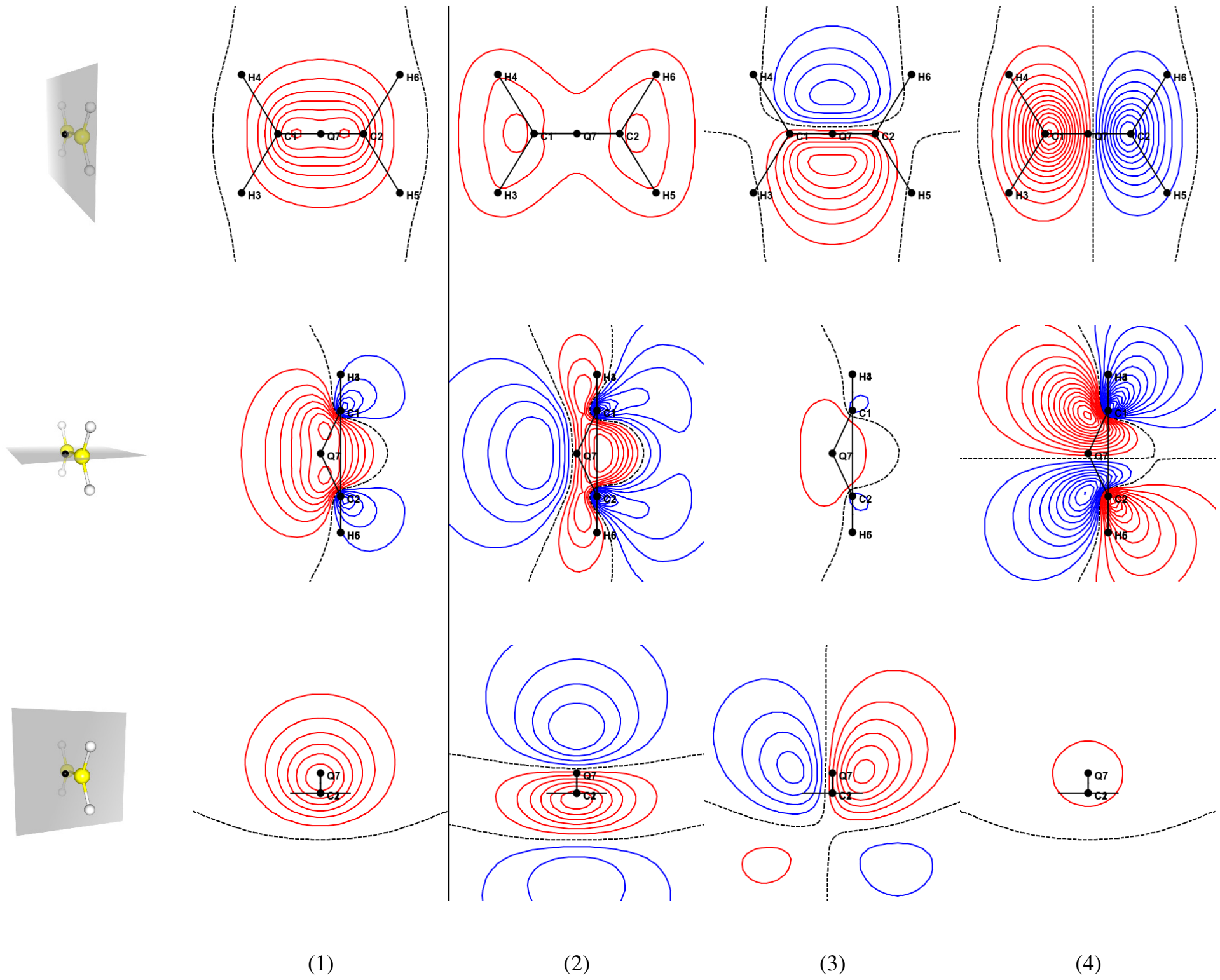

FIGURE 4.3: Isocontours des orbitales oscillantes projetées (POO, colonnes (2), (3) et (4)) construites à partir d'une orbitale localisée occupée (LMO, colonne (1)) correspondant à une liaison C-C de l'éthylène. On montre (tout à gauche) un schéma indiquant les plans de coupe sur lesquels sont calculés les isocontours ; on représente par un *point* noir labelisé "Q" le centroïde de la LMO.

---

Je montre figure 4.4 un exemple d'une telle approximation pour un point de référence $\mathbf{r}_A$ au niveau du centroïde d'une orbitale localisée correspondant à un doublet libre de l'oxygène de la molécule d'eau et figure 4.5 pour un point de référence au niveau du centroïde d'une orbitale localisée correspondant à une liaison C-C de l'éthylène. On est enclins à penser que la bonne correspondance entre les représentations des objets des deux côtés du signe égal de l'équation (4.3.3) indique la qualité de la localisation des orbitales.

## 4.4 Fonction de réponse

[NOTE: **ce travail a depuis fait l'objet d'une publication :** ***Relationships between charge density response functions, exchange holes and localized orbitals.*** **B. Mussard, J.G. Ángyán, Comp. Theor. Chem. 1053 44-52 (2015)**]

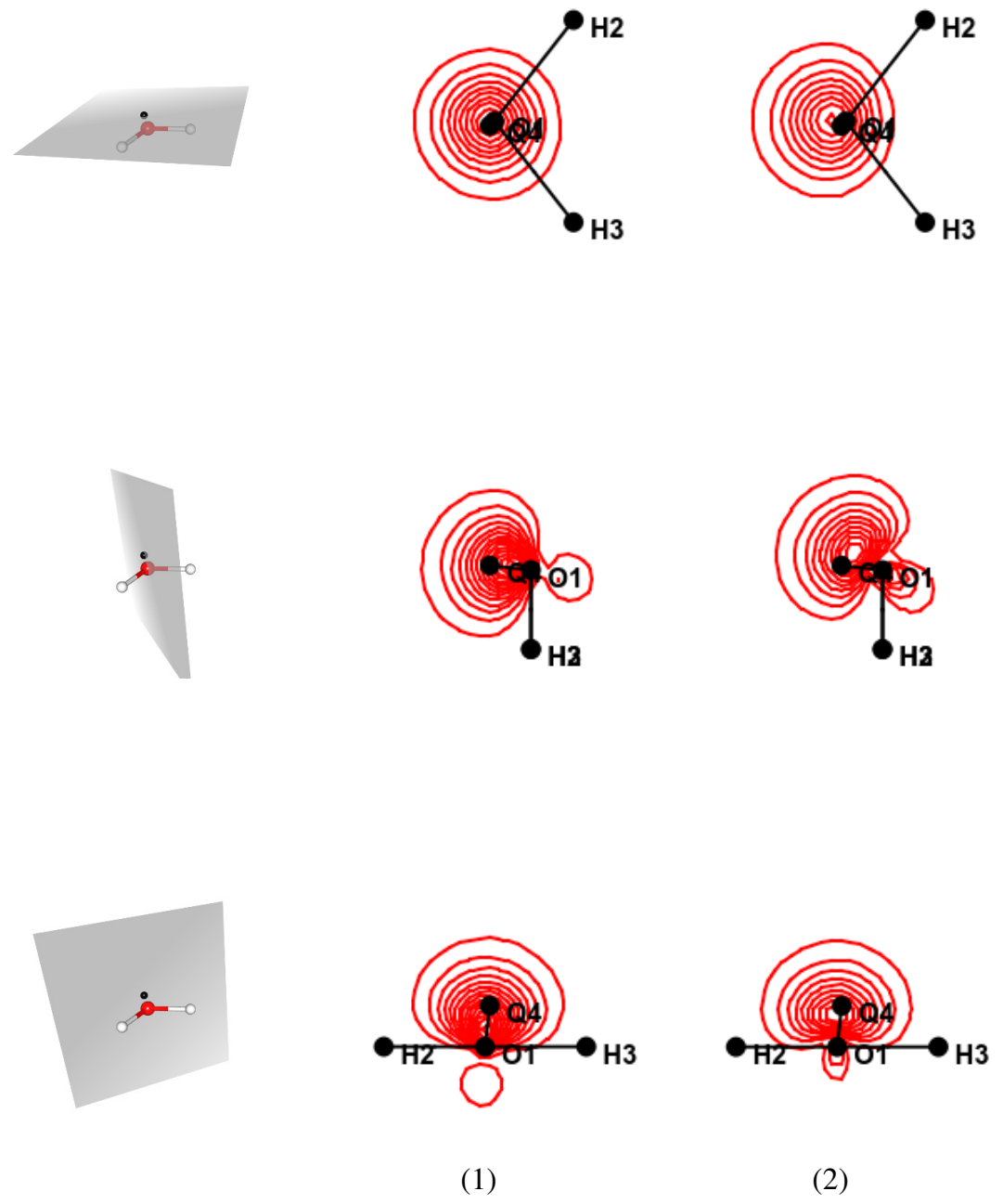

FIGURE 4.4: Représentation sur trois plans de coupe des isocontours (1) du trou d'échange correspondant au point de référence $\mathbf{r}_A$ montré par le *point* noir labelisé "Q" et (2) du carré de l'orbitale localisée dont le centroïde est le point de référence, $\mathbf{r}_A$.

---

Des définitions de la densité à deux particules $\hat{n}_2(\mathbf{r}_1, \mathbf{r}_2) = \hat{n}(\mathbf{r}_1)\hat{n}(\mathbf{r}_2) - \delta(\mathbf{r}_1, \mathbf{r}_2)\hat{n}(\mathbf{r}_1)$ et du trou d'échange-corrélation $h_{xc}(\mathbf{r}_1, \mathbf{r}_2) = \dfrac{n_2(\mathbf{r}_1, \mathbf{r}_2)}{n(\mathbf{r}_1)} - n(\mathbf{r}_2)$, on peut déduire la relation qui suit :

$$\begin{aligned}
n(\mathbf{r}_1)h_{xc}(\mathbf{r}_1, \mathbf{r}_2) &= n_2(\mathbf{r}_1, \mathbf{r}_2) - n(\mathbf{r}_1)n(\mathbf{r}_2) \\
&= \langle \hat{n}_2(\mathbf{r}_1, \mathbf{r}_2) \rangle - \langle \hat{n}(\mathbf{r}_1) \rangle \langle \hat{n}(\mathbf{r}_2) \rangle \\
&= \langle \hat{n}(\mathbf{r}_1)\hat{n}(\mathbf{r}_2) \rangle - \delta(\mathbf{r}_1, \mathbf{r}_2) \langle \hat{n}(\mathbf{r}_1) \rangle - \langle \hat{n}(\mathbf{r}_1) \rangle \langle \hat{n}(\mathbf{r}_2) \rangle \\
&= \langle \delta\hat{n}(\mathbf{r}_1)\delta\hat{n}(\mathbf{r}_2) \rangle - \delta(\mathbf{r}_1, \mathbf{r}_2) \langle \hat{n}(\mathbf{r}_1) \rangle ,
\end{aligned} \tag{4.4.1}$$

où $\langle \sqcup \rangle$ désigne une valeur moyenne sur l'état fondamental, $\langle 0|\sqcup|0 \rangle$ et où on utilise $\hat{n} = n + \delta\hat{n}$ (voir 2.2). Ainsi :

$$n(\mathbf{r}_1)h_{xc}(\mathbf{r}_1, \mathbf{r}_2) + \delta(\mathbf{r}_1, \mathbf{r}_2)n(\mathbf{r}_1) = \langle \delta\hat{n}(\mathbf{r}_1)\delta\hat{n}(\mathbf{r}_2) \rangle \tag{4.4.2}$$

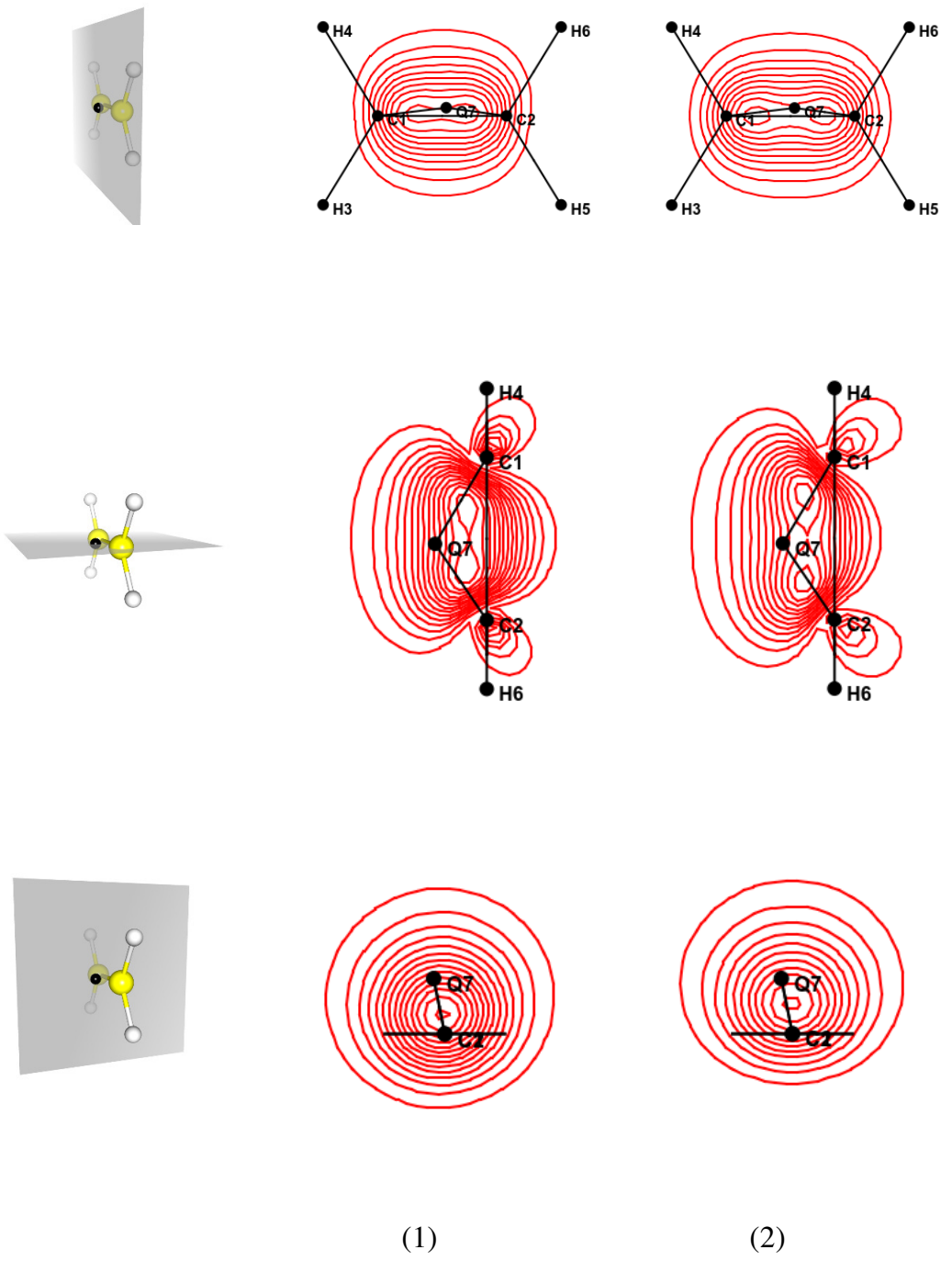

FIGURE 4.5: Représentation sur trois plans de coupe des isocontours (1) du trou d'échange correspondant au point de référence $\mathbf{r}_A$ montré par le *point* noir labelisé "Q" et (2) du carré de l'orbitale localisée dont le centroïde est le point de référence, $\mathbf{r}_A$.

On retrouve une expression qui met en jeu les *fluctuations* de densité de charge. Comme on a vu dans le chapitre sur la RPA, et notamment équation (2.2.4), ces fluctuations sont liées à la fonction de réponse par le théorème de fluctuation-dissipation. On va d'ailleurs voir à présent que l'on peut lier l'équation (4.4.2) à la fonction de réponse *statique*. Ainsi, comme il sera rendu abondamment clair dans la suite (voir chapitre 5), on peut approximer la fonction de réponse en utilisant des dénominateurs effectifs de sorte à « faire monter » la sommation sur les états excités au numérateur, y utiliser une résolution de l'identité et se débarrasser du besoin de connaître le spectre d'excitation. Ceci appliqué simplement à la fonction de réponse *statique* donne :

$$\chi(\mathbf{r}_1, \mathbf{r}_2; 0) = \sum_{\alpha \neq 0} \frac{n_\alpha(\mathbf{r}_1) n^*_\alpha(\mathbf{r}_2)}{i\eta^+ - \Omega_\alpha} + \frac{n^*_\alpha(\mathbf{r}_1) n_\alpha(\mathbf{r}_2)}{i\eta^+ - \Omega_\alpha} \tag{4.4.3}$$

$$= 2 \sum_{\alpha \neq 0} \frac{n_\alpha(\mathbf{r}_1) n_\alpha(\mathbf{r}_2)}{i\eta^+ - \Omega_\alpha}$$

$$\doteq 2 \frac{\sum_{\alpha \neq 0} n_\alpha(\mathbf{r}_1) n_\alpha(\mathbf{r}_2)}{\Omega(\mathbf{r}_1, \mathbf{r}_2)} \tag{4.4.4}$$

$$\approx 2 \frac{\sum_{\alpha \neq 0} n_\alpha(\mathbf{r}_1) n_\alpha(\mathbf{r}_2)}{\overline{\Omega}} \propto \sum_{\alpha \neq 0} n_\alpha(\mathbf{r}_1) n_\alpha(\mathbf{r}_2), \tag{4.4.5}$$

où, à part la supposition que les orbitales soient réelles, seule la dernière étape est une approximation, c'est-à-dire seule l'étape qui réduit la fonction $\Omega(\mathbf{r}_1, \mathbf{r}_2)$ définie ligne (4.4.4) à une simple constante $\overline{\Omega}$ est une approximation. On fait le lien entre la forme de la fonction de réponse vue ligne (4.4.3) et celles vues dans le chapitre sur la RPA dans l'Annexe A.6.3. L'approximation que l'on voit ici est une esquisse d'une technique largement développée dans la suite, et d'une certaine manière vue dans les références[31, 153]. L'idée pour le moment est simplement de pouvoir montrer qu'au final, une comparaison des équations (4.4.2) et (4.4.5) permet d'écrire :

$$\chi(\mathbf{r}_1, \mathbf{r}_2; 0) \propto n(\mathbf{r}_1) h_{xc}(\mathbf{r}_1, \mathbf{r}_2) + \delta(\mathbf{r}_1, \mathbf{r}_2) n(\mathbf{r}_1) \tag{4.4.6}$$

Les portées de cette équation sont multiples. Elle permet de comprendre que la réponse de la densité de charge d'un système au point $\mathbf{r}_2$ à une perturbation extérieure appliquée en $\mathbf{r}_1$ peut être prédite à partir de la connaissance du trou d'échange-corrélation du point de référence $\mathbf{r}_1$. Ainsi la réponse du système (la redistribution des électrons ; c'est-à-dire la polarisation des électrons) ne sera non-nulle que dans l'espace où le trou est non-nul. Ceci lie intimement les notions de trou d'échange-corrélation, de fonction de réponse, et de localisation des électrons[153].

Avec le programme décrit section 4.1, on peut calculer sur une grille les fonctions $\chi(\mathbf{r}_A, \mathbf{r}_B)$, $h_{xc}(\mathbf{r}_A, \mathbf{r}_B)$ et $\delta(\mathbf{r}_A, \mathbf{r}_B)$, de sorte que l'on peut représenter visuellement les objets à gauche et à droite du signe égal de l'équation (4.4.6).

De telles représentations sont montrées figure 4.6 pour un point de référence $\mathbf{r}_A$ placé au niveau du centroïde d'une orbitale localisée correspondant à un doublet libre de l'oxygène de la molécule d'eau et figure 4.7 pour un point de référence placé au niveau de la liaison C-C de l'éthylène. On voit que la structure primaire de la fonction de réponse est récupérée par l'approximation vue équation (4.4.6).

## 4.5 Conclusion

Le travail présenté ici rend compte de pistes de reflexions autour de l'utilisation de grilles de l'espace réel pour des calculs de corrélation RPA. Cette possibilité a toujours été repoussée à cause de la résolution nécessaire lorsque l'on utilise des grilles parallélépipédiques, et donc du nombre rédhibitoire de points de grilles sur lesquels faire des sommations. L'utilisation (et à terme : l'adaptation) des grilles de type "DFT" devrait ouvrir la voie à d'intéressants développements à ce sujet. Avec le

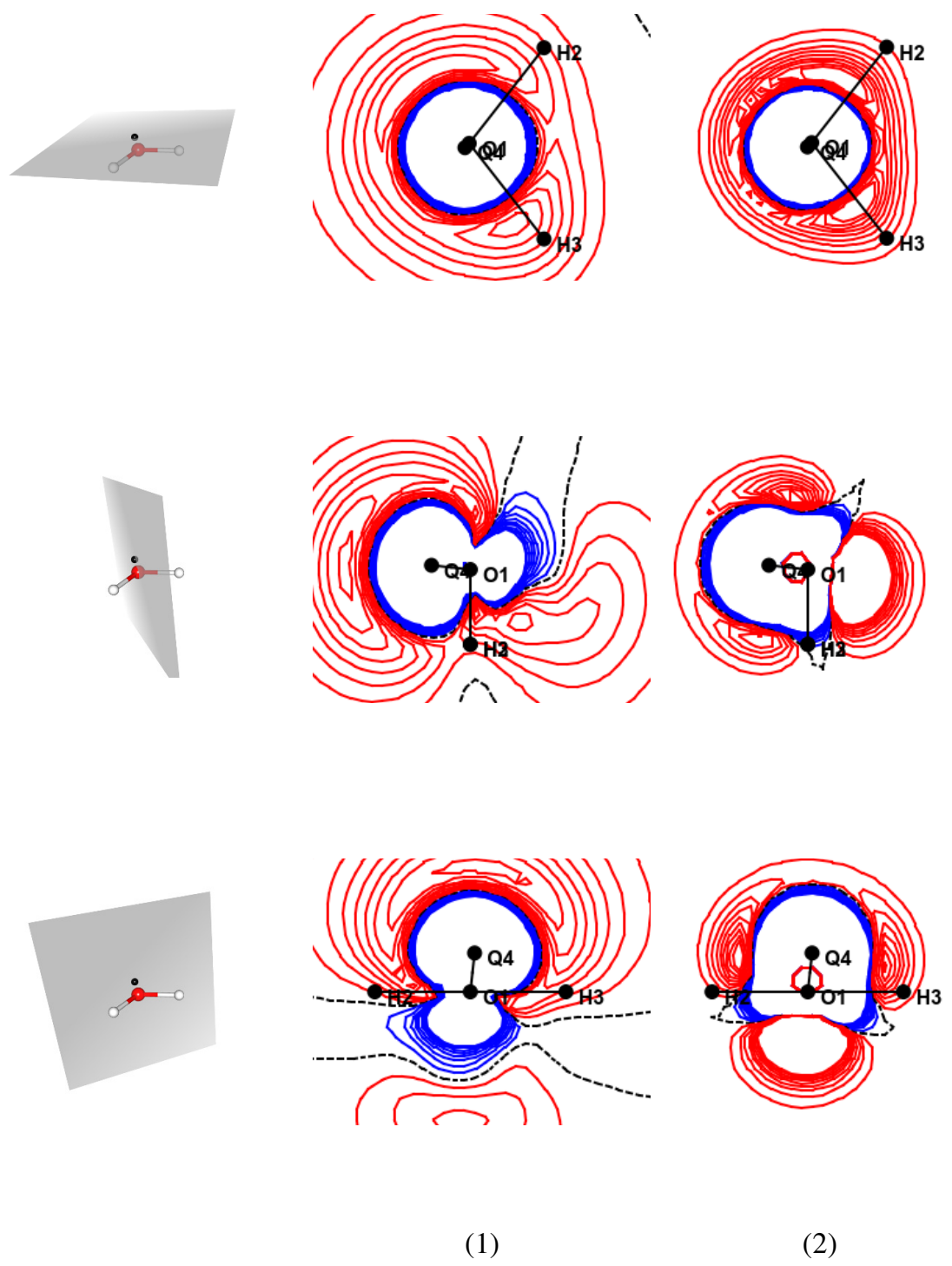

FIGURE 4.6: Représentation sur trois plans de coupe des isocontours de (1) la fonction de réponse correspondant au point de référence $\mathbf{r}_A$ montré par le *point* noir labelisé "Q" et de (2) l'objet vu à droite du signe égal de l'équation (4.4.6), également au point de référence $\mathbf{r}_A$.

---

programme que j'ai écrit lors de ces réflexions, il est tout de même d'ores et déjà possible d'obtenir des visualisations, entre autres : de fonctions de réponse. Ceci est un outil assez peu répandu malgré son intérêt certain pour atteindre une meilleure compréhension des phénomènes liés à la corrélation.

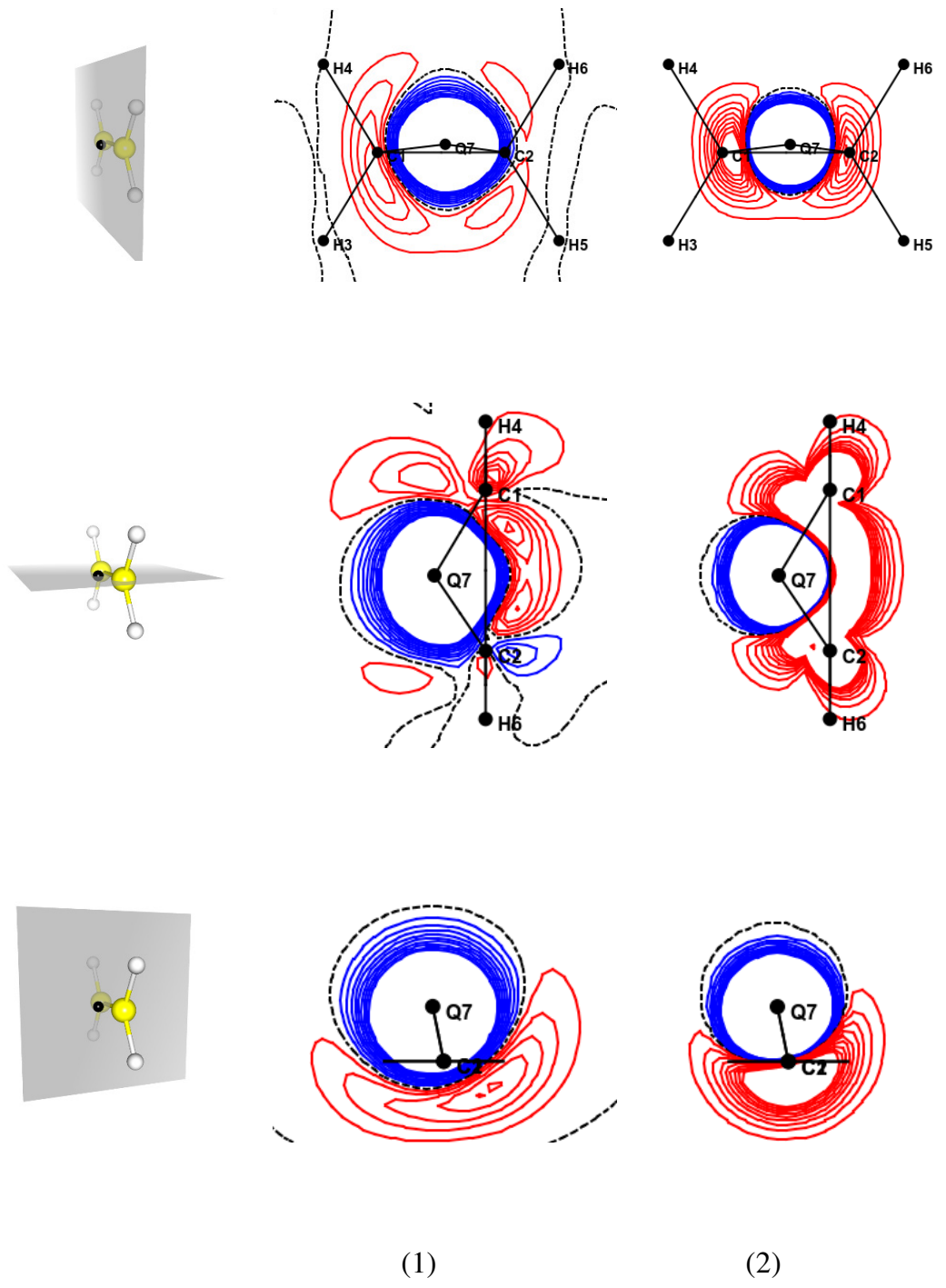

FIGURE 4.7: Représentation sur trois plans de coupe des isocontours de (1) la fonction de réponse correspondant au point de référence $\mathbf{r}_A$ montré par le *point* noir labelisé "Q" et de (2) l'objet vu à droite du signe égal de l'équation (4.4.6), également au point de référence $\mathbf{r}_A$.

# Chapitre 5

# EED : Approximation du dénominateur effectif

Le but de ce chapitre est de montrer un développement d'une approximation appelée EED (pour *Effective Energy Denominator*) et qui consiste à écrire la "sommation de fractions" qui apparaît dans les expressions des fonctions de réponse comme une "fraction d'une sommation", c'est-à-dire que l'on écrit un objet qui autorise l'application d'une résolution de l'identité au numérateur, et permet ainsi de s'affranchir de la connaissance du spectre de l'hamiltonien. Cette technique est une généralisation de l'approximation de Unsöld où l'énergie moyenne dépend de la fréquence. On trouve en Annexe E les détails des dérivations.

Ce développement se place dans l'espace direct, c'est-à-dire que l'on manipule des objets avec leurs dépendances aux coordonnées spatiales, par opposition aux habitudes de présenter des développements dans les espaces de Hilbert, c'est-à-dire sur des objets écrits avec des éléments de matrice sur une base d'orbitales.

## 5.1 Contexte et motivation

Habituellement, on travaille en chimie quantique avec une représentation matricielle des objets qui nous intéressent, où les dimensions du problème sont déterminées par celles de l'espace des fonctions de base (par exemple orbitales atomiques, ondes planes, *etc.*.. ). On réfère souvent à une telle technique comme une approche *spectrale*, étant donné qu'il s'agit d'une expansion de type Fourier de la fonction cible ; ce que l'on peut mettre en opposition avec des méthodes qui représentent cette fonction cible directement sur une grille de points soit dans l'espace direct, soit dans l'espace réciproque. Typiquement, en DFT, on utilise une représentation sur grille de la densité électronique, qui est une fonction relativement simple à 3 variables. On a vu qu'en RPA la quantité fondamentale est la fonction de réponse (et les fonctions de corrélation de paire), qui dépendent de 2 fois 3 variables d'espace, ce qui entraîne une complexité considérablement plus importante.

Dans les expressions de la fonction de réponse que l'on rencontre dans les autres sections, on a besoin du spectre complet des excitations de l'hamiltonien, c'est-à-dire que l'on a besoin d'une lourde sommation sur les états virtuels. On montre ici une sorte de généralisation de l'approximation de Unsöld[190], autrement connue sous le nom d'« approximation du dénominateur effectif ». L'idée de base, très simple, est de sortir le dénominateur de la sommation en définissant un dénominateur effectif. La sommation ne porte plus ainsi que sur le numérateur, sur lequel on applique une simple résolution de l'identité pour se débarrasser de la sommation sur les états excités.

Lors des applications de l'approximation d'Unsöld, le dénominateur effectif peut être choisi de différentes manières : à partir du premier potentiel d'ionisation[191], pour reproduire une valeur expérimentale, ou à partir des considérations de perturbatives[192]. Ces choix présentent leurs limites lorsque l'on considère la quantité de la fonction de réponse approximée sur l'ensemble de la gamme de fréquence.

## 5.2 Technique de l'énergie effective

Une approche alternative, potentiellement exacte, a été proposée par Berger *et. al.* [193, 194] sous la forme d'un dénominateur effectif qui est fréquence-dépendant. Cette technique porte le nom d'approximation du dénominateur effectif (*Effective Energy Denominator* : EED). La dérivation de Berger *et. al.* est faite dans le cadre de la physique du solide, donc en espace réciproque. Je montre ici une étude du sujet en espace réel.

On utilise une forme de la fonction de réponse $\chi$ qui nous sera pratique dans la suite. Cette forme est dérivée dans la section A.6.3, et s'écrit (se souvenir que la quantité $i\eta^+$ que l'on voit dans l'Annexe tend *in fine* vers zéro et n'est présente que pour justifier certains raisonnements, voir A.6.1 et surtout A.6.2) :

$$\chi(\mathbf{r}_1, \mathbf{r}_2; i\omega) = \sum_{\alpha\neq 0}\left\{\frac{n_\alpha(\mathbf{r}_1)n_\alpha^*(\mathbf{r}_2)}{i\omega - \Omega_\alpha} + \frac{n_\alpha^*(\mathbf{r}_1)n_\alpha(\mathbf{r}_2)}{-i\omega - \Omega_\alpha}\right\} = \chi_+(\mathbf{r}_1, \mathbf{r}_2; i\omega) + \chi_-(\mathbf{r}_1, \mathbf{r}_2; i\omega), \tag{5.2.1}$$

où l'on observe la propriété suivante : $\chi_+^*(\mathbf{r}_1, \mathbf{r}_2; -\omega) = \chi_-(\mathbf{r}_1, \mathbf{r}_2; \omega)$. Ainsi, la procédure que l'on va décrire dans la suite peut être appliquée à $\chi_+$ et $\chi_-$. Ceci générerait, comme on va le voir dans la suite, une série d'objets $\Omega_\pm^{nn}$, $\Omega_\pm^{nj}$, *etc...* qui sont tous liés par le même type de relation qui lie $\chi_\pm$. Il semble plus pédagogique de travailler sous l'hypothèse d'orbitales réelles, où on a $n_\alpha^*(\mathbf{r}_2) = n_\alpha(\mathbf{r}_2)$ et où on peut écrire :

$$\chi(\mathbf{r}_1, \mathbf{r}_2; i\omega) = \sum_{\alpha\neq 0} n_\alpha(\mathbf{r}_1)n_\alpha(\mathbf{r}_2)\left\{\frac{1}{i\omega - \Omega_\alpha} + \frac{1}{-i\omega - \Omega_\alpha}\right\} = 2\mathrm{Re}\left\{\sum_{\alpha\neq 0}\frac{n_\alpha(\mathbf{r}_1)n_\alpha(\mathbf{r}_2)}{i\omega - \Omega_\alpha}\right\}, \tag{5.2.2}$$

et d'appliquer le procédure directement à $\chi \doteq \sum_{\alpha\neq 0}\frac{n_\alpha(\mathbf{r}_1)n_\alpha(\mathbf{r}_2)}{i\omega - \Omega_\alpha}$.

La procédure est donc la suivante : on définit le dénominateur effectif (l'EED) $\Omega^{nn}(\mathbf{r}_1, \mathbf{r}_2, i\omega)$ par la relation :

$$\chi(\mathbf{r}_1, \mathbf{r}_2; i\omega) = \frac{\sum_{\alpha\neq 0} n_\alpha(\mathbf{r}_1)n_\alpha(\mathbf{r}_2)}{i\omega - \Omega^{nn}(\mathbf{r}_1, \mathbf{r}_2, i\omega)}, \tag{5.2.3}$$

qui donne une nouvelle expression (en principe exacte) de la fonction de réponse. Grâce à cette formulation, on peut utiliser la résolution de l'identité au numérateur et obtenir la fonction de réponse comme une valeur moyenne sur l'état fondamental (voir section 5.3) pourvu, bien sur, que $\Omega^{nn}$ soit connu.

Notons que l'énergie effective $\Omega^{nn}(\mathbf{r}_1, \mathbf{r}_2, i\omega)$ dépend à la fois des deux coordonnées qui sont impliquées dans la fonction réponse, ainsi que de la fréquence $y$. Cette dépendance peut être considérée comme une généralisation naturelle d'une observation rapportée dans la littérature à propos des utilisations diverses de l'approximation de Unsöld : il n'existe pas *une* énergie d'excitation moyenne pour un système donné ; il faut dériver ou ajuster des énergies effectives séparément pour les différentes composantes de la polarisabilité statique, pour la polarisabilité moyenne, pour l'estimation du coefficient de dispersion, *etc.*.. [192].

Avec cette méthode, ce qu'il faut trouver à présent c'est une expression exacte pour l'EED, et un cadre pour l'obtenir par des approximations successives. Une relation intéressante est obtenue en multipliant l'équation (5.2.3) de chaque côté par $(i\omega - \Omega^{nn}(\mathbf{r}_1, \mathbf{r}_2, i\omega))$ puis en multipliant chaque terme de la somme sur les états excités à droite du signe égal par $1 = \frac{i\omega-\Omega_\alpha}{i\omega-\Omega_\alpha}$ :

$$(i\omega - \Omega^{nn}(\mathbf{r}_1, \mathbf{r}_2, i\omega))\ \chi(\mathbf{r}_1, \mathbf{r}_2; i\omega) = \sum_{\alpha\neq 0} \frac{n_\alpha(\mathbf{r}_1) n_\alpha(\mathbf{r}_2)}{i\omega - \Omega_\alpha} (i\omega - \Omega_\alpha), \tag{5.2.4}$$

qui se simplifie en :

$$\Omega^{nn}(\mathbf{r}_1, \mathbf{r}_2, i\omega)\ \chi^{nn}(\mathbf{r}_1, \mathbf{r}_2; i\omega) = \sum_{\alpha\neq 0} \frac{n_\alpha(\mathbf{r}_1) n_\alpha(\mathbf{r}_2)\Omega_\alpha}{i\omega - \Omega_\alpha} \doteq \chi^{nj} \tag{5.2.5}$$

La nouvelle quantité obtenue à la droite de l'équation (5.2.5) ressemble à une fonction de réponse : on la désigne par $\chi^{nj}$, tandis que la fonction réponse de densité de charge a été désignée comme $\chi^{nn}$.

**L'idée est** d'appliquer la même procédure de manière répétée pour obtenir une hiérarchie d'objets $\chi^{nn}, \chi^{nj} \ldots, \Omega^{nn}, \Omega^{nj} \ldots$, ouvrant ainsi la porte à un cadre précis où approximer $\chi^{nn}$.

Ainsi on définit un deuxième EED sur la « quasi-fonction de réponse » $\chi^{nj}$ :

$$\chi^{nj}(\mathbf{r}_1, \mathbf{r}_2; i\omega) = \frac{\sum_{\alpha\neq 0} n_\alpha(\mathbf{r}_1) n_\alpha(\mathbf{r}_2)\Omega_\alpha}{i\omega - \Omega^{nj}(\mathbf{r}_1, \mathbf{r}_2, i\omega)}, \tag{5.2.6}$$

conduisant au même type de relation que nous avons vu dans l'équation (5.2.5) :

$$\Omega^{nj}(\mathbf{r}_1, \mathbf{r}_2, i\omega)\chi^{nj}(\mathbf{r}_1, \mathbf{r}_2; i\omega) = \sum_{\alpha\neq 0} \frac{n_\alpha(\mathbf{r}_1) n_\alpha(\mathbf{r}_2)\Omega_\alpha\Omega_\alpha}{i\omega - \Omega_\alpha}, \tag{5.2.7}$$

qui, à son tour, définit une nouvelle fonction $\chi^{jj}$ pour laquelle on peut écrire :

$$\chi^{jj}(\mathbf{r}_1, \mathbf{r}_2; i\omega) = \sum_{\alpha\neq 0} \frac{n_\alpha(\mathbf{r}_1)n_\alpha(\mathbf{r}_2)\Omega_\alpha\Omega_\alpha}{i\omega - \Omega_\alpha} = \frac{\sum_{\alpha\neq 0} n_\alpha(\mathbf{r}_1)n_\alpha(\mathbf{r}_2)\Omega_\alpha\Omega_\alpha}{i\omega - \Omega^{jj}(\mathbf{r}_1, \mathbf{r}_2; i\omega)}, \tag{5.2.8}$$

et ainsi de suite : la procédure itérative peut continuer *ad infinitum*.

## 5.3 Expression des numérateurs

Une observation attentive des numérateurs des équations (5.2.3) (5.2.6) et (5.2.8) va nous permettre de montrer que l'on peut les exprimer comme des valeurs moyennes sur l'état fondamental, c'est-à-dire que l'on peut s'affranchir de la connaissance du spectre de l'hamiltonien pour les calculer. Je fais également émerger des relations avec les règles de somme[195], un aspect qui est plus développé Annexe E.5.

Comme il a déjà été vu, par exemple équation (4.4.1), le numérateur de la définition du premier EED, équation (5.2.3), va s'écrire :

$$\begin{aligned}
\sum_{\alpha\neq 0} n_\alpha(\mathbf{r}_1)n_\alpha(\mathbf{r}_2) &= \frac{1}{2}\sum_{\alpha\neq 0} \langle 0|\hat{n}(\mathbf{r}_1)|\alpha\rangle \langle\alpha|\hat{n}(\mathbf{r}_2)|0\rangle + \langle 0|\hat{n}(\mathbf{r}_2)|\alpha\rangle \langle\alpha|\hat{n}(\mathbf{r}_1)|0\rangle \\
&= \frac{1}{2}\left(\langle\hat{n}(\mathbf{r}_1)\hat{n}(\mathbf{r}_2)\rangle - n(\mathbf{r}_1)n(\mathbf{r}_2) + \langle\hat{n}(\mathbf{r}_2)\hat{n}(\mathbf{r}_1)\rangle - n(\mathbf{r}_2)n(\mathbf{r}_1)\right) \\
&= \frac{1}{2}\left(\langle\delta\hat{n}(\mathbf{r}_1)\delta\hat{n}(\mathbf{r}_2)\rangle + \langle\delta\hat{n}(\mathbf{r}_2)\delta\hat{n}(\mathbf{r}_1)\rangle\right),
\end{aligned} \tag{5.3.1}$$

où $\langle\sqcup\rangle$ désigne une valeur moyenne sur l'état fondamental, $\langle 0|\sqcup|0\rangle$. Ainsi le numérateur s'écrit en fait (voir Annexe E.5) :

$$S^{nn}(\mathbf{r}_1, \mathbf{r}_2) = \frac{1}{2}\Big(S_{-1}(\mathbf{r}_1, \mathbf{r}_2) + S_{-1}(\mathbf{r}_2, \mathbf{r}_1)\Big) \tag{5.3.2}$$

J'ai désigné ce numérateur par $S^{nn}$, de sorte que l'on peut écrire l'équation (5.2.3) d'une manière abrégée :

$$\chi^{nn} = \frac{S^{nn}}{i\omega - \Omega^{nn}} \tag{5.3.3}$$

Concernant le numérateur de l'équation (5.2.6), on considère que :

$$\begin{aligned}
\sum_{\alpha\neq 0} n_\alpha(\mathbf{r}_1)n_\alpha(\mathbf{r}_2)\Omega_\alpha = \frac{1}{4}\sum_{\alpha\neq 0} &\langle 0|\hat{n}(\mathbf{r}_1)|\alpha\rangle \Big(\langle\alpha|\hat{n}(\mathbf{r}_2)|0\rangle\,\Omega_\alpha\Big) + \Big(\Omega_\alpha\,\langle 0|\hat{n}(\mathbf{r}_1)|\alpha\rangle\Big)\langle\alpha|\hat{n}(\mathbf{r}_2)|0\rangle \\
&+ \langle 0|\hat{n}(\mathbf{r}_2)|\alpha\rangle \Big(\langle\alpha|\hat{n}(\mathbf{r}_1)|0\rangle\,\Omega_\alpha\Big) + \Big(\Omega_\alpha\,\langle 0|\hat{n}(\mathbf{r}_2)|\alpha\rangle\Big)\langle\alpha|\hat{n}(\mathbf{r}_1)|0\rangle,
\end{aligned} \tag{5.3.4}$$

où l'on utilise le théorème hyperviriel[196, 197] (voir en Annexe E.1 les quelques lignes qui permettent de démontrer cette relation) pour écrire :

$$\begin{aligned}\sum_{\alpha\neq 0} n_\alpha(\mathbf{r}_1)n_\alpha(\mathbf{r}_2)\Omega_\alpha &= \frac{1}{4}\sum_{\alpha\neq 0} \langle 0|\hat{n}(\mathbf{r}_1)|\alpha\rangle \left\langle \alpha\middle|\left[\hat{H},\hat{n}(\mathbf{r}_2)\right]\middle|0\right\rangle - \left\langle 0\middle|\left[\hat{H},\hat{n}(\mathbf{r}_1)\right]\middle|\alpha\right\rangle \langle \alpha|\hat{n}(\mathbf{r}_2)|0\rangle \\ &\qquad + \langle 0|\hat{n}(\mathbf{r}_2)|\alpha\rangle \left\langle \alpha\middle|\left[\hat{H},\hat{n}(\mathbf{r}_1)\right]\middle|0\right\rangle - \left\langle 0\middle|\left[\hat{H},\hat{n}(\mathbf{r}_2)\right]\middle|\alpha\right\rangle \langle \alpha|\hat{n}(\mathbf{r}_1)|0\rangle \\ &= \frac{1}{4}\left(\left\langle\left[\hat{n}(\mathbf{r}_1),\left[\hat{H},\hat{n}(\mathbf{r}_2)\right]\right] + \left[\hat{n}(\mathbf{r}_2),\left[\hat{H},\hat{n}(\mathbf{r}_1)\right]\right]\right\rangle\right), \end{aligned} \tag{5.3.5}$$

où $[\sqcup,\sqcup]$ est un commutateur. On utilise, pour passer de la première à la deuxième ligne, le fait que $\left\langle\left[\hat{H},\hat{n}(\mathbf{r})\right]\right\rangle = 0$ pour appliquer une résolution de l'identité. On montre dans les Annexes E.2 et E.3 que le commutateur est lié à la divergence de la densité de courant : $\left[\hat{H},\hat{n}(\mathbf{r}_2)\right] = i\nabla_{\mathbf{r}_2}\cdot\hat{\mathbf{j}}(\mathbf{r}_2)$. Ceci explique la notation $\sqcup^{nj}$ utilisée pour la fonction de réponse $\chi^{nj}$ ; je désigne d'ailleurs ce numérateur par $S^{nj}$ :

$$S^{nj}(\mathbf{r}_1,\mathbf{r}_2) = \frac{1}{2}\Big(S_0(\mathbf{r}_1,\mathbf{r}_2) + S_0(\mathbf{r}_2,\mathbf{r}_1)\Big), \tag{5.3.6}$$

et on écrit :

$$\chi^{nj} = \frac{S^{nj}}{i\omega - \Omega^{nj}} \tag{5.3.7}$$

Dans les développements de l'EED dans l'espace réciproque, ce commutateur est travaillé de sorte à faire sortir deux termes différents (voir par exemple les équations (6) à (9) de [193] et (6) à (10) de [194]) : le lecteur trouvera dans l'Annexe E.4 une discussion sur le lien entre ces dérivations et l'expression de notre commutateur en fonction de la densité de courant.

De manière analogue, le théorème hyperviriel sur le numérateur de l'équation (5.2.8) nous permet d'écrire :

$$\begin{aligned}\sum_{\alpha\neq 0} n_\alpha(\mathbf{r}_1)n_\alpha(\mathbf{r}_2)\Omega_\alpha\Omega_\alpha &= \frac{1}{2}\sum_{\alpha\neq 0}\Big(\Omega_\alpha \langle 0|\hat{n}(\mathbf{r}_1)|\alpha\rangle\Big)\Big(\langle \alpha|\hat{n}(\mathbf{r}_2)|0\rangle\,\Omega_\alpha\Big) + \Big(\Omega_\alpha \langle 0|\hat{n}(\mathbf{r}_2)|\alpha\rangle\Big)\Big(\langle \alpha|\hat{n}(\mathbf{r}_1)|0\rangle\,\Omega_\alpha\Big) \\ &= -\frac{1}{2}\sum_{\alpha\neq 0}\left\langle 0\middle|\left[\hat{H},\hat{n}(\mathbf{r}_1)\right]\middle|\alpha\right\rangle\left\langle \alpha\middle|\left[\hat{H},\hat{n}(\mathbf{r}_2)\right]\middle|0\right\rangle + \left\langle 0\middle|\left[\hat{H},\hat{n}(\mathbf{r}_2)\right]\middle|\alpha\right\rangle\left\langle \alpha\middle|\left[\hat{H},\hat{n}(\mathbf{r}_1)\right]\middle|0\right\rangle \\ &= -\frac{1}{2}\left\langle\left[\hat{H},\hat{n}(\mathbf{r}_1)\right]\left[\hat{H},\hat{n}(\mathbf{r}_2)\right] + \left[\hat{H},\hat{n}(\mathbf{r}_2)\right]\left[\hat{H},\hat{n}(\mathbf{r}_1)\right]\right\rangle, \end{aligned} \tag{5.3.8}$$

qui se réduit à une moyenne sur l'état fondamental pour les mêmes raisons que précédemment. Ce numérateur est appelé $S^{jj}$ pour des raisons qui sont à présent évidentes, et on peut écrire (voir Annexe E.5) :

$$S^{jj}(\mathbf{r}_1,\mathbf{r}_2) = \frac{1}{2}\Big(S_1(\mathbf{r}_1,\mathbf{r}_2) + S_1(\mathbf{r}_2,\mathbf{r}_1)\Big), \tag{5.3.9}$$

d'où :

$$\chi^{jj} = \frac{S^{jj}}{i\omega - \Omega^{jj}} \tag{5.3.10}$$

On a dérivé une série d'objets connus ($S^{nn}$, $S^{nj}$ et $S^{jj}$) qui permettent d'exprimer les premières approximations d'une séquence hiérarchique potentiellement exacte par de simples moyennes sur l'état fondamental, c'est-à-dire sans connaître l'ensemble des états excités (mais pourvu que l'on soit capable de calculer les EED $\Omega^{nn}$, *etc.*.. ).

Toutefois, il serait trop tôt pour se réjouir : d'une part, le théorème hyperviriel, que nous avons évoqué lors des dérivations n'est valable que pour les solutions exactes de l'équation de Schrödinger, d'autre part les calculs de ces moyennes sur l'état fondamental sont loin d'être triviaux à effectuer, car nous avons affaire à des opérateurs à deux électrons, venant de produits d'opérateurs mono-électroniques. Avant aborder la question d'un cadre pratique de l'utilisation de ces moyennes sur l'état fondamental, je vais montrer comment clore cette série des approximations successives de l'EED.

## 5.4 Approximations des EED

L'objet d'intérêt ici est $\chi^{nn}$ ; les numérateurs $S^{nn}$, $S^{nj}$ et $S^{jj}$ sont connus et les dénominateurs effectifs $\Omega^{\sqcup}$ sont les éléments qu'il nous faut approximer. Les définitions les plus explicites des EED sont obtenues lorsque l'on considère, par exemple pour $\Omega^{nn}$ :

$$\Omega^{nn} = \frac{\chi^{nj}}{\chi^{nn}} = \frac{S^{nj}}{S^{nn}} \frac{i\omega - \Omega^{nn}}{i\omega - \Omega^{nj}}, \tag{5.4.1}$$

dans laquelle, à nouveau, on peut exprimer explicitement $\Omega^{nn}$ et $\Omega^{nj}$ pour écrire :

$$\Omega^{nn} = \frac{S^{nj}}{S^{nn}} \frac{i\omega - \frac{S^{nj}}{S^{nn}} \frac{i\omega - \Omega^{nn}}{i\omega - \Omega^{nj}}}{i\omega - \frac{S^{jj}}{S^{nj}} \frac{i\omega - \Omega^{nj}}{i\omega - \Omega^{jj}}}, \tag{5.4.2}$$

dans une procédure que l'on peut répéter à l'envi pour obtenir une hiérarchie d'expressions (exactes) pour $\Omega^{nn}$. Une approximation de premier ordre pour $\Omega^{nn}$ est obtenue en posant $\Omega^{nn} \approx \Omega^{nj}$ dans (5.4.1) :

$$\Omega^{nn(1)} = \frac{S^{nj}}{S^{nn}}, \tag{5.4.3}$$

et une approximation de deuxième ordre est obtenue en posant $\Omega^{nn} \approx \Omega^{nj} \approx \Omega^{jj}$ dans (5.4.2) :

$$\Omega^{nn(2)} = \frac{S^{nj}}{S^{nn}} \frac{i\omega - \frac{S^{nj}}{S^{nn}}}{i\omega - \frac{S^{jj}}{S^{nj}}} \tag{5.4.4}$$

Remarquons que dans cette approximation d'ordre deux, l'EED comporte une dépendance explicite et non triviale à la fréquence.

## 5.5 Évaluation de la fonction réponse non-interagissante

D'un point de vue plus pratique, on peut chercher à mettre en place des expressions calculables pour les numérateurs, et pour les fonctions de réponse. Dans une perspective d'applications pour calculer l'énergie de corrélation RPA par une approche matrice diélectrique, l'objet central qui nous intéresse est la fonction réponse non-interagissante. Ainsi, l'hamiltonien devient l'hamiltonien Kohn-Sham ou Kohn-Sham généralisé, éventuellement l'opérateur Hartree-Fock ; la somme sur les états excités se simplifie comme une sommation sur des orbitales virtuelles.

Deux approches s'offrent à nous : dans une première approche, on travaille sur la fonction de réponse avec sa dépendance explicite aux coordonnées $\mathbf{r}_1$ et $\mathbf{r}_2$ et on dégage des EED qui dépendent des coordonnées : $\Omega(\mathbf{r}_1, \mathbf{r}_2)$ ; dans une deuxième approche, on applique la procédure des EED directement à un objet qui est la fonction de réponse écrite sur la base des orbitales atomiques et on obtient des EED qui dépendent des bases : $\Omega_{\mu\nu,\sigma\rho}$ (au lieu de dépendre des coordonnées).

On considère dans les deux cas la fonction de réponse suivante :

$$\chi(\mathbf{r}_1, \mathbf{r}_2; i\omega) = 2\mathrm{Re}\left\{\sum_{ia} \frac{\phi_i^*(\mathbf{r}_1)\phi_a(\mathbf{r}_1)\phi_a^*(\mathbf{r}_2)\phi_i(\mathbf{r}_2)}{i\omega - \Omega_{ia}}\right\} \tag{5.5.1}$$

### 5.5.1 Approche avec une EED dépendante des coordonnées

Considérons l'objet $\chi^{(i)}$ défini de la manière suivante :

$$\chi(\mathbf{r}_1, \mathbf{r}_2; i\omega) = 2\mathrm{Re}\left\{\sum_i \underbrace{\sum_a \frac{\phi_i^*(\mathbf{r}_1)\phi_a(\mathbf{r}_1)\phi_a^*(\mathbf{r}_2)\phi_i(\mathbf{r}_2)}{i\omega - \Omega_{ia}}}_{\chi^{(i)}}\right\} \tag{5.5.2}$$

La procédure EED que l'on connaît bien à présent, appliquée à $\chi^{(i)}$, se résume aux équations suivantes :

$$
\begin{aligned}
\chi^{(i)}(\mathbf{r}_1, \mathbf{r}_2; i\omega) &= \frac{\phi_i^*(\mathbf{r}_1)\left(\sum_a \phi_a(\mathbf{r}_1)\phi_a^*(\mathbf{r}_2)\right)\phi_i(\mathbf{r}_2)}{i\omega - \Omega^{(i)}(\mathbf{r}_1, \mathbf{r}_2, i\omega)} \\
\Omega^{(i)}(\mathbf{r}_1, \mathbf{r}_2, i\omega)\chi^{(i)}(\mathbf{r}_1, \mathbf{r}_2; i\omega) &= \sum_a \frac{\phi_i^*(\mathbf{r}_1)\phi_a(\mathbf{r}_1)\phi_a^*(\mathbf{r}_2)\phi_i(\mathbf{r}_2)\Omega_{ia}}{i\omega - \Omega_{ia}} = Y^{(i)}(\mathbf{r}_1, \mathbf{r}_2; i\omega) \\
Y^{(i)}(\mathbf{r}_1, \mathbf{r}_2; i\omega) &= \frac{\phi_i^*(\mathbf{r}_1)\left(\sum_a \phi_a(\mathbf{r}_1)\Omega_{ia}\phi_a^*(\mathbf{r}_2)\right)\phi_i(\mathbf{r}_2)}{i\omega - \tilde{\Omega}^{(i)}(\mathbf{r}_1, \mathbf{r}_2, i\omega)} \\
& \quad etc\ldots,
\end{aligned} \tag{5.5.3}
$$

où l'on évite soigneusement d'utiliser les dénominations « *nn* », « *nj* », *etc...* Comme précédemment, une première approximation de l'EED est :

$$
\Omega^{(i)}(\mathbf{r}_1, \mathbf{r}_2, i\omega) = \frac{\phi_i^*(\mathbf{r}_1)\left(\sum_a \phi_a(\mathbf{r}_1)\Omega_{ia}\phi_a^*(\mathbf{r}_2)\right)\phi_i(\mathbf{r}_2)}{\phi_i^*(\mathbf{r}_1)\left(\sum_a \phi_a(\mathbf{r}_1)\phi_a^*(\mathbf{r}_2)\right)\phi_i(\mathbf{r}_2)}, \tag{5.5.4}
$$

ce qui produit une première approximation de la fonction de réponse :

$$
\chi^{(i)}(\mathbf{r}_1, \mathbf{r}_2; i\omega) = \frac{\phi_i^*(\mathbf{r}_1)\left(\sum_a \phi_a(\mathbf{r}_1)\phi_a^*(\mathbf{r}_2)\right)\phi_i(\mathbf{r}_2)}{i\omega - \dfrac{\phi_i^*(\mathbf{r}_1)\left(\sum_a \phi_a(\mathbf{r}_1)\Omega_{ia}\phi_a^*(\mathbf{r}_2)\right)\phi_i(\mathbf{r}_2)}{\phi_i^*(\mathbf{r}_1)\left(\sum_a \phi_a(\mathbf{r}_1)\phi_a^*(\mathbf{r}_2)\right)\phi_i(\mathbf{r}_2)}} = \phi_i^*(\mathbf{r}_1)\left\{\frac{\sum_a \phi_a(\mathbf{r}_1)\phi_a^*(\mathbf{r}_2)}{i\omega - \dfrac{\sum_a \phi_a(\mathbf{r}_1)\Omega_{ia}\phi_a^*(\mathbf{r}_2)}{\sum_a \phi_a(\mathbf{r}_1)\phi_a^*(\mathbf{r}_2)}}\right\}\phi_i(\mathbf{r}_2) \tag{5.5.5}
$$

Les quantités qui contiennent les sommations sur les états virtuels peuvent être écrites dans une base d'orbitales atomiques :

$$
\sum_a \phi_a(\mathbf{r}_1)\phi_a^*(\mathbf{r}_2) = \sum_{\mu\nu}\sum_a c_{a\mu}c_{a\nu}\chi_\mu(\mathbf{r}_1)\chi_\nu^*(\mathbf{r}_2) = \sum_{\mu\nu} Q_{\nu\mu}\chi_\mu(\mathbf{r}_1)\chi_\nu^*(\mathbf{r}_2), \tag{5.5.6}
$$

et :

$$
\begin{aligned}
\sum_a \phi_a(\mathbf{r}_1)\left(\epsilon_a - \epsilon_i\right)\phi_a^*(\mathbf{r}_2) &= \sum_{\mu\nu}\sum_a c_{a\mu}c_{a\nu}\epsilon_a\chi_\mu(\mathbf{r}_1)\chi_\nu^*(\mathbf{r}_2) - \epsilon_i\sum_{\mu\nu}\sum_a c_{a\mu}c_{a\nu}\chi_\mu(\mathbf{r}_1)\chi_\nu^*(\mathbf{r}_2) \\
&= \sum_{\mu\nu}\left(\mathbf{S}^{-1}\mathbf{F}\mathbf{Q} - \epsilon_i\mathbf{Q}\right)_{\nu\mu}\chi_\mu(\mathbf{r}_1)\chi_\nu^*(\mathbf{r}_2)
\end{aligned} \tag{5.5.7}
$$

L'expression finale de la première approximation de la fonction de réponse est avec cette approche :

$$\chi^{(i)}(\mathbf{r}_1,\mathbf{r}_2;i\omega) = \sum_{\mu\nu}\sum_{\lambda\sigma}\chi^*_\lambda(\mathbf{r}_1)\chi_\mu(\mathbf{r}_1)c^*_{i\lambda} \left\{ \frac{\sum_{\mu\nu} Q_{\nu\mu}\chi_\mu(\mathbf{r}_1)\chi^*_\nu(\mathbf{r}_2)}{i\omega - \dfrac{\sum_{\mu\nu}\left(\mathbf{S}^{-1}\mathbf{F}\mathbf{Q} - \epsilon_i\mathbf{Q}\right)_{\nu\mu}\chi_\mu(\mathbf{r}_1)\chi^*_\nu(\mathbf{r}_2)}{\sum_{\mu\nu} Q_{\nu\mu}\chi_\mu(\mathbf{r}_1)\chi^*_\nu(\mathbf{r}_2)}} \right\} c_{i\sigma}\chi^*_\nu(\mathbf{r}_2)\chi_\sigma(\mathbf{r}_2) \tag{5.5.8}$$

On peut constater que cette stratégie n'est pas très adaptée aux applications numériques car la quantité définie par l'équation (5.5.8) ne se prête pas aux techniques d'intégrations analytiques et l'inévitable intégration numérique risque d'être coûteuse.

### 5.5.2 Approche avec fonctions de base atomiques

Avec cette approche, on cherche directement à faire émerger une expression de la fonction de réponse dans la base des orbitales atomiques :

$$\chi(\mathbf{r}_1,\mathbf{r}_2;i\omega) = 2\mathrm{Re}\left\{\sum_{\mu\nu}\sum_{\lambda\sigma}\underbrace{\left(\sum_{ia}\frac{c^*_{i\mu}c_{a\nu}c^*_{a\lambda}c_{i\sigma}}{i\omega - \Omega_{ia}}\right)}_{\Pi_{\mu\nu,\lambda\sigma}(i\omega)}\chi^*_\mu(\mathbf{r}_1)\chi_\nu(\mathbf{r}_1)\chi^*_\lambda(\mathbf{r}_2)\chi_\sigma(\mathbf{r}_2)\right\} \tag{5.5.9}$$

On applique la procédure des EED directement à $\Pi_{\mu\nu,\lambda\sigma}(i\omega)$ :

$$\begin{aligned}
\Pi_{\mu\nu,\lambda\sigma}(i\omega) &= \frac{\sum_{ia} c^*_{i\mu}c_{a\nu}c^*_{a\lambda}c_{i\sigma}}{i\omega - \tilde{\Omega}_{\mu\nu,\lambda\sigma}(i\omega)} \\
\tilde{\Omega}_{\mu\nu,\lambda\sigma}(i\omega)\Pi_{\mu\nu,\lambda\sigma}(i\omega) &= \sum_{ia}\frac{c^*_{i\mu}c_{a\nu}c^*_{a\lambda}c_{i\sigma}\Omega_{ia}}{i\omega - \Omega_{ia}} = \Upsilon_{\mu\nu,\lambda\sigma}(i\omega) \\
\Upsilon_{\mu\nu,\lambda\sigma}(i\omega) &= \frac{\sum_{ia} c^*_{i\mu}c_{a\nu}c^*_{a\lambda}c_{i\sigma}\Omega_{ia}}{i\omega - \tilde{\tilde{\Omega}}_{\mu\nu,\lambda\sigma}(i\omega)} \\
& \textit{etc}\ldots
\end{aligned} \tag{5.5.10}$$

Ainsi la première approximation de l'élément de matrice de la fonction de réponse est :

$$\Pi_{\mu\nu,\lambda\sigma}(i\omega) = \frac{\sum_{ia} c^*_{i\mu}c_{a\nu}c^*_{a\lambda}c_{i\sigma}}{i\omega - \dfrac{\sum_{ia} c^*_{i\mu}c_{a\nu}c^*_{a\lambda}c_{i\sigma}\Omega_{ia}}{\sum_{ia} c^*_{i\mu}c_{a\nu}c^*_{a\lambda}c_{i\sigma}}} \tag{5.5.11}$$

À nouveau, on peut écrire les quantités qui impliquent les sommations dans la base des orbitales atomiques :

$$\sum_i c_{i\sigma} c^*_{i\mu} \sum_a c_{a\nu} c^*_{a\lambda} = P_{\sigma\mu} Q_{\nu\lambda}$$

$$\sum_i c_{i\sigma} c^*_{i\mu} \sum_a c_{a\nu} c^*_{a\lambda} \epsilon_a - \sum_i c_{i\sigma} c^*_{i\mu} \epsilon_i \sum_a c_{a\nu} c^*_{a\lambda} = P_{\sigma\mu} \left(\mathbf{S}^{-1}\mathbf{FQ}\right)_{\nu\lambda} - \left(\mathbf{S}^{-1}\mathbf{FP}\right)_{\sigma\mu} Q_{\nu\lambda} \quad (5.5.12)$$

Ainsi on écrit pour $\Pi$ :

$$\Pi_{\mu\nu,\lambda\sigma}(i\omega) = \frac{P_{\sigma\mu} Q_{\nu\lambda}}{i\omega - \dfrac{P_{\sigma\mu} \left(\mathbf{S}^{-1}\mathbf{FQ}\right)_{\nu\lambda} - \left(\mathbf{S}^{-1}\mathbf{FP}\right)_{\sigma\mu} Q_{\nu\lambda}}{P_{\sigma\mu} Q_{\nu\lambda}}} \quad (5.5.13)$$

### 5.5.3 Illustration numérique : polarisabilité dynamique à fréquence imaginaire

Une fois que l'on a établi cette équation, on peut éventuellement l'utiliser dans l'expression de la polarisabilité dynamique dépendante de la fréquence, à une fréquence imaginaire, c'est-à-dire dans :

$$\begin{aligned} \alpha_{\alpha\beta}(i\omega) &= 2\mathrm{Re}\left\{\sum_{\mu\nu}\sum_{\lambda\sigma}\left(\sum_i\sum_a \frac{c^*_{i\mu} c_{a\nu} c^*_{a\lambda} c_{i\sigma}}{\omega - \Omega_{ia}}\right) \langle \mu | r_\alpha | \nu \rangle \left\langle \lambda \middle| r_\beta \middle| \sigma \right\rangle \right\} \\ &= 2 \sum_{\mu\nu}\sum_{\lambda\sigma} \mathrm{Re}\left\{\Pi_{\mu\nu,\lambda\sigma}(i\omega) \langle \mu | r_\alpha | \nu \rangle \left\langle \lambda \middle| r_\beta \middle| \sigma \right\rangle\right\}, \end{aligned} \quad (5.5.14)$$

On propose plutôt la procédure suivante : considérons l'expression de la polarisabilité :

$$\alpha_{\alpha\beta}(i\omega) = 2\mathrm{Re} \sum_i \underbrace{\left\{\sum_a \frac{\langle i | \hat{r}_\alpha | a \rangle \left\langle a \middle| \hat{r}_\beta \middle| i \right\rangle}{i\omega - \Omega_{ia}}\right\}}_{\alpha_{i,\alpha\beta}(i\omega)} \quad (5.5.15)$$

On obtient, en appliquant la procédure EED à $\alpha_{i,\alpha\beta}(i\omega)$, l'expression suivante :

$$\alpha_{i,\alpha\beta}(i\omega) = \frac{\tilde{S}^{nn}_{i,\alpha\beta}}{i\omega - \tilde{\Omega}_{i,\alpha\beta}}, \quad (5.5.16)$$

où $\tilde{S}^{nn}_{i,\alpha\beta}$ n'est pas tout à fait similaire au numérateur $S^{nn}$ que l'on a rencontré précédemment (on contracte ici directement avec les opérateurs position $\hat{r}_\alpha$ une quantité qui dépend des orbitales occupées) ; la similitude est cependant évidente.

Les approximations d'ordre un et deux de $\tilde{\Omega}_{i,\alpha\beta}$ sont :

$$\tilde{\Omega}^{(1)}_{i,\alpha\beta} = \frac{\tilde{S}^{nj}_{i,\alpha\beta}}{\tilde{S}^{nn}_{i,\alpha\beta}} \qquad \text{et :} \qquad \tilde{\Omega}^{(2)}_{i,\alpha\beta} = \frac{\tilde{S}^{nj}_{i,\alpha\beta}}{\tilde{S}^{nn}_{i,\alpha\beta}} \frac{\omega - \dfrac{\tilde{S}^{nj}_{i,\alpha\beta}}{\tilde{S}^{nn}_{i,\alpha\beta}}}{\omega - \dfrac{\tilde{S}^{jj}_{i,\alpha\beta}}{\tilde{S}^{nj}_{i,\alpha\beta}}} \tag{5.5.17}$$

On est en mesure de calculer les objets $\tilde{S}^{nn}_{i,\alpha\beta}$, $\tilde{S}^{nj}_{i,\alpha\beta}$, et $\tilde{S}^{jj}_{i,\alpha\beta}$ par des opérations matricielles simples à partir des matrices **P**, $\mathbf{S}^{-1}\mathbf{F}$, et **X**, **Y** et **Z**.

On présente ainsi aux figures 5.1 et 5.2 des applications numériques de polarisabilité dynamique dépendante de la fréquence, à une fréquence imaginaire, c'est-à-dire des applications numériques de l'objet $\alpha(i\omega)$. On montre des calculs d'approximation à l'ordre un et à l'ordre deux avec des orbitales occupées canoniques et localisées par le critère de Boys ; ces calculs sont faits avec `MOLPRO` à partir de la procédure décrite juste au-dessus.

On compare dans chaque figure les comportements en fonction de la fréquence des composants *xx* (en bleu), *yy* (en rouge) et *zz* (en vert) ainsi que de la moyenne sphérique (en noir) des polarisabilités de l'éthylène $C_2H_4$ calculées dans la base vDz. On trouve des résultats similaires avec le benzène $C_6H_6$. On calcule également une polarisabilité que l'on considère référente et qui est obtenue par sommation explicite sur les états virtuels dans un cadre *Uncoupled Hartree-Fock*.

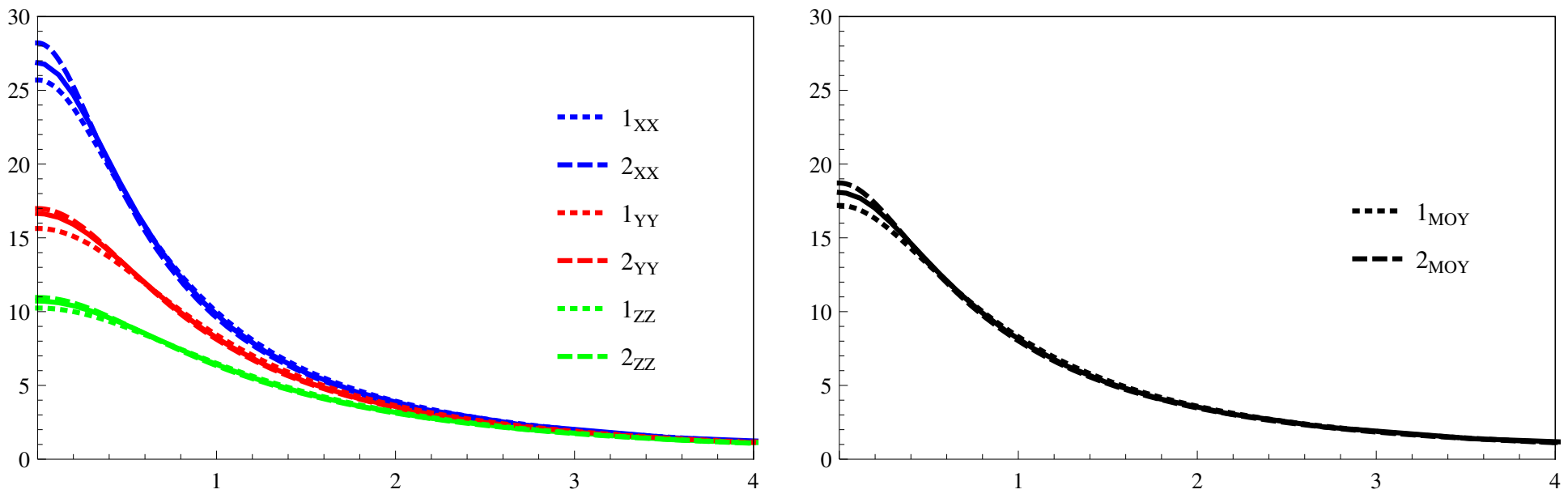

TABLE 5.1: Comparaison des composantes *xx* (en bleu), *yy* (en rouge) et *zz* (en vert) ainsi que des moyennes sphériques (en noir, à droite) des polarisabilités dépendantes de la fréquence, à fréquence imaginaire, approximée à l'ordre un (en pointillé fin) et deux (en pointillé plus large) selon l'approximation EED. Toutes les comparaisons sont à lire en fonction de la polarisabilité référente correspondante, montrée en trait plein.

Dans la figure 5.1 sont montrés les comportements de : l'approximation d'ordre un (en pointillé fin), l'approximation d'ordre deux (en pointillé plus large) et la polarisabilité référente (en trait plein). On voit que ces deux approximations reproduisent très bien la forme de la polarisabilité tout au long de la gamme de fréquence. Notons que, comme on l'a vu dans les développements de l'approximation EED, l'approximation de l'ordre deux du dénominateur effectif dépend explicitement de la fréquence.

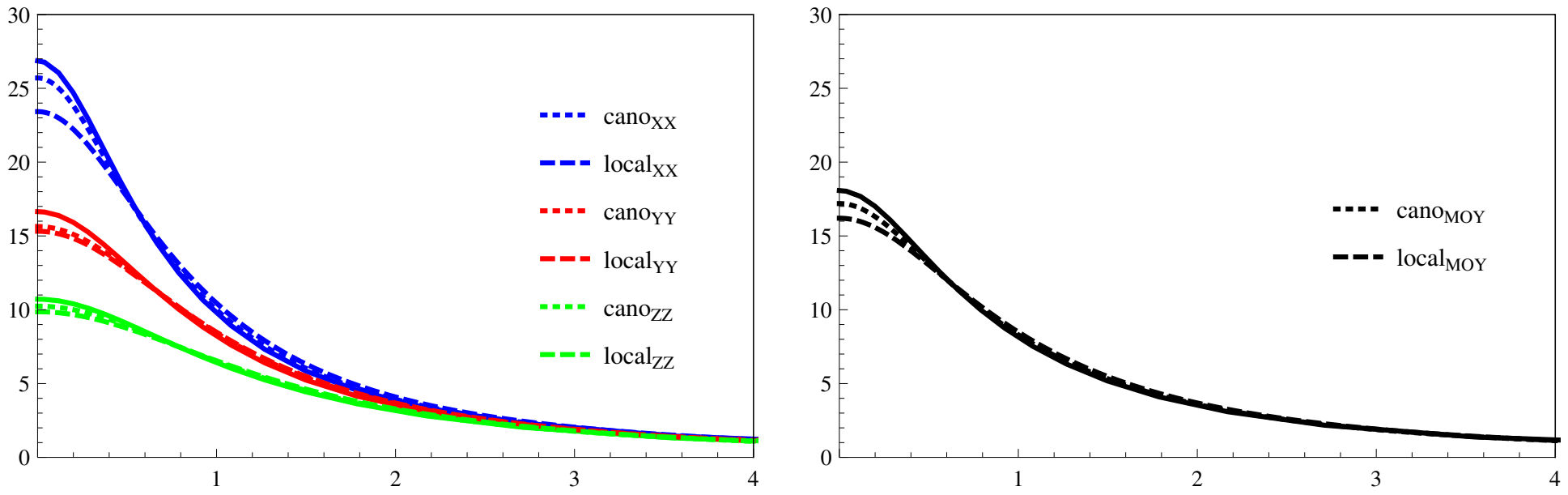

TABLE 5.2: Comparaison des composantes *xx* (en bleu), *yy* (en rouge) et *zz* (en vert) ainsi que des moyennes sphériques (en noir, à droite) des polarisabilités dépendantes de la fréquence, à fréquence imaginaire, approximée à l'ordre un selon l'approximation EED avec des orbitales canoniques (en pointillé fin) et localisées selon le critère de Boys (en pointillé plus large). Toutes les comparaisons sont à lire en fonction de la polarisabilité référente correspondante, montrée en trait plein.

---

On montre dans la figure 5.2 les résultats de calculs approximés à l'ordre un réalisés avec des orbitales canoniques (en pointillé fin) ou locales (en pointillé plus large). À nouveau la polarisabilité référente est donnée en trait plein, et à nouveau on voit que le choix des orbitales influe peu sur la qualité des résultats.

## 5.6 Conclusion

Dans ce chapitre on montre une adaptation à l'espace réel de la technique dite du dénominateur effectif (EED), qui est développée dans l'espace réciproque par Berger *et. al.*. Au cours de cette adaptation, on montre que les numérateurs qui émergent sont des règles de sommes qui sont bien connues dans la littérature. On dérive également des moyens de calculer la fonction de réponse approximée par la méthode de l'EED et on montre quelques résultats numériques sur le sujet. Le but à terme est bien entendu le calcul de l'énergie de corrélation sans avoir à utiliser explicitement d'orbitales virtuelles ; en cela on rejoint l'objectif poursuivi dans le chapitre 3 lorsque l'on a dérivé les équations RPA dans la base des POO.

# Chapitre 6

# Gradients analytiques des énergies RSH-RPA

On dérive ici les équations qui permettent de calculer le gradient de l'énergie RSH+RPA, c'est-à-dire de l'énergie courte-portée RSH *et* longue-portée RPA. La dérivation d'un gradient RPA est une nouveauté de cette thèse, tout comme la dérivation tout-en-un d'un gradient d'une énergie mêlant courte- et longue-portée. Le travail de Chabbal, Leininger et Stoll portant sur la dérivation d'un gradient RSH-(L)MP2, et qui semble être le seul dans la littérature à donner un gradient d'une énergie à séparation de portée, propose une dérivation basée sur une modification du gradient MP2 *sans* séparation de portée. Notre formulation offre une dérivation claire et tout-en-un des couplages entre les contributions courte- et longue-portée de l'énergie RSH-RPA. Notons que, parallèlement à notre travail, le groupe de Helgaker a dérivé des gradients RPA pour une fonction d'onde de référence de type Hartree-Fock (voir [198]).

On rappelle pour commencer les bases du formalisme Lagrangien, que l'on utilise pour dériver le gradient RSH-RPA : à partir de l'énergie RSH-RPA qui n'est pas variationnelle, l'idée est simplement de se rapporter à un objet variationnel, et donc plus facile à manier. On cherche à établir des formules analytiques du gradient de l'énergie RSH-RPA, où l'énergie RPA est exprimée dans sa formulation avec équations de "Riccati " (ring CCD), que l'on rappelle ici. Au cours de la dérivation, de nouvelles équations émergent pour le calcul des différentes composantes du gradient, notamment une équation CP-RPA, similaire aux équations *Coupled Perturbed* habituellement rencontrées lorsque l'on dérive des gradients. À nouveau, ce chapitre est à lire en coordination avec l'Annexe F, qui offre des détails des dérivations.

On peut mettre en avant un parallèle entre les gradients RSH-RPA et les gradients RSH-MP2 . Une façon de voir ce parallèle est de dire que les gradients RSH-RPA sont une généralisation des gradients RSH-MP2, ou les gradients RSH-MP2 un cas particulier des gradients RSH-RPA. Cette observation est utilisée pour implémenter les gradients RSH-RPA dans MOLPRO, dont on présente quelques premiers résultats.

[NOTE: **Ce travail a depuis fait l'objet d'une publication :** ***Analytical Energy Gradients in Range-Separated Hybrid Density Functional Theory with Random Phase Approximation.*** **B. Mussard, P. Szalay, J.G. Ángyán, J. Chem. Theory Comput. 10 1968-1979 (2014)**]

## 6.1 Formalisme Lagrangien

Le calcul de gradients analytiques d'énergies de corrélation a été initié par les travaux de Pople *et. al.* [199] et on peut retrouver les avancées théoriques les plus importantes sur le sujet aux références[200–205]. On introduit ici notamment le formalisme Lagrangien[206–210] utilisé dans la suite, et qui donne le cadre général pour la dérivation de gradients d'énergies de méthodes non variationnelles.

Considérons une énergie $E$ qui dépend de paramètres $\mathbf{V}$ et $\mathbf{T}$ et qui, aux paramètres corrects $\mathbf{V}^*$ et $\mathbf{T}^*$, prend une valeur voulue $E[\mathbf{V}^*, \mathbf{T}^*] = \mathcal{E}$. On a distingué ici les paramètres stationnaires $\mathbf{V}$ et les paramètres non stationnaires $\mathbf{T}$. Les paramètres stationnaires sont déterminés en annulant la dérivée de l'énergie $\frac{\partial E}{\partial \mathbf{V}} = \mathbf{0}$, c'est-à-dire que les paramètres corrects $\mathbf{V}^*$ avec lesquels il faut calculer l'énergie sont ceux qui donnent : $\left.\frac{\partial E}{\partial \mathbf{V}}\right|_{\mathbf{V}=\mathbf{V}^*} = \mathbf{0}$. Les paramètres non stationnaires sont déterminés par d'autres conditions $\mathbf{R}[\mathbf{T}] = \mathbf{0}$, c'est-à-dire que les paramètres corrects avec lesquels il faut calculer l'énergie, $\mathbf{T}^*$, sont ceux qui annulent les règles : $\mathbf{R}[\mathbf{T}^*] = 0$.

On cherche à obtenir la dérivée de l'énergie par rapport à une perturbation $x$, $\frac{\mathrm{d}E}{\mathrm{d}x}$. Il nous faut alors bien comprendre de quelles manières l'énergie dépend de la perturbation. L'énergie dépend de la perturbation de manière directe, à travers l'hamiltonien, et de manière implicite à travers les paramètres de la fonction d'onde ($\mathbf{V}$ et $\mathbf{T}$). On écrit donc l'énergie $E \doteq E[x, \mathbf{V}(x), \mathbf{T}(x)]$, et la dérivée :

$$\frac{\mathrm{d}E}{\mathrm{d}x} = \frac{\partial E}{\partial x} + \frac{\partial E}{\partial \mathbf{V}}\frac{\partial \mathbf{V}}{\partial x} + \frac{\partial E}{\partial \mathbf{T}}\frac{\partial \mathbf{T}}{\partial x} \tag{6.1.1}$$

La dépendance de l'hamiltonien est en fait une dépendance des fonctions de bases, c'est-à-dire de $h_{\mu\nu}$, $(\mu\nu|\sigma\rho)$, et des recouvrements $S_{\mu\nu}$. Considérant ces dépendances, le gradient s'écrit :

$$\frac{\mathrm{d}E}{\mathrm{d}x} = \frac{\partial E}{\partial \mathbf{h}}\mathbf{h}^{(x)} + \frac{\partial E}{\partial(\mu\nu|\sigma\rho)}(\mu\nu|\sigma\rho)^{(x)} + \frac{\partial E}{\partial \mathbf{S}}\mathbf{S}^{(x)} + \frac{\partial E}{\partial \mathbf{V}}\mathbf{V}^{(x)} + \frac{\partial E}{\partial \mathbf{T}}\mathbf{T}^{(x)}, \tag{6.1.2}$$

où la notation en exposant $A^{(x)}$ dénote la dérivée par rapport à une perturbation, évaluée au point où la perturbation est nulle (au point où la fonction d'onde est optimisée, où les quantités telles que les intégrales, les paramètres... sont calculées) :

$$A^{(x)} = \left.\frac{\mathrm{d}A}{\mathrm{d}x}\right|_{x=0} \tag{6.1.3}$$

Le calcul de $\mathbf{V}^{(x)}$ n'est jamais nécessaire ($\frac{\partial E}{\partial \mathbf{V}} = \mathbf{0}$) et , avec une méthode de calcul de l'énergie totalement stationnaire (pas de paramètre $\mathbf{T}$), le gradient s'écrit simplement :

$$\frac{\mathrm{d}E}{\mathrm{d}x} = \frac{\partial E}{\partial \mathbf{h}}\mathbf{h}^{(x)} + \frac{\partial E}{\partial(\mu\nu|\sigma\rho)}(\mu\nu|\sigma\rho)^{(x)} + \frac{\partial E}{\partial \mathbf{S}}\mathbf{S}^{(x)}, \tag{6.1.4}$$

où $\mathbf{h}^{(x)}$, $(\mu\nu|\sigma\rho)^{(x)}$ et $\mathbf{S}^{(x)}$ sont des dérivées de l'hamiltonien de cœur, des intégrales bi-électroniques et de la matrice de recouvrement. L'équation (6.1.4) est connue sous le nom de théorème de Hellmann-Feynman généralisé[211, 212], et n'est bien sûr valable que pour des méthodes de calcul de l'énergie totalement stationnaires, comme mentionné plus haut. Avec une méthode de calcul de l'énergie contenant des paramètres **T**, en revanche, la non stationnarité $\frac{\partial E}{\partial \mathbf{T}} \neq \mathbf{0}$ oblige à effectuer d'une manière ou d'une autre le calcul de $\mathbf{T}^{(x)}$. Le mieux est alors d'utiliser un formalisme Lagrangien[206, 207, 209] (voir figure 6.1) :

$$\mathcal{L}[\mathbf{V}, \mathbf{T}, \mathbf{z}] = E[\mathbf{V}, \mathbf{T}] + \mathbf{z}\mathbf{R}[\mathbf{T}], \tag{6.1.5}$$

où chaque règle pour déterminer les paramètres non stationnaires est introduite avec un multiplicateur de Lagrange inconnu, **z**. Tous les paramètres du Lagrangien (les paramètres de l'énergie **V** et **T** et le nouveau paramètre **z**) sont trouvés en imposant la stationnarité du Lagrangien :

$$\frac{\partial \mathcal{L}}{\partial \mathbf{T}} = \frac{\partial E}{\partial \mathbf{T}} + \mathbf{z}\frac{\partial R}{\partial \mathbf{T}} = 0 \tag{6.1.6a}$$

$$\frac{\partial \mathcal{L}}{\partial \mathbf{V}} = \frac{\partial E}{\partial \mathbf{V}} = 0 \tag{6.1.6b}$$

$$\frac{\partial \mathcal{L}}{\partial \mathbf{z}} = \mathbf{R}[\mathbf{T}] = 0 \tag{6.1.6c}$$

Ces trois équations stationnaires apportent chacune des informations importantes sur le formalisme Lagrangien : les équations (6.1.6b) et (6.1.6c) montrent que les paramètres corrects de l'énergie sont calculés de la même manière dans le formalisme Lagrangien que lorsque l'on travaille avec la fonctionnelle $E[\mathbf{V}, \mathbf{T}]$, c'est-à-dire que le Lagrangien est une fonctionnelle de la même théorie que l'énergie, et ne définit pas une nouvelle théorie. Ainsi, la stationnarité des paramètres du Lagrangien fournira les même paramètres $\mathbf{V}^*$ et $\mathbf{T}^*$ du calcul de l'énergie.

**L'équation** (6.1.6a) est une nouvelle équation qui émerge du formalisme Lagrangien ; elle permet de calculer le multiplicateur de Lagrange de telle manière que le Lagrangien soit stationnaire par rapport aux paramètres **T**.

Ainsi, aux paramètres corrects ($\mathbf{V}^*$ et $\mathbf{T}^*$, qui proviennent de la fonctionnelle de l'énergie, et $\mathbf{z}^*$, qui assurent la stationnarité (6.1.6a)), on a :

$$\mathcal{L}[\mathbf{V}^*, \mathbf{T}^*, \mathbf{z}^*] = E[\mathbf{V}^*, \mathbf{T}^*] + \mathbf{z}^*\mathbf{R}[\mathbf{T}^*] = \mathcal{E} \tag{6.1.7}$$

Et le gradient est simplement :

$$\begin{aligned}\frac{\mathrm{d}\mathcal{L}}{\mathrm{d}x} = \frac{\mathrm{d}\mathcal{E}}{\mathrm{d}x} &= \frac{\partial \mathcal{L}}{\partial \mathbf{h}}\mathbf{h}^{(x)} + \frac{\partial \mathcal{L}}{\partial(\mu\nu|\sigma\rho)}(\mu\nu|\sigma\rho)^{(x)} + \frac{\partial \mathcal{L}}{\partial \mathbf{S}}\mathbf{S}^{(x)} + \frac{\partial \mathcal{L}}{\partial \mathbf{V}}\mathbf{V}^{(x)} + \frac{\partial \mathcal{L}}{\partial \mathbf{T}}\mathbf{T}^{(x)} + \frac{\partial \mathcal{L}}{\partial \mathbf{z}}\mathbf{z}^{(x)} \\ &= \frac{\partial \mathcal{L}}{\partial \mathbf{h}}\mathbf{h}^{(x)} + \frac{\partial \mathcal{L}}{\partial(\mu\nu|\sigma\rho)}(\mu\nu|\sigma\rho)^{(x)} + \frac{\partial \mathcal{L}}{\partial \mathbf{S}}\mathbf{S}^{(x)},\end{aligned} \tag{6.1.8}$$

où, comme on l'a vu plus haut, $\mathbf{h}^{(x)}$, $(\mu\nu|\sigma\rho)^{(x)}$ et $\mathbf{S}^{(x)}$ sont des quantités bien connues.

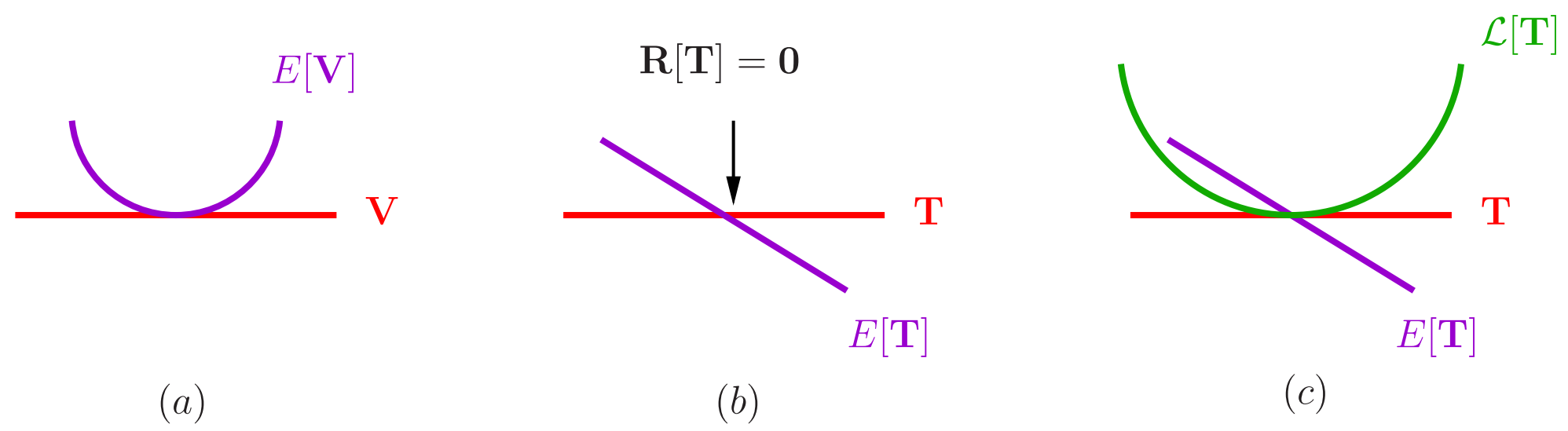

Figure 6.1: Comportement de l'énergie $E$ d'une méthode avec des paramètres ($a$) stationnaires (les paramètres corrects pour le calcul de l'énergie sont ceux qui annulent la dérivée) et ($b$) non stationnaires (les paramètres corrects pour le calcul de l'énergie doivent être définis par une règle autre qu'une procédure d'annulation de la dérivée, $\mathbf{R}[\mathbf{T}] = \mathbf{0}$). Je montre en ($c$) la construction du Lagrangien $\mathcal{L}[\mathbf{T}]$, qui est égal à l'énergie aux paramètres corrects pour le calcul de l'énergie, et qui est stationnaire par rapport aux paramètres de l'énergie.

## 6.2 Lagrangien RSH-RPA

Dans la suite, je dérive un gradient de l'énergie totale RSH-RPA $E$ par un formalisme Lagrangien. La dérivation d'un gradient RPA est un travail original de cette thèse ; de plus la dérivation d'un gradient d'une énergie « RSH+énergie de corrélation longue-portée » n'a jamais été faite en utilisant un Lagrangien contenant par construction tous les termes courte- et longue-portée.

On rappelle que l'énergie totale dans un cadre RSH s'écrit :

$$E = E_{\mathrm{RSH}}^{\mathrm{SR,LR}} + E_c^{\mathrm{LR}} \tag{6.2.1}$$

### 6.2.1 Expression de l'énergie RSH

On trouvera qu'il est pratique, dans un tel contexte, d'exprimer l'énergie RSH avec des fockiennes courte- et longue-portée :

$$E_{\mathrm{RSH}} = \left\langle \mathbf{d}^{(0)}\mathbf{f} \right\rangle + \Delta_{\mathrm{DC}} = \left\langle \mathbf{d}^{(0)}\mathbf{f}^{\mathrm{LR}} \right\rangle + \Delta_{\mathrm{DC}}^{\mathrm{LR}} + \left\langle \mathbf{d}^{(0)}\mathbf{f}^{\mathrm{SR}} \right\rangle + \Delta_{\mathrm{DC}}^{\mathrm{SR}}, \tag{6.2.2}$$

où les termes « $\Delta_{\mathrm{DC}}$ » sont les termes de double comptage. Dans ce chapitre, on ne rencontrera plus de problématique où il est bon de distinguer les traces « $\mathrm{Tr}(\sqcup)$ » et « $\mathrm{tr}(\sqcup)$ ». On utilise la notation plus compacte vue ci-dessus : « $\langle \sqcup \rangle \doteq \mathrm{tr}(\sqcup)$ ». On a :

$$\begin{aligned}
\mathbf{f}^{\mathrm{LR}} &= \mathbf{h} + \mathbf{g}^{\mathrm{LR}}\left[\mathbf{d}^{(0)}\right] \\
\Delta_{\mathrm{DC}}^{\mathrm{LR}} &= -\frac{1}{2}\left\langle \mathbf{d}^{(0)}\mathbf{g}^{\mathrm{LR}}\left[\mathbf{d}^{(0)}\right]\right\rangle \\
\text{avec :} & \\
g^{\mathrm{LR}}\left[\mathbf{d}^{(0)}\right]_{pq} &= d_{rs}^{(0)}\left(\langle pr|qs\rangle^{\mathrm{LR}} - \frac{1}{2}\langle ps|rq\rangle^{\mathrm{LR}}\right)
\end{aligned} \tag{6.2.3}$$

Et où on peut dériver une $f^{\mathrm{SR}}$ en considérant que :

$$E_{\mathrm{Hxc}}^{\mathrm{SR}} = \int dr\ F(\boldsymbol{\xi}(r)), \tag{6.2.4}$$

où $\boldsymbol{\xi}$ est un vecteur contenant toutes les dépendances envisageables pour une fonctionnelle de la densité, c'est-à-dire où $\boldsymbol{\xi} = \{\xi_A\} = \{\rho_\alpha, \nabla\rho_\alpha\nabla\rho_\beta, \ldots\}$. On a alors :

$$f_{ab}^{\mathrm{SR}} = \sum_A \int dr\ \frac{\partial F}{\partial \xi_A}\frac{\partial \xi_A}{\partial d_{ab}^{(0)}} \tag{6.2.5}$$

Observons le terme $\left\langle \mathbf{d}^{(0)}\mathbf{f}^{\mathrm{SR}}\right\rangle$ :

$$\left\langle \mathbf{d}^{(0)}\mathbf{f}^{\mathrm{SR}}\right\rangle = \sum_A \int dr\ \frac{\partial F}{\partial \xi_A}\left(\frac{\partial \xi_A}{\partial d_{ab}^{(0)}} d_{ab}^{(0)}\right), \tag{6.2.6}$$

qui n'est pas en général égal à $E_{\mathrm{Hxc}}^{\mathrm{SR}}$ : on a donc $\Delta_{\mathrm{DC}}^{\mathrm{SR}} = E_{\mathrm{Hxc}}^{\mathrm{SR}} - \left\langle \mathbf{d}^{(0)}\mathbf{f}^{\mathrm{SR}}\right\rangle$. La définition en terme de fockienne courte-portée est donc purement formelle, mais est utile dans la suite de la dérivation.

### 6.2.2 Expression de l'énergie RPA

Dans l'équation (6.2.1), l'énergie de corrélation longue-portée est ici une énergie RPA, et plus spécifiquement une énergie RPA exprimée dans la formulation avec équations de "Riccati ". On rappelle que dans cette formulation, on dispose des énergies suivantes :

$$E_c^{\mathrm{dRPA\text{-}I\text{-}Riccati}} = \frac{1}{2}\,\mathrm{tr}\left\{{}^1\mathbf{B}^{\mathrm{dRPA}}\ {}^1\mathbf{T}\right\} \tag{6.2.7}$$

$$E_c^{\mathrm{dRPA\text{-}I\text{-}SOSEX}} = \frac{1}{2}\,\mathrm{tr}\left\{{}^1\mathbf{B}^{\mathrm{RPAx}}\ {}^1\mathbf{T}\right\}, \tag{6.2.8}$$

avec les amplitudes :

$$\mathbf{0} = {}^1\mathbf{B}^{\mathrm{dRPA}} + {}^1\mathbf{B}^{\mathrm{dRPA}}\ {}^1\mathbf{T} + {}^1\mathbf{T}\ {}^1\mathbf{B}^{\mathrm{dRPA}} + {}^1\mathbf{T}\ {}^1\mathbf{B}^{\mathrm{dRPA}}\ {}^1\mathbf{T} + \boldsymbol{\varepsilon}\ {}^1\mathbf{T} + {}^1\mathbf{T}\boldsymbol{\varepsilon}, \tag{6.2.9}$$

et :

$$E_c^{\text{RPAx-II-Riccati}} = \frac{1}{4}\,\text{tr}\left\{ {}^1\mathbf{B}^{\text{RPAx}}\,{}^1\mathbf{T} + 3\,{}^3\mathbf{B}^{\text{RPAx}}\,{}^3\mathbf{T}\right\} \tag{6.2.10}$$

$$E_c^{\text{RPAx-II-SO2}} = \frac{1}{2}\,\text{tr}\left\{ {}^1\mathbf{B}^{\text{dRPA}}\,{}^1\mathbf{T}\right\} \tag{6.2.11}$$

$$E_c^{\text{RPAx-II-SO1}} = \frac{1}{2}\,\text{tr}\left\{ {}^1\mathbf{B}^{\text{RPAx}}\left( {}^1\mathbf{T} - {}^3\mathbf{T}\right)\right\}, \tag{6.2.12}$$

avec les amplitudes :

$$\begin{aligned} \mathbf{0} &= {}^1\mathbf{B}^{\text{RPAx}} + {}^1\mathbf{A}'^{\text{RPAx}}\,{}^1\mathbf{T} + {}^1\mathbf{T}\,{}^1\mathbf{A}'^{\text{RPAx}} + {}^1\mathbf{T}\,{}^1\mathbf{B}^{\text{RPAx}}\,{}^1\mathbf{T} + \boldsymbol{\varepsilon}\,{}^1\mathbf{T} + {}^1\mathbf{T}\boldsymbol{\varepsilon} \\ \mathbf{0} &= {}^3\mathbf{B}^{\text{RPAx}} + {}^3\mathbf{A}'^{\text{RPAx}}\,{}^3\mathbf{T} + {}^3\mathbf{T}\,{}^3\mathbf{A}'^{\text{RPAx}} + {}^3\mathbf{T}\,{}^3\mathbf{B}^{\text{RPAx}}\,{}^3\mathbf{T} + \boldsymbol{\varepsilon}\,{}^3\mathbf{T} + {}^3\mathbf{T}\boldsymbol{\varepsilon} \end{aligned} \tag{6.2.13}$$

On peut aussi considérer la version RPAX2 (voir section 2.6) :

$$E_c^{\text{RPAX2}} = \frac{1}{2}\,\text{tr}\left\{ {}^1\mathbf{B}^{\text{dRPA}}\,{}^1\mathbf{T}\right\}, \tag{6.2.14}$$

avec les amplitudes :

$$\mathbf{0} = {}^1\mathbf{B}^{\text{RPAx}} + {}^1\mathbf{B}^{\text{RPAx}}\,{}^1\mathbf{T} + {}^1\mathbf{T}\,{}^1\mathbf{B}^{\text{RPAx}} + {}^1\mathbf{T}\,{}^1\mathbf{B}^{\text{RPAx}}\,{}^1\mathbf{T} + \boldsymbol{\varepsilon}\,{}^1\mathbf{T} + {}^1\mathbf{T}\boldsymbol{\varepsilon} \tag{6.2.15}$$

### 6.2.3 Lagrangien total

La structure du Lagrangien associé à l'énergie totale RSH-RPA dans un cadre Riccati est :

$$\mathcal{L} = E_{\text{RSH}}^{\text{SR,LR}} + E_{\text{RPA}}^{\text{LR}} + \text{règles} \tag{6.2.16}$$

Les règles à ajouter au Lagrangien sont :
- les équations de Riccati correspondant à la version de RPA choisie : $\mathbf{R}(\mathbf{C},\mathbf{T}) = \mathbf{0}$,
- l'orthogonalité des orbitales moléculaires : $\mathbf{C}^\dagger\mathbf{S}\mathbf{C} - \mathbf{1} = \mathbf{0}$, où $\mathbf{C}$ sont les coefficients orbitalaires,
- le théorème de Brillouin, qui ne concerne que les éléments $ai$ (orbitales virtuelles-occupées) de la matrice de Fock : $(\mathbf{f})_{ai} = 0$,

Ces règles sont respectivement associées aux multiplicateurs de Lagrange (inconnus) $\boldsymbol{\lambda}$, $\mathbf{x}$ et $\mathbf{z}$. Le Lagrangien s'écrit donc :

$$\mathcal{L}[\mathbf{C},\mathbf{x},\mathbf{z},\mathbf{T},\boldsymbol{\lambda}] = \left\langle \mathbf{d}^{(0)}\mathbf{f}\right\rangle + \Delta_{\text{DC}} + E_c^{\text{RPA}} + \langle \boldsymbol{\lambda}\mathbf{R}(\mathbf{C},\mathbf{T})\rangle + \left\langle \mathbf{x}(\mathbf{C}^\dagger\mathbf{S}\mathbf{C} - \mathbf{1})\right\rangle + \langle \mathbf{z}\mathbf{f}\rangle \tag{6.2.17}$$

(on définit $z_{ij} = z_{ab} = 0$ pour pouvoir écrire une formulation matricielle compacte).

Comme voulu dans le formalisme du Lagrangien, les conditions stationnaires du Lagrangien par rapport à ses multiplicateurs :

$$\begin{aligned}
\frac{\partial \mathcal{L}}{\partial \boldsymbol{\lambda}} &= \mathbf{R}(\mathbf{C}, \mathbf{T}) = \mathbf{0} \\
\frac{\partial \mathcal{L}}{\partial \mathbf{x}} &= \mathbf{C}^{\dagger}\mathbf{S}\mathbf{C} - \mathbf{1} = \mathbf{0} \\
\frac{\partial \mathcal{L}}{\partial z_{ai}} &= f_{ai} = 0,
\end{aligned} \tag{6.2.18}$$

assurent que l'on a bien défini un objet de la méthode RSH-RPA, et non pas une autre. Les conditions stationnaires par rapport aux paramètres non stationnaires de l'énergie vont fournir de nouvelles équations déterminant les multiplicateurs de Lagrange.

## 6.3 Conditions stationnaires

Les conditions stationnaires par rapport aux amplitudes $\mathbf{T}$ sont très simples à dériver (peu de termes dépendent de $\mathbf{T}$ dans (6.2.17)) :

$$\frac{\partial \mathcal{L}}{\partial \mathbf{T}} = \mathbf{Q}(\mathbf{T})\boldsymbol{\lambda} + \boldsymbol{\lambda}\mathbf{Q}(\mathbf{T})^{\dagger} = -\mathbf{I}^{\dagger}, \tag{6.3.1}$$

et fournissent $\boldsymbol{\lambda}$ avec des équations très proches des équations de Riccati qui définissent $\mathbf{T}$. L'expression de la matrice $\mathbf{Q}(\mathbf{T})$ dépend des équations de Riccati choisies pour le calcul et la matrice $\mathbf{I}$ de la formule de l'énergie considérée (voir la section 6.2.2).

Les conditions stationnaires par rapport aux coefficients des orbitales $\mathbf{C}$, en revanche, demandent une longue dérivation. Dans toutes les versions des équations de Riccati et de l'énergie, on peut toujours réécrire les termes suivants comme :

$$E_c^{\text{RPA}} + \langle \boldsymbol{\lambda}\mathbf{R}(\mathbf{C}, \mathbf{T}) \rangle = \langle \mathbf{K}\mathbf{M} \rangle + \langle \mathbf{K}'\mathbf{N} \rangle + \langle \mathbf{J}\mathbf{O} \rangle + \langle \boldsymbol{\lambda}\boldsymbol{\varepsilon}\mathbf{T} + \boldsymbol{\lambda}\mathbf{T}\boldsymbol{\varepsilon} \rangle, \tag{6.3.2}$$

où je distingue les contributions liées aux intégrales $\mathbf{K}$, $\mathbf{K}'$, et $\mathbf{J}$ des contributions liées à la matrice de Fock présentent dans la matrice $\boldsymbol{\varepsilon}$ :

$$\begin{aligned}
\langle \boldsymbol{\lambda}\boldsymbol{\varepsilon}\mathbf{T} + \boldsymbol{\lambda}\mathbf{T}\boldsymbol{\varepsilon} \rangle &= \lambda_{ia,jb}(f_{bc}\delta_{jk} - f_{jk}\delta_{bc})T_{kc,ia} + \lambda_{ia,jb}T_{jb,kc}(f_{ca}\delta_{ki} - f_{ki}\delta_{ca}) \\
&= \lambda_{ia,jb}T_{jc,ia}f_{bc} - \lambda_{ia,jb}T_{kb,ia}f_{jk} + \lambda_{ia,jb}T_{jb,ic}f_{ca} - \lambda_{ia,jb}T_{jb,ka}f_{ki} \\
&= \{\mathbf{T}, \boldsymbol{\lambda}\}_{cb}\, f_{bc} - \{\mathbf{T}, \boldsymbol{\lambda}\}_{kj}\, f_{jk} + \{\boldsymbol{\lambda}, \mathbf{T}\}_{ac}\, f_{ca} - \{\boldsymbol{\lambda}, \mathbf{T}\}_{ik}\, f_{ki} \qquad (6.3.3) \\
&= \left\langle \mathbf{d}^{(2)}\mathbf{f} \right\rangle, \qquad (6.3.4)
\end{aligned}$$

où j'introduis à la fois $\mathbf{d}^{(2)}$ :

$$
\begin{aligned}
\left(\mathbf{d}^{(2)}\right)_{ij} &= -\left(\mathbf{T}\boldsymbol{\lambda} + \boldsymbol{\lambda}\mathbf{T}\right)_{ia,ja} = -\{\mathbf{T},\boldsymbol{\lambda}\}_{ij} - \{\boldsymbol{\lambda},\mathbf{T}\}_{ij} \\
\left(\mathbf{d}^{(2)}\right)_{ab} &= \left(\mathbf{T}\boldsymbol{\lambda} + \boldsymbol{\lambda}\mathbf{T}\right)_{ia,ib} = \{\mathbf{T},\boldsymbol{\lambda}\}_{ab} + \{\boldsymbol{\lambda},\mathbf{T}\}_{ab} \\
\left(\mathbf{d}^{(2)}\right)_{ai} &= 0,
\end{aligned}
\tag{6.3.5}
$$

et la notation d'une sorte de « trace incomplète », qui dépend encore de deux indices parmi ceux qui composent les super-indices de deux matrices **X** et **Y** :

$$
\begin{aligned}
\{\mathbf{X},\mathbf{Y}\}_{ij} &= X_{\mathbf{i}a,kc}Y_{kc,\mathbf{j}a} \\
\{\mathbf{X},\mathbf{Y}\}_{ab} &= X_{i\mathbf{a},kc}Y_{kc,i\mathbf{b}}
\end{aligned}
\tag{6.3.6}
$$

Cette réécriture n'est pas utile en amont, car elle rendrait plus compliquée la dérivation des conditions de stationnarité par rapport aux amplitudes **T**, mais elle permet ici d'écrire le Lagrangien en factorisant les termes qui dépendent des coefficients **C**, facilitant la dérivation des conditions stationnaires. Le Lagrangien s'écrit en effet à présent :

$$
\mathcal{L} = \left\langle\left(\mathbf{d}^{(0)} + \mathbf{z} + \mathbf{d}^{(2)}\right)\mathbf{f}\right\rangle + \Delta_{\mathrm{DC}} + \langle\mathbf{KM}\rangle + \langle\mathbf{K}'\mathbf{N}\rangle + \langle\mathbf{JO}\rangle + \left\langle\mathbf{x}(\mathbf{C}^{\dagger}\mathbf{SC} - \mathbf{1})\right\rangle
\tag{6.3.7}
$$

Une dérivation minutieuse de chacun de ces termes par rapport à une rotation **V** des coefficients **C**, détaillée dans l'Annexe F, permet d'écrire :

$$
\begin{aligned}
\frac{\partial\mathcal{L}}{\partial V_{ij}} &= 2\Big(\mathbf{fd}^{(2)} + \mathbf{fd}^{(0)} + \mathbf{g}^{\mathrm{LR}}\left[\mathbf{z} + \mathbf{d}^{(2)}\right]\mathbf{d}^{(0)} + \mathbf{W}^{\mathrm{SR}}\left[\mathbf{z} + \mathbf{d}^{(2)}\right]\mathbf{d}^{(0)} + \{\mathbf{K},\mathbf{M}\} + \{\mathbf{K}',\mathbf{N}\} + \{\mathbf{J},\mathbf{O}\} + x\Big)_{ij} \\
\frac{\partial\mathcal{L}}{\partial V_{aj}} &= 2\Big(\mathbf{fz} + \mathbf{g}^{\mathrm{LR}}\left[\mathbf{z} + \mathbf{d}^{(2)}\right]\mathbf{d}^{(0)} + \mathbf{W}^{\mathrm{SR}}\left[\mathbf{z} + \mathbf{d}^{(2)}\right]\mathbf{d}^{(0)} + \left\{\overline{\mathbf{K}},\mathbf{M}\right\} + \left\{\overline{\mathbf{K}}',\mathbf{N}\right\} + \left\{\overline{\mathbf{J}},\mathbf{O}\right\} + x\Big)_{aj} \\
\frac{\partial\mathcal{L}}{\partial V_{ib}} &= 2\Big(\mathbf{fz} + \left\{\overline{\mathbf{K}},\mathbf{M}\right\} + \left\{\overline{\mathbf{K}}',\mathbf{N}\right\} + \left\{\overline{\mathbf{J}},\mathbf{O}\right\} + x\Big)_{ib} \\
\frac{\partial\mathcal{L}}{\partial V_{ab}} &= 2\Big(\mathbf{fd}^{(2)} + \{\mathbf{K},\mathbf{M}\} + \{\mathbf{K}',\mathbf{N}\} + \{\mathbf{J},\mathbf{O}\} + x\Big)_{ab},
\end{aligned}
\tag{6.3.8}
$$

où, dans un souci d'exhaustivité, on montre les structures de chaque élément de matrice ($ij$, $aj$, $ib$, et $ab$). Les différences dans les expressions s'expliquent par la structure des matrices qui composent le Lagrangien (voir Annexe F.5). Dans la suite, on rassemble les éléments qui ne dépendent ni de **x** ni de **z** dans la matrice $\boldsymbol{\Theta}$ et les éléments qui dépendent de **z** dans la matrice $\widetilde{\boldsymbol{\Theta}}(\mathbf{z})$, de sorte que la condition de stationnarité du Lagrangien par rapport à la rotation des coefficients **C** s'écrit :

$$
\frac{1}{2}\frac{\partial\mathcal{L}}{\partial\mathbf{V}} = \boldsymbol{\Theta} + \widetilde{\boldsymbol{\Theta}}(\mathbf{z}) + 2\mathbf{x} = \mathbf{0}
\tag{6.3.9}
$$

C'est l'équation qui nous servira pour déterminer les multiplicateurs **z** et **x**.

## 6.4 Multiplicateurs z et x

Le multiplicateur de Lagrange $\mathbf{x}$ est hermitien, on peut donc écrire à partir de (6.3.9) :

$$\begin{cases} \boldsymbol{\Theta} - \boldsymbol{\Theta}^\dagger + \widetilde{\boldsymbol{\Theta}}(\mathbf{z}) - \widetilde{\boldsymbol{\Theta}}(\mathbf{z})^\dagger = \mathbf{0} & (6.4.1a) \\ \boldsymbol{\Theta} + \boldsymbol{\Theta}^\dagger + \widetilde{\boldsymbol{\Theta}}(\mathbf{z}) + \widetilde{\boldsymbol{\Theta}}(\mathbf{z})^\dagger = -4\mathbf{x} & (6.4.1b) \end{cases}$$

La première équation est une sorte d'équation CP-RPA[200, 213–215] (*Coupled Perturbed RPA*) dont uniquement la partie virtuelle-occupée ($ai$) est à considérer. En effet, on peut montrer que les blocs occupée-occupée $ij$ et virtuelle-virtuelle $ab$ de $\boldsymbol{\Theta}$ et $\widetilde{\boldsymbol{\Theta}}(\mathbf{z})$ sont hermitiens ; et de toutes les manières, seul $z_{ai}$ nous intéresse. On ne doit donc résoudre que $\left(\boldsymbol{\Theta} - \boldsymbol{\Theta}^\dagger + \widetilde{\boldsymbol{\Theta}}(\mathbf{z}) - \widetilde{\boldsymbol{\Theta}}(\mathbf{z})^\dagger\right)_{ai} = 0$, c'est-à-dire :

$$\left(\boldsymbol{\Theta} - \boldsymbol{\Theta}^\dagger + \mathbf{fz} - \mathbf{zf} + 4\mathbf{g}^{\mathrm{LR}}[\mathbf{z}] + 4\mathbf{W}^{\mathrm{SR}}[\mathbf{z}]\right)_{ai} = 0 \qquad (6.4.2)$$

Une fois $\mathbf{z}$ connu, l'équation (6.4.1b) fournit $\mathbf{x}$.

Le formalisme qui est développé ici contraste avec la démarche suivie par Chabbal *et. al.* [216], où les gradients de méthodes de type séparation de portée (RSH+(L)MP2) ont été dérivés à partir du gradient MP2 *sans* séparation de portée , et ensuite accommodés pour la longue-portée. La partie courte-portée correspondant à un gradient DFT n'était rajoutée qu'ultérieurement, terme par terme. La théorie que je développe ici prend le chemin d'une dérivation tout-en-un du gradient total courte- et longue-portée. Notamment, l'émergence de l'objet $\mathbf{W}^{\mathrm{SR}}[\mathbf{z}]$, l'analogue courte-portée de $\mathbf{g}^{\mathrm{LR}}[\mathbf{z}]$, montre une synergie claire entre la dérivation des parties courte- et longue-portée du gradient de l'énergie (voir l'Annexe F.4).

De plus, on peut séparer dans le bloc $ij$ de $\mathbf{x}$ (1) les termes provenant de la dérivation de la partie du gradient provenant de l'énergie RPA longue-portée et (2) les termes émergeant de la dérivation de la partie du gradient qui vient de l'énergie RSH de référence. En d'autres termes, on peut écrire (voir l'équation (6.4.1b) pour trouver la formule pour $\mathbf{x}$ et l'équation (6.3.8) pour l'origine des termes $2(\mathbf{fd}^{(0)})_{ij} + 2(\mathbf{d}^{(0)}\mathbf{f})_{ij}$) :

$$\begin{aligned} -4(\mathbf{x})_{ij} &= -4(\mathbf{x}^{\mathrm{RPA}})_{ij} + 2(\mathbf{fd}^{(0)})_{ij} + 2(\mathbf{d}^{(0)}\mathbf{f})_{ij} \\ (\mathbf{x})_{ij} &= (\mathbf{x}^{\mathrm{RPA}})_{ij} - 2(\mathbf{f})_{ij}, \end{aligned} \qquad (6.4.3)$$

où l'on reconnaît le terme du multiplicateur de Lagrange trouvé dans les dérivations d'énergies de référence.

Une fois les équations (6.3.1), (6.4.2) et (6.4.1b) résolues pour $\boldsymbol{\lambda}$, $\mathbf{z}$ et $\mathbf{x}$, on connaît entièrement le Lagrangien, stationnaire par rapport à tous les paramètres de l'énergie RSH-RPA, et égal à l'énergie RSH-RPA aux paramètres $\mathbf{T}^*$ et $\mathbf{C}^*$ de l'énergie RSH-RPA.

## 6.5 Gradient analytique RSH-RPA

J'obtiens alors un gradient total (voir Annexe F.6) :

$$E^{(x)}_{\text{RSH+RPA}} = \mathcal{L}^{(x)} = \left\langle \mathbf{D}^1\mathbf{H}^{(x)} \right\rangle + \left(\mathbf{D}^2 + \mathbf{\Gamma}^2\right)_{\mu\nu,\rho\sigma} (\mu\nu|\rho\sigma)^{\text{LR}(x)} + \left\langle \mathbf{X}\mathbf{S}^{(x)} \right\rangle + \text{SR}^{(x)}, \tag{6.5.1}$$

où :

$$\begin{aligned}
(\mathbf{D}^1)_{\mu\nu} &= C_{\mu p}\left(\mathbf{d}^{(0)} + \mathbf{d}^{(2)} + \mathbf{z}\right)_{pq} C^\dagger_{q\nu} = \left(\mathbf{D}^{(0)} + \mathbf{D}^{(2)} + \mathbf{Z}\right)_{\mu\nu} \\
(\mathbf{D}^2)_{\mu\nu,\sigma\rho} &= \left(\tfrac{1}{2}\mathbf{D}^{(0)} + \mathbf{D}^{(2)} + \mathbf{Z}\right)_{\mu\nu} D^{(0)}_{\rho\sigma} - \frac{1}{2}\left(\tfrac{1}{2}\mathbf{D}^{(0)} + \mathbf{D}^{(2)} + \mathbf{Z}\right)_{\mu\rho} D^{(0)}_{\nu\sigma} \\
(\mathbf{\Gamma}^2)_{\mu\nu,\sigma\rho} &= C_{\mu k}C_{\nu j}C^\dagger_{c\rho}C^\dagger_{b\sigma}(\mathbf{M})_{ia,kc} + C_{\mu k}C_{\nu j}C^\dagger_{b\rho}C^\dagger_{c\sigma}(\mathbf{N})_{ia,kc} + C_{\mu k}C_{\nu b}C^\dagger_{j\rho}C^\dagger_{c\sigma}(\mathbf{O})_{ia,kc} \\
(\mathbf{X})_{\mu\nu} &= C_{\mu p}(\mathbf{x})_{pq}C^\dagger_{q\nu} \\
\text{SR}^{(x)} &= \sum_A \omega^{(x)}_\lambda \left(F(\xi_A) + \frac{\partial F}{\partial \xi_A}\left(\xi^{\mathbf{d}^{(2)}}_A + \xi^{\mathbf{z}}_A\right)\right) \\
&\quad + \sum_A \omega_\lambda \frac{\partial F}{\partial \xi_A}\left(\xi^{\mathbf{d}^{(0)}(x)}_A + \xi^{\mathbf{d}^{(2)}(x)}_A + \xi^{\mathbf{z}(x)}_A\right) + \sum_{AB} \omega_\lambda \frac{\partial^2 F}{\partial \xi_B \partial \xi_A}\left(\xi^{\mathbf{d}^{(2)}}_A + \xi^{\mathbf{z}}_A\right)\xi^{(x)}_B
\end{aligned} \tag{6.5.2}$$

## 6.6 Discussion autour du parallèle avec le gradient RSH-MP2

La dérivation présentée dans les sections précédentes montre d'intéressants parallèles avec la dérivation du gradient de l'énergie RSH-MP2. Concernant l'énergie MP2 sans séparation de portée, considérons la fonctionnelle Hylleraas [156, 217] suivante :

$$\mathcal{H} = \left\langle \psi^{(1)} \middle| \hat{H}^{(0)} - E^{(0)} \middle| \psi^{(1)} \right\rangle + 2\left\langle \psi^{(1)} \middle| \hat{H} \middle| \psi^{(0)} \right\rangle, \tag{6.6.1}$$

où :

$$\begin{aligned}
\psi^{(1)} &= \frac{1}{2}T_{ia,jb}\phi_{ia,jb} = \widetilde{T}_{ia,jb}\widetilde{\phi}_{ia,jb} \\
\phi_{ia,jb} &= \hat{E}_{ai}\hat{E}_{bj}\psi^{(0)} \\
\widetilde{\phi}_{ia,jb} &= \frac{1}{6}\left(2\phi_{ia,jb} + \phi_{ib,ja}\right),
\end{aligned} \tag{6.6.2}$$

où $\hat{E}_{ai}$ sont les opérateurs d'excitations à une particule bien connus, et où les objets *contravariants* $\widetilde{\mathbf{T}}$ et $\widetilde{\phi}$[218, 219] sont introduits pour des raisons qui ne sont pas développées ici.

Les conditions de stationnarité de la fonctionnelle Hylleraas par rapport à $\widetilde{\mathbf{T}}$ s'écrivent :

$$\frac{\partial \mathcal{H}}{\partial \widetilde{T}_{ia,jb}} = 2\left\langle \widetilde{\phi}_{ia,jb} \middle| \hat{H}^{(0)} - E^{(0)} \middle| \widetilde{T}_{kl,cd}\widetilde{\phi}_{kl,cd} \right\rangle + 2\left\langle \widetilde{\phi}_{ia,jb} \middle| \hat{H} \middle| \psi^{(0)} \right\rangle = 2\widetilde{R}_{ia,jb} = 0 \tag{6.6.3}$$

Une dérivation attentive montre que :

$$\widetilde{\mathbf{R}} = \mathbf{K} + \mathbf{T}\boldsymbol{\varepsilon} + \boldsymbol{\varepsilon}\mathbf{T}, \tag{6.6.4}$$

et, avec $E^{(2)} = \widetilde{T}_{ia,jb} \left\langle \widetilde{\phi}_{ia,jb} \middle| \hat{H} \middle| \psi^{(0)} \right\rangle = \widetilde{T}_{ia,jb} K_{ia,jb}$ , que :

$$\mathcal{H} = E^{(2)} + \widetilde{T}_{ia,jb} \widetilde{R}_{ia,jb} \tag{6.6.5}$$

Ainsi, lorsque les conditions stationnaires de la fonctionnelle Hylleraas par rapport aux amplitudes $\widetilde{\mathbf{T}}$ sont respectées (*i.e.* lorsque $\widetilde{\mathbf{R}} = \mathbf{0}$), la fonctionnelle Hylleraas est à la fois égale à l'énergie MP2, $E^{(2)}$, et stationnaire par rapport à $\widetilde{\mathbf{T}}$. Ceci n'est bien évidemment rien d'autre qu'une conséquence de la stationnarité de la fonctionnelle Hylleraas, mais peut être vu dans un formalisme Lagrangien comme :

$$\mathcal{H} \doteq \mathcal{L} = E^{(2)} + \left\langle \boldsymbol{\lambda} \widetilde{\mathbf{R}} \right\rangle, \tag{6.6.6}$$

où $\boldsymbol{\lambda} = \widetilde{\mathbf{T}}$ est *le* multiplicateur qui assure la stationnarité du Lagrangien par rapport aux amplitudes.

**En d'autres termes** : là où, dans le cas RPA, les équations de Riccati permettent de calculer les amplitudes, et les conditions stationnaires d'un Lagrangien sont imposées dans un deuxième temps à travers un multiplicateur de Lagrange non trivial, $\boldsymbol{\lambda}$, ici, dans le cas MP2, les amplitudes sont déterminées par la stationnarité de la fonctionnelle Hylleraas elle-même, c'est-à-dire que dans un parallèle avec un formalisme Lagrangien, on introduit pour s'assurer de la stationnarité un multiplicateur de Lagrange trivial (redondant) : $\boldsymbol{\lambda} = \widetilde{\mathbf{T}}$.

C'est pourquoi on retrouve dans des dérivations du gradient MP2[220, 221] la quantité suivante :

$$\mathcal{L} = E^{(2)} + \left\langle \widetilde{\mathbf{T}} \widetilde{\mathbf{R}} \right\rangle = 2 \left\langle \widetilde{\mathbf{T}} \mathbf{K} \right\rangle + \left\langle \mathbf{d}^{(2)} \mathbf{f} \right\rangle, \tag{6.6.7}$$

avec :

$$\begin{aligned} \left(\mathbf{d}^{(2)}\right)_{ij} &= -2 \left\{ \widetilde{\mathbf{T}}, \mathbf{T} \right\}_{ij} \\ \left(\mathbf{d}^{(2)}\right)_{ab} &= 2 \left\{ \widetilde{\mathbf{T}}, \mathbf{T} \right\}_{ab} \\ \left(\mathbf{d}^{(2)}\right)_{ai} &= 0, \end{aligned} \tag{6.6.8}$$

qui est à comparer à la définition de $\mathbf{d}^{(2)}$ dans le cas RPA, équation (6.3.5), en se souvenant qu'ici $\boldsymbol{\lambda} = \widetilde{\mathbf{T}}$. Le Lagrangien total RSH+MP2 est :

$$\mathcal{L} = \left\langle \left( \mathbf{d}^{(0)} + \mathbf{z} + \mathbf{d}^{(2)} \right) \mathbf{f} \right\rangle + \Delta_{\mathrm{DC}} + 2 \left\langle \mathbf{K} \widetilde{\mathbf{T}} \right\rangle + \left\langle \mathbf{x} (\mathbf{C}^{\dagger} \mathbf{S} \mathbf{C} - \mathbf{1}) \right\rangle, \tag{6.6.9}$$

où, mise à part la définition de $\mathbf{d}^{(2)}$, seuls les termes impliquant les matrices **M**, **N** et **O** sont différents : c'est-à-dire que la totalité de la dérivation du gradient de l'énergie RSH-MP2 *peut* être vue comme un cas particulier de RSH-RPA, ou la dérivation du gradient de l'énergie RSH-RPA comme une généralisation de RSH-MP2.

## 6.7 Implémentation et Validation

Cette (nouvelle) dérivation de gradient RSH-RPA a été implémentée dans MOLPRO [131]. Un gros effort a été fait pour implémenter les gradients dans le « cœur » de MOLPRO (par opposition à l'écriture d'un module en périphérie de MOLPRO) de sorte que, à la suite de cette implémentation, toutes les possibilités de MOLPRO soient disponibles pour les gradients RSH-RPA (notamment : optimisation de géométrie, calcul de dipôle, *etc.*. . ). Le parallèle entre les gradients RSH-MP2 et RSH-RPA a été utilisé, c'est-à-dire qu'une implémentation préexistante des gradients RSH-MP2[216] a été modifiée pour permettre le calcul de gradient RSH-RPA.

On cherche ainsi à construire les objets vus équation (6.5.2), après avoir calculé les amplitudes RPA par résolution d'une équation de Riccati et les multiplicateurs de Lagrange *via* les équations (6.3.1), (6.4.2) et (6.4.1b). Chronologiquement : les matrices $\mathbf{T}$ et $\boldsymbol{\lambda}$ sont obtenues par résolution itérative des équations de Riccati et de stationnarité du Lagrangien par rapport aux amplitudes (équation (6.3.1)). Ces deux résolutions sont très semblables et ne posent pas de problème majeur. Une fois connus ces deux éléments, on peut calculer les matrices $\mathbf{M}$, $\mathbf{N}$ et $\mathbf{O}$ ainsi que la matrice $\mathbf{d}^{(2)}$. À ce moment, il reste à calculer les multiplicateurs $\mathbf{z}$ et $\mathbf{x}$, c'est-à-dire à résoudre l'équation CP-RPA (6.4.2). L'étape capitale est la contraction des objets $\{\mathbf{K}, \mathbf{M}\}$, *etc.*. . , et, surtout, $\left\{\overline{\mathbf{K}}, \mathbf{M}\right\}$ *etc.*. . vus équation (6.3.8) et qui composent la matrice $\boldsymbol{\Theta} - \boldsymbol{\Theta}^{\dagger}$ utilisée dans l'équation CP-RPA. En pratique, tous ces objets sont calculés en amont d'un calcul de gradient et sauvegardés dans des *records* que MOLPRO utilise dans un calcul très classique de gradient, comme montré équation (6.5.1).

Dans l'état actuel de l'implémentation, toutes les versions de RPA qui ne contiennent pas les intégrales bi-électroniques $\mathbf{J}$, dans l'expression de l'énergie ou des équations de Riccati sont fonctionnelles (c'est-à-dire : dRPA-I, SOSEX, et RPAX2). Les autres versions ne sont pas théoriquement plus compliquées, et sont d'ailleurs prévues dans le programme, mais demandent la contraction d'objets de la forme $\{\mathbf{J}, \mathbf{O}\}$ et surtout $\left\{\overline{\mathbf{J}}, \mathbf{O}\right\}$ dont la construction efficace reste à mettre en œuvre dans le programme. La résolution de ce problème est une tâche prioritaire pour rendre notre implémentation la plus générale possible.

L'implémentation en elle-même a été testée par des comparaisons à des gradients numériques calculés avec des formules à 3- et 5-points ; les gradients analytiques coïncident tous avec une précision d'au moins $10^{-6}$ Hartree aux gradients numériques. Le temps de calcul nécessaire est, comme attendu, grossièrement le double du temps de calcul nécessaire pour un calcul d'énergie RPA. En terme de pourcentage du temps de calcul d'une énergie, la même différence de temps de calcul est observée en RPA et en MP2 entre le calcul d'un gradient et le calcul d'une énergie ; également : la même différence est observée entre un coût de calcul RPA ou MP2, qu'il s'agisse d'un calcul de gradient ou d'énergie. Ceci tend à montrer que l'implémentation dans le « cœur » de MOLPRO permet de s'assurer que l'on ne perd pas de temps de calcul sur un point ou l'autre de la procédure. La majorité du temps de calcul est passé à construire les objets $\left\{\overline{\mathbf{K}}, \mathbf{M}\right\}$ et à la résolution des équations CP-RPA (comme dans le cas d'un calcul MP2). On remarque que les convergences avec la base sont meilleures en RSH que sans séparation de portée, en accord avec les observations faites dans la littérature[222].

## 6.8 Premiers résultats

### 6.8.1 Densités corrélées

Une première exploitation du développement de ces gradients est à trouver dans l'étude des densités corrélées qui émergent dès l'équation (6.3.7), et que l'on trouve dans l'expression du gradient total, équation (6.5.1) :

$$\left(\mathbf{D}^1\right)_{\mu\nu} = C_{\mu p}\left(\mathbf{d}^{(0)} + \mathbf{d}^{(2)} + \mathbf{z}\right)_{pq} C^{\dagger}_{q\nu} = \left(\mathbf{D}^{(0)} + \mathbf{D}^{(2)} + \mathbf{Z}\right)_{\mu\nu} \tag{6.8.1}$$

Vu que l'on dispose de tous les outils pour visualiser des objets sur tout type de grille (voir section 4.1), on peut montrer ici les structures des corrections aux densités de référence dues à la corrélation, c'est-à-dire les structures des objets :

$$n^{\text{corrélée}} = n^{(0)} + \underbrace{n^{(2)}}_{\Delta^{(2)}} \quad \text{et :} \quad n^{\text{corrélée}} = n^{(0)} + \underbrace{n^{(2)} + n^{(z)}}_{\Delta^{(2+z)}}, \tag{6.8.2}$$

où $n^{(0)}$, $n^{(2)}$ et $n^{(z)}$ sont formées à partir des matrices densité $\mathbf{d}^{(0)}$, $d^{(2)}$ et $\mathbf{z}$. On montre figures 6.2 et 6.3 les isocontours des objets $\Delta^{(2)}$ et $\Delta^{(2+z)}$ pour un calcul RSH+RPA réalisé sur les molécules d'eau et d'éthylène. On a vérifié que ces densités s'intègrent bien à zéro. À propos du signe de ces densités, on écrit la chose suivante : si $\Delta > 0$ (en bleu dans la figure), cela signifie que cette zone de l'espace est munie de « plus d'électrons » (d'une densité électronique plus importante) que dans la description de référence ; si $\Delta < 0$ (en rouge sur la figure), la zone subit une déplétion du nombre d'électron. On voit donc sur la figure 6.2, qui concerne la molécule d'eau, que les électrons sont, relativement à la référence RSH, repoussés par rapport aux noyaux et redistribués à l'extérieur de l'espace des liaisons O-H. On trouve des conclusions similaires aux références[223, 224]. Concernant l'éthylène, on voit également que les électrons sont éloignés des noyaux, mais ils semblent se concentrer sur la liaison C-C dans le cas de la représentation de $\Delta^{(2)}$ et plutôt hors de la liaison dans le cas de $\Delta^{(2+z)}$.

Des calculs sur des dipôles et quadrupoles moléculaires au niveau RSH-dRPA-I sont en cours sur un ensemble de systèmes étudiés récemment au niveau coupled cluster [225].

### 6.8.2 Optimisation de géométrie

Une bonne façon d'exploiter ces nouveaux gradients RSH+RPA est de les utiliser dans le cadre d'optimisation de géométrie de molécules ou de complexes intermoléculaires qui présentent des caractéristiques qui exigent un traitement fin des corrélations. Pour critiquer les résultats obtenus on peut (1) comparer les géométries en elles-mêmes (c'est-à-dire comparer les $3N - 6$ variables indépendantes du système une à une) ou (2) dans un cadre d'interaction intermoléculaire : comparer les énergies d'interaction obtenues *après* optimisation de géométrie.

Pour des raisons techniques l'optimisation de géométrie est réalisée par un optimiseur externe qui, à chaque itération, produit de nouvelles coordonnées à partir des énergies et gradients d'un calcul `MOLPRO` en utilisant des algorithmes d'optimisation quasi-Newton avec contrainte sur les coordonnées internes. Les scripts de l'optimiseur sont rassemblés dans un programme appelé `GADGET` développé par Bučko *et. al.* [226]. Ce programme est pensé dans l'esprit d'une interface avec `VASP`,

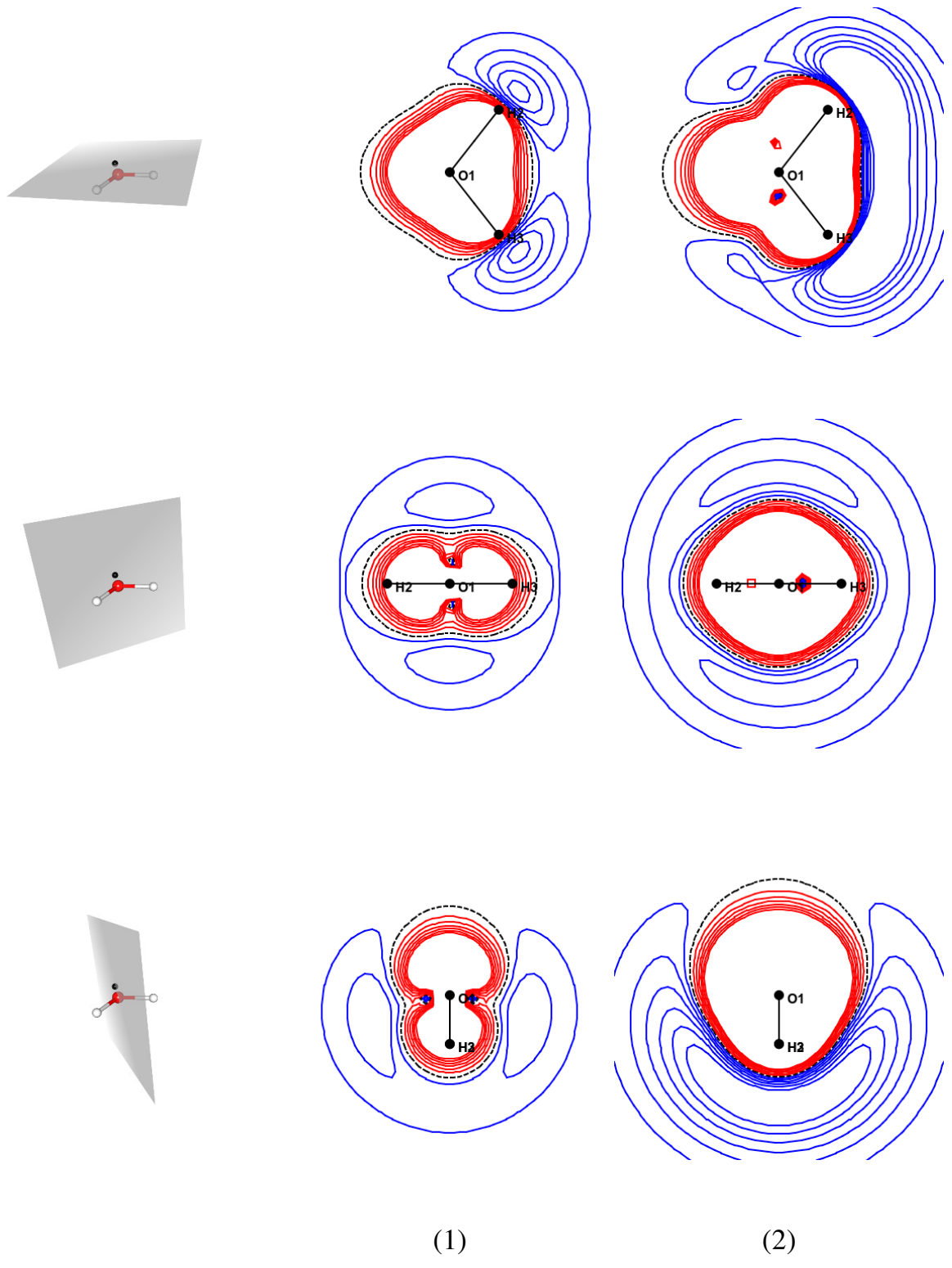

FIGURE 6.2: Isocontours sur trois plans de coupe des densités (1) $\Delta^{(2)}$ et (2) $\Delta^{(2+z)}$ formées à partir des matrices densité $\mathbf{d}^{(2)}$ et $\mathbf{d}^{(2)} + \mathbf{z}$ pour la molécule d'eau.

---

des scripts ont donc été écrits pendant cette thèse, en profitant des conseils de Tomáš Bučko, pour traduire des *input/output* de sorte à pouvoir utiliser `GADGET` avec `MOLPRO`.

De manière générale j'utilise pour présenter les résultats des outils simples qu'il faut définir ici. Pour comparer un ensemble de données $\{a\}$ à un ensemble de référence $\{a^{\text{ref}}\}$, j'utilise :

$$\begin{aligned}
&\textit{Mean Absolute Error} : && \text{MAE} = \frac{1}{N}\sum_i \left|a_i - a_i^{\text{ref}}\right| \\
&\textit{Mean Signed Error} : && \text{MSE} = \frac{1}{N}\sum_i a_i - a_i^{\text{ref}} \\
&\textit{Mean Absolute percentage Error} : && \text{MA\%E} = \frac{1}{N}\sum_i \left|\frac{a_i - a_i^{\text{ref}}}{a_i^{\text{ref}}}\right| \\
&\textit{Mean Signed pperentage Error} : && \text{MS\%E} = \frac{1}{N}\sum_i \frac{a_i - a_i^{\text{ref}}}{a_i^{\text{ref}}}
\end{aligned} \tag{6.8.3}$$

Je propose des résultats d'optimisation de géométrie sur une série de petites molécules proposées,

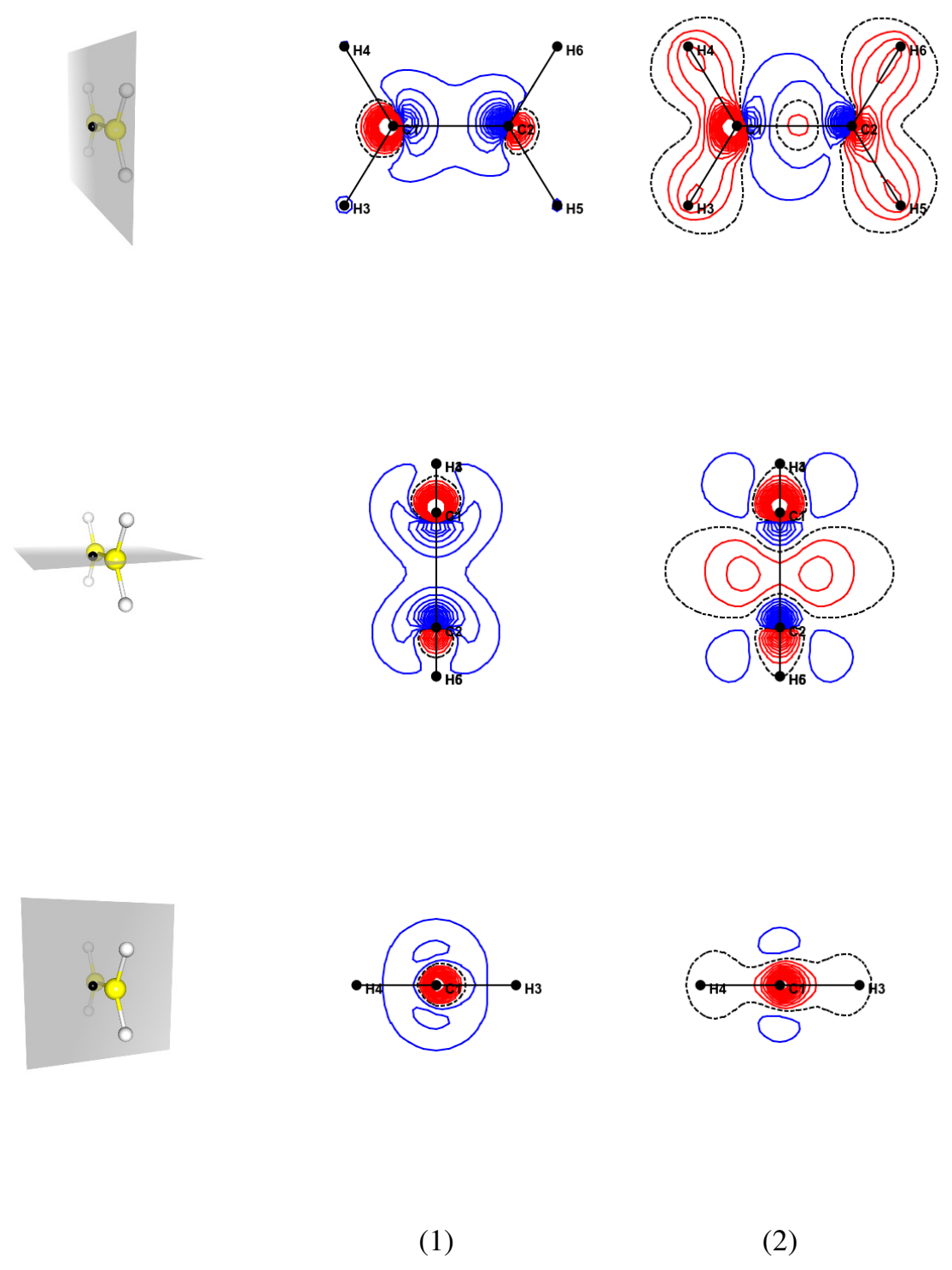

FIGURE 6.3: Isocontours sur trois plans de coupe des densités (1) $\Delta^{(2)}$ et (2) $\Delta^{(2+z)}$ formées à partir des matrices densité $\mathbf{d}^{(2)}$ et $\mathbf{d}^{(2)} + \mathbf{z}$ pour l'éthylène.

entre autres, par Helgaker *et. al.* [227]. Les géométries ont été optimisées au niveau RSH+RPA, avec la fonctionnelle d'échange-corrélation courte-portée srLDA[58, 60] et avec les versions de dRPA-I et SOSEX (voir section 2.6). On dispose également des résultats publiés récemment par Rekkedal *et. al.* [198] concernant leur dérivation des gradients dRPA-I sans séparation de portée (notons que j'ai pu reproduire ces résultats exactement, ce qui d'une certaine manière valide également la théorie et l'implémentation développées ici). Par soucis de comparaison homogène entre ces résultats et les optimisations que je présente ici, tous les calculs présentés ici sont faits avec la base cc-pvQz. Ceci étant dit, la comparaison de matériel fourni par Rekkedal *et. al.* tend à montrer que la convergence des résultats des optimisations avec la base utilisée est plus rapide dans le cas des calculs RSH, comme attendu (ceci pourrait faire l'objet d'une étude plus systématique).

On présente dans les figures 6.4 à 6.8 les longueurs de liaisons des géométries optimisées avec ces différentes méthodes, relativement à des longueurs de liaisons considérées comme référentes, obtenues par des calculs de type CCSD(T)[228]. Les liaisons montrées dans les graphes sont classées par longueurs référentes croissantes ; toutes les erreurs relatives sont inférieures à 0.1 Ångström. On présente figure 6.4 les longueurs de liaisons des géométries optimisées aux niveaux RHF-MP2, RHF-dRPA et RHF-SOSEX (c'est-à-dire sans séparation de portée). On voit que les longueurs RHF-MP2 sont quelque peu meilleures que les longueurs RHF-RPA, notamment pour les liaisons longues. Les optimisations au niveau RHF-SOSEX sont légèrement inférieures aux optimisations RHF-dRPA. Aux figures 6.5, 6.6 et 6.7 sont montrées des comparaisons entre les longueurs de liaisons des géométries optimisées avec et sans séparation de portée, pour les méthodes MP2, dRPA et SOSEX. La MP2 voit sa performance réduite par l'utilisation de la séparation de portée, surtout pour les liaisons longues, alors que les résultats dRPA et SOSEX sont tous améliorés dans le cadre RSH, notamment pour les liaisons courtes. Les méthodes dRPA et SOSEX ont exactement le même comportement. Enfin, on présente figure 6.8 les résultats compilés des longueurs de liaisons des géométries optimisées aux niveaux LDA-MP2, LDA-dRPA et LDA-SOSEX. On voit, comme observé par Rekkedal *et. al.*, que les longueurs des liaisons $F-F$ de $F_2$ et $O-F$ de $HOF$ sont les plus éloignées de la référence ; elles contribuent à un haut niveau aux moyennes qui seront montrées plus loin. On n'observe plus la différence que l'on voyait dans les calculs sans séparation de portée entre MP2 et dRPA/SOSEX. La perte de performance de MP2 et le gain de dRPA/SOSEX résultent en des qualités de longueurs de liaisons comparables pour les trois méthodes dans le cadre de la séparation de portée. La physique de ces liaisons s'apparente à des corrélations plutôt de courte portée, on n'est donc pas étonné de voir que la performance des méthodes de type séparation de portée n'est pas largement meilleure que celles des méthodes sans séparation de portée. Le but *in fine* sera plutôt d'utiliser les optimisations dans le cadre d'interaction inter-moléculaires. La figure 6.9, qui montre les *Mean Absolute Error* et *Mean Signed Error* en Ångström, ainsi que les *Mean Absolute pourcentage Error* et *Mean Signed pourcentage Error* des longueurs de liaisons de toutes les méthodes confirment les remarques précédemment faites. Notons que malgré les différences que l'on énumère ici, toutes les géométries sont au final très proches les unes des autres.

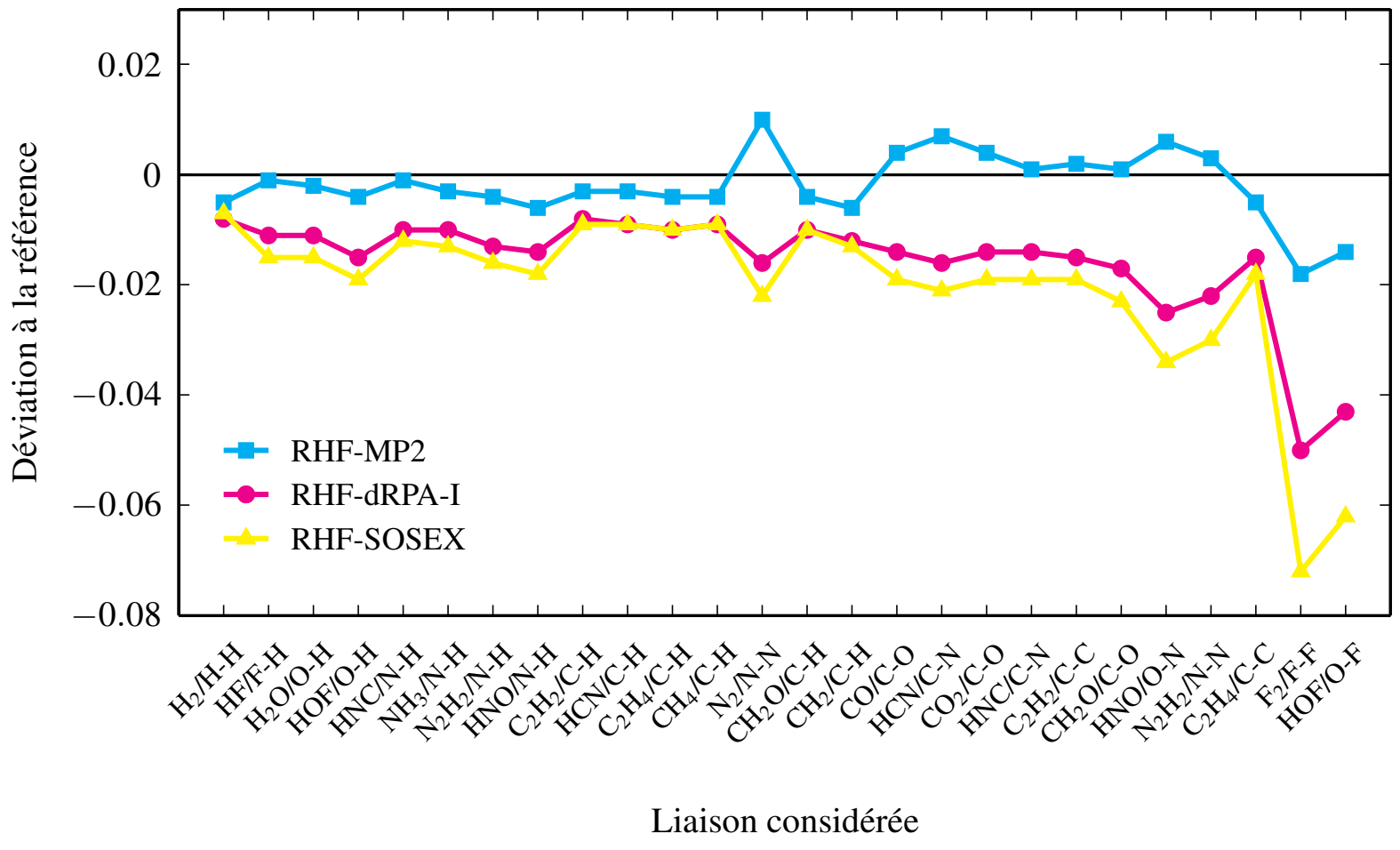

FIGURE 6.4: Déviations (Å) à la référence[228] des longueurs de liaison de 16 molécules optimisées aux niveaux RHF-MP2 (en cyan, carré plein), RHF-dRPA (en magenta, cercle plein) et RHF-SOSEX (en jaune, triangle plein). Les liaisons sont classées par longueurs référentes croissantes.

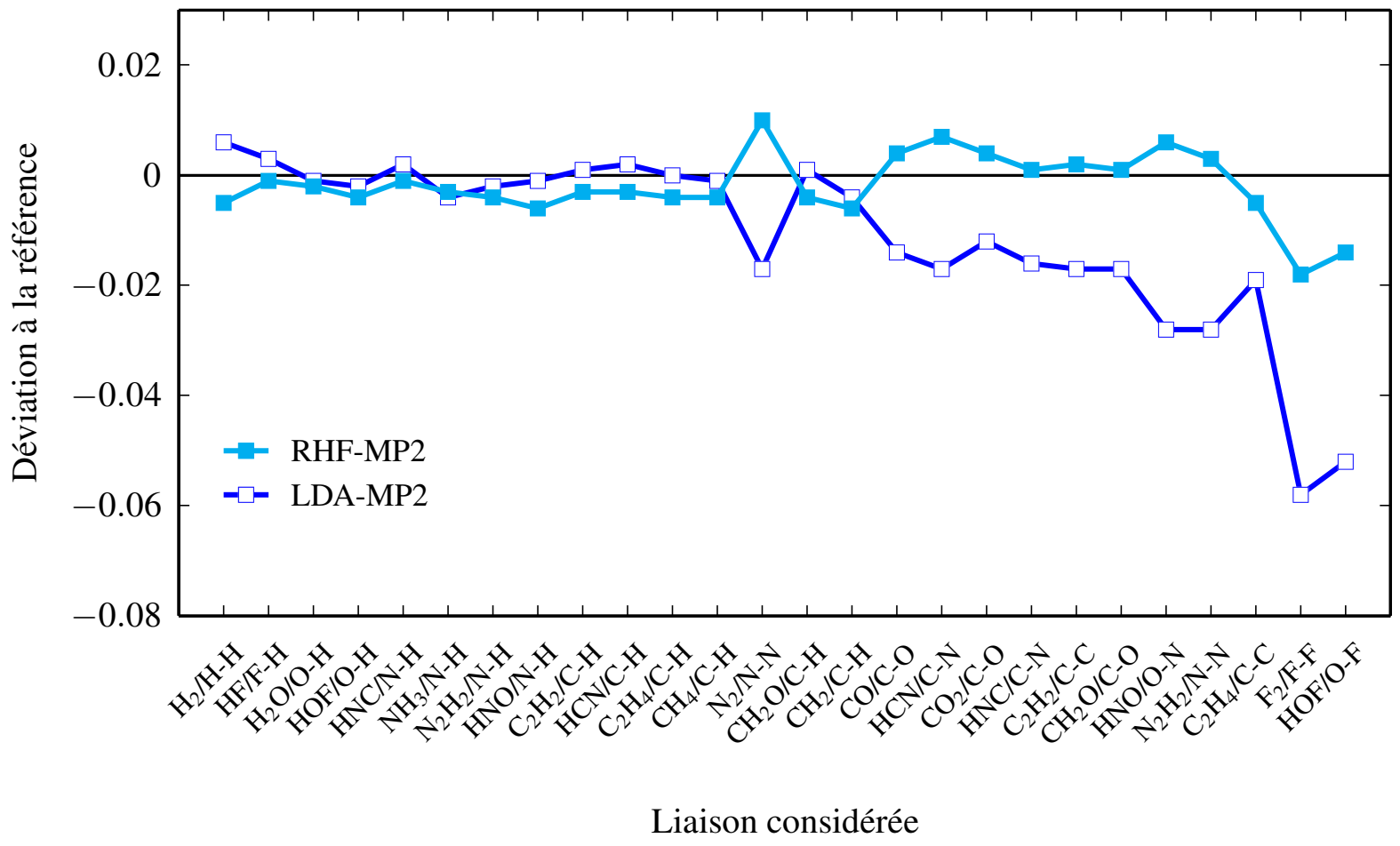

FIGURE 6.5: Déviations (Å) à la référence[228] des longueurs de liaison de 16 molécules optimisées aux niveaux RHF-MP2 (en cyan, carré plein), et LDA-MP2 (en bleu, carré vide). Les liaisons sont classées par longueurs référentes croissantes.

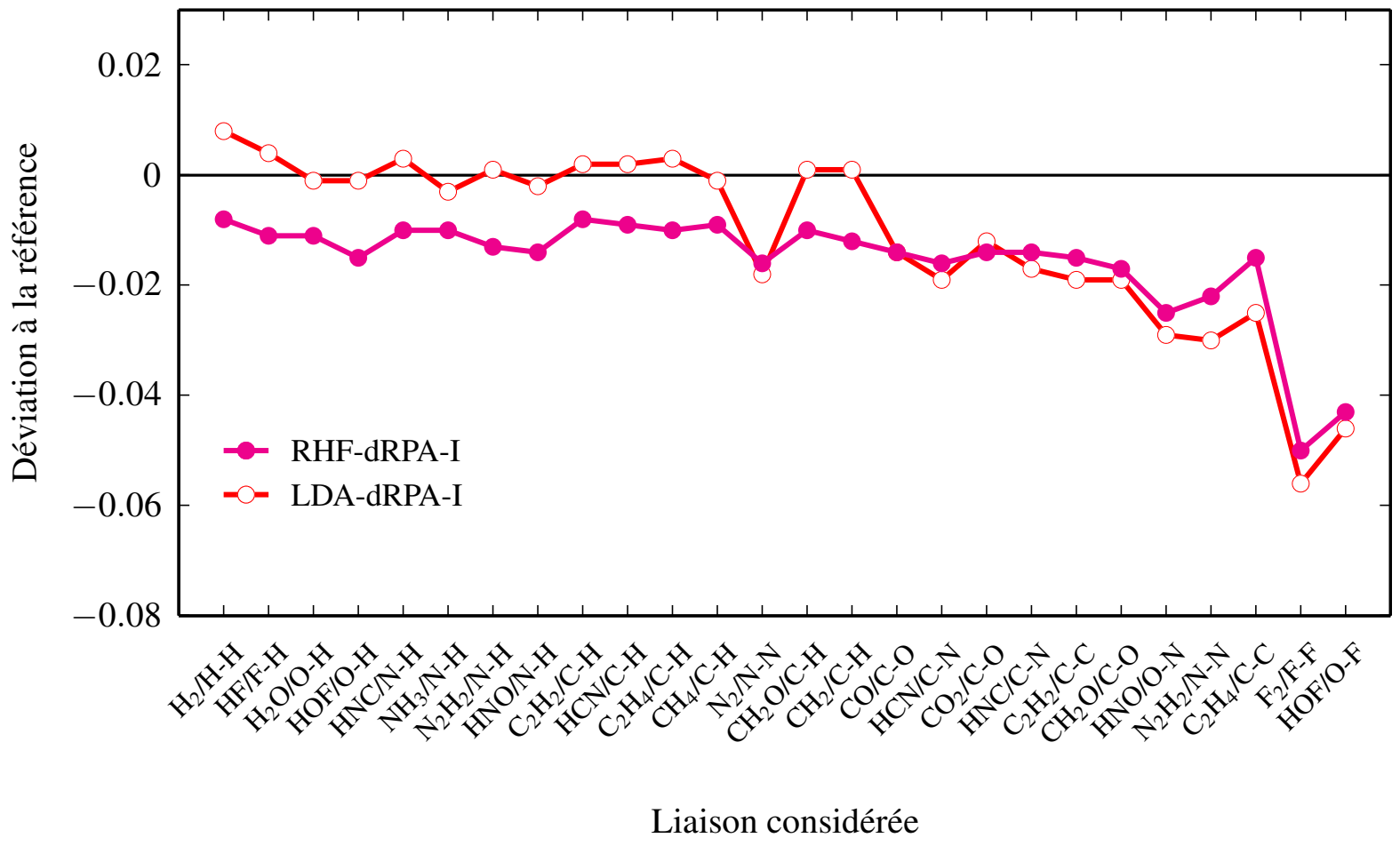

FIGURE 6.6: Déviations (Å) à la référence[228] des longueurs de liaison de 16 molécules optimisées aux niveaux RHF-dRPA (en magenta, cercle plein), et LDA-dRPA (en rouge, cercle vide). Les liaisons sont classées par longueurs référentes croissantes.

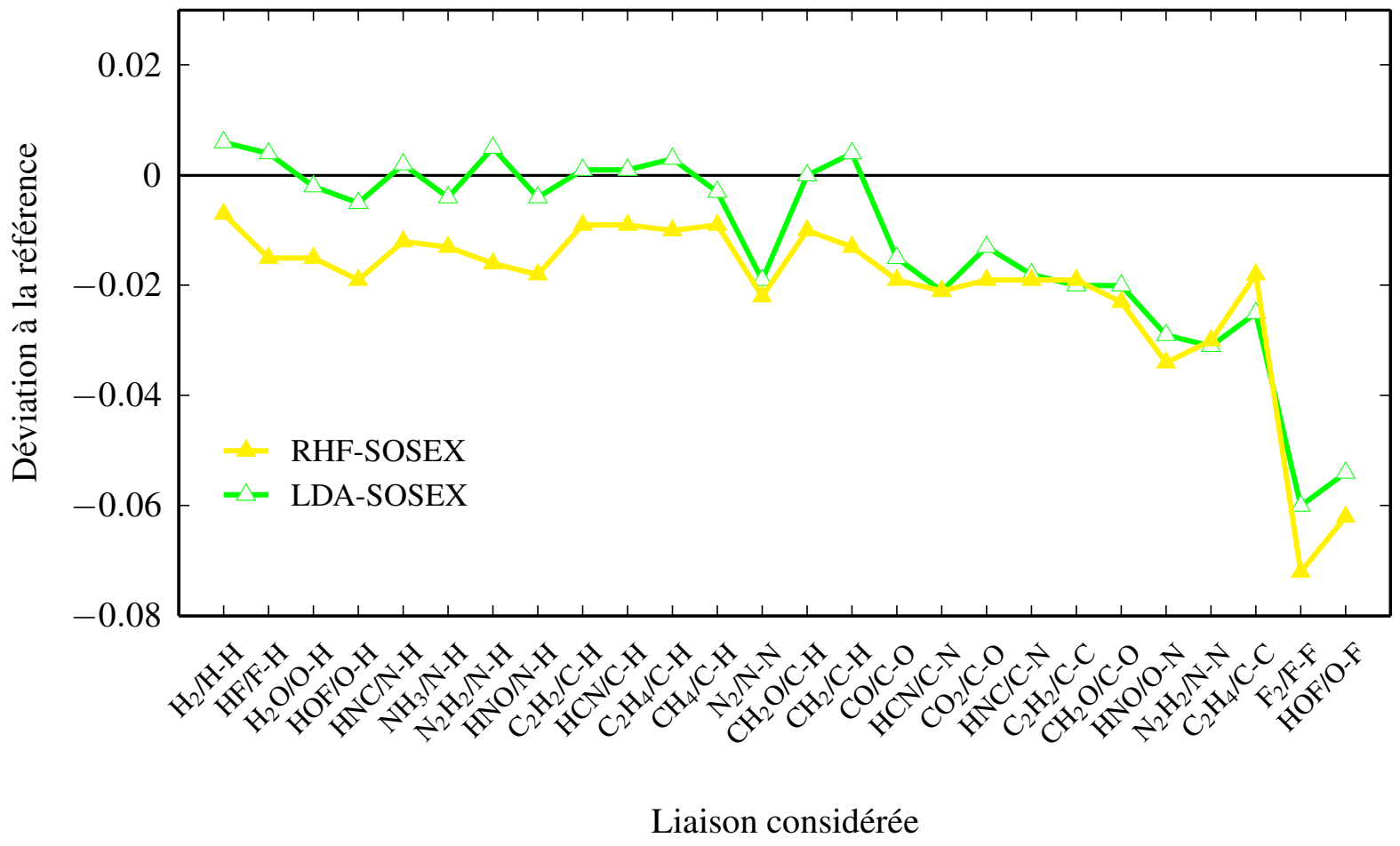

FIGURE 6.7: Déviations (Å) à la référence[228] des longueurs de liaison de 16 molécules optimisées aux niveaux RHF-SOSEX (en jaune, triangle plein), et LDA-SOSEX (en vert, triangle vide). Les liaisons sont classées par longueurs référentes croissantes.

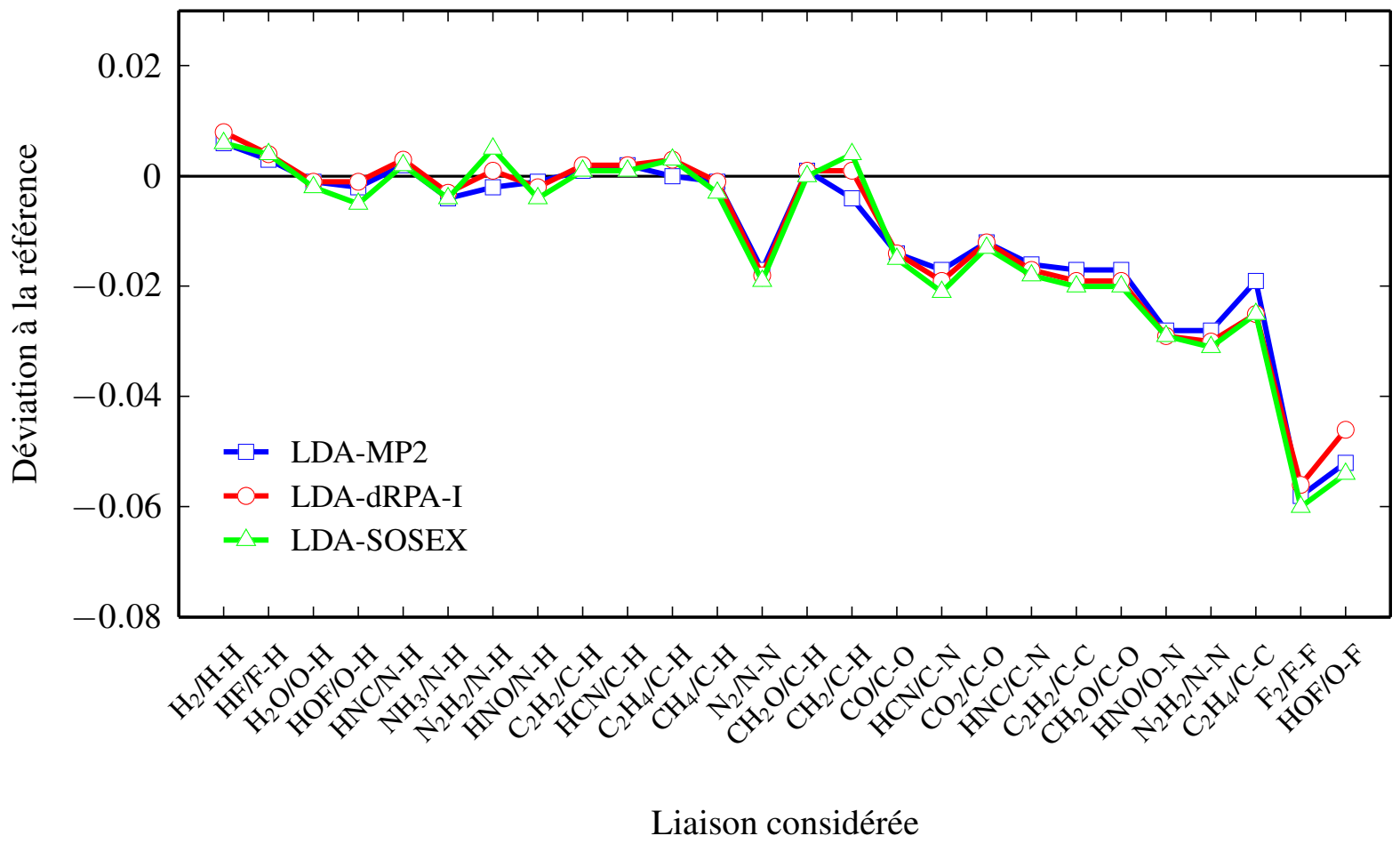

FIGURE 6.8: Déviations (Å) à la référence[228] des longueurs de liaison de 16 molécules optimisées aux niveaux LDA-MP2 (en bleu, carré vide), LDA-dRPA (en rouge, cercle vide) et LDA-SOSEX (en vert, triangle vide). Les liaisons sont classées par longueurs référentes croissantes.

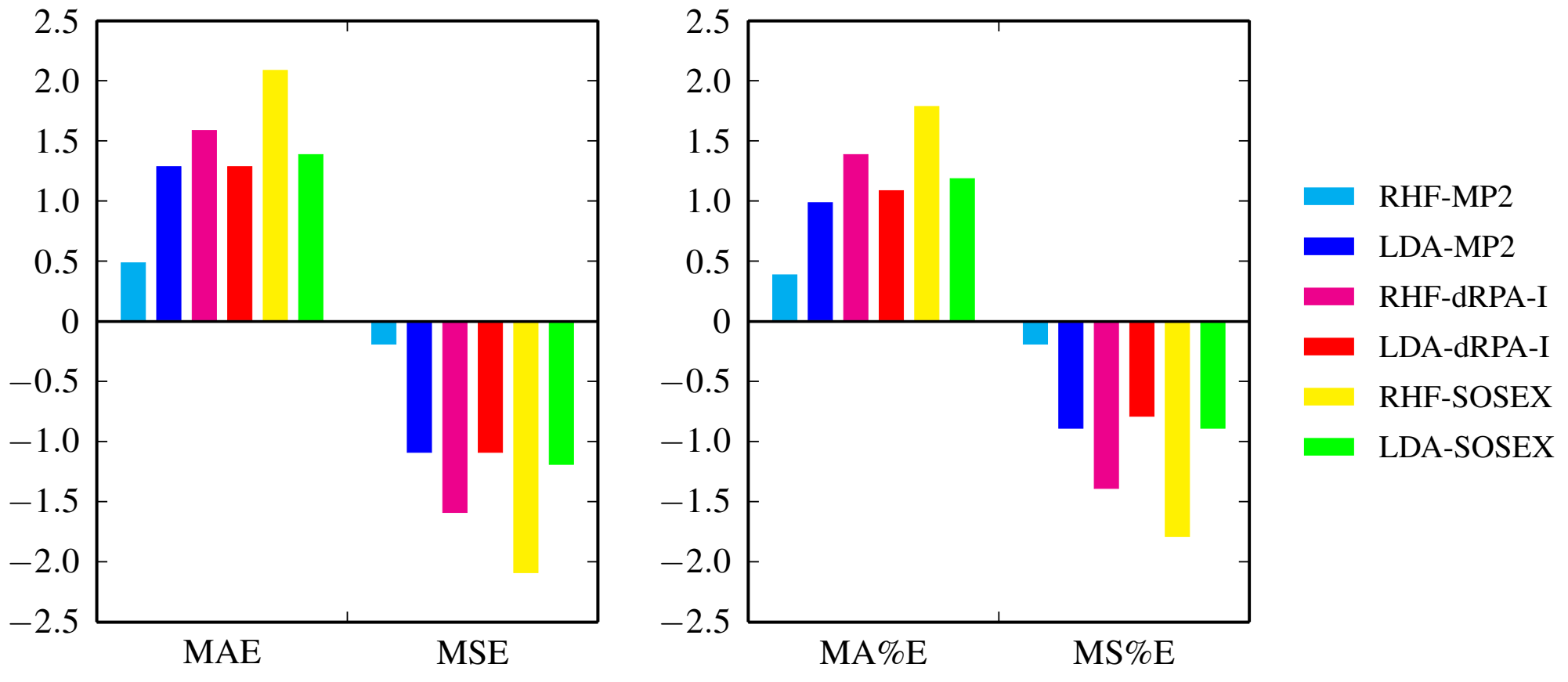

FIGURE 6.9: Moyennes telles que définies (6.8.3) concernant les longueurs de liaisons de géométries optimisées avec les méthodes RHF-MP2, LDA-MP2 ; RHF-dRPA, LDA-dRPA et RHF-SOSEX, LDA-SOSEX. (Si le lecteur lit une version noir et blanc : les barres sont dans le même ordre que dans la légende.)

# Conclusion

Le travail de cette thèse a mené à divers résultats de natures très différentes, toujours autour de développements de l'approximation de la phase aléatoire (RPA).

Du point de vue du formalisme, il a été fait une certaine unification dans la manière de voir ce que l'on appelle les *flavors* de RPA (dRPA-I, dRPA-II, RPAx-I et RPAx-II) avec l'introduction de l'« interrupteur » $\xi$. Également, j'ai présenté des explorations du formalisme "matrice diélectrique" et des approximations qui peuvent y être faites. Ceci est dans la continuité de ce qui a été fait par Ángyán *et. al.* [2] à propos du formalisme "connexion adiabatique". Côté programmation, cet effort d'unification s'est traduit par un effort pour rassembler dans MOLPRO en un seul et même script (et une seule et même syntaxe) toutes les possibilités de calculs RPA qui sont décrits dans le manuscrit. D'un point de vue des détails de dérivations, j'ai démontré des relations qui n'avaient à notre connaissance jamais vraiment été montrées en détail dans la littérature, comme par exemple l'intégration de la formulation "matrice diélectrique", proposée par McLachlan et que le lecteur peut voir dans l'Annexe B.3. Lors des développements des expressions de type "matrice diélectrique" de RPA avec échange, nous sommes arrivés à une expression qui s'est avérée très proche à la méthode RPAX2, proposée récemment par Hesselmann, et qui lui a permis d'avoir des résultats numériques excellents sans séparation de portée. Étant donné qu'il s'agit d'un thème intéressant qui a surgit d'une manière inattendue (comme c'est souvent le cas dans la recherche) nous n'avons pas trouvé encore le temps pour approfondir la question du statut de la formule que nous avons identifiée et sa performance dans un cadre *avec* séparation de portée. Les premiers essais n'étaient pas très convaincant, mais nous n'avons certainement pas fait encore le tour de la question.

On peut discerner deux travaux principaux en rapport avec les orbitales locales. D'une part, un travail a été conduit à l'occasion d'une collaboration avec le Laboratoire de Chimie Théorique de l'Université Pierre et Marie Curie, à Paris. Là-bas a été développée une procédure pour construire des orbitales localisées d'un dimère à partir d'orbitales qui ont été localisées pour les monomères. De cette manière il est possible de discerner les di-excitations qui contribuent significativement à la contribution de la RPA à l'énergie d'interaction. À cette occasion, un programme a été écrit à partir de codes mentionnés précédemment. Ce programme est pensé pour être interfacé avec MOLPRO et avec des scripts écrits à Paris. Il permet de calculer une énergie RPA à partir d'une liste de di-excitations importantes dans des bases d'orbitales locales. D'autre part, j'ai montré des développements d'orbitales dites « orbitales oscillantes projetées » (POO). Ces développements inédits utilisent les pro-priétés de ces orbitales virtuelles localisées pour écrire des versions locales des équations RPA. De plus, grâce à une structure analytique intéressante des POO, dans le cadre d'un formalisme détaillée dans la thèse, il est possible de s'affranchir de la connaissance explicite des orbitales virtuelles et d'écrire une énergie de corrélation approximative, mais fort vraisemblablement bien adaptée pour le calcul de l'énergie de dispersion, uniquement en fonction des orbitales occupées. C'est un résultat intéressant en soi, et qui de plus ouvre la possibilité de l'élaboration de fonctionnelles de corrélation de qualité RPA, mais qui ne nécessitent pas l'utilisation des orbitales virtuelles.

Un des volets de cette thèse concerne des réflexions dans l'espace direct, c'est-à-dire sur des fonctions vues avec leurs dépendances aux coordonnées spatiales. Les travaux dans ce domaine ont été doubles : d'un côté, il a été mis en place un programme capable de lire des orbitales d'un calcul `MOLPRO` et de calculer des fonctions telles que la fonction de réponse $\chi$, le trou d'échange $h_x$, la fonction de Dirac $\delta$, *etc.*.. Les orbitales peuvent être fournies sur des grilles parallélépipédique, que l'on dit "régulières", ou sur des grilles de type "DFT", et dans des fichiers formatés ou non. À court terme, ce programme n'a été utilisé que pour générer les visualisations dans l'espace direct qui sont montrées dans cette thèse. Un développement futur pourrait consister à repenser la construction des grilles de type "DFT" pour échantillonner l'espace de sorte à correspondre plus aux besoins des fonctions rencontrées dans les calculs de corrélation (c'est-à-dire l'espace *entre* les atomes).

D'un autre côté, je présente une adaptation à la chimie quantique de l'approche EED, développée à l'origine par Berger *et. al.* dans l'espace réciproque et avec des ondes planes. Lors de ce travail d'adaptation, j'ai pu dégager des relations intéressantes entre certains objets qui émergent lors des dérivations et les règles de sommes des fonctions de réponse, ainsi que – d'un point de vue plus pragmatique – des pistes pour calculer la fonction de réponse avec des éléments de matrices simples. Cette méthode peut trouver des applications futures dans un formalisme RPA dans l'espace direct, où le calcul de l'énergie de corrélation se ferait sans utiliser explicitement les orbitales virtuelles. On retrouve en cela une visée déjà explorée lors la dérivation des équations RPA dans la base des POO.

Ce que l'on peut considérer comme le travail majeur de cette thèse a consisté à développer les gradients analytiques de l'énergie de corrélation RPA dans un contexte de séparation de portée, c'est-à-dire les gradients analytiques de l'énergie $E = E_{\mathrm{RSH}}^{\mathrm{SR,LR}} + E_{\mathrm{RPA}}^{\mathrm{LR}}$. Lorsque nous avons commencé ce projet, il n'existait pas dans la littérature de gradients analytiques pour la RPA, et il n'existait pas non plus de dérivation tout-en-un de gradients d'énergies avec séparation de portée. Une Communication vient d'être publiée par le groupe de Helgaker sur les gradients RPA, avec des orbitales de référence Hartree-Fock, qui est du point de vue formalisme le cas le plus simple, mais présente probablement le moins d'intérêt du point de vue des applications. En utilisant le formalisme Lagrangien, j'ai pu mettre en place une dérivation qui permet d'obtenir en une fois un gradient qui mêle des contributions courte- et longue-portée. Au cours de la dérivation, le parallèle entre les comportements des termes bi-électroniques courte- et longue-portée est clair et les couplages émergent naturellement. Cette théorie a pu être implémentée dans le « cœur » de `MOLPRO`, c'est-à-dire d'une manière qui profite des outils de programmation mis en place lors de précédentes implémentations des gradients corrélés, notamment des gradients RSH-MP2. À la suite de cette programmation, les calculs de dipôle au niveau RSH-RPA ainsi que les optimisations de géométrie sont du coup disponibles immédiatement. J'ai présenté des résultats d'optimisation de géométrie aux niveaux RSH-dRPA-I et RSH-SOSEX, ainsi que des densités corrélées au niveau RSH-dRPA-I. Il faut voir ces résultats comme une solide validation de l'implémentation, mais le but à terme est de calculer des énergies d'interaction sur géométries optimisées sur des complexes présentant des interactions faibles (complexes de l'ensemble

S22, par exemple). Également, quelques premiers tests semblent indiquer que des résultats intéressants pourraient être obtenus en faisant des calculs de dipôles.

Il est certain que beaucoup reste à faire en relation des travaux présentés dans cette thèse. Je mentionnerais la programmation des gradients analytiques pour les variantes RPA avec échange, notamment les deux versions proposées par Szabo et Ostlund, qui se montrent particulièrement efficace pour calculer des forces intermoléculaires dans un cadre de séparation de portée. Également, il serait important de poursuivre les travaux basés sur les orbitales localisées, notamment implémenter et tester le formalisme POO numériquement. La liste des pistes à explorer pourrait être rallongée.

# Annexe A

# Un contexte pour la RPA, des discussions autour des fonctions de réponse

Cette Annexe offre au lecteur des discussions et développements autour de notions que l'on manipule lorsque l'on dérive les équations RPA, et qui sont à la limite des habitudes d'un chimiste théoricien. On introduit notamment le formalisme des fonctions de Green, qui offre un cadre particulièrement adapté à la RPA et s'impose comme une méthode de choix pour comprendre les implications physiques de l'approximation RPA. Sont succinctement montrés des développements tels que l'approximation GW (que l'on trouve plutôt en physique du solide) et l'équation de Bethe-Salpeter, centrale dans notre façon de dériver les équations RPA. On montre également une discussion sur la nature de l'objet $\chi$ que l'on peut rencontrer dans de nombreux contextes.

Ces développements sont indiqués dans un souci de positionner la RPA, et cette thèse, dans un contexte plus général, mais ne sont pas démontrés dans les plus grands détails : cette Annexe est un effort de vulgarisation d'un domaine limitrophe à la chimie théorique.

## A.1 Théorème de Fluctuation-Dissipation

L'étude d'un système se fait très souvent par l'étude de sa réponse à une perturbation[229, 230]. Un système soumis à une perturbation extérieure (à une force extérieure) voit les valeurs de ses observables dévier de leur valeur moyenne. En théorie de la réponse linéaire, on écrit la variation de la valeur moyenne d'une observable $\hat{B}$ dans un système soumis à une perturbation $F(t)\hat{A}$ ($F$ est l'amplitude de la perturbation, $\hat{A}$ est l'opérateur du système auquel est couplée la perturbation) comme :

$$\langle B \rangle (t) - \langle B \rangle_0 = \int_{-\infty}^{t} \chi_{BA}(t - t')F(t')\, dt', \tag{A.1.1}$$

où l'on a défini $\chi_{BA}$, dont la transformée de Fourier est $\chi_{BA}(\omega)$. Cette fonction de réponse dépend de l'hamiltonien à l'équilibre $\hat{H}_0$ : elle décrit la dynamique du système à l'équilibre mais, dans une approximation linéaire, permet de connaître le comportement du système hors de son équilibre.

Le système soumis à une perturbation extérieure perd de l'énergie par *dissipation* et les observables retournent à leur valeur d'équilibre. Une façon de voir cela est de dire que les degrés de liberté du système sont couplés les uns aux autres et jouent les uns pour les autres le rôle de réservoir (thermique) qui offre une résistance à la force extérieure. Le lecteur trouvera cette explication, et les démonstrations des équations qui sont esquissées dans la suite, par exemple dans la référence [231]. On peut montrer[231] que cette variation de l'énergie moyenne du système soumis à une perturbation extérieure (cette *dissipation*) est liée à la partie imaginaire de la fonction de réponse :

$$\overline{\frac{\partial E}{\partial t}} \propto \omega \,\mathrm{Im}\,(\chi(\omega)) \tag{A.1.2}$$

D'autres sources d'écart à la moyenne sont les fluctuations statistiques autour de la valeur moyenne, mesurées par les fonctions d'auto-corrélation telles que celles qui apparaissent équation (2.2.3). On peut montrer[231] que ces fluctuations sont également liées à la partie imaginaire de $\chi$ :

$$G(\omega) \propto \mathrm{Im}\,(\chi(\omega)) \tag{A.1.3}$$

En d'autres mots, avec les équations (A.1.2) et (A.1.3), on montre que la réponse d'un système à une force extérieure (que l'on exprime en terme de dissipation) est identique à la réponse d'un système à une fluctuation autour de l'équilibre ; c'est-à-dire qu'un système « ne sait pas » ce qui l'a mené à un état hors équilibre (une force extérieure ou une fluctuation spontanée autour de la moyenne) et son évolution de retour à l'équilibre sera la même dans les deux cas : c'est ce que l'on exprime dans le théorème de Fluctuation-Dissipation.

## A.2 Introduction au formalisme des fonctions de Green

L'étude du problème à $N$-corps est l'étude des effets des interactions entre les $N$ corps sur le comportement du système à $N$-corps. Il s'agit d'un problème d'une grande complexité, qui est en général insoluble. Bien souvent, les premières approches de résolution passent par la supposition de l'absence d'interaction pure et simple, c'est-à-dire que l'on substitue au problème à $N$-corps compliqué une superposition de $N$ problèmes à un corps, très simples. Ces suppositions, dites de « champ moyen », donnent étonnement de bon résultats, malgré la sévérité de l'approximation. Cependant, ce qu'il reste à décrire, la partie qui est perdue dans l'approximation, est physiquement très importante.

La théorie quantique des champs offre une manière unifiée et systématique pour attaquer le problème à $N$-corps, et ce dans tous les domaines où il peut émerger. Elle propose une nouvelle description, c'est-à-dire propose de substituer l'étude du problème de $N$ particules réelles en interaction par l'étude de $N$ particules fictives, que l'on appelle « quasi-particules », qui sont (presque) sans interaction. En d'autres termes : on observe que le système des corps réels interagissant fortement est bien décrit par un système de corps fictif interagissant faiblement. (Rappelons-nous que l'on a mentionné plus haut que, en effet, les résultats obtenus avec l'approximation (drastique) de particule sans interaction sont « bons ».)

Une bonne manière de comprendre l'efficacité de cette description est de considérer le comportement et les évènements entourant le déplacement d'une particule réelle. Lorsqu'une particule réelle se déplace dans le système à $N$-corps en interaction attractive (on peut faire globalement le même raisonnement pour une interaction répulsive, bien évidemment), elle emporte avec elle un « nuage » de ses plus proches voisins, avec lesquels elle est en forte interaction. Ce nuage écrante la particule à la vue des autres particules du système. Il est donc plus pertinent d'étudier directement le comportement des quasi-particules (les particules *et* leur nuage), qui interagissent entre elles bien plus faiblement que les particules réelles : on peut considérer que leurs comportements sont indépendants les uns des autres. Dans divers domaines de la physique, l'interaction entre les particules réelles est appelée interaction « nue », là où l'interaction entre quasi-particules est appelée interaction « effective », « habillée », ou : « renormalisée ».

Dans un système à $N$ particules, et dans une description en terme de quasi particules, chaque particule est à la fois au cœur d'une quasi particule et membre des nuages de plusieurs autres quasi particules. Ainsi, une étude du système des quasi particules risque de prendre en compte les particules réelles plus d'une fois. Il est plus simple de définir une description en terme de quasi particules dans un contexte où l'on ajoute une particule réelle au système et où l'on observe et décrit le mouvement de cette particule supplémentaire dans le système. On peut aussi étudier la propagation d'un « trou », c'est-à-dire l'évolution d'une situation où l'on a enlevé une particule au système.

Dans une telle description, les quasi-particules ont des propriétés propres : elles ont, ou peuvent avoir, une masse effective, une charge effective, *etc.*.. et ont, donc, également une énergie propre. On donne le nom de *self-energy* ($\Sigma$) à la différence entre l'énergie d'une quasi-particule et l'énergie d'une particule « nue ». Une explication que l'on peut donner pour le nom de *self-energy* est la suivante : la particule nue interagit avec le système à $N$-corps, génère ainsi le nuage qui l'entoure, et le nuage interagit avec la particule nue, modifiant son comportement. En un sens : la particule interagit avec elle-même *via* le système à $N$-corps et ce faisant change sa propre énergie d'une quantité $\Sigma$.

Dans la suite je présente ces formalismes peut-être peu connus du lecteur chimiste théoricien. De bonnes références à lire sur le sujet seront, par exemple, les chapitres des livres [229, 230, 232–236] et les références [237, 238].

## A.3 Fonctions de Green

On introduit la fonction de Green à une particule[234, 237–240] :

$$
\begin{aligned}
iG_1(1,2) &= \left\langle {}^N_0 \right| \mathrm{T}\left[\hat{\psi}_1 \hat{\psi}_2^\dagger\right] \left| {}^N_0 \right\rangle \\
&= \Theta(t_1 > t_2) \left\langle {}^N_0 \right| \hat{\psi}_1 \hat{\psi}_2^\dagger \left| {}^N_0 \right\rangle - \Theta(t_1 < t_2) \left\langle {}^N_0 \right| \hat{\psi}_2^\dagger \hat{\psi}_1 \left| {}^N_0 \right\rangle ,
\end{aligned}
\tag{A.3.1}
$$

où $\left| {}^N_0 \right\rangle$ est l'état fondamental à $N$ particules du système, l'opérateur de Wick T ordonne les opérateurs qui le composent en temps décroissants (l'opérateur qui agit temporellement en premier est à droite). Les opérateurs sont des opérateurs en seconde quantification, de création ($\hat{\psi}^\dagger$) et d'annihilation ($\hat{\psi}$) ; les variables d'espace-temps sont contractées en $1 \doteq (\mathbf{x}_1, t_1)$, et on note $\Theta(t_1 > t_2)$ la fonction de Heaviside qui vaut un pour $t_1 > t_2$. Cette formulation permet de traiter les électrons et les trous de la même manière.

On peut atteindre une interprétation de la fonction de Green en utilisant la représentation en interaction des opérateurs[234, 237], où la dépendance temporelle est exprimée explicitement comme $\hat{\psi}_1 = e^{i\hat{H}t_1}\hat{\psi}_{\mathbf{x}_1}e^{-i\hat{H}t_1}$ :

$$\begin{aligned} iG_1(1,2) &= \Theta(t_1 > t_2)\langle {}^N_0 | e^{i\hat{H}t_1}\hat{\psi}_{\mathbf{x}_1}e^{-i\hat{H}t_1}e^{i\hat{H}t_2}\hat{\psi}^\dagger_{\mathbf{x}_2}e^{-i\hat{H}t_2} | {}^N_0 \rangle \\ &\quad - \Theta(t_1 < t_2)\langle {}^N_0 | e^{i\hat{H}t_2}\hat{\psi}^\dagger_{\mathbf{x}_2}e^{-i\hat{H}t_2}e^{i\hat{H}t_1}\hat{\psi}_{\mathbf{x}_1}e^{-i\hat{H}t_1} | {}^N_0 \rangle \\ &= \Theta(t_1 > t_2)e^{iE_{N,0}(t_1-t_2)}\langle {}^N_0 | \hat{\psi}_{\mathbf{x}_1}e^{-i\hat{H}(t_1-t_2)}\hat{\psi}^\dagger_{\mathbf{x}_2} | {}^N_0 \rangle \\ &\quad - \Theta(t_1 < t_2)e^{-iE_{N,0}(t_1-t_2)}\langle {}^N_0 | \hat{\psi}^\dagger_{\mathbf{x}_2}e^{i\hat{H}(t_1-t_2)}\hat{\psi}_{\mathbf{x}_1} | {}^N_0 \rangle, \end{aligned} \tag{A.3.2}$$

où l'on voit que la fonction de Green représente, pour un temps $t_1$ supérieur à $t_2$, l'amplitude de probabilité de l'événement suivant : création d'un électron en $\mathbf{x}_2$ ($\hat{\psi}^\dagger_{\mathbf{x}_2}$), propagation de cet électron de $t_2$ à $t_1$ ($e^{-i\hat{H}(t_1-t_2)}$) et destruction de l'électron en $\mathbf{x}_1$ ($\hat{\psi}_{\mathbf{x}_1}$). Il s'agit donc de l'amplitude de probabilité de trouver au temps $t_1$ et en $\mathbf{x}_1$ un électron crée en $\mathbf{x}_2$ au temps $t_2$. De la même manière, pour un temps $t_2$ supérieur à $t_1$, la fonction de Green est la probabilité de trouver au temps $t_2$ et en $\mathbf{x}_2$ un trou crée en $\mathbf{x}_1$ au temps $t_1$. Avec cette réécriture, on voit bien également que la fonction de Green ne dépend que de la différence $\tau$ des deux temps $t_1$ et $t_2$ :

$$G_1(1,2) \doteq G_1(\mathbf{x}_1,\mathbf{x}_2;\tau) \tag{A.3.3}$$

On peut prouver que la fonction de Green à une particule est intimement liée à la densité à une particule[234, 237, 238], et que l'on peut donc obtenir les observables de l'état fondamental comme des moyennes de fonction de Green $\langle G_1(1,2)\rangle$. Mais toutes les informations du système ne sont pas contenues dans la fonction de Green à une particule : les phénomènes dûs à des couplages électron-trou sont perdus, les phénomènes dûs aux corrélations entre particules sont perdus. On peut écrire de manière plus générale des fonctions de Green à $N$ particules, liées aux densités à $N$ particules :

$$G_n(1 \to 2n) \doteq (-i)^n \langle {}^N_0 | \mathrm{T}\left[\hat{\psi}_1 \ldots \hat{\psi}_n\hat{\psi}^\dagger_{n+1} \ldots \hat{\psi}^\dagger_{2n}\right] | {}^N_0 \rangle \tag{A.3.4}$$

## A.4 Équations de Dyson et de Hedin, Approximation GW

### A.4.1 Hiérarchie des fonctions de Green

Les équations de propagation des opérateurs de création et d'annihilation, appliquées aux $G_n$ nous donnent une hiérarchie d'équations de propagation des fonctions de Green[232, 238, 239], où la fonction de Green à une particule dépend de la fonction de Green à deux particules, qui dépend de la fonction de Green à trois particules, *etc.*.. On cherche à tronquer cette hiérarchie en exprimant la fonction de Green à deux particules en fonction de la fonction de Green à une particule. Plus précisément : on modifie l'équation de propagation de $G_1$ (A.4.1) pour obtenir l'équation de Dyson (A.4.2)[232, 233, 237–240] :

$$[i\partial_{t_1} - h(\mathbf{x}_1)]\, G_1(1,2) + i \int v(1,3) G_2(1,3;2,3) = \delta(1,2) \tag{A.4.1}$$

$$\Leftrightarrow [i\partial_{t_1} - h(\mathbf{x}_1)]\, G_1(1,2) - \int \Sigma(1,3) G_1(3,2) = \delta(1,2), \tag{A.4.2}$$

où la « self-energy » $\Sigma$ rend compte de tous les effets portés par $G_2$. Une interprétation de $\Sigma$ émerge si l'on considère un système sans effet de couplage de deux particules, dont la fonction de Green à une particule $G_1^{\text{tronquée}}$ obéit à l'équation de propagation :

$$[i\partial_{t_1} - h(\mathbf{x}_1)]\, G_1^{\text{tronquée}}(1,2) = \delta(1,2) \tag{A.4.3}$$

On retrouve ici une définition des fonctions de Green plus classique dans d'autre domaine de la physique : $G_1^{\text{tronquée}}(1,2)^{-1} = [i\partial_{t_1} - h(\mathbf{x}_1)]$. L'équation de Dyson se réécrit alors schématiquement :

$$G_1 = G_1^{\text{tronquée}} + G_1^{\text{tronquée}}\, \Sigma\, G_1, \tag{A.4.4}$$

et la *self-energy* se comprend comme la connexion entre le système sans effet de couplage de deux particules et le système réel décrit par $G_1$. Tous les effets d'interactions sont inclus dans $\Sigma$, c'est-à-dire les effets d'interaction de Coulomb, les effets d'échange et de corrélation : $\Sigma \doteq \Sigma_{\text{Hxc}} = v_h + \Sigma_{\text{XC}}$.

### A.4.2 Approximation GW

L'enjeu est ensuite bien sûr de trouver une expression de la *self-energy* exacte et un cadre où l'approximer. On trouve satisfaction dans une théorie des perturbations, que je vais présenter dans la suite de manière très succincte : les notations ne seront pas détaillées en profondeur, les lois d'intégrations ne seront pas redémontrées, les dépendances ne seront pas surveillées avec attention. Le but ici est simplement de comprendre d'où vient l'approximation faite sur $\Sigma$, et ses conséquences dans le cadre de la RPA. Le lecteur intéressé trouvera de bonnes revues sur le sujet dans la littérature, entre autres dans la référence [237]. On ajoute au système une perturbation $U$ dépendante du temps, et on définit le potentiel :

$$V = U + v_h = U - i \int v\, G_1, \tag{A.4.5}$$

dont la présence force à redéfinir les fonctions de Green (non détaillé ici). Une dérivation attentive des nouvelles fonctions de Green permet rapidement d'obtenir la formule de Schwinger :

$$\frac{\partial G_1}{\partial U} = -G_2 + G_1 G_1, \tag{A.4.6}$$

dont on comprend immédiatement la portée : cette équation permet d'exprimer la fonction de Green à deux particules exclusivement en fonction de fonctions de Green à une particule. Insérée dans l'équation (A.4.1) et (A.4.2), elle donne en effet :

$$\Sigma_{XC} = i\int v\,\frac{\partial G_1}{\partial U}\,G_1^{-1} \overset{\overset{1}{\downarrow}}{=} -i\int v\,G_1\,\frac{\partial G_1^{-1}}{\partial U} \overset{\overset{2}{\downarrow}}{=} -i\int v\,G_1\,\frac{\partial G_1^{-1}}{\partial V}\,\frac{\partial V}{\partial U} = i\int G_1\,W\,\widetilde{\Gamma}, \tag{A.4.7}$$

où la loi 1 est la « loi de l'inverse » et la loi 2 est la « chain-rule », deux lois d'analyse fonctionnelle bien connues. Cette équation définit les objets $\widetilde{\Gamma}$ et $W$ :

$$\widetilde{\Gamma} = -\frac{\partial G_1^{-1}}{\partial V} \overset{\overset{\text{Dyson avec}}{\overset{\Sigma \doteq V+\Sigma_{XC}}{\downarrow}}}{=} -\frac{\partial G_1^{\text{tronquée},-1}}{\partial V} + \frac{\partial V}{\partial V} + \frac{\partial \Sigma_{XC}}{\partial V} \overset{\overset{1+2}{\downarrow}}{=} \delta\delta + \int \frac{\partial \Sigma_{XC}}{\partial G_1}\,G_1\,\widetilde{\Gamma}\,G_1, \tag{A.4.8}$$

et :

$$W = \int v\,\frac{\partial V}{\partial U} \overset{\overset{\text{def de}}{\overset{V+2}{\downarrow}}}{=} v - i\int v\,\frac{\partial G_1}{\partial V}\,v\,\frac{\partial V}{\partial U} = v + \int v\,\widetilde{\chi}\,W, \tag{A.4.9}$$

où :

$$\widetilde{\chi} = -i\frac{\partial G_1}{\partial V} \overset{\overset{1}{\downarrow}}{=} i\int G_1\,\frac{\partial G_1^{-1}}{\partial V}\,G_1 = -i\int G_1\,\widetilde{\Gamma}\,G_1, \tag{A.4.10}$$

est la polarisabilité irréductible du système, c'est-à-dire la réponse du système -de $G_1$- à une perturbation $V$. La dénomination « irréductible » souligne le fait que $\widetilde{\chi}$ est définie par rapport à $V$, là où une polarisabilité « réductible » peut être définie comme $\chi = -i\frac{\partial G_1}{\partial U}$.

Il est intéressant de se pencher sur la définition de $W$. Il s'agit du potentiel $v$ écranté par l'inverse de la fonction diélectrique : $\varepsilon^{-1} = \frac{\partial V}{\partial U}$. Comme on l'a vu dans l'introduction, l'interaction entre les quasi particules est bien plus faible (elle est quasi nulle) que l'interaction entre les particules réelles : $W$ est plus faible que le $v$.

**On a donc** une collection de cinq objets ($\Sigma_{XC}$ et $G_1$, et $\widetilde{\Gamma}$, $\widetilde{\chi}$ et $W$) qui ont émergés dans la dérivation et de cinq équations ((A.4.4), (A.4.8), (A.4.10), (A.4.9), (A.4.7)) que l'on peut faire tourner dans une procédure SCF jusqu'à convergence, c'est-à-dire jusqu'à obtenir l'exacte *self-energy*.

Ce schéma, proposé par Hedin[241], est une théorie des perturbations dont la composante de base est le potentiel écranté $W$, qui est beaucoup plus faible que le potentiel $v$ : une série en puissance de $W$ converge beaucoup plus rapidement qu'une théorie des perturbations en puissance de $v$. (Cette idée n'est pas sans lien avec le modèle de Hubbard[242] en physique du solide). De fait, l'approximation la plus répandue consiste à faire un unique cycle $\left(\Sigma_{XC}^{\text{guess}}, G_1^{\text{guess}}\right) \rightarrow \left(\widetilde{\Gamma}, \widetilde{\chi}, W\right) \rightarrow \left(\Sigma_{XC}^{\text{new}}, G_1^{\text{new}}\right)$. Avec une *self-energy* de départ $\Sigma_{XC}^{\text{guess}} = 0$, on obtient :

$$\begin{cases} G_1 &= G_1^{\text{tronquée}} \\ \widetilde{\Gamma} &= \delta\delta \\ \widetilde{\chi} &= -iG_1\, G_1 = \chi^{\text{IP}} \\ W &= v - i \displaystyle\int v\, G_1\, G_1\, W \\ \Sigma_{\text{XC}}^{\text{new}} &= iG\, W, \end{cases} \tag{A.4.11}$$

où l'expression de la *self-energy* ainsi obtenue donne son nom à l'approximation : $GW$.

## A.5 Équation de Bethe-Salpeter, Approximation du noyau

Dans les cas où la fonction de Green à deux particules est effectivement nécessaire (dans le cas de couplage électron-trou par exemple), on utilise une équation similaire à l'équation de Dyson : l'équation de Bethe-Salpeter[243]. Considérons la polarisabilité « réductible », en portant attention cette fois aux dépendances :

$$\chi(1,2;1',2') = -i\frac{\partial G_1(1,1')}{\partial U(2',2)} = iG_2(1,2;1',2') - iG_1(1,1')G_1(2,2') \tag{A.5.1}$$

**Cette équation** est du plus grand intérêt d'un point de vue physique : $\chi$ est un objet à 4 points, définit à la fois (voir l'équation (A.4.6)) comme la réponse (linéaire) d'un système à une perturbation extérieure et (voir l'équation (A.5.1)) comme la différence entre le mouvement *couplé* d'un électron et d'un trou et leur mouvement *indépendant*. On voit ici les prémisses d'une utilisation dans un cadre de théorème de fluctuation dissipation

On peut procéder comme précédemment et dériver l'équation de Bethe-Salpeter[80, 237, 240] de la manière simplifiée suivante :

$$\begin{aligned} \chi &\overset{\overset{1}{\downarrow}}{\doteq} i\int G_1\, \frac{\partial G_1^{-1}}{\partial U}\, G_1 \overset{\substack{\text{Dyson avec} \\ \Sigma \doteq U + v_h + \Sigma_{\text{XC}} \\ \downarrow}}{\doteq} i\int G_1\, \frac{\partial\left(G_1^{\text{tronquée},-1} - U - v_h - \Sigma_{\text{XC}}\right)}{\partial U}\, G_1 \\ &\overset{\overset{2}{\downarrow}}{\doteq} -iG_1\, G_1 - i\int G_1\, \frac{\partial\,(v_h + \Sigma_{\text{XC}})}{\partial G_1}\, \frac{\partial G_1}{\partial U}\, G_1 = \chi^{\text{IP}} + \int \chi^{\text{IP}}\, \mathcal{K}\, \chi, \end{aligned} \tag{A.5.2}$$

où l'on retrouve la polarisabilité à particules indépendantes $\chi^{\text{IP}}$ qui apparaît (à deux points) dans l'équation (A.4.11), et où l'on définit le noyau $\mathcal{K} = i\frac{\partial(v_h+\Sigma_{\text{XC}})}{\partial G_1}$ à partie de la *self-energy* totale, c'est-à-dire du potentiel de Hartree et de la *self-energy d'échange-corrélation*. On voit que l'équation de Bethe-Salpeter est une sorte d'équation de Dyson pour la polarisabilité à 4 points[80], et que le noyau $\mathcal{K}$ joue le même rôle que la *self-energy* : il connecte le système à particules indépendantes représenté par $\chi^{\text{IP}}$ au système réel. On représente schématiquement l'équation de Bethe-Salpeter :

$$\chi = \chi^{\text{IP}} + \chi^{\text{IP}} \mathcal{K} \chi \tag{A.5.3}$$

Comme précédemment avec la *self-energy*, le noyau doit être approximé, c'est-à-dire la dérivée de $\Sigma_{\text{XC}}$ doit être approximée : il semble naturel d'utiliser l'approximation mise en place pour $\Sigma_{\text{XC}}$, *i.e.* l'approximation $GW$, où $\Sigma_{\text{XC}} = 0$. Ceci mène à un noyau qui ne contient que le terme de Hartree, et est appelé TD-H, ou : RPA, que l'on appelle ici direct-RPA (dRPA).

$$\mathcal{K}^{\text{RPA}}(1,2;3,4) = i\frac{\partial v_h(3)\delta(3,4)}{\partial G_1(4,2)} = v(1,4)\,\delta(1,3)\delta(2,4) \tag{A.5.4}$$

(voir définition de $v_h$, équation (A.4.5)). Dans ce cas, les « $\delta$ » qui émergent vont réduire l'équation de Bethe-Salpeter (A.5.2) à une équation d'objets à deux points. Une approximation moins sévère consiste à inclure un terme d'échange, ce qui donne un noyau :

$$\mathcal{K}^{\text{RPAx}}(1,2;3,4) = i\frac{\partial\,(v_h + \Sigma_{\text{X}})}{\partial G_1} = v(1,4)\,\delta(1,3)\delta(2,4) - v(1,3)\,\delta(1,4)\delta(2,3) \tag{A.5.5}$$

Il s'agit de l'approximation TD-HF, ou : RPA-échange (RPAx).

**Ainsi** l'approximation RPA sur le noyau de l'équation de Bethe-Salpeter est similaire à l'approximation $GW$ sur la *self-energy* de l'équation de Dyson et consiste à négliger totalement toutes contributions autres que Hartree. L'approximation RPA-échange (RPAx) consiste à inclure un terme d'échange dans la *self-energy*, c'est-à-dire dans le noyau de l'équation de Bethe-Salpeter.

## A.6 Fonctions de réponse

### A.6.1 Représentations de Lehmann

La représentation spectrale dans le contexte du formalisme des fonctions de Green est appelé « représentation de Lehmann ». Pour obtenir la représentation de Lehmann de la fonction de Green à une particule[234, 237, 240], on insère dans l'équation (A.3.1) la complétude d'une base d'états à $M$ particules dans leurs états fondamental ou excités $n$, $\sum_{M,n} |{}^{M}_{n}\rangle\langle{}^{M}_{n}| = 1$. Du fait des opérateurs de création/annihilation, seuls les états $|{}^{N+1}_{n}\rangle$ (*resp.* $|{}^{N-1}_{n}\rangle$) survivent pour $\tau > 0$ (*resp.* $\tau < 0$) :

$$\begin{aligned} iG_1(1,2) = \sum_n & \Theta(\tau > 0)\,\langle{}^{N}_{0}|\hat{\psi}_1|{}^{N+1}_{n}\rangle\langle{}^{N+1}_{n}|\hat{\psi}^\dagger_2|{}^{N}_{0}\rangle \\ & - \Theta(\tau < 0)\,\langle{}^{N}_{0}|\hat{\psi}^\dagger_2|{}^{N-1}_{n}\rangle\langle{}^{N-1}_{n}|\hat{\psi}_1|{}^{N}_{0}\rangle \end{aligned} \tag{A.6.1}$$

La représentation en interaction des opérateurs donne :

$$iG_1(\mathbf{x}_1,\mathbf{x}_2;\tau) = \sum_n \Theta(\tau>0)e^{-i\left(E_{N+1,n}-E_{N,0}\right)\tau}\left\langle {}^N_0 \middle| \hat{\psi}_{\mathbf{x}_1} \middle| {}^{N+1}_n \right\rangle \left\langle {}^{N+1}_n \middle| \hat{\psi}^\dagger_{\mathbf{x}_2} \middle| {}^N_0 \right\rangle - \Theta(\tau<0)e^{-i\left(E_{N,0}-E_{N-1,n}\right)\tau}\left\langle {}^N_0 \middle| \hat{\psi}^\dagger_{\mathbf{x}_2} \middle| {}^{N-1}_n \right\rangle \left\langle {}^{N-1}_n \middle| \hat{\psi}_{\mathbf{x}_1} \middle| {}^N_0 \right\rangle, \tag{A.6.2}$$

dont la transformée de Fourier est :

$$G_1(\mathbf{x}_1,\mathbf{x}_2;\omega) = \sum_n \frac{\left\langle {}^N_0 \middle| \hat{\psi}_{\mathbf{x}_1} \middle| {}^{N+1}_n \right\rangle \left\langle {}^{N+1}_n \middle| \hat{\psi}^\dagger_{\mathbf{x}_2} \middle| {}^N_0 \right\rangle}{\omega-(E_{N+1,n}-E_{N,0})+i\eta^+} + \frac{\left\langle {}^N_0 \middle| \hat{\psi}^\dagger_{\mathbf{x}_2} \middle| {}^{N-1}_n \right\rangle \left\langle {}^{N-1}_n \middle| \hat{\psi}_{\mathbf{x}_1} \middle| {}^N_0 \right\rangle}{\omega-(E_{N,0}-E_{N-1,n})-i\eta^+} \tag{A.6.3}$$

On a utilisé :

$$\int_{-\infty}^{\infty} dt\;\Theta(\tau>0)e^{-i\varepsilon\tau}e^{i\omega\tau} = \frac{i}{\omega-\varepsilon+i\eta^+}$$
$$\text{et : } \int_{-\infty}^{\infty} dt\;\Theta(\tau<0)e^{-i\varepsilon\tau}e^{i\omega\tau} = \frac{-i}{\omega-\varepsilon-i\eta^+}, \tag{A.6.4}$$

qui s'explique lorsque l'on comprend, par exemple pour la première relation, que $e^{i(\omega-\varepsilon)\tau} = e^{i\mathrm{Re}(\omega-\varepsilon)\tau}$ $e^{-\mathrm{Im}(\omega-\varepsilon)\tau}$ est intégrable analytiquement si $\mathrm{Im}\,(\omega-\varepsilon)$ et $\tau$ ont le même signe. Ici, vu la présence de la fonction de Heaviside, si $\mathrm{Im}\,(\omega-\varepsilon)>0$. Ainsi l'apparition des éléments $\pm i\eta^+$ est due à la nature physique de $G_1$ et à la scission entre le traitement de la propagation d'un électron pour $\tau$ supérieur à 0 et le traitement de la propagation d'un trou pour $\tau$ inférieur à 0[229].

On adopte la notation de Feynman, où $\eta^+$ implique la présence d'une limite : $\eta^+ \doteq \lim_{\eta^+\to 0}$. Il faut ainsi comprendre que les pôles de la première fraction de l'équation (A.6.3) sont $E_{N+1,n}-E_{N,0}-i\eta^+$, c'est-à-dire sont les différences d'énergie $E_{N+1,n}-E_{N,0}$ décalées à partir de l'axe des réels dans le plan complexe inférieur, et ceci simplement pour justifier certains raisonnements, avant de prendre $\eta^+\to 0$. Une plus ample discussion sur la présence et la signification de $\eta^+$ est donnée section A.6.2.

Considérons à présent la représentation de Lehmann de $\chi^{\mathrm{IP}}(1,2;1',2') = -iG_1(1,2')G_1(2,1')$. On procède de la même manière que précédemment, c'est-à-dire que l'on insère des bases complètes pour obtenir (on peut aussi utiliser directement les représentations de Lehmann dérivées plus haut) :

$$i\chi^{\mathrm{IP}}(1,2;1',2') = \Theta(\tau>0)\left\langle {}^N_0 \middle| \hat{\psi}_1\hat{\psi}^\dagger_{2'} \middle| {}^N_0 \right\rangle \left\langle {}^N_0 \middle| \hat{\psi}^\dagger_{1'}\hat{\psi}_2 \middle| {}^N_0 \right\rangle + \Theta(\tau<0)\left\langle {}^N_0 \middle| \hat{\psi}^\dagger_{2'}\hat{\psi}_1 \middle| {}^N_0 \right\rangle \left\langle {}^N_0 \middle| \hat{\psi}_2\hat{\psi}^\dagger_{1'} \middle| {}^N_0 \right\rangle \tag{A.6.5}$$

En explicitant les coordonnées d'espace et de temps, on a :

$$\begin{aligned} i\chi^{\mathrm{IP}}(\mathbf{x}_1,\mathbf{x}_2;\mathbf{x}_{1'},\mathbf{x}_{2'};\tau) = {} & \Theta(\tau>0)e^{-i\left(E_{N+1,n}-E_{N,0}-E_{N,0}+E_{N-1,0}\right)\tau} \\ & \left\langle {}^N_0 \middle| \hat{\psi}_{\mathbf{x}_1} \middle| {}^{N+1}_n \right\rangle \left\langle {}^{N+1}_n \middle| \hat{\psi}^\dagger_{\mathbf{x}_{2'}} \middle| {}^N_0 \right\rangle \left\langle {}^N_0 \middle| \hat{\psi}^\dagger_{\mathbf{x}_{1'}} \middle| {}^{N-1}_n \right\rangle \left\langle {}^{N-1}_n \middle| \hat{\psi}_{\mathbf{x}_2} \middle| {}^N_0 \right\rangle \\ & + \Theta(\tau<0)e^{i\left(E_{N+1,n}-E_{N,0}-E_{N,0}+E_{N-1,0}\right)\tau} \\ & \left\langle {}^N_0 \middle| \hat{\psi}_{\mathbf{x}_{2'}} \middle| {}^{N-1}_n \right\rangle \left\langle {}^{N-1}_n \middle| \hat{\psi}^\dagger_{\mathbf{x}_1} \middle| {}^N_0 \right\rangle \left\langle {}^N_0 \middle| \hat{\psi}^\dagger_{\mathbf{x}_2} \middle| {}^{N+1}_n \right\rangle \left\langle {}^{N+1}_n \middle| \hat{\psi}_{\mathbf{x}_{1'}} \middle| {}^N_0 \right\rangle \end{aligned} \tag{A.6.6}$$

L'équation (A.6.5) montre que $\chi^{\text{IP}}$ décrit la propagation simultanée d'un électron *et* d'un trou. La transformée de Fourier de (A.6.6) est (avec des définitions évidentes pour $f$) :

$$\chi^{\text{IP}}(\mathbf{x}_1,\mathbf{x}_2;\mathbf{x}_{1'},\mathbf{x}_{2'};\omega) = \frac{f_{N+1}(\mathbf{x}_1)f_{N-1}(\mathbf{x}_{1'})f^*_{N+1}(\mathbf{x}_{2'})f^*_{N-1}(\mathbf{x}_2)}{\omega-(E_{N+1,n}-E_{N,0})-(E_{N,0}-E_{N-1,0})+i\eta^+} - \frac{f_{N+1}(\mathbf{x}_2)f_{N-1}(\mathbf{x}_{2'})f^*_{N+1}(\mathbf{x}_{1'})f^*_{N-1}(\mathbf{x}_1)}{\omega+(E_{N+1,n}-E_{N,0})-(E_{N,0}-E_{N-1,0})-i\eta^+}, \tag{A.6.7}$$

où, à nouveau, les $\pm i\eta^+$ sont une conséquence de la nature de $\chi^{\text{IP}}$.

Notons que $\chi^{\text{IP}} = -iG_1G_1$ n'est pas encore bien définie puisque l'on a pas choisi de fonction de Green à une particule en particulier. En choisissant les fonctions de Green sans interaction $G_1^0$ d'un calcul KS ou HF, on obtient $\chi^{\text{IP}} = -iG_1^0G_1^0 = \chi^0$. Ainsi, dans ce contexte de fonctions d'onde mono-déterminentales, on retrouve pour la représentation de Lehmann la définition de l'équation (2.3.2).

### A.6.2 Des propagateurs "causal", "avancé" et "retardé"

L'expression de Lehmann, équation (A.6.7), introduit une quantité positive $\eta^+$ que l'on comprend lorsque l'on considère les conditions d'analyticité des transformés de Fourier vues équations (A.6.4). Une plus importante discussion peut être faite sur les $\eta^+$ des fonctions de réponse $\chi$.

La fonction de corrélation temporelle $K$ qui émerge naturellement dans un contexte de réponse linéaire[229], où l'on cherche à exprimer l'évolution de la valeur moyenne $\langle B\rangle$ d'un système d'hamiltonien perturbé $\hat{H} = \hat{H}_0 + F(t)\hat{A}$ avec les fonctions d'onde perturbées écrites $|0'\rangle = |0\rangle + c_n(t)e^{-i\omega_{0,n}t}\,|n\rangle$, est :

$$\langle B\rangle - \langle B\rangle_0 = \int_{-\infty}^{t} K(t-t')F(t')\,dt' \tag{A.6.8}$$

et :

$$K(t-t') = \sum\left(\langle 0|B|n\rangle\,\langle n|A|0\rangle\,e^{-i\omega_{0,n}(t-t')} - \langle 0|A|n\rangle\,\langle n|B|0\rangle\,e^{i\omega_{0,n}(t-t')}\right), \tag{A.6.9}$$

où c'est la causalité qui dicte le choix des bornes d'intégration : la réponse de $\langle B\rangle$ au temps $t$ dépend de la perturbation aux temps inférieurs à $t$. Cette fonction de corrélation temporelle est la transformé de Fourier d'une fonction de réponse (d'un propagateur) $\chi$. Plus exactement, étant donné que $K$ n'est définie que pour $\tau = t - t' > 0$, on a :

$$\chi(\omega) = \int_{-\infty}^{\infty} \Theta(\tau)K(\tau)e^{i\omega\tau}d\tau \tag{A.6.10}$$

et :

$$\Theta(\tau)K(\tau) \propto \int_{-\infty}^{\infty} \chi(\omega)e^{-i\omega\tau}d\omega \tag{A.6.11}$$

Pour obtenir une expression de $\chi$ à partir de $K$, équation (A.6.9), on ajoute à la perturbation un facteur de convergence $e^{\eta^+ t}$ qui assure une apparition graduelle à $t = -\infty$ (essentiellement, cela revient à travailler avec une fonction plus lisse que $\Theta$ dans l'équation (A.6.10)), on résout l'équation de Schrödinger pour les $c_n$ (avec les bons choix pour $\hat{H}'$), pour finalement pouvoir écrire[229] :

$$\chi(\omega) = \sum \frac{\langle 0|B|n\rangle \langle n|A|0\rangle}{\omega - \omega_{0,n} + i\eta^+} - \frac{\langle n|A|0\rangle \langle 0|B|n\rangle}{\omega + \omega_{0,n} + i\eta^+} \tag{A.6.12}$$

(Noter les signes des termes de convergence : $+i\eta^+$, là où l'équation (A.6.7) montrait des termes $i\eta^+$ et $-i\eta^+$. Je reviendrai sur cette question dans la suite). C'est à ce moment, il me semble, que l'on comprend le mieux le lien entre la causalité (c'est-à-dire : le fait que $\tau > 0$) et la présence de $\eta^+$. La fonction $\chi$ à des pôles sur l'axe des réels $\pm\omega_{0,n}$, déplacés dans le plan complexe inférieur d'une quantité $i\eta^+$. L'intégrale (A.6.11) est donc : analytique dans le plan complexe inférieur (sauf en ces pôles) pour $\tau > 0$ (on applique alors le théorème des résidus pour obtenir une intégration par contour non nulle, voir par exemple B.2.2) ; analytique, pour $\tau < 0$, dans la totalité du plan complexe supérieur, qui ne contient pas de pôle et dans lequel l'intégration par contour donne un résultat nul en vertu du théorème de Cauchy (voir B.1.3).

**Une manière** de présenter les choses est de dire que l'on impose la causalité soit par le biais de fonction de Heaviside, soit par la présence du facteur de convergence qui décale les pôles de $\chi$ de manière à forcer le signe de $\tau$ dans la transformée de Fourier (A.6.11).

Cette formulation du propagateur $\chi$ est appelée « retardée »[232]. Il décrit un phénomène physique réel, où la réponse est déterminée après l'application de la perturbation, et le facteur de convergence $e^{\eta^+ t}$ est inclus pour respecter la causalité. Une autre formulation du propagateur peut être écrite, en remplaçant les termes $+i\eta^+$ par $-i\eta^+$ dans (A.6.12), pour obtenir :

$$\chi(\omega) = \sum \frac{\langle 0|B|n\rangle \langle n|A|0\rangle}{\omega - \omega_{0,n} - i\eta^+} - \frac{\langle n|A|0\rangle \langle 0|B|n\rangle}{\omega + \omega_{0,n} - i\eta^+} \tag{A.6.13}$$

Ce propagateur a des pôles décalés dans le plan complexe supérieur, et avec le même raisonnement que précédemment, cela mène à une fonction de corrélation temporelle pour laquelle $\tau < 0$. En d'autres termes, ce propagateur résulte d'une vision où l'on a ajouté un facteur de convergence inverse $e^{-\eta^+ t}$ et décrit un phénomène imaginaire où le temps « descend » de $+\infty$, et où la réponse est déterminée avant la perturbation. On appelle ce propagateur : propagateur « avancé »[232].

Le propagateur de l'équation (A.6.7), obtenu dans un formalisme de fonction de Green avec l'opérateur de Wick, est appelé propagateur « causal »[232]. On rappelle ici son expression :

$$\chi(\omega) = \sum \frac{\langle 0|B|n\rangle \langle n|A|0\rangle}{\omega - \omega_{0,n} + i\eta^+} - \frac{\langle n|A|0\rangle \langle 0|B|n\rangle}{\omega + \omega_{0,n} - i\eta^+} \tag{A.6.14}$$

Pour résumer : le propagateur retardé est analytique dans le plan complexe supérieur, a tous ses pôles dans le plan complexe inférieur et donne donc une intégrale non nulle dans le plan complexe inférieur. Le propagateur avancé est analytique dans le plan complexe inférieur, a tous ses pôles dans le plan complexe supérieur et donne une intégrale non nulle dans le plan complexe supérieur. Le propagateur causal n'est analytique dans aucun des deux plans, mais a (donc) une transformée de Fourier bien définie et non nulle pour toute valeur de $\tau$ (sauf $\tau = 0$). Notons que dans la section B.2.2, on est amené à intégrer la fonction de réponse causale et à discuter de la position de ses pôles.

### A.6.3 Une expression alternative

Concernant la fonction de réponse $\chi$ à 4 points vue par exemple équation (A.5.1), ayant quatre opérateurs, elle peut décrire le comportement d'une paire d'électron, d'une paire de trous, ou, ce qui nous intéresse ici : d'un électron et d'un trou. On choisit donc judicieusement l'ordonnement des temps :

$$\chi(1,2;1',2') \doteq \chi(\mathbf{x}_1,t_1,\mathbf{x}_2,t_2;\mathbf{x}_{1'},t_1^+,\mathbf{x}_{2'},t_2^+), \tag{A.6.15}$$

où les temps $t_1^+$ et $t_2^+$ sont légèrement supérieurs à $t_1$ et $t_2$. Ainsi l'opérateur de Wick de la fonction de Green à deux particules de l'équation (A.5.1) ne fournit-il que deux composants, qui sont :

$$-G_2(\mathbf{x}_1,t_1,\mathbf{x}_2,t_2;\mathbf{x}_{1'},t_1^+,\mathbf{x}_{2'},t_2^+) = \Theta(t_1>t_2)\langle {}^N_0|\hat\psi^\dagger_{1'}\hat\psi_1\hat\psi^\dagger_{2'}\hat\psi_2|{}^N_0\rangle - \Theta(t_2>t_1)\langle {}^N_0|\hat\psi^\dagger_{2'}\hat\psi_2\hat\psi^\dagger_{1'}\hat\psi_1|{}^N_0\rangle \tag{A.6.16}$$

Si l'on considère la définition (A.5.1) de la fonction de réponse $\chi$, on se retrouve à écrire la représentation de Lehmann suivante :

$$\chi(\mathbf{x}_1,\mathbf{x}_2;\mathbf{x}_{1'},\mathbf{x}_{2'};\omega) = \sum_{n\neq 0}\left\{\frac{\langle {}^N_0|\hat\psi^\dagger_{\mathbf{x}_{1'}}\hat\psi_{\mathbf{x}_1}|{}^N_n\rangle\langle {}^N_n|\hat\psi^\dagger_{\mathbf{x}_{2'}}\hat\psi_{\mathbf{x}_2}|{}^N_0\rangle}{\omega+i\eta^+-\Omega_n} + \frac{\langle {}^N_0|\hat\psi^\dagger_{\mathbf{x}_{2'}}\hat\psi_{\mathbf{x}_2}|{}^N_n\rangle\langle {}^N_n|\hat\psi^\dagger_{\mathbf{x}_{1'}}\hat\psi_{\mathbf{x}_1}|{}^N_0\rangle}{-\omega+i\eta^+-\Omega_n}\right\}, \tag{A.6.17}$$

que l'on obtient en insérant la complétude de la base $\sum_n |{}^N_n\rangle\langle {}^N_n| = 1$ et où la somme sur $n \neq 0$ s'explique par la présence du terme $-iG_1G_1$ dans (A.5.1). $\Omega_n$ est la différence entre les énergies de l'état excité $n$ et de l'état fondamental. L'objet à 2 points, $\chi(\mathbf{r}_1,\mathbf{r}_2)$ (on ne s'occupe pas ici de la coordonnée de spin, qui est intégrée), s'écrit donc :

$$\chi(\mathbf{r}_1,\mathbf{r}_2;\omega) = \sum_{\alpha\neq 0}^{\infty}\left\{\frac{n_\alpha(\mathbf{r}_1)n_\alpha^*(\mathbf{r}_2)}{\omega+i\eta^+-\Omega_\alpha} + \frac{n_\alpha^*(\mathbf{r}_1)n_\alpha(\mathbf{r}_2)}{-\omega+i\eta^+-\Omega_\alpha}\right\} = \chi_+(\mathbf{r}_1,\mathbf{r}_2;\omega) + \chi_-(\mathbf{r}_1,\mathbf{r}_2;\omega), \tag{A.6.18}$$

où l'on préfère écrire $|0\rangle \doteq |{}^N_0\rangle$ et $|\alpha\rangle \doteq |{}^N_n\rangle$ (on n'a plus besoin de mentionner explicitement le nombre de particule, qui est toujours $N$). On définit $n_\alpha(\mathbf{r}) = \langle 0|\hat\psi^\dagger_{\mathbf{r}}\hat\psi_{\mathbf{r}}|\alpha\rangle = \langle 0|\hat n(\mathbf{r})|\alpha\rangle$.

# Annexe B

# Éléments d'intégration complexe

Le but de cette Annexe se situe dans les sections B.2.2 et B.3 : il s'agit de démontrer clairement des relations liées à des intégrations dans le chapitre sur la RPA. On présente en amont de ces sections des notions de base d'intégration complexe, dans le but d'éclairer le lecteur sur les méthodes et le vocabulaire utilisé lorsque que l'on intègre les équations, notamment (2.4.6) et (2.7.29). Tous les outils nécessaires à la compréhension du théorème des résidus, relativement bien connu des chimistes théoriciens, et du principe de l'argument, qui n'est qu'une application un peu exotique du théorème des résidus, sont présentés ici. À nouveau, il s'agit d'une Annexe de vulgarisation et les démonstrations sont simplement indicatives ; la plupart des dérivations sont une simplification de la référence [244].

## B.1 Notions de bases

Afin de dissiper d'éventuels doutes dans la suite, je présente ici des notations et quelques définitions de base en intégration complexe. L'idée, bien sûr, n'est pas de rentrer dans la totalité des détails que l'on pourrait rigoureusement réclamer (on suppose notamment acquises les existences des objets lorsqu'elles sont nécessaires), mais de se convaincre du fonctionnement des points clés qui mènent aux formulations importantes des sections B.2.2 et B.3. On définit l'intégrale sur un arc $\gamma$ paramétré par $z = z(t) \in \mathbb{C}, a \leq t \leq b$ de la fonction complexe $f(t)$ comme :

$$\int_\gamma f(z)\, dz = \int_a^b f(z(t))z'(t)dt, \tag{B.1.1}$$

dont la valeur est invariante par changement de paramétrisation de l'arc. On accepte également une généralisation des arcs en chaîne d'arcs, et on peut notamment écrire symboliquement :

$$\int_{\gamma_1+\gamma_2+\cdots+\gamma_n} f\, dz = \int_{\gamma_1} f\, dz + \int_{\gamma_2} f\, dz + \cdots + \int_{\gamma_n} f\, dz, \tag{B.1.2}$$

en particulier : $\int_{-\gamma} f\ dz = -\int_{\gamma} f\ dz$. Une chaîne d'arc qui forme un contour fermé est appelée un cycle.

Dans la suite, on dit que $a$ est un zéro d'ordre $n$ de la fonction $f$ si $f(a)$ ainsi que les $n-1$ premières dérivées, $f^{(i)}(a)$, sont nulles. C'est-à-dire que l'on peut écrire $f(z) = (z-a)^n f_n(z)$ où $f_n(a) \neq 0$. De la même manière un pôle d'ordre $n$ de $f$ est un zéro d'ordre $n$ de $g(z) = 1/f(z)$, c'est-à-dire que l'on peut écrire $f(z) = (z-a)^{-n} f_n(z)$, avec $f_n(z) = 1/g_n(z)$.

### B.1.1 Index

Une notion importante à considérer en intégration complexe est l'*index d'un point par rapport à un contour fermé*. De manière simple, l'index indique le nombre de fois où un cycle tourne autour d'un point qui n'est pas sur le cycle. Observons l'intégrale de $1/(z-a)$ sur un cycle $\gamma$ :

$$\int_\gamma \frac{dz}{z-a} = \int_\gamma d\mathrm{Log}(z-a) = \int_\gamma d\mathrm{Log}|z-a| + i\int_\gamma d\mathrm{Arg}(z-a) \tag{B.1.3}$$

Le premier terme est nul (sur un cycle, $\mathrm{Log}|z-a|$ retrouve sa valeur initiale), et le second terme augmente ou diminue d'un multiple de $2\pi i$ . On définit justement l'index $n(\gamma, a)$ d'un point $a$ par rapport à un cycle $\gamma$ comme ce multiple :

$$n(\gamma, a) = \frac{1}{2\pi i}\int_\gamma \frac{dz}{z-a}, \tag{B.1.4}$$

qui indique donc le nombre de fois où un cycle tourne dans le sens trigonométrique autour d'un point qui n'est pas sur la cycle (l'index prend une valeur négative lorsque le cycle tourne dans le sens anti-trigonométrique autour d'un point). Ceci n'est pas une démonstration rigoureuse mais donne « avec les mains » l'esprit de la justification complète. On trouve une démonstration plus rigoureuse dans la section 2.1 de la référence [244].

### B.1.2 Homologie

Une fois la notion d'index définie, une propriété des cycles émerge : l'*homologie*. En quelques mots : dire que des cycles $\gamma_1$ et $\gamma_2$ sont homologues par rapport a une région $\Omega$, c'est dire qu'ils tournent autour de tous les points hors de cette région le même nombre de fois (voir la figure B.1). En effet, si $\gamma$ est un cycle dans une région $\Omega$ pouvant contenir des trous, on dit que $\gamma$ est *homologue à zéro par rapport à* $\Omega$ si $n(\gamma, a) = 0$ pour tous les points $a$ dans le complément de $\Omega$. On écrit $\gamma \sim 0[\Omega]$ et on comprend que $\gamma_1 \sim \gamma_2[\Omega]$ est équivalent à $\gamma_1 - \gamma_2 \sim 0[\Omega]$.

### B.1.3 Théorème de Cauchy

Un théorème fondamental pour nous est le théorème de Cauchy, qui dit que : si $f$ est analytique dans une région $\Omega$, alors :

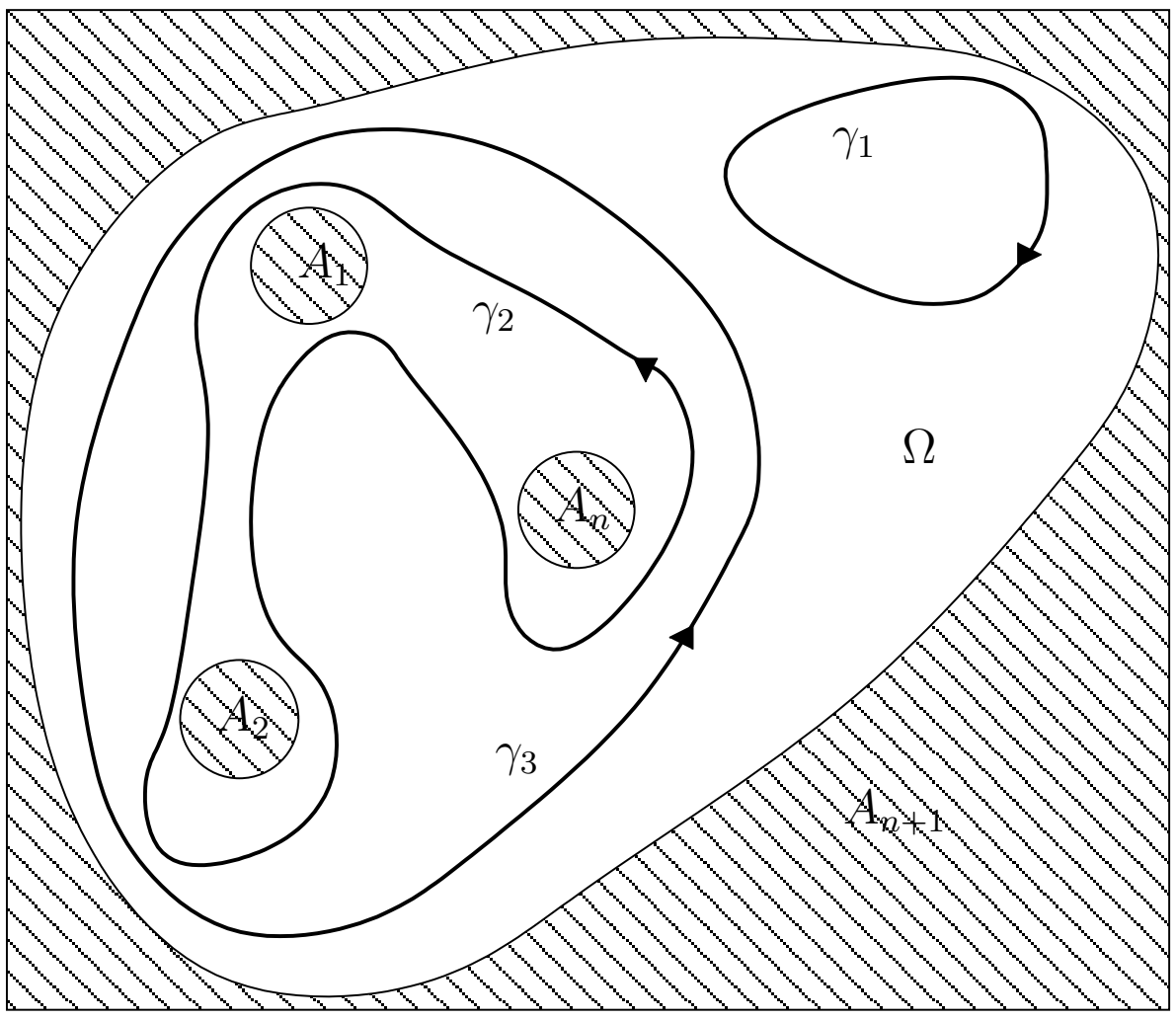

FIGURE B.1: $\Omega$ est une région contenant les trous $A_1, A_2, ..., A_n, A_{n+1}$ ($A_{n+1}$ est l'« extérieur » de $\Omega$). On a : $\gamma_1 \sim 0[\Omega]$, *i.e.* le cycle ne tourne autour d'aucun point hors de $\Omega$. Et : $\gamma_2 \sim \gamma_3[\Omega]$, *i.e.* les deux cycles tournent autour de tous les points hors de $\Omega$ de la même manière.

---

$$\int_\gamma f(z)\,dz = 0, \tag{B.1.5}$$

pour tout $\gamma$ tel que $\gamma \sim 0[\Omega]$. Ce théorème est intimement connecté avec la propriété qu'ont certaines intégrales de ne dépendre que des points d'extrémités de l'arc $\gamma$, c'est-à-dire d'être nulles pour un contour fermé. Je ne démontrerai pas ce théorème, le lecteur intéressé trouvera une démonstration claire dans la partie 4 de [244].

### B.1.4 Théorème des résidus

Considérons une région $\Omega$ munie de trous $A_1, A_2, ..., A_n, A_{n+1}$ (on dit que $\Omega$ est multi-connectée). Étant donnée un cycle $\gamma$ de $\Omega$, on peut montrer que l'index $n(\gamma, a)$ est constant quand $a$ varie dans un même $A_i$. En d'autres termes : le cycle tourne autour des trous, et l'index est le même pour tous les points d'un trou. On appelle $c_i$ la valeur constante de l'index des points d'un trou $A_i$. On est en mesure de construire des cycles $\gamma_1, \gamma_2, ..., \gamma_n$ autour des trous tels que $n(\gamma_i, a) = 1$ pour $a \in A_i$ et $n(\gamma_i, a) = 0$ pour tout autre point hors de $\Omega$. (Notons que $A_{n+1}$ est l'« extérieur » de la région, et n'est pas à strictement parlé un « trou », voir figure B.1 ; de toutes les manières : $n(\gamma, a) = 0$ pour tout cycle $\gamma$ de $\Omega$ et tout point $a$ de $A_{n+1}$). Ainsi pour tout cycle $\gamma$ de $\Omega$, il existe des cycles $\gamma_1, \gamma_2, ..., \gamma_n$ tels que :

$$\gamma \sim (c_1\gamma_1 + c_2\gamma_2 + ... + c_n\gamma_n)[\Omega], \tag{B.1.6}$$

c'est-à-dire : les cycles de chaque côté du signe ~ tournent autour des trous le même nombre de fois. Les cycles $\gamma_i$ sont parfois appelés une « base d'homologie pour $\Omega$ ».

Considérons une fonction $f$ analytique sur toute la région, sauf en des pôles isolés $a_1, a_2, ..., a_n$ qui sont situés dans les trous $A_1, A_2, ..., A_n$ ($f$ est dite méromorphe). Par le théorème de Cauchy, on a :

$$\begin{aligned} \int_\gamma f\, dz &= c_1 \int_{\gamma_1} f\, dz + c_2 \int_{\gamma_2} f\, dz + ... + c_n \int_{\gamma_n} f\, dz \\ &= 2\pi i (c_1 R_{a_1} + c_2 R_{a_2} + ... + c_n R_{a_n}), \end{aligned} \tag{B.1.7}$$

qui définit les $R_{a_i}$, les *résidus de $f$ aux points $a_i$*. L'équation (B.1.7) est appelée le théorème des résidus, et fournit une manière de calculer l'intégrale $\int_\gamma f\, dz$ pourvu que l'on dispose d'un moyen de calculer les résidus $R_{a_i}$. Dans le cas particulier d'un pôle simple en $a_i$, on obtient le résidu $R_{a_i}$ comme la limite $(z - a_i) f(z)$ pour $z \to a_i$.

### B.1.5 Principe de l'argument

Une application particulière du théorème des résidus est d'un intérêt spécial pour nous : il s'agit du principe de l'argument. Considérons un fonction méromorphe $f$ qui a des zéros $a_i$ d'ordre $h_i$ et des pôles $b_j$ d'ordre $h_j$. Pour chaque zéros, on peut écrire :

$$\begin{aligned} f(z) &= (z - a_i)^{h_i} f_{h_i}(z) \qquad \text{avec :} \qquad f_{h_i}(a_i) \neq 0 \\ f'(z) &= h_i (z - a_i)^{h_i - 1} f_{h_i}(z) + (z - a_i)^{h_i} f'_{h_i}(z) \end{aligned} \tag{B.1.8}$$

La fonction $f'/f$ s'écrit alors :

$$\frac{f'(z)}{f(z)} = \frac{h_i}{z - a_i} + \frac{f'_{h_i}(z)}{f_{h_i}(z)}, \tag{B.1.9}$$

et a des pôles simples $a_i$ de résidus $h_i$. On peut écrire pareillement, pour chaque pôle de $f$ :

$$\frac{f'(z)}{f(z)} = \frac{-h_j}{z - b_i} + \frac{f'_{h_i}(z)}{f_{h_i}(z)} \tag{B.1.10}$$

En résumé : si $f$ est une fonction méromorphe avec des zéros $a_i$ d'ordre $h_i$ et des pôles $b_j$ d'ordre $h_j$, alors la fonction $f'/f$ a des pôles simples $a_i$ et $b_j$ de résidus $h_i$ et $-h_j$. Le théorème des résidus appliqué à $f'/f$ donne :

$$\frac{1}{2\pi i} \int_\gamma \frac{f'(z)}{f(z)}\, dz = \sum_i n(\gamma, a_i) h_i - \sum_j n(\gamma, b_j) h_j \tag{B.1.11}$$

Cette équation est appelée principe de l'argument, et peut être généralisée à une fonction $g\frac{f'}{f}$ :

$$\frac{1}{2\pi i}\int_\gamma g(z)\frac{f'(z)}{f(z)}\,dz = \sum_i n(\gamma, a_i)g(a_i)h_i - \sum_j n(\gamma, b_j)g(b_j)h_j \tag{B.1.12}$$

Le lecteur trouvera dans la littérature le principe de l'argument écrit avec une convention où les sommations sur les zéros et les pôles sont répétées autant de fois que l'exige leur ordre respectif, ce qui permet une expression compacte :

$$\frac{1}{2\pi i}\int_\gamma g(z)\frac{f'(z)}{f(z)}\,dz = \sum_i n(\gamma, a_i)g(a_i) - \sum_j n(\gamma, b_j)g(b_j), \tag{B.1.13}$$

et même, avec un contour choisi tel que les index sont tous égaux à 1 :

$$\frac{1}{2\pi i}\int_\gamma g(z)\frac{f'(z)}{f(z)}\,dz = \sum_i g(a_i) - \sum_j g(b_j) \tag{B.1.14}$$

## B.2 Intégration curviligne des fonctions de réponse

### B.2.1 Représentation spectrale de la représentation matricielle de la fonction de réponse

On rappelle que l'on résout $\mathbb{A}_\alpha \mathbb{C}_{\alpha,n} = \omega_{\alpha,n}\mathbb{\Delta}\mathbb{C}_{\alpha,n}$ ; que $\mathbb{\Pi}_\alpha^{-1} = \omega\mathbb{\Delta} - \mathbb{A}_\alpha$ et que $\mathbb{\Delta} = \begin{pmatrix} \mathbf{1} & \mathbf{0} \\ \mathbf{0} & -\mathbf{1} \end{pmatrix}$. Considérant les symétries du problème (voir 2.4.1), on écrit :

$$\begin{aligned}\mathbb{A}_\alpha &= \sum_n \omega_{\alpha,n}\mathbb{C}_{\alpha,n}\mathbb{C}^\dagger_{\alpha,n} = \sum_{n>0}\omega_{\alpha,n}\mathbb{C}_{\alpha,n}\mathbb{C}^\dagger_{\alpha,n} + \sum_{n<0}\omega_{\alpha,n}\mathbb{C}_{\alpha,n}\mathbb{C}^\dagger_{\alpha,n} \\ &= \sum_{n>0}\omega_{\alpha,n}\mathbb{C}_{\alpha,n}\mathbb{C}^\dagger_{\alpha,n} - \omega_{\alpha,-n}\mathbb{C}_{\alpha,-n}\mathbb{C}^\dagger_{\alpha,-n}\end{aligned} \tag{B.2.1}$$

$$\begin{aligned}\mathbb{\Delta} &= \sum_n \Delta_n\mathbb{C}_{\alpha,n}\mathbb{C}^\dagger_{\alpha,n} = \sum_{n>0}\Delta_n\mathbb{C}_{\alpha,n}\mathbb{C}^\dagger_{\alpha,n} + \sum_{n<0}\Delta_n\mathbb{C}_{\alpha,n}\mathbb{C}^\dagger_{\alpha,n} \\ &= \sum_{n>0}\mathbb{C}_{\alpha,n}\mathbb{C}^\dagger_{\alpha,n} - \mathbb{C}_{\alpha,-n}\mathbb{C}^\dagger_{\alpha,-n},\end{aligned} \tag{B.2.2}$$

où $\Delta_n$, bien sur, sont les valeurs propres qui permettent de reconstruire $\mathbb{\Delta}$. On a donc l'expression suivante pour $\mathbb{\Pi}_\alpha^{-1}$ :

$$(\mathbb{\Pi}_\alpha)^{-1} = \omega\mathbb{\Delta} - \mathbb{A}_\alpha = \sum_{n>0}(\omega - \omega_{\alpha,n})\mathbb{C}_{\alpha,n}\mathbb{C}^\dagger_{\alpha,n} + (-\omega + \omega_{\alpha,-n})\mathbb{C}_{\alpha,-n}\mathbb{C}^\dagger_{\alpha,-n}, \tag{B.2.3}$$

et, en étant attentif à rajouter les quantités $i\eta^+$ (voir discussion section A.6) :

$$\Pi_\alpha = \sum_{n>0} \frac{\mathbb{C}_{\alpha,n}\mathbb{C}^\dagger_{\alpha,n}}{\omega - \omega_{\alpha,n} + i\eta^+} + \frac{\mathbb{C}_{\alpha,-n}\mathbb{C}^\dagger_{\alpha,-n}}{-\omega + \omega_{\alpha,-n} + i\eta^+} \tag{B.2.4}$$

### B.2.2 L'intégration

L'intégration de l'équation (2.4.6) se fait par une méthode appelée « décalage du contour d'intégration ». Rappelons que :

$$\mathbb{P}^{\mathrm{RPA}}_{c,\alpha} = \int_{-\infty}^{\infty} \frac{-d\omega}{2\pi i}\left[\Pi^{\mathrm{RPA}}_\alpha(\omega) - \Pi_0(\omega)\right] \tag{2.4.6}$$

On cherche donc à intégrer une fonction $f$ sur l'axe réel, de $-\infty$ à $\infty$. Une telle intégration peut s'écrire :

$$\frac{1}{2\pi i}\int_{-\infty}^{\infty} f(\omega)d\omega = \lim_{R\to\infty} \frac{1}{2\pi i}\int_{-R}^{+R} f(\omega)d\omega = \lim_{R\to\infty} \frac{1}{2\pi i}\int_{\gamma_R} f(z)\,dz, \tag{B.2.5}$$

où $\gamma_R$ est le segment de l'axe réel allant de $-R$ à $R$. Le décalage du contour d'intégration consiste à exprimer l'intégrale sur une variable réelle par une intégrale complexe que l'on peut aisément calculer. On considère l'intégrale suivante, où l'on ajoute à $\gamma_R$ un arc $C_R$ qui clôt un contour dans le plan complexe supérieur :

$$\frac{1}{2\pi i}\left(\int_{\gamma_R} f(z)\,dz + \int_{C_R} f(z)\,dz\right) = \frac{1}{2\pi i}\oint_{\gamma_R + C_R} f(z)\,dz \tag{B.2.6}$$

Le choix le plus simple pour fermer le contour est un demi-cercle, allant du point réel $R$ au point réel $-R$, en passant par le plan complexe supérieur (voir figure B.2). En considérant que $f$ tend vers zéro à l'infini, on peut considérer que l'intégrale sur l'arc $C_R$ ne contribue pas :

$$\lim_{R\to\infty} \frac{1}{2\pi i}\int_{\gamma_R} f(z)\,dz = \lim_{R\to\infty} \frac{1}{2\pi i}\oint_{\gamma_R + C_R} f(z)\,dz \tag{B.2.7}$$

On est capable de calculer l'intégrale sur un contour fermé en utilisant le théorème des résidus. Observons la fonction $f$ :

$$f(z) = \sum_n \left\{ \frac{-\mathbb{C}_{\alpha,n}\mathbb{C}^\dagger_{\alpha,n}}{z - \omega_{\alpha,n} + i\eta^+} + \frac{+\mathbb{C}_{\alpha,-n}\mathbb{C}^\dagger_{\alpha,-n}}{z + \omega_{\alpha,n} - i\eta^+} + \frac{+\mathbb{C}_{0,n}\mathbb{C}^\dagger_{0,n}}{z - \omega_{0,n} + i\eta^+} + \frac{-\mathbb{C}_{0,-n}\mathbb{C}^\dagger_{0,-n}}{z + \omega_{0,n} - i\eta^+} \right\} \tag{B.2.8}$$

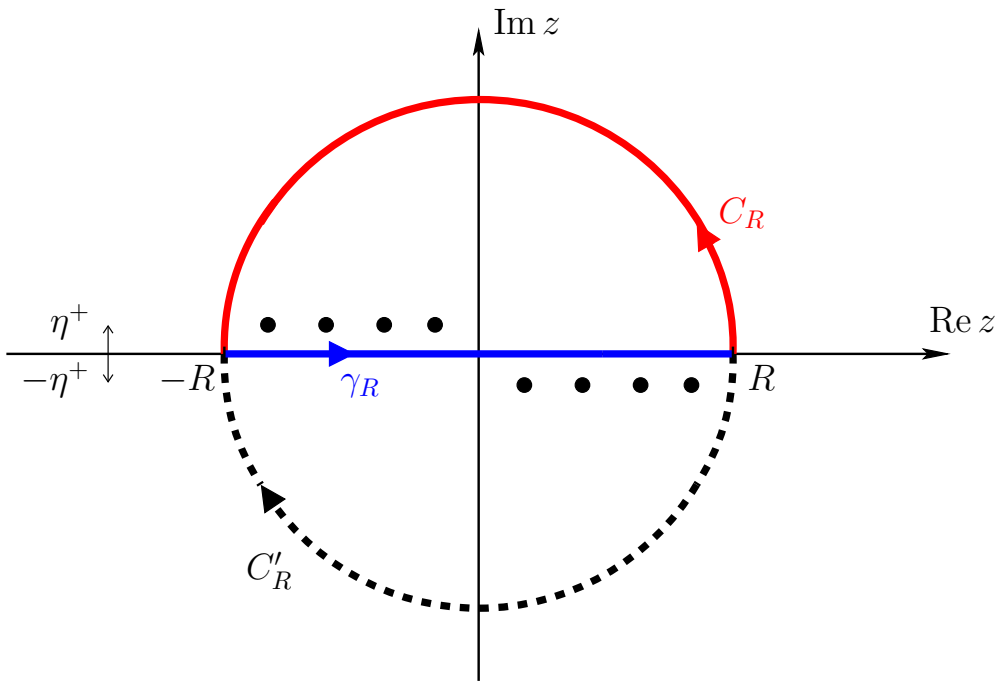

FIGURE B.2: Schéma du plan complexe $z$. Les pôles de $f$ (points noirs, voir équation (B.2.8)) sont légèrement au-dessus de l'axe des réels négatifs ($-\omega_{\alpha,n} + i\eta^+$ et $-\omega_{0,n} + i\eta^+$ ; on rappelle que $\omega_{\alpha,n}$ et $\omega_{0,n}$ sont positifs) et légèrement au-dessous de l'axe des réels positifs ($\omega_{\alpha,n} - i\eta^+$ et $\omega_{0,n} - i\eta^+$). On intègre sur le contour trigonométrique $\gamma_R + C_R$, mais l'intégration sur le contour $\gamma_R + C'_R$ est équivalente.

Cette fonction possède des pôles dans le plan complexe supérieur, légèrement au-dessus de l'axe réel : aux points $z = -\omega_{\alpha,n} + i\eta^+$, de résidus $\mathbb{C}_{\alpha,-n}\mathbb{C}^\dagger_{\alpha,-n}$ et $z = -\omega_{0,n} + i\eta^+$, de résidus $-\mathbb{C}_{0,-n}\mathbb{C}^\dagger_{0,-n}$. Tous les index sont égaux à +1. Ainsi :

$$\frac{1}{2\pi i}\int_{-\infty}^{\infty} f(\omega)d\omega = \lim_{R\to\infty}\frac{1}{2\pi i}\oint_{\gamma_R + C_R} f(z)\,dz = \sum_n \left\{\mathbb{C}_{\alpha,-n}\mathbb{C}^\dagger_{\alpha,-n} - \mathbb{C}_{0,-n}\mathbb{C}^\dagger_{0,-n}\right\} \tag{B.2.9}$$

Pour être complet, on peut remarquer que la fonction $f$ possède également des pôles dans le plan complexe inférieur, aux points $z = \omega_{\alpha,n} - i\eta^+$, de résidus $-\mathbb{C}_{\alpha,n}\mathbb{C}^\dagger_{\alpha,n}$ et $z = \omega_{0,n} - i\eta^+$, de résidus $\mathbb{C}_{0,n}\mathbb{C}^\dagger_{0,n}$ ; on peut intégrer dans le plan complexe inférieur : le sens du contour est alors anti-trigonométrique et les index des singularités sont tous égaux à −1. L'intégration donne un résultat qui à terme produit la même expression de l'énergie (c'est-à-dire qui a la même trace).

## B.3 Intégration de la formulation "matrice diélectrique"

Ayant à présent tous les outils à disposition, on est en mesure ici d'expliquer en détail la méthode esquissée par McLachlan *et al.*[245]. L'intégrale en fréquence (2.7.28) est calculée par intégration par parties du logarithme et application du principe de l'argument, pour obtenir la formulation (2.7.29) après intégration curviligne.

Considérons, donc, l'équation (2.7.28) :

$$E_c^{\text{dRPA-I}} = \frac{1}{2}\int_{-\infty}^{\infty}\frac{d\omega}{2\pi}\sum_{ia}\left\{\text{Log}\left(\frac{\Omega_{ia}^2+\omega^2}{\varepsilon_{ia}^2+\omega^2}\right) - \frac{M_{1,ia,ia}-\varepsilon_{ia}^2}{\varepsilon_{ia}^2+\omega^2}\right\} \tag{2.7.28}$$

Le second terme s'intègre en $K_{ia,ia}$, et le logarithme peut être intégré par partie :

$$\frac{1}{2\pi}\int_{-\infty}^{\infty}\text{Log}(f(z))\,dz = \frac{1}{2\pi}\Big[\text{Log}(f(z))\,z\Big]_{-\infty}^{\infty} - \frac{1}{2\pi}\int_{-\infty}^{\infty}\frac{f'(z)}{f(z)}\,z, \tag{B.3.1}$$

où le premier terme est nul et où on peut appliquer le principe de l'argument de l'équation (B.1.14) au second terme, avec : $g(z) = z$ et $f(z) = \frac{\Omega_{ia}^2+z^2}{\varepsilon_{ia}^2+z^2}$. La fonction $f$ a deux zéros simples en $\pm i\Omega_{ia}$ et deux pôles simples en $\pm i\varepsilon_{ia}$. En utilisant le même raisonnement que dans la section B.2.2 sur les contours de la figure B.2 où cette fois les éléments intéressants sont sur l'axe imaginaire, on intègre de manière équivalente sur le plan complexe supérieur ou sur le plan complexe inférieur pour obtenir :

$$-\frac{1}{2\pi}\int_{-\infty}^{\infty}\frac{f'(z)}{f(z)}\,z = -i\frac{1}{2\pi i}\int_{-\infty}^{\infty}\frac{f'(z)}{f(z)}\,z = -i\left(i\Omega_{ia} - i\varepsilon_{ia}\right), \tag{B.3.2}$$

c'est-à-dire pour obtenir une énergie de plasmon :

$$E_c^{\text{dRPA-I}} = \frac{1}{2}\left\{\sum_{ia}(\Omega_{ia}-\varepsilon_{ia}) - (K_{ia,ia})\right\} \tag{B.3.3}$$

# Annexe C

# Détails de l'adaptation de spin

Dans cette Annexe, on se convainc de l'adaptation de spin des matrices d'intégrales bi-électroniques, qui sont diagonales par bloc avec des composants singulets et triplets[2, 80, 246]. Une fois ceci établi, il est aisé de montrer que les composants triplets ne sont à prendre en compte que dans la formulation RPAx-II[2].

## C.1 Structure des matrices bi-électroniques

Une intégrale bi-électronique entre des spin-orbitales $p, q, r, s$ s'écrit :

$$\begin{aligned}\langle pq|rs\rangle &= \int p^*(\mathbf{x}_1)q^*(\mathbf{x}_2)w(\mathbf{x}_1,\mathbf{x}_2)r(\mathbf{x}_1)s(\mathbf{x}_2)\\ &= \int p^*(\mathbf{r}_1)q^*(\mathbf{r}_2)w(\mathbf{r}_1,\mathbf{r}_2)r(\mathbf{r}_1)s(\mathbf{r}_2)\int s_p^*(s_1)s_r(s_1)\int s_q^*(s_2)s_s(s_2),\end{aligned} \quad \text{(C.1.1)}$$

et ne peut être non nulle que si les fonctions de spin respectent $s_p = s_r$ et $s_q = s_s$. Ainsi, sur les $2^4$ combinaisons de spin possible pour les orbitales $p, q, r, s$, seules $2^2$ sont non nulles. Plus particulièrement, on a pour les matrices $\mathbf{K}$, $\mathbf{K}'$ et $\mathbf{J}$ les structures suivantes :

$$K_{ia,jb} = \langle ij|ab\rangle \qquad K'_{ia,jb} = \langle ij|ba\rangle \qquad J_{ia,jb} = \langle ib|ja\rangle$$

$$s_i = s_a \text{ et } s_j = s_b \qquad s_i = s_b \text{ et } s_j = s_a \qquad s_i = s_j \text{ et } s_b = s_a$$

$$ia\left[\overbrace{\begin{pmatrix} \uparrow\uparrow,\uparrow\uparrow & \uparrow\uparrow,\downarrow\downarrow & \mathbf{0} & \mathbf{0} \\ \downarrow\downarrow,\uparrow\uparrow & \downarrow\downarrow,\downarrow\downarrow & \mathbf{0} & \mathbf{0} \\ \mathbf{0} & \mathbf{0} & \mathbf{0} & \mathbf{0} \\ \mathbf{0} & \mathbf{0} & \mathbf{0} & \mathbf{0} \end{pmatrix}}^{jb} \quad \overbrace{\begin{pmatrix} \uparrow\uparrow,\uparrow\uparrow & \mathbf{0} & \mathbf{0} & \mathbf{0} \\ \mathbf{0} & \downarrow\downarrow,\downarrow\downarrow & \mathbf{0} & \mathbf{0} \\ \mathbf{0} & \mathbf{0} & \mathbf{0} & \uparrow\downarrow,\downarrow\uparrow \\ \mathbf{0} & \mathbf{0} & \downarrow\uparrow,\uparrow\downarrow & \mathbf{0} \end{pmatrix}}^{jb} \quad \overbrace{\begin{pmatrix} \uparrow\uparrow,\uparrow\uparrow & \mathbf{0} & \mathbf{0} & \mathbf{0} \\ \mathbf{0} & \downarrow\downarrow,\downarrow\downarrow & \mathbf{0} & \mathbf{0} \\ \mathbf{0} & \mathbf{0} & \uparrow\downarrow,\uparrow\downarrow & \mathbf{0} \\ \mathbf{0} & \mathbf{0} & \mathbf{0} & \downarrow\uparrow,\downarrow\uparrow \end{pmatrix}}^{jb}\right.$$

La transformation de chacune de ces matrices selon $\widetilde{\mathbf{X}} = \mathbf{U}^{\dagger}\mathbf{X}\mathbf{U}$ (voir (2.6.2)) donne :

$$\widetilde{K}_{ia,jb} = \begin{pmatrix} K_1 & K_3 & \mathbf{0} & \mathbf{0} \\ K_2 & K_4 & \mathbf{0} & \mathbf{0} \\ \mathbf{0} & \mathbf{0} & \mathbf{0} & \mathbf{0} \\ \mathbf{0} & \mathbf{0} & \mathbf{0} & \mathbf{0} \end{pmatrix} \qquad \widetilde{K'}_{ia,jb} = \begin{pmatrix} K'_1 & K'_3 & \mathbf{0} & \mathbf{0} \\ K'_2 & K'_4 & \mathbf{0} & \mathbf{0} \\ \mathbf{0} & \mathbf{0} & K'_5 & K'_7 \\ \mathbf{0} & \mathbf{0} & K'_6 & K'_8 \end{pmatrix} \qquad \widetilde{J}_{ia,jb} = \begin{pmatrix} J_1 & J_3 & \mathbf{0} & \mathbf{0} \\ J_2 & J_4 & \mathbf{0} & \mathbf{0} \\ \mathbf{0} & \mathbf{0} & J_5 & J_7 \\ \mathbf{0} & \mathbf{0} & J_6 & J_8 \end{pmatrix}$$

où :

$$\begin{aligned} K_1 &= \tfrac{1}{2}\,(\uparrow\uparrow,\uparrow\uparrow + \downarrow\downarrow,\uparrow\uparrow \\ &\quad + \uparrow\uparrow,\downarrow\downarrow + \downarrow\downarrow,\downarrow\downarrow) = 2K_{ia,jb} \\ K_2 &= \tfrac{1}{2}\,(\uparrow\uparrow,\uparrow\uparrow - \downarrow\downarrow,\uparrow\uparrow \\ &\quad + \uparrow\uparrow,\downarrow\downarrow - \downarrow\downarrow,\downarrow\downarrow) = 0 \\ K_3 &= \tfrac{1}{2}\,(\uparrow\uparrow,\uparrow\uparrow + \downarrow\downarrow,\uparrow\uparrow \\ &\quad - \uparrow\uparrow,\downarrow\downarrow - \downarrow\downarrow,\downarrow\downarrow) = 0 \\ K_4 &= \tfrac{1}{2}\,(\uparrow\uparrow,\uparrow\uparrow - \downarrow\downarrow,\uparrow\uparrow \\ &\quad - \uparrow\uparrow,\downarrow\downarrow + \downarrow\downarrow,\downarrow\downarrow) = 0 \end{aligned}$$

$$\begin{aligned} K'_1 &= \tfrac{1}{2}\,(\uparrow\uparrow,\uparrow\uparrow + \downarrow\downarrow,\downarrow\downarrow) = K'_{ia,jb} \\ K'_2 &= \tfrac{1}{2}\,(\uparrow\uparrow,\uparrow\uparrow - \downarrow\downarrow,\downarrow\downarrow) = 0 \\ K'_3 &= \tfrac{1}{2}\,(\uparrow\uparrow,\uparrow\uparrow - \downarrow\downarrow,\downarrow\downarrow) = 0 \\ K'_4 &= \tfrac{1}{2}\,(\uparrow\uparrow,\uparrow\uparrow + \downarrow\downarrow,\downarrow\downarrow) = K'_{ia,jb} \\ K'_5 &= \tfrac{1}{2}\,(\downarrow\uparrow,\uparrow\downarrow + \uparrow\downarrow,\downarrow\uparrow) = K'_{ia,jb} \\ K'_6 &= \tfrac{1}{2}\,(\downarrow\uparrow,\uparrow\downarrow - \uparrow\downarrow,\downarrow\uparrow) = 0 \\ K'_7 &= \tfrac{1}{2}\,(\uparrow\downarrow,\downarrow\uparrow - \downarrow\uparrow,\uparrow\downarrow) = 0 \\ K'_8 &= -\tfrac{1}{2}\,(\downarrow\uparrow,\uparrow\downarrow + \uparrow\downarrow,\downarrow\uparrow) = -K'_{ia,jb} \end{aligned}$$

$$\begin{aligned} J_1 &= \tfrac{1}{2}\,(\uparrow\uparrow,\uparrow\uparrow + \downarrow\downarrow,\downarrow\downarrow) = J_{ia,jb} \\ J_2 &= \tfrac{1}{2}\,(\uparrow\uparrow,\uparrow\uparrow - \downarrow\downarrow,\downarrow\downarrow) = 0 \\ J_3 &= \tfrac{1}{2}\,(\uparrow\uparrow,\uparrow\uparrow - \downarrow\downarrow,\downarrow\downarrow) = 0 \\ J_4 &= \tfrac{1}{2}\,(\uparrow\uparrow,\uparrow\uparrow + \downarrow\downarrow,\downarrow\downarrow) = J_{ia,jb} \\ J_5 &= \tfrac{1}{2}\,(\uparrow\downarrow,\uparrow\downarrow + \downarrow\uparrow,\downarrow\uparrow) = J_{ia,jb} \\ J_6 &= \tfrac{1}{2}\,(\uparrow\downarrow,\uparrow\downarrow - \downarrow\uparrow,\downarrow\uparrow) = 0 \\ J_7 &= \tfrac{1}{2}\,(\uparrow\downarrow,\uparrow\downarrow - \downarrow\uparrow,\downarrow\uparrow) = 0 \\ J_8 &= \tfrac{1}{2}\,(\uparrow\downarrow,\uparrow\downarrow + \downarrow\uparrow,\downarrow\uparrow) = J_{ia,jb} \end{aligned}$$

Les matrices $\widetilde{K}_{ia,jb}$, $\widetilde{K'}_{ia,jb}$ et $\widetilde{J}_{ia,jb}$ sont donc bien bloc-diagonales, et on construit les matrices $\mathbf{A}'$, et $\mathbf{B}$ vues équation (2.6.5).

## C.2 Contributions triplets

On veut montrer ici que des contributions triplets n'apparaissent que dans le cas de figure de la RPAx-II. Rappelons l'équation générale des énergies RPA AC-FDT :

$$E_c = \frac{1}{2}\int_0^1 d\alpha \,\mathrm{tr}\left\{\frac{1}{2}(\mathbf{A}' \overset{\text{I/II}}{\downarrow}{+} \mathbf{B})\underset{\text{d/x}}{\overset{}{\mathbf{Q}_\alpha}} + \frac{1}{2}(\mathbf{A}' \overset{\text{I/II}}{\downarrow}{-} \mathbf{B})\underset{\text{d/x}}{\mathbf{Q}_\alpha^{-1}} - \overset{\text{I/II}}{\mathbf{A}'}\right\}, \qquad (2.4.22)$$

où les matrices $\mathbf{Q}_\alpha$ (et $\mathbf{M}_\alpha$) sont définies aux équations (2.4.27) et (2.4.25). On rappelle que les matrices sont construites dans un cadre *simple-barre* ou *double-barre* ; direct-RPA ou RPA-échange. Il est clair que les formulations *simple-barre* ne font pas émerger de contribution triplet (les matrices ${}^3\mathbf{A}'^{\mathrm{I}}$ et ${}^3\mathbf{B}^{\mathrm{I}}$ sont nulles). Pour le cas des formulations direct-RPA, on a :

$${}^3\mathbf{M}_\alpha^{\mathrm{dRPA}} = (\boldsymbol{\varepsilon} + \alpha\mathbf{0} - \alpha\mathbf{0})^{\frac{1}{2}}\,(\boldsymbol{\varepsilon} + \alpha\mathbf{0} + \alpha\mathbf{0})\,(\boldsymbol{\varepsilon} + \alpha\mathbf{0} - \alpha\mathbf{0})^{\frac{1}{2}} = \boldsymbol{\varepsilon}^2, \qquad (\text{C.2.1})$$

et :

$${}^3\mathbf{Q}_\alpha^{\mathrm{dRPA}} = (\boldsymbol{\varepsilon} + \alpha\mathbf{0} - \alpha\mathbf{0})^{\frac{1}{2}}\left(\boldsymbol{\varepsilon}^2\right)^{-\frac{1}{2}}(\boldsymbol{\varepsilon} + \alpha\mathbf{0} - \alpha\mathbf{0})^{\frac{1}{2}} = \mathbf{1} \qquad (\text{C.2.2})$$

Ce qui mène à écrire la contribution triplet de l'intégrande de l'équation (2.4.22), dans le cas direct-RPA (indépendamment de la construction *simple-barre* ou *double-barre* des matrices $\mathbf{A}'$ et $\mathbf{B}$) :

$$\frac{1}{2}(\mathbf{A}' + \mathbf{B}) + \frac{1}{2}(\mathbf{A}' - \mathbf{B}) - \mathbf{A}' = \mathbf{0} \qquad (\text{C.2.3})$$

Ainsi, en effet, seule la combinaison de matrices construites dans un cadre RPAx-II font émerger des contributions triplets.

# Annexe D

# Dérivations pour les orbitales localisées

On dérive ici tout ce qui est nécessaire aux développements concernant les POO dans le manuscrit, c'est-à-dire que l'on écrit : les différentes procédures itératives pour calculer les matrices **T** selon le niveau d'approximation (approximation des excitations locales, moyenne sphérique), les objets qui émergent lors de l'expansion multipolaire des intégrales bi-électroniques, et les éléments de matrice de Fock dans la base des POO.

## D.1 Résolution itérative des équations de Riccati locales

### D.1.1 Équation de Riccati dans la base des POO

Du fait de la non orthogonalité des POO et la structure non diagonales de la matrice de Fock, le schéma de résolution habituel des équations de Riccati doit être revu. On sépare dans (3.5.10) les contributions des paires $ij$ pures, et les contributions d'autres paires $ik$ ou $kj$ :

$$\begin{aligned}\mathbf{R}^{ij} &= \mathbf{B}^{ij} + \left(\mathbf{f} - f_{ii}\mathbf{S} + \mathbf{A}'^{ii}\right)\mathbf{T}^{ij}\mathbf{S} + \mathbf{S}\mathbf{T}^{ij}\left(\mathbf{f} - \mathbf{S}f_{jj} + \mathbf{A}'^{jj}\right) \\ &- \sum_{k\neq i} f_{ik}\mathbf{S}\mathbf{T}^{kj}\mathbf{S} - \sum_{k\neq j}\mathbf{S}\mathbf{T}^{ik}\mathbf{S}f_{kj} + \sum_{k\neq i}\mathbf{A}'^{ik}\mathbf{T}^{kj}\mathbf{S} + \sum_{k\neq j}\mathbf{S}\mathbf{T}^{ik}\mathbf{A}'^{kj} + \sum_{kl}\mathbf{S}\mathbf{T}^{ik}\mathbf{B}^{kl}\mathbf{T}^{lj}\mathbf{S} = \mathbf{0}\end{aligned} \quad \text{(D.1.1)}$$

Dans ces équations, on ne peut pas traiter les termes impliquant la matrice $(\mathbf{f})_{p_\alpha q_\beta}$, que l'on va diagonaliser en utilisant la matrice $\mathbf{X}$ solution du problème aux valeurs propres généralisé $\mathbf{fX} = \mathbf{SX}\boldsymbol{\varepsilon}$. On va donc écrire des objets tels que, par exemple, $\mathbf{X}^\dagger\mathbf{fX}$ : j'attire l'attention du lecteur sur le fait que cette transformation n'est *pas* un retour vers les orbitales virtuelles VMO. On a :

$$\mathbf{f}_{\text{POO}} = \mathbf{V}^\dagger\mathbf{f}_{\text{VMO}}\mathbf{V} \quad \text{c'est-à-dire :} \quad \left(\mathbf{V}^\dagger\right)^{-1}\mathbf{f}_{\text{POO}}\left(\mathbf{V}\right)^{-1} = \mathbf{f}_{\text{VMO}}, \quad \text{(D.1.2)}$$

où la relation qui lie $(\mathbf{V})^{-1}$ et $\mathbf{V}$ n'est pas évidente ($\mathbf{V}$ n'est *pas* une matrice orthogonale). La transformation avec la matrice $\mathbf{X}$ s'écrit :

$$X^{\dagger}_{\widetilde{a}p_\alpha} f_{p_\alpha q_\beta} X_{q_\beta \widetilde{b}} = \left(\mathbf{X}^{\dagger}\mathbf{S}\mathbf{X}\right)_{\widetilde{a}\widetilde{b}} \varepsilon_{\widetilde{b}} = \delta_{\widetilde{a}\widetilde{b}} \varepsilon_{\widetilde{b}}, \tag{D.1.3}$$

où les orbitales $\widetilde{a}$ ne sont *pas* les orbitales virtuelles VMO, mais des orbitales virtuelles pseudo-canoniques qui diagonalisent la matrice de fock virtuelle exprimée en POO. Ainsi, on transforme les équations de Riccati dans une base d'orbitales virtuelles pseudo-canoniques qui diagonalisent la matrice $\mathbf{f}_{\text{POO}}$, et ce séparément pour chaque paire $[ij]$.

On multiplie donc à gauche par $\mathbf{X}^{\dagger}$ et à droite par $\mathbf{X}$ :

$$\begin{aligned}
\mathbf{X}^{\dagger}\mathbf{R}^{ij}\mathbf{X} = \mathbf{X}^{\dagger}\mathbf{B}^{ij}\mathbf{X} &+ \left(\mathbf{X}^{\dagger}\mathbf{f} - f_{ii}\mathbf{X}^{\dagger}\mathbf{S} + \mathbf{X}^{\dagger}\mathbf{A}'^{ii}\right)\mathbf{T}^{ij}\mathbf{S}\mathbf{X} + \mathbf{X}^{\dagger}\mathbf{S}\mathbf{T}^{ij}\left(\mathbf{f}\mathbf{X} - \mathbf{S}\mathbf{X}f_{jj} + \mathbf{A}'^{jj}\mathbf{X}\right) \\
&- \sum_{k\neq i} f_{ik}\mathbf{X}^{\dagger}\mathbf{S}\mathbf{T}^{kj}\mathbf{S}\mathbf{X} - \sum_{k\neq j}\mathbf{X}^{\dagger}\mathbf{S}\mathbf{T}^{ik}\mathbf{S}\mathbf{X}f_{kj} + \sum_{k\neq i}\mathbf{X}^{\dagger}\mathbf{A}'^{ik}\mathbf{T}^{kj}\mathbf{S}\mathbf{X} + \sum_{k\neq j}\mathbf{X}^{\dagger}\mathbf{S}\mathbf{T}^{ik}\mathbf{A}'^{kj}\mathbf{X} \\
&+ \sum_{kl}\mathbf{X}^{\dagger}\mathbf{S}\mathbf{T}^{ik}\mathbf{B}^{kl}\mathbf{T}^{lj}\mathbf{S}\mathbf{X} = \mathbf{0}
\end{aligned} \tag{D.1.4}$$

On applique l'équation aux valeurs propres, ainsi que les relations $\mathbf{I} = \mathbf{S}\mathbf{X}\mathbf{X}^{\dagger} = \mathbf{X}\mathbf{X}^{\dagger}\mathbf{S}$, pour obtenir :

$$\begin{aligned}
\mathbf{X}^{\dagger}\mathbf{R}^{ij}\mathbf{X} = \mathbf{X}^{\dagger}\mathbf{B}^{ij}\mathbf{X} &+ \left(\boldsymbol{\varepsilon} - f_{ii}\mathbf{I} + \mathbf{X}^{\dagger}\mathbf{A}'^{ii}\mathbf{X}\right)\mathbf{X}^{\dagger}\mathbf{S}\mathbf{T}^{ij}\mathbf{S}\mathbf{X} + \mathbf{X}^{\dagger}\mathbf{S}\mathbf{T}^{ij}\mathbf{S}\mathbf{X}\left(\boldsymbol{\varepsilon} - f_{jj}\mathbf{I} + \mathbf{X}^{\dagger}\mathbf{A}'^{jj}\mathbf{X}\right) \\
&- \sum_{k\neq i} f_{ik}\mathbf{X}^{\dagger}\mathbf{S}\mathbf{T}^{kj}\mathbf{S}\mathbf{X} - \sum_{k\neq j}\mathbf{X}^{\dagger}\mathbf{S}\mathbf{T}^{ik}\mathbf{S}\mathbf{X}f_{kj} + \sum_{k\neq i}\mathbf{X}^{\dagger}\mathbf{A}'^{ik}\mathbf{X}\mathbf{X}^{\dagger}\mathbf{S}\mathbf{T}^{kj}\mathbf{S}\mathbf{X} + \sum_{k\neq j}\mathbf{X}^{\dagger}\mathbf{S}\mathbf{T}^{ik}\mathbf{S}\mathbf{X}\mathbf{X}^{\dagger}\mathbf{A}'^{kj}\mathbf{X} \\
&+ \sum_{kl}\mathbf{X}^{\dagger}\mathbf{S}\mathbf{T}^{ik}\mathbf{S}\mathbf{X}\mathbf{X}^{\dagger}\mathbf{B}^{kl}\mathbf{X}\mathbf{X}^{\dagger}\mathbf{S}\mathbf{T}^{lj}\mathbf{S}\mathbf{X} = \mathbf{0},
\end{aligned} \tag{D.1.5}$$

qui peut s'écrire de manière plus compacte :

$$\begin{aligned}
\overline{\mathbf{R}}^{ij} = \overline{\mathbf{B}}^{ij} &+ \left(\boldsymbol{\varepsilon} - f_{ii}\mathbf{I} + \overline{\mathbf{A}'}^{ii}\right)\overline{\mathbf{T}}^{ij} + \overline{\mathbf{T}}^{ij}\left(\boldsymbol{\varepsilon} - f_{jj}\mathbf{I} + \overline{\mathbf{A}'}^{jj}\right) \\
&- \sum_{k\neq i} f_{ik}\overline{\mathbf{T}}^{kj} - \sum_{k\neq j}\overline{\mathbf{T}}^{ik} f_{kj} + \sum_{k\neq i}\overline{\mathbf{A}'}^{ik}\overline{\mathbf{T}}^{kj} + \sum_{k\neq j}\overline{\mathbf{T}}^{ik}\overline{\mathbf{A}'}^{kj} + \sum_{kl}\overline{\mathbf{T}}^{ik}\overline{\mathbf{B}}^{kl}\overline{\mathbf{T}}^{lj} = \mathbf{0},
\end{aligned} \tag{D.1.6}$$

où l'on a introduit les notations :

$$\begin{aligned}
\overline{\mathbf{R}}^{ij} &= \mathbf{X}^{\dagger}\mathbf{R}^{ij}\mathbf{X} \\
\overline{\mathbf{A}'}^{ij} &= \mathbf{X}^{\dagger}\mathbf{A}'^{ij}\mathbf{X} \\
\overline{\mathbf{B}}^{ij} &= \mathbf{X}^{\dagger}\mathbf{B}^{ij}\mathbf{X} \\
\overline{\mathbf{T}}^{ij} &= \mathbf{X}^{\dagger}\mathbf{S}\mathbf{T}^{ij}\mathbf{S}\mathbf{X}
\end{aligned} \tag{D.1.7}$$

On résout l'équation (D.1.6) par la formule itérative :

$$\overline{T}_{\widetilde{ab}}^{ij\,(n)} = -\frac{\overline{B}_{\widetilde{ab}}^{ij} + \Delta\overline{R}_{\widetilde{ab}}^{ij}\left(\overline{\mathbf{T}}^{(n-1)}\right)}{\left(\varepsilon_{\widetilde{a}} - f_{ii} + A'^{ii}_{\widetilde{aa}}\right) + \left(\varepsilon_{\widetilde{b}} - f_{jj} + A'^{jj}_{\widetilde{bb}}\right)}, \tag{D.1.8}$$

où $\Delta\overline{\mathbf{R}}^{ij}\left(\overline{\mathbf{T}}\right)$ est :

$$\Delta\overline{\mathbf{R}}^{ij}\left(\overline{\mathbf{T}}\right) = -\sum_{k\neq i} f_{ik}\overline{\mathbf{T}}^{kj} - \sum_{k\neq j}\overline{\mathbf{T}}^{ik} f_{kj} + \sum_{k\neq i}\overline{\mathbf{A}'}^{ik}\overline{\mathbf{T}}^{kj} + \sum_{k\neq j}\overline{\mathbf{T}}^{ik}\overline{\mathbf{A}'}^{kj} + \sum_{kl}\overline{\mathbf{T}}^{ik}\overline{\mathbf{B}}^{kl}\overline{\mathbf{T}}^{lj} \tag{D.1.9}$$

Après convergence, les amplitudes obtenues peuvent être transformées vers la base POO d'origine par la simple transformation $\mathbf{T}^{ij} = \mathbf{X}\overline{\mathbf{T}}^{ij}\mathbf{X}^{\dagger}$, en effet on a :

$$\mathbf{T}^{ij} = \left(\mathbf{X}^{\dagger}\mathbf{S}\right)^{-1}\overline{\mathbf{T}}^{ij}(\mathbf{S}\mathbf{X})^{-1} = \mathbf{S}^{-1}(\mathbf{X}^{\dagger})^{-1}\overline{\mathbf{T}}^{ij}\mathbf{X}^{-1}\mathbf{S}^{-1} = \mathbf{X}\mathbf{X}^{\dagger}(\mathbf{X}^{\dagger})^{-1}\overline{\mathbf{T}}^{ij}\mathbf{X}^{-1}\mathbf{X}\mathbf{X}^{\dagger} = \mathbf{X}\overline{\mathbf{T}}^{ij}\mathbf{X}^{\dagger} \tag{D.1.10}$$

En fait, l'énergie de corrélation peut tout à fait être obtenue sans transformation, par exemple :

$$\begin{aligned}
\sum_{ij}\operatorname{tr}\left\{\overline{\mathbf{K}}^{ij}\overline{\mathbf{T}}^{ij}\right\} &= \sum_{ij}\operatorname{tr}\left\{\mathbf{X}^{\dagger}\mathbf{K}^{ij}\mathbf{X}\mathbf{X}^{\dagger}\mathbf{S}\mathbf{T}^{ij}\mathbf{S}\mathbf{X}\right\} \\
&= \sum_{ij}\operatorname{tr}\left\{\mathbf{K}^{ij}\mathbf{S}^{-1}\mathbf{S}\mathbf{T}^{ij}\mathbf{S}\mathbf{X}\mathbf{X}^{\dagger}\right\} \\
&= \sum_{ij}\operatorname{tr}\left\{\mathbf{K}^{ij}\mathbf{T}^{ij}\mathbf{S}\mathbf{S}^{-1}\right\} \\
&= \sum_{ij}\operatorname{tr}\left\{\mathbf{K}^{ij}\mathbf{T}^{ij}\right\}
\end{aligned} \tag{D.1.11}$$

### D.1.2 Équation de Riccati dans le modèle des excitations locales

Pour ce qui est de la résolution des équations de Riccati dans le modèle des excitations locales, équations (3.5.15), le même raisonnement aboutit au même genre de procédure :

$$\overline{T}_{\widetilde{ab}}^{ij\,(n)} = -\frac{\overline{B}_{\widetilde{ab}}^{ij} + \Delta\overline{R}_{\widetilde{ab}}^{ij}\left(\overline{\mathbf{T}}^{(n-1)}\right)}{\left(\varepsilon_{\widetilde{a}} - f_{ii} + A'^{ii}_{\widetilde{aa}}\right) + \left(\varepsilon_{\widetilde{b}} - f_{jj} + A'^{jj}_{\widetilde{bb}}\right)}, \tag{D.1.12}$$

où cette fois :

$$\Delta\overline{\mathbf{R}}^{ij}\left(\overline{\mathbf{T}}\right) = \sum_{k\neq i}\overline{\mathbf{A}'}^{ik}\overline{\mathbf{T}}^{kj} + \sum_{k\neq j}\overline{\mathbf{T}}^{ik}\overline{\mathbf{A}'}^{kj} + \sum_{kl}\overline{\mathbf{T}}^{ik}\overline{\mathbf{B}}^{kl}\overline{\mathbf{T}}^{lj}, \tag{D.1.13}$$

et :

$$\overline{\mathbf{T}}^{ij} = \mathbf{X}^{\dagger}\mathbf{s}^{ii}\mathbf{T}^{ij}\mathbf{s}^{jj}\mathbf{X} \tag{D.1.14}$$

Les amplitudes trouvées peuvent être retransformées dans la base POO comme précédemment.

### D.1.3 Équation de Riccati dans l'approximation des moyennes sphériques

Quant aux équations trouvées dans le cadre d'approximation avec moyenne sphérique, (3.5.28) : elles peuvent être résolues directement, *i.e.* sans transformation pseudo-canonique. On trouve la procédure d'itération suivante :

$$T^{ij\,(n)}_{\alpha\beta} = -\frac{s^i s^j L_{\mathrm{B}}{}^{ij}_{\alpha\beta} + \Delta R^{ij}_{\alpha\beta}\left(\mathbf{T}^{(n-1)}\right)}{\Delta_{ij} + \Delta_{ji}}, \tag{D.1.15}$$

où :

$$\Delta_{ij} = f^i s^j - f_{ii} s^i s^j + \tfrac{1}{3} s^i s^j s^i L^{ii}_{\alpha\beta} - 3 s^j L_{\mathrm{A}}{}^{ii}_{\alpha\alpha}, \tag{D.1.16}$$

et où les seules quantités nécessaires sont les moyennes sphériques $s^i$ et $f^i$ ainsi que les tenseurs dipôle-dipôle $L^{ij}$. On a :

$$\Delta\mathbf{R}^{ij}(\mathbf{T}) = \tfrac{1}{3}\sum_{k\neq i} s^i s^j s^k \mathbf{L}^{ik}\mathbf{T}^{kj} - 3 s^j \mathbf{L}_{\mathrm{A}}{}^{ii}_{\substack{\text{hors-}\\\text{diag}}}\mathbf{T}^{ij} + \tfrac{1}{3}\sum_{k\neq j} s^i s^j s^k \mathbf{T}^{ik}\mathbf{L}^{kj} - 3 s^i \mathbf{T}^{ij}\mathbf{L}_{\mathrm{A}}{}^{jj}_{\substack{\text{hors-}\\\text{diag}}} + \tfrac{1}{3^2}\sum_{kl} s^i s^j s^k s^l \mathbf{T}^{ik}\mathbf{L}_{\mathrm{B}}^{kl}\mathbf{T}^{lj} \tag{D.1.17}$$

## D.2 Expansions multipolaires

### D.2.1 Généralités

On écrit les expansions multipolaires des intégrales bi-électroniques en choisissant les centroïdes des orbitales occupées $\mathbf{D}^i$ et $\mathbf{D}^j$ comme centres d'expansion. On dénote le vecteur connectant les centres d'expansion par $\mathbf{R}^{ij}$ de sorte à écrire, pour deux points quelconque $\mathbf{r}^i$ et $\mathbf{r}^j$ autour de $\mathbf{D}^i$ et $\mathbf{D}^j$ : $\mathbf{R} = \mathbf{r}^i - \mathbf{r}^j = \left(\mathbf{r}^i - \mathbf{D}^i\right) + \mathbf{R}^{ij} - \left(\mathbf{r}^j - \mathbf{D}^j\right)$. On écrit la double expansion de Taylor de l'interaction longue-portée :

$$L(\mathbf{R}) = L^{ij} + (r^i_\alpha - D^i_\alpha)L^{ij}_\alpha + (r^j_\beta - D^j_\beta)L^{ij}_\beta + (r^i_\alpha - D^i_\alpha)(r^j_\beta - D^j_\beta)L^{ij}_{\alpha\beta} + \ldots, \tag{D.2.1}$$

où les définitions de $L^{ij}_\alpha$, $L^{ij}_{\alpha\beta}$, *etc...* sont évidentes. Par exemple, pour $L = \mathrm{erf}(\mu R)/R$, on a :

$$L^{ij}_\alpha(\mathbf{R}) = -\frac{R_\alpha}{R^3}\left(1 - \frac{2}{\sqrt{\pi}}\mu R \mathrm{e}^{-\mu^2 R^2} - \mathrm{erf}(\mu R)\right) \tag{D.2.2}$$

$$\begin{aligned} L^{ij}_{\alpha\beta}(\mathbf{R}) = {} & \frac{3R_\alpha R_\beta}{R^5}\left(\mathrm{erf}(\mu R) - \frac{2}{\sqrt{\pi}}\mu R \mathrm{e}^{-\mu^2 R^2}\left(1 + \tfrac{2}{3}\mu^2 R^2\right)\right) \\ & - \frac{\delta_{\alpha\beta} R^2}{R^5}\left(\mathrm{erf}(\mu R) - \frac{2}{\sqrt{\pi}}\mu R \mathrm{e}^{-\mu^2 R^2}\right) \end{aligned} \tag{D.2.3}$$

En se rappelant que le tenseur de l'interaction Coulombienne pleine portée s'écrit $T^{ij}_{\alpha\beta} = \frac{3R_\alpha R_\beta}{R^5} - \frac{\delta_{\alpha\beta} R^2}{R^5}$, on peut écrire $L^{ij}_{\alpha\beta}$ sous une forme alternative qui montre bien la contribution amortie de l'interaction dipôle-dipôle :

$$L^{ij}_{\alpha\beta}(\mathbf{R}) = T^{ij}_{\alpha\beta}\left(\mathrm{erf}(\mu R) - \frac{2}{\sqrt{\pi}}\mu R \mathrm{e}^{-\mu^2 R^2}\left(1 + \tfrac{2}{3}\mu^2 R^2\right)\right) - \delta_{\alpha\beta}\frac{4\mu^3}{3\sqrt{\pi}}e^{-\mu^2 R^2} \tag{D.2.4}$$

La trace du produit de deux tenseurs de second ordre (que l'on utilisera au moment de dériver des coefficients $\mathrm{C}_6$ approximés, voir section 3.5.5) s'écrit alors :

$$L^{ij}_{\alpha\beta}L^{ij}_{\beta\alpha} = \frac{6}{R^6}\underbrace{\left(4\mathrm{e}^{-2\mu^2 R^2}\mu R\left(\frac{\mu R\left(3 + 4\mu^2 R^2 + 2\mu^4 R^4\right)}{3\pi} - \frac{\left(3 + 2\mu^2 R^2\right)\mathrm{erf}(\mu R)}{3\sqrt{\pi}}\right) + \mathrm{erf}(\mu R)^2\right)}_{F^\mu(R)} \tag{D.2.5}$$

### D.2.2 Intégrales Coulombiennes

Lorsque l'on veut écrire l'expansion multipolaire des intégrales de type $J^{ij}_{m_\alpha n_\beta}$, on est amené à écrire :

$$J^{ij}_{m_\alpha n_\beta} = \left(ij\middle|m_\alpha n_\beta\right) = \left\langle i\middle|\hat{r}_\gamma\middle|j\right\rangle L\left\langle m\middle|\hat{r}_\alpha\left(1 - \hat{P}\right)\hat{r}_\delta\left(1 - \hat{P}\right)\hat{r}_\beta\middle|n\right\rangle \tag{D.2.6}$$

Ces intégrales représentent l'interaction entre des densités provenant de sites éloignés les uns des autres : on ne doit considérer comme non nuls que les éléments $\left(ii\middle|i_\alpha i_\beta\right)$ :

$$J^{ii}_{i_\alpha i_\beta} = \left\langle i\middle|\hat{r}_\gamma\middle|i\right\rangle L^{ii}_{\gamma\delta} \left\langle m\middle|\hat{r}_\alpha \left(1-\hat{P}\right) \hat{r}_\delta \left(1-\hat{P}\right) \hat{r}_\beta\middle|m\right\rangle \tag{D.2.7}$$

Le deuxième *bra-ket* s'écrit :

$$\begin{aligned}\left\langle m\middle|\hat{r}_\alpha \left(1-\hat{P}\right) \hat{r}_\delta \left(1-\hat{P}\right) \hat{r}_\beta\middle|m\right\rangle &= \left\langle m\middle|\hat{r}_\alpha\hat{r}_\delta\hat{r}_\beta\middle|m\right\rangle - \sum_k \langle m|\hat{r}_\alpha|k\rangle \left\langle k\middle|\hat{r}_\delta\hat{r}_\beta\middle|m\right\rangle - \sum_k \langle m|\hat{r}_\alpha\hat{r}_\delta|k\rangle \left\langle k\middle|\hat{r}_\beta\middle|m\right\rangle \\ &\quad + \sum_{kl} \langle m|\hat{r}_\alpha|k\rangle \langle k|\hat{r}_\delta|l\rangle \left\langle l\middle|\hat{r}_\beta\middle|m\right\rangle \\ &= \left\langle m\middle|\hat{r}_\alpha\hat{r}_\delta\hat{r}_\beta\middle|m\right\rangle - \langle m|\hat{r}_\alpha|m\rangle \left\langle m\middle|\hat{r}_\delta\hat{r}_\beta\middle|m\right\rangle - \langle m|\hat{r}_\alpha\hat{r}_\delta|m\rangle \left\langle m\middle|\hat{r}_\beta\middle|m\right\rangle \\ &\quad + \langle m|\hat{r}_\alpha|m\rangle \langle m|\hat{r}_\delta|m\rangle \left\langle m\middle|\hat{r}_\beta\middle|m\right\rangle, \end{aligned} \tag{D.2.8}$$

où on néglige les éléments hors diagonaux de $\langle m|\hat{r}_\alpha|k\rangle$, qui sont d'une certaine manière minimisés dans une localisation selon Boys. On se souvenant que $D^i_\alpha = \langle i|r_\alpha|i\rangle$, on a :

$$J^{ii}_{i_\alpha i_\beta} = D^i_\gamma L^{ii}_{\gamma\delta} \left\{\left\langle m\middle|\hat{r}_\alpha\hat{r}_\delta\hat{r}_\beta\middle|m\right\rangle - D^m_\alpha \left\langle m\middle|\hat{r}_\delta\hat{r}_\beta\middle|m\right\rangle - \langle m|\hat{r}_\alpha\hat{r}_\delta|m\rangle D^m_\beta + D^m_\alpha D^m_\delta D^m_\beta\right\} \tag{D.2.9}$$

On va donc écrire la matrice $\mathbf{A}'$ dans l'approximation des excitations locales :

$$A'^{ij}_{\alpha\beta} = K^{ij}_{\alpha\beta} - \delta_{ij} L_{\mathrm{A}}{}^{ii}_{\alpha\beta}, \tag{D.2.10}$$

où je définis les matrices $\mathbf{L}^{ii}_{\mathrm{A}}$ :

$$L_{\mathrm{A}}{}^{ii}_{\alpha\beta} = \zeta \sum_{\gamma\delta} D^i_\gamma L^{ii}_{\gamma\delta} \left\{\left\langle m\middle|\hat{r}_\alpha\hat{r}_\delta\hat{r}_\beta\middle|m\right\rangle - D^m_\alpha \left\langle m\middle|\hat{r}_\delta\hat{r}_\beta\middle|m\right\rangle - \langle m|\hat{r}_\alpha\hat{r}_\delta|m\rangle D^m_\beta + D^m_\alpha D^m_\delta D^m_\beta\right\}, \tag{D.2.11}$$

qui incluent par soucis de brièveté l'« interrupteur » qui sert à passer d'une formulation dRPA à une formulation RPAx. On veut, en donnant ce nom $\mathbf{L}_{\mathrm{A}}$, mimer l'objet $\mathbf{L}_{\mathrm{B}}$ qui émerge dans la dérivation de $\mathbf{B}$.

## D.3 Éléments de la matrice de fock

Il nous faut expliciter les éléments de matrices dans la base des POO, c'est-à-dire les éléments $f_{i_\alpha j_\beta}$. On peut facilement écrire :

$$\mathbf{f}_{\mathrm{POO}} = \mathbf{V}^\dagger \mathbf{f}_{\mathrm{VMO}} \mathbf{V}, \tag{D.3.1}$$

qui implique explicitement le bloc virtuel/virtuel de la matrice de fock. On préférerait exprimer tous les éléments de matrices avec des orbitales occupées, on procède donc d'une manière alternative, en écrivant :

$$\left\langle i_\alpha \middle| \hat{f} \middle| j_\beta \right\rangle = \left\langle i \middle| \hat{r}_\alpha \left(1-\hat{P}\right) \hat{f} \left(1-\hat{P}\right) \hat{r}_\beta \middle| j \right\rangle$$
$$= \left\langle i \middle| \hat{r}_\alpha \hat{f} \hat{r}_\beta \middle| j \right\rangle - \sum_{kl} \langle i|\hat{r}_\alpha|k\rangle f_{kl} \left\langle l \middle| \hat{r}_\beta \middle| j \right\rangle \tag{D.3.2}$$

Le triple produit d'opérateurs est transformé selon :

$$\hat{r}_\alpha \hat{f} \hat{r}_\beta = \frac{1}{2}(\underbrace{\left[\hat{r}_\alpha, \hat{f}\right] \hat{r}_\beta + \hat{r}_\alpha \left[\hat{f}, \hat{r}_\beta\right]}_{(a)} + \underbrace{\hat{f} \hat{r}_\alpha \hat{r}_\beta + \hat{r}_\alpha \hat{r}_\beta \hat{f}}_{(b)}) \tag{D.3.3}$$

Pour ce qui est du terme (b) dans l'équation (D.3.3), on utilise la complétude de la base et le théorème de Brillouin local pour écrire les éléments de matrice :

$$\left\langle i \middle| \hat{f} \hat{r}_\alpha \hat{r}_\beta \middle| j \right\rangle = \left\langle i \middle| \hat{f} \middle| p \right\rangle \left\langle p \middle| \hat{r}_\alpha \hat{r}_\beta \middle| j \right\rangle = f_{ik} \left\langle k \middle| \hat{r}_\alpha \hat{r}_\beta \middle| j \right\rangle$$

et :

$$\left\langle i \middle| \hat{r}_\alpha \hat{r}_\beta \hat{f} \middle| j \right\rangle = \left\langle i \middle| \hat{r}_\alpha \hat{r}_\beta \middle| k \right\rangle f_{kj} \tag{D.3.4}$$

D'une manière générale (avec un opérateur d'échange non local), le commutateur de l'opérateur de position avec la fockienne trouvée dans le terme (a) est :

$$\left[\hat{r}_\alpha, \hat{f}\right] = \frac{1}{2}\left[\hat{r}_\alpha, \hat{\nabla}^2\right] + \left[\hat{r}_\alpha, \hat{K}\right] = \frac{1}{2}\nabla_\alpha + \left[\hat{r}_\alpha, \hat{K}\right], \tag{D.3.5}$$

c'est-à-dire que les contributions des deux commutateurs dans l'équation (D.3.3), (a), sont :

$$(a) \qquad \frac{1}{2}\left(\nabla_\alpha \hat{r}_\beta - \hat{r}_\beta \nabla_\alpha\right) + \left(\left[\hat{r}_\alpha, \hat{K}\right] \hat{r}_\beta - \hat{r}_\alpha \left[\hat{r}_\beta, \hat{K}\right]\right) \tag{D.3.6}$$

Si l'on utilise la définition de l'échange non local :

$$\hat{K} = \sum_k^{\mathrm{LMO}} \int d\mathbf{r}' \; \phi_k^*(\mathbf{r}') w(\mathbf{r}, \mathbf{r}') \hat{P} \phi_k(\mathbf{r}'), \tag{D.3.7}$$

on écrit les éléments de matrice suivant :

$$\left\langle i \middle| \hat{r}_\alpha \hat{K} \middle| j \right\rangle = \sum_k^{\mathrm{LMO}} \iint d\mathbf{r} d\mathbf{r}' \phi_i^*(\mathbf{r}) \phi_k^*(\mathbf{r}') r_\alpha w(\mathbf{r}, \mathbf{r}') \phi_k(\mathbf{r}) \phi_j(\mathbf{r}')$$
$$= \sum_k^{\mathrm{LMO}} \langle ik | r_\alpha w(\mathbf{r}, \mathbf{r}') | kj \rangle$$

et :

$$\left\langle i \middle| \hat{K} \hat{r}_\alpha \middle| j \right\rangle = \sum_k^{\mathrm{LMO}} \langle ik | r'_\alpha w(\mathbf{r}, \mathbf{r}') | kj \rangle \tag{D.3.8}$$

Muni de la relation canonique $\left\langle i\middle|\hat{\nabla}_\alpha \hat{r}_\beta - \hat{r}_\alpha \hat{\nabla}_\beta\middle|j\right\rangle = \delta_{\alpha\beta}\delta_{ij}$, on exprime l'élément de matrice de (a) (équation (D.3.6)) sous la forme simple :

$$\frac{1}{2}\delta_{\alpha\beta}\delta_{ij} + \sum_{k}^{\mathrm{LMO}} \left\langle ik\middle| w(\mathbf{r},\mathbf{r}')\left(\left(\hat{r}_\alpha - \hat{r}'_\alpha\right)\hat{r}'_\beta - \hat{r}_\alpha\left(\hat{r}_\beta - \hat{r}'_\beta\right)\right)\middle|kj\right\rangle \tag{D.3.9}$$

On obtient donc finalement :

$$\begin{aligned}\left\langle i_\alpha\middle|\hat{f}\middle|j_\beta\right\rangle = &\frac{1}{4}\delta_{ij}\delta_{\alpha\beta} + \frac{1}{2}\sum_{k}^{\mathrm{LMO}} \left\langle ik\middle| w(\mathbf{r},\mathbf{r}')\left(\left(\hat{r}_\alpha - \hat{r}'_\alpha\right)\hat{r}'_\beta - \hat{r}_\alpha\left(\hat{r}_\beta - \hat{r}'_\beta\right)\right)\middle|kj\right\rangle \\ &+\frac{1}{2}\sum_{k}\left(f_{ik}\left\langle k\middle|\hat{r}_\alpha\hat{r}_\beta\middle|j\right\rangle + \left\langle i\middle|\hat{r}_\alpha\hat{r}_\beta\middle|k\right\rangle f_{kj}\right) - \sum_{kl}\langle i|\hat{r}_\alpha|k\rangle f_{kl}\left\langle l\middle|\hat{r}_\beta\middle|j\right\rangle\end{aligned} \tag{D.3.10}$$

On peut négliger le commutateur d'échange dans un premier temps. Une bonne manière de la traiter pourrait être d'utiliser la même approximation multipolaire que pour les intégrales bi-électroniques.

# Annexe E

# Dérivations autour de l'approximation EED

Ici on montre les détails des dérivations liées aux développements de l'approximation EED. On remontre notamment, suivant deux points de vue complémentaires, le lien qui existe entre le commutateur de l'hamiltonien avec l'opérateur densité et la densité de courant. Une discussion courte permet de faire le pont entre cette expression du commutateur et celle qui est vue dans les développements faits par d'autres dans l'espace réciproque. Finalement, on montre que les numérateurs qui émergent dans les procédures EED sont bien des règles de somme.

## E.1 Théorème Hyperviriel

Pour montrer la version du théorème hyperviriel que l'on utilise ici, il suffit d'appliquer les définitions de $n_\alpha(\mathbf{r})$ et $\Omega_\alpha$ :

$$\begin{aligned} n^*_\alpha(\mathbf{r})\Omega_\alpha &= \langle\alpha|\,\hat{n}(\mathbf{r})\,(E_\alpha - E_0)\,|0\rangle \\ &= \langle\alpha|\,\hat{H}\hat{n}(\mathbf{r}) - \hat{n}(\mathbf{r})\hat{H}\,|0\rangle \\ &= \langle\alpha|\left[\hat{H}, \hat{n}(\mathbf{r})\right]|0\rangle, \end{aligned} \tag{E.1.1}$$

là où :

$$\begin{aligned} -n_\alpha(\mathbf{r})\Omega_\alpha &= \langle 0|\,\hat{n}(\mathbf{r})\,(E_0 - E_\alpha)\,|\alpha\rangle \\ &= \langle 0|\,\hat{H}\hat{n}(\mathbf{r}) - \hat{n}(\mathbf{r})\hat{H}\,|\alpha\rangle \\ &= \langle 0|\left[\hat{H}, \hat{n}(\mathbf{r})\right]|\alpha\rangle \end{aligned} \tag{E.1.2}$$

## E.2 Lien entre commutateur et densité de courant

On voit émerger, équation (5.3.5), le commutateur $\left[\hat{H}, \hat{n}(\mathbf{r}_1)\right]$. Les opérateurs de potentiel commutent avec la densité, ainsi seule la partie énergie cinétique de l'hamiltonien contribue. On écrit :

$$\left[\hat{H}, \hat{n}(\mathbf{r}_1)\right] = -\frac{1}{2} \sum_i \sum_j \left[\nabla_{\mathbf{r}_i} \cdot \nabla_{\mathbf{r}_i}, \delta(\mathbf{r}_1 - \mathbf{r}_j)\right] \tag{E.2.1}$$

On utilise dans la suite l'identité des commutateurs $[AB, C] = [A, C]\, B + A\, [B, C]$, c'est-à-dire :

$$\left[\nabla_{\mathbf{r}_i} \cdot \nabla_{\mathbf{r}_i}, \delta(\mathbf{r}_1 - \mathbf{r}_j)\right] = \left[\nabla_{\mathbf{r}_i}, \delta(\mathbf{r}_1 - \mathbf{r}_j)\right] \cdot \nabla_{\mathbf{r}_i} + \nabla_{\mathbf{r}_i} \cdot \left[\nabla_{\mathbf{r}_i}, \delta(\mathbf{r}_1 - \mathbf{r}_j)\right] \tag{E.2.2}$$

Il est facile alors de montrer que le terme bi-électronique ($i \neq j$) ne contribue pas : le commutateur que l'on trouve ici à droite, appliqué à une fonction test, produit :

$$\left[\nabla_{\mathbf{r}_i}, \delta(\mathbf{r}_1 - \mathbf{r}_j)\right] f(\mathbf{r}_i) = \nabla_{\mathbf{r}_i} \left(\delta(\mathbf{r}_1 - \mathbf{r}_j) f(\mathbf{r}_i)\right) - \delta(\mathbf{r}_1 - \mathbf{r}_j) \nabla_{\mathbf{r}_i} f(\mathbf{r}_i) = 0 \tag{E.2.3}$$

Le même commutateur dans le cas mono-électronique($i = j$) donne en revanche :

$$\begin{aligned} [\nabla_{\mathbf{r}_i}, \delta(\mathbf{r}_1 - \mathbf{r}_i)]\, f(\mathbf{r}_i) &= (\nabla_{\mathbf{r}_i} \delta(\mathbf{r}_1 - \mathbf{r}_i))\, f(\mathbf{r}_i) + \delta(\mathbf{r}_1 - \mathbf{r}_i)\, (\nabla_{\mathbf{r}_i} f(\mathbf{r}_i)) - \delta(\mathbf{r}_1 - \mathbf{r}_i)\, (\nabla_{\mathbf{r}_i} f(\mathbf{r}_i)) \\ &= (\nabla_{\mathbf{r}_i} \delta(\mathbf{r}_1 - \mathbf{r}_i))\, f(\mathbf{r}_i) \end{aligned} \tag{E.2.4}$$

Une *chain-rule* permet d'échanger les variables de dérivation, selon $\nabla_{\mathbf{r}_i} \delta(\mathbf{r}_1 - \mathbf{r}_i) = -\nabla_{\mathbf{r}_1} \delta(\mathbf{r}_1 - \mathbf{r}_i)$, pour écrire la contribution mono-électronique de l'équation (E.2.1) :

$$\begin{aligned} -\frac{1}{2} \sum_i [\nabla_{\mathbf{r}_i} \cdot \nabla_{\mathbf{r}_i}, \delta(\mathbf{r}_1 - \mathbf{r}_i)] &= -\frac{1}{2} \sum_i (-\nabla_{\mathbf{r}_1} \delta(\mathbf{r}_1 - \mathbf{r}_i) \cdot \nabla_{\mathbf{r}_i} - \nabla_{\mathbf{r}_i} \cdot \nabla_{\mathbf{r}_1} \delta(\mathbf{r}_1 - \mathbf{r}_i)) \\ &= \frac{1}{2} \nabla_{\mathbf{r}_1} \cdot \sum_i \{\delta(\mathbf{r}_1 - \mathbf{r}_i) \cdot \nabla_{\mathbf{r}_i} + \nabla_{\mathbf{r}_i} \delta(\mathbf{r}_1 - \mathbf{r}_i)\}, \end{aligned} \tag{E.2.5}$$

ou l'on reconnaît en effet les éléments de la densité de courant $\hat{\mathbf{j}}(\mathbf{r}_1) = -\frac{i}{2} \sum_i \{\delta(\mathbf{r}_1 - \mathbf{r}_i) \cdot \nabla_{\mathbf{r}_i} + \nabla_{\mathbf{r}_i} \delta(\mathbf{r}_1 - \mathbf{r}_i)\}$, ainsi :

$$\left[\hat{H}, \hat{n}(\mathbf{r}_1)\right] = i \nabla_{\mathbf{r}_1} \cdot \hat{\mathbf{j}}(\mathbf{r}_1) \tag{E.2.6}$$

Cette équation peut être considérée comme la forme opérateur de l'équation de continuité de mécanique quantique.

## E.3 Lien entre commutateur et densité de courant : point de vue des éléments de matrices

On met au jour la relation entre le commutateur et la densité de courant d'une autre manière : en s'intéressant à leurs éléments de matrices. Cette démonstration alternative est complémentaire et permet de comparer ce résultat au résultat dérivé dans l'espace réciproque par Berger *et. al.* (voir section E.4).

### E.3.1 Éléments de matrice du commutateur

On va utiliser dans la suite les relations suivantes concernant la fonction Dirac et ses dérivées (dans les formules générales de gauche, les $x_i$ sont les zéros de la fonction $g$) :

$$
\begin{aligned}
&\int dx\ f(x)\delta(g(x)) = \sum_i \frac{f(x_i)}{|g'(x_i)|} && \text{en part. :} \int d\mathbf{r}_j\ f(\mathbf{r}_j)\delta(\mathbf{r}_1 - \mathbf{r}_j) = f(\mathbf{r}_1) \\
&\int dx\ f(x)\delta^{(n)}(g(x)) = \sum_i \frac{(-1)^n f^{(n)}(x_i)}{|g'(x_i)|} && \text{en part. :} \int d\mathbf{r}_j\ f(\mathbf{r}_j)\delta^{(n)}(\mathbf{r}_1 - \mathbf{r}_j) = (-1)^n f^{(n)}(\mathbf{r}_1)
\end{aligned}
\tag{E.3.1}
$$

On cherche à écrire les éléments de matrices de la partie mono-électronique du commutateur de l'opérateur énergie cinétique avec l'opérateur densité, c'est-à-dire que l'on cherche à expliciter les objets :

$$
C_j = -\frac{1}{2}\int d\mathbf{r}_j\ \phi_p(\mathbf{r}_j)\left[\nabla^2_{\mathbf{r}_j}, \delta(\mathbf{r}_1 - \mathbf{r}_j)\right]\phi_q(\mathbf{r}_j) \tag{E.3.2}
$$

En développant le commutateur et en utilisant la dérivée d'un produit, on obtient :

$$
\begin{aligned}
C_j &= -\frac{1}{2}\int d\mathbf{r}_j\ \phi_p(\mathbf{r}_j)\nabla^2_{\mathbf{r}_j}\delta(\mathbf{r}_1 - \mathbf{r}_j)\phi_q(\mathbf{r}_j) + \frac{1}{2}\int d\mathbf{r}_j\ \phi_p(\mathbf{r}_j)\delta(\mathbf{r}_1 - \mathbf{r}_j)\nabla^2_{\mathbf{r}_j}\phi_q(\mathbf{r}_j) \\
&= \underbrace{-\frac{1}{2}\int d\mathbf{r}_j\ \phi_p(\mathbf{r}_j)\left(\nabla^2_{\mathbf{r}_j}\delta(\mathbf{r}_1 - \mathbf{r}_j)\right)\phi_q(\mathbf{r}_j)}_{(1)} \underbrace{-\frac{1}{2}2\int d\mathbf{r}_j\ \phi_p(\mathbf{r}_j)\left(\nabla_{\mathbf{r}_j}\delta(\mathbf{r}_1 - \mathbf{r}_j)\right)\left(\nabla_{\mathbf{r}_j}\phi_q(\mathbf{r}_j)\right)}_{(2)} \\
&\quad -\frac{1}{2}\int d\mathbf{r}_j\ \phi_p(\mathbf{r}_j)\delta(\mathbf{r}_1 - \mathbf{r}_j)\left(\nabla^2_{\mathbf{r}_j}\phi_q(\mathbf{r}_j)\right) + \frac{1}{2}\int d\mathbf{r}_j\ \phi_p(\mathbf{r}_j)\delta(\mathbf{r}_1 - \mathbf{r}_j)\left(\nabla^2_{\mathbf{r}_j}\phi_q(\mathbf{r}_j)\right)
\end{aligned}
\tag{E.3.3}
$$

Les deux derniers termes s'annulent ; vues les relations (E.3.1) concernant les dérivées de la fonction de Dirac, le premier terme donne :

$$
\begin{aligned}
(1) &= -\frac{1}{2}(-1)^2\left\{\nabla^2_{\mathbf{r}_j}\left(\phi_p(\mathbf{r}_j)\phi_q(\mathbf{r}_j)\right)\right\}\Big|_{\mathbf{r}_1} \\
&= \left\{-\frac{1}{2}\left(\nabla^2_{\mathbf{r}_j}\phi_p(\mathbf{r}_j)\right)\phi_q(\mathbf{r}_j) - \frac{1}{2}2\left(\nabla_{\mathbf{r}_j}\phi_p(\mathbf{r}_j)\right)\left(\nabla_{\mathbf{r}_j}\phi_q(\mathbf{r}_j)\right) - \frac{1}{2}\phi_p(\mathbf{r}_j)\left(\nabla^2_{\mathbf{r}_j}\phi_q(\mathbf{r}_j)\right)\right\}\Bigg|_{\mathbf{r}_1},
\end{aligned}
\tag{E.3.4}
$$

et le deuxième :

$$(2) = -\frac{1}{2}2(-1)\left\{\nabla_{\mathbf{r}_j}\left(\phi_p(\mathbf{r}_j)\left(\nabla_{\mathbf{r}_j}\phi_q(\mathbf{r}_j)\right)\right)\right\}\Big|_{\mathbf{r}_1}$$
$$= \left\{\left(\nabla_{\mathbf{r}_j}\phi_p(\mathbf{r}_j)\right)\left(\nabla_{\mathbf{r}_j}\phi_q(\mathbf{r}_j)\right) + \phi_p(\mathbf{r}_j)\left(\nabla^2_{\mathbf{r}_j}\phi_q(\mathbf{r}_j)\right)\right\}\Big|_{\mathbf{r}_1}, \quad \text{(E.3.5)}$$

si bien que les éléments de matrice du commutateur s'écrivent :

$$\left\langle p\left|\left[\hat{H}, n(\hat{\mathbf{r}}_1)\right]\right|q\right\rangle = \frac{1}{2}\left\{\phi_p(\mathbf{r}_1)\left(\nabla^2_{\mathbf{r}_1}\phi_q(\mathbf{r}_1)\right) - \left(\nabla^2_{\mathbf{r}_1}\phi_p(\mathbf{r}_1)\right)\phi_q(\mathbf{r}_1)\right\} \quad \text{(E.3.6)}$$

### E.3.2 Éléments de matrice de la densité de de courant

Cherchons d'abord à retrouver les éléments de matrice de l'opérateur densité de courant :

$$\hat{\mathbf{j}}(\mathbf{r}_1) = -\frac{i}{2}\sum_j\left(\delta(\mathbf{r}_1 - \mathbf{r}_j)\nabla_{\mathbf{r}_j} + \nabla_{\mathbf{r}_j}\delta(\mathbf{r}_1 - \mathbf{r}_j)\right), \quad \text{(E.3.7)}$$

c'est-à-dire cherchons à travailler les objets :

$$J_j = -\frac{i}{2}\int d\mathbf{r}_j\ \phi_p(\mathbf{r}_j)\left(\delta(\mathbf{r}_1 - \mathbf{r}_j)\nabla_{\mathbf{r}_j} + \nabla_{\mathbf{r}_j}\delta(\mathbf{r}_1 - \mathbf{r}_j)\right)\phi_q(\mathbf{r}_j) \quad \text{(E.3.8)}$$

De la même manière que précédemment, le développement de dérivées d'un produit et l'utilisation des relations (E.3.1) concernant les dérivées de la fonction de Dirac donnent :

$$J_j = -\frac{i}{2}\int d\mathbf{r}_j\ \phi_p(\mathbf{r}_j)\delta(\mathbf{r}_1 - \mathbf{r}_j)\left(\nabla_{\mathbf{r}_j}\phi_q(\mathbf{r}_j)\right)$$
$$- \frac{i}{2}\int d\mathbf{r}_j\ \phi_p(\mathbf{r}_j)\left(\nabla_{\mathbf{r}_j}\delta(\mathbf{r}_1 - \mathbf{r}_j)\right)\phi_q(\mathbf{r}_j) - \frac{i}{2}\int d\mathbf{r}_j\ \phi_p(\mathbf{r}_j)\delta(\mathbf{r}_1 - \mathbf{r}_j)\left(\nabla_{\mathbf{r}_j}\phi_q(\mathbf{r}_j)\right)$$
$$= -\frac{i}{2}2\left\{\phi_p(\mathbf{r}_j)\left(\nabla_{\mathbf{r}_j}\phi_q(\mathbf{r}_j)\right)\right\}\Big|_{\mathbf{r}_1} - \frac{i}{2}(-1)\left\{\nabla_{\mathbf{r}_j}\left(\phi_p(\mathbf{r}_j)\phi_q(\mathbf{r}_j)\right)\right\}\Big|_{\mathbf{r}_1}$$
$$= -i\left\{\phi_p(\mathbf{r}_j)\left(\nabla_{\mathbf{r}_j}\phi_q(\mathbf{r}_j)\right)\right\}\Big|_{\mathbf{r}_1} + \frac{i}{2}\left\{\left(\nabla_{\mathbf{r}_j}\phi_p(\mathbf{r}_j)\right)\phi_q(\mathbf{r}_j)\right\}\Big|_{\mathbf{r}_1} + \frac{i}{2}\left\{\phi_p(\mathbf{r}_j)\left(\nabla_{\mathbf{r}_j}\phi_q(\mathbf{r}_j)\right)\right\}\Big|_{\mathbf{r}_1} \quad \text{(E.3.9)}$$

On retrouve finalement l'expression bien connue :

$$\left\langle p\left|\hat{\mathbf{j}}(\mathbf{r}_1)\right|q\right\rangle = \frac{i}{2}\left\{\left(\nabla_{\mathbf{r}_1}\phi_p(\mathbf{r}_1)\right)\phi_q(\mathbf{r}_1) - \phi_p(\mathbf{r}_1)\left(\nabla_{\mathbf{r}_1}\phi_q(\mathbf{r}_1)\right)\right\} \quad \text{(E.3.10)}$$

Quant à démontrer le lien entre les éléments de matrice du commutateur et de la densité de courant, on voit que :

$$i\nabla_{\mathbf{r}_1}\hat{\mathbf{j}}(\mathbf{r}_1) = \frac{1}{2}\nabla_{\mathbf{r}_1}\cdot\left(\delta(\mathbf{r}_1-\mathbf{r}_j)\nabla_{\mathbf{r}_j} + \nabla_{\mathbf{r}_j}\delta(\mathbf{r}_1-\mathbf{r}_j)\right), \tag{E.3.11}$$

dont les éléments de matrice s'écrivent (tout simplement à partir de l'équation (E.3.10)) :

$$\begin{aligned}
-\frac{1}{2}\nabla_{\mathbf{r}_1}\left\{\left(\nabla_{\mathbf{r}_1}\phi_p(\mathbf{r}_1)\right)\phi_q(\mathbf{r}_1) - \phi_p(\mathbf{r}_1)\left(\nabla_{\mathbf{r}_1}\phi_q(\mathbf{r}_1)\right)\right\} &= -\frac{1}{2}\Big\{\left(\nabla^2_{\mathbf{r}_1}\phi_p(\mathbf{r}_1)\right)\phi_q(\mathbf{r}_1) + \left(\nabla_{\mathbf{r}_1}\phi_p(\mathbf{r}_1)\right)\left(\nabla_{\mathbf{r}_1}\phi_q(\mathbf{r}_1)\right) \\
&\quad - \left(\nabla_{\mathbf{r}_1}\phi_p(\mathbf{r}_1)\right)\left(\nabla_{\mathbf{r}_1}\phi_q(\mathbf{r}_1)\right) - \phi_p(\mathbf{r}_1)\left(\nabla^2_{\mathbf{r}_1}\phi_q(\mathbf{r}_1)\right)\Big\} \\
&= \frac{1}{2}\left\{\phi_p(\mathbf{r}_1)\left(\nabla^2_{\mathbf{r}_1}\phi_q(\mathbf{r}_1)\right) - \left(\nabla^2_{\mathbf{r}_1}\phi_p(\mathbf{r}_1)\right)\phi_q(\mathbf{r}_1)\right\},
\end{aligned} \tag{E.3.12}$$

qui est l'équation (E.3.6).

## E.4 Commentaire sur l'espace réciproque

Au cours de dérivations de l'approximation EED dans l'espace réciproque, Berger *et. al.* [194] sont amenés à écrire une équation similaire à notre équation (5.2.5), et qui s'écrit :

$$\Omega^{nn}(\mathbf{k}_1,\mathbf{k}_2;\omega)\chi(\mathbf{k}_1,\mathbf{k}_2;\omega) = \frac{1}{2}\sum_{v,c} n_v \frac{\tilde{\rho}^*_{cv}(\mathbf{k}_1)\left\langle c\middle|\left[\hat{H}, e^{-i\mathbf{k}_2\cdot\mathbf{r}'}\right]\middle|v\right\rangle + h.c.}{\omega - \Omega_{cv} + i\eta^+}, \tag{E.4.1}$$

où je montre une version partiellement « traduite » vers nos notations. Les sommations sur les états de valence $v$ et de conduction $c$ correspondent à nos sommations sur les états occupés et virtuels. Le commutateur à travailler ici est le même que celui qui a été travaillé dans les sections précédentes : traduire l'équation (E.3.3) en espace réciproque donne (se souvenir que les deux derniers termes s'annulent) :

$$\begin{aligned}
J_j &= -\frac{1}{2}\int d\mathbf{r}_j\ \phi_c(\mathbf{r}_j)\left(\nabla^2_{\mathbf{r}_j}e^{-i\mathbf{k}_2\mathbf{r}_j}\right)\phi_v(\mathbf{r}_j) - \frac{1}{2}2\int d\mathbf{r}_j\ \phi_c(\mathbf{r}_j)\left(\nabla_{\mathbf{r}_j}e^{-i\mathbf{k}_2\mathbf{r}_j}\right)\left(\nabla_{\mathbf{r}_j}\phi_v(\mathbf{r}_j)\right) \\
&= \frac{\mathbf{k}_2^2}{2}\int d\mathbf{r}_j\ \phi_c(\mathbf{r}_j)e^{-i\mathbf{k}_2\mathbf{r}_j}\phi_v(\mathbf{r}_j) + \int d\mathbf{r}_j\ \phi_c(\mathbf{r}_j)e^{-i\mathbf{k}_2\mathbf{r}_j}\left(i\nabla_{\mathbf{r}_j}\phi_v(\mathbf{r}_j)\right)\mathbf{k}_2 \\
&= \frac{\mathbf{k}_2^2}{2}\tilde{\rho}_{cv}(\mathbf{k}_2) + \left\langle c\middle|e^{-i\mathbf{k}_2\mathbf{r}_j}\left(i\nabla_{\mathbf{r}_j}\right)\middle|v\right\rangle\cdot\mathbf{k}_2
\end{aligned} \tag{E.4.2}$$

Ainsi l'équation (E.4.1) s'écrit-elle :

$$\begin{aligned}
\Omega^{nn}(\mathbf{k}_1,\mathbf{k}_2;\omega)\chi(\mathbf{k}_1,\mathbf{k}_2;\omega) &= \frac{1}{2}\left(\frac{\mathbf{k}_2^2}{2}+\frac{\mathbf{k}_1^2}{2}\right)\sum_{v,c} n_v\frac{\tilde{\rho}^*_{cv}(\mathbf{k}_1)\tilde{\rho}_{cv}(\mathbf{k}_2)}{\omega-\Omega_{cv}+i\eta^+} \\
&\quad + \frac{1}{2}\sum_{v,c} n_v\frac{\tilde{\rho}^*_{cv}(\mathbf{k}_1)\left\langle c\middle|e^{-i\mathbf{k}_2\mathbf{r}_j}\left(i\nabla_{\mathbf{r}_j}\right)\middle|v\right\rangle\cdot\mathbf{k}_2 + h.c.}{\omega-\Omega_{cv}+i\eta^+}
\end{aligned} \tag{E.4.3}$$

La division par $\chi$ va faire émerger le terme $Q$ vu équation (7) de la référence [194] (l'apparition de $\mathbf{k}_1^2/2$ est un effet du terme $+h.c.$ dans l'équation (E.4.1)). Cette manipulation n'est possible que dans le contexte de l'espace réciproque, où la dérivé seconde de l'exponentielle laisse une expression de l'élément de matrice densité intouchée par une dérivation, ce qui permet de reformer la fonction $\chi$. Il semble (voir sections précédentes) que pour former la densité de courant l'on ait besoin des *deux* termes de l'équation (E.4.2). Ainsi l'objet appelé $\tilde{j}_{cv}(\mathbf{k}_2)$ dans l'équation (9) de la référence [194] n'est *pas* la densité de courant.

## E.5 Règles de somme

On cherche ici à confirmer les liens entre les numérateurs des équations (5.2.3), (5.2.6) et (5.2.8), c'est-à-dire entre $\sum_{\alpha\neq 0} n_\alpha(\mathbf{r}_1)n_\alpha(\mathbf{r}_2)$, $\sum_{\alpha\neq 0} n_\alpha(\mathbf{r}_1)n_\alpha(\mathbf{r}_2)\Omega_\alpha$, et $\sum_{\alpha\neq 0} n_\alpha(\mathbf{r}_1)n_\alpha(\mathbf{r}_2)\Omega_\alpha\Omega_\alpha$ et les règles de sommes, que l'on peut définir de la manière suivante (voir [247] et surtout [248]) :

$$\tilde{S}_k(\mathbf{r}_1,\mathbf{r}_2) = \sum_\alpha \Omega_\alpha^{k+1} n_\alpha(\mathbf{r}_1)n_\alpha(\mathbf{r}_2), \tag{E.5.1}$$

et dériver (voir toujours référence [248]) :

$$\begin{aligned} \tilde{S}_{2k-1}(\mathbf{r}_1,\mathbf{r}_2) &= (-1)^k \left\langle \hat{n}_{(k)}(\mathbf{r}_1)\hat{n}_{(k)}(\mathbf{r}_2)\right\rangle \\ \tilde{S}_{2k}(\mathbf{r}_1,\mathbf{r}_2) &= \frac{1}{2}(-1)^k \left\langle \hat{n}_{(k)}(\mathbf{r}_1)\hat{n}_{(k+1)}(\mathbf{r}_2) - \hat{n}_{(k+1)}(\mathbf{r}_1)\hat{n}_{(k)}(\mathbf{r}_2)\right\rangle, \end{aligned} \tag{E.5.2}$$

où $\hat{n}_{(k)}(\mathbf{r}_1)$ est un commutateur défini de la manière suivante :

$$\begin{cases} \hat{n}_{(0)}(\mathbf{r}_1) = \hat{n}(\mathbf{r}_1) \\ \hat{n}_{(k)}(\mathbf{r}_1) = \left[\hat{H}, \hat{n}_{(k-1)}(\mathbf{r}_1)\right] \end{cases} \quad \text{c'est-à-dire :} \quad \hat{n}_{(k)}(\mathbf{r}_1) = \left[\hat{H},\left[\hat{H},\left[\ldots,\left[\hat{H},\hat{n}(\mathbf{r}_1)\right]\right]\right]\right] \tag{E.5.3}$$

Appliquer ces formules pour nos besoins donne :

$$\tilde{S}_{-1}(\mathbf{r}_1,\mathbf{r}_2) = \langle \hat{n}(\mathbf{r}_1)\hat{n}(\mathbf{r}_2)\rangle \tag{E.5.4}$$

$$\tilde{S}_0(\mathbf{r}_1,\mathbf{r}_2) = \frac{1}{2}\left\langle \hat{n}(\mathbf{r}_1)\left[\hat{H},\hat{n}(\mathbf{r}_2)\right] - \left[\hat{H},\hat{n}(\mathbf{r}_1)\right]\hat{n}(\mathbf{r}_2)\right\rangle \tag{E.5.5}$$

$$\tilde{S}_1(\mathbf{r}_1,\mathbf{r}_2) = -\left\langle \left[\hat{H},\hat{n}(\mathbf{r}_1)\right]\left[\hat{H},\hat{n}(\mathbf{r}_2)\right]\right\rangle \tag{E.5.6}$$

Ce qui a été fait ici est une application *stricte* des règles de somme telles qu'on les trouve dans les références citées plus haut. Dans notre cas de figure, la sommation n'inclut pas l'état fondamental. On définit donc en fait :

$$
\begin{aligned}
S_{-1}(\mathbf{r}_1,\mathbf{r}_2) &= \langle \hat{n}(\mathbf{r}_1)\hat{n}(\mathbf{r}_2)\rangle - n_0(\mathbf{r}_1)n_0(\mathbf{r}_2) = \langle \delta\hat{n}(\mathbf{r}_1)\delta\hat{n}(\mathbf{r}_2)\rangle \\
S_0(\mathbf{r}_1,\mathbf{r}_2) &= \frac{1}{2}\left\langle \hat{n}(\mathbf{r}_1)\left[\hat{H},\hat{n}(\mathbf{r}_2)\right] - \left[\hat{H},\hat{n}(\mathbf{r}_1)\right]\hat{n}(\mathbf{r}_2)\right\rangle \\
S_1(\mathbf{r}_1,\mathbf{r}_2) &= -\left\langle\left[\hat{H},\hat{n}(\mathbf{r}_1)\right]\left[\hat{H},\hat{n}(\mathbf{r}_2)\right]\right\rangle,
\end{aligned}
\tag{E.5.7}
$$

où l'on utilise $\left\langle\left[\hat{H},\hat{n}(\mathbf{r}_1)\right]\right\rangle = 0$. Passer de ces relations aux équations (5.3.2), (5.3.6) et (5.3.9) est aisé, et montre que les numérateurs qui émergent dans une procédure EED sont bien des règles de somme.

# Annexe F

# Détails des équations du gradient RSH-RPA

Dans cette Annexe sont explicitées les équations principales impliquées dans la dérivation du gradient de l'énergie RSH-RPA. Mise à part la dernière section, toute l'Annexe consiste à dériver les conditions stationnaires de tous les éléments du Lagrangien par rapport aux coefficients orbitalaires. En particulier, la dérivation des termes liés à la fockienne courte- et longue-portée font tout l'intérêt de notre formulation, qui montre naturellement un parallèle remarquable entre les deux dérivations.

Au cours de la dérivation des conditions stationnaires des termes liés à la fockienne courte-portée par rapport aux coefficients orbitalaires, des notations inédites sont introduites, qui permettent un traitement très général des termes dérivant de la fonctionnelle d'échange-corrélation.

L'objet de ces sections est principalement de dériver les conditions stationnaires du Lagrangien de l'équation (6.3.7) par rapport aux coefficients orbitalaires. Une modification des orbitales moléculaires peut s'exprimer comme une rotation des coefficients $\mathbf{C}$ :

$$|p\rangle = C^{\dagger}_{p\mu}\,|\mu\rangle \qquad \text{devient au premier ordre :} \qquad |\overset{*}{p}\rangle = \left(\mathbf{1} + \mathbf{V}^{\dagger}\right)_{pt} C^{\dagger}_{t\mu}\,|\mu\rangle \tag{F.0.1}$$

Autrement dit, les coefficients subissent la transformation :

$$\overset{*}{\mathbf{C}} \leftarrow \mathbf{C} + \mathbf{C}\mathbf{V} \tag{F.0.2}$$

On cherche donc à obtenir ici :

$$\left.\frac{\partial \mathcal{X}}{\partial \mathbf{V}}\right|_{\mathbf{V}=\mathbf{0}}, \tag{F.0.3}$$

pour tous les termes $\mathcal{X}$ du Lagrangien. Pour certains de ces termes, on dérive directement l'équation (F.0.3) (voir section F.3). Dans d'autres cas il est plus simple de considérer une expansion au premier ordre en $\mathbf{V}$ :

$$\mathcal{X}(\overset{*}{\mathbf{C}}) \underset{\mathbf{V}\to 0}{=} \mathcal{X}(\mathbf{C}) + \left.\frac{d\mathcal{X}}{d\mathbf{V}}\right|_{\mathbf{V}\to\mathbf{0}} \mathbf{V} = \mathcal{X}^{(0)} + \mathcal{X}^{(1)} \tag{F.0.4}$$

On atteint alors la dérivée de ces termes par rapport à une modification des coefficients en considérant $\mathcal{X}^{(1)}$ (voir sections F.1 et F.2).

J'attire l'attention du lecteur sur le fait que, particulièrement dans cette Annexe, j'utilise extensivement la convention de Einstein concernant les indices implicitement sommés.

## F.1 Intégrales bi-électroniques

Considérons une intégrale bi-électronique quelconque $\langle pq|rs\rangle$. On cherche à écrire une expression des intégrales bi-électroniques calculées avec des orbitales ayant subi une rotation $\overset{*}{\mathbf{C}} = \mathbf{C} + \mathbf{CV}$. Le développement en série par rapport à $\mathbf{V}$ de $\langle p\overset{*}{q}|rs\rangle$ est :

$$\begin{aligned}\langle \overset{*}{pq}|rs\rangle &= \overset{*}{C}_{\mu p}\overset{*}{C}_{\nu q}\langle\mu\nu|\rho\sigma\rangle \overset{*}{C}^{\dagger}_{rp}\overset{*}{C}^{\dagger}_{s\sigma}\\ &= \left(C_{\mu p} + C_{\mu t}V_{tp}\right)\left(C_{\nu q} + C_{\nu t}V_{tq}\right)\langle\mu\nu|\rho\sigma\rangle\left(C^{\dagger}_{r\rho} + V^{\dagger}_{rt}C^{\dagger}_{t\rho}\right)\left(C^{\dagger}_{s\sigma} + V^{\dagger}_{st}C^{\dagger}_{t\sigma}\right)\\ &= \langle pq|rs\rangle + V_{tp}\langle tq|rs\rangle + V_{tq}\langle pt|rs\rangle\\ &\quad + V_{tr}\langle pq|ts\rangle + V_{ts}\langle pq|rt\rangle + O(V^2)\end{aligned} \tag{F.1.1}$$

On travaille dans la suite avec les intégrales $(\mathbf{K})_{ia,jb} = \langle ij|ab\rangle$ ; le raisonnement est inchangé pour les intégrales de type $\mathbf{K}'$ et $\mathbf{J}$. Les intégrales bi-électroniques apparaissent toujours dans le contexte d'une trace avec un autre objet, c'est-à-dire dans des sommations du type $X_{ia,jb}K_{jb,ia}$ : nous travaillerons directement sur l'expansion de ces traces. Les termes du premier ordre des traces sont :

$$\begin{aligned}\left(X_{ia,jb}\overset{*}{K}_{jb,ia}\right)^{(1)} &= X_{ia,jb}V_{tj}\langle ti|ba\rangle + X_{ia,jb}V_{ti}\langle jt|ba\rangle\\ &\quad + X_{ia,jb}V_{tb}\langle ji|ta\rangle + X_{ia,jb}V_{ta}\langle ji|bt\rangle\\ &= X_{ia,jb}V_{tj}\langle ti|ba\rangle + X_{jb,ia}V_{tj}\langle it|ab\rangle\\ &\quad + X_{ia,jb}V_{tb}\langle ji|ta\rangle + X_{jb,ia}V_{tb}\langle ij|at\rangle\\ &= \left(\mathbf{X} + \mathbf{X}^{\dagger}\right)_{ia,jb}V_{tj}\langle ti|ba\rangle\\ &\quad + \left(\mathbf{X} + \mathbf{X}^{\dagger}\right)_{ia,jb}V_{tb}\langle ji|ta\rangle,\end{aligned} \tag{F.1.2}$$

où c'est en échangeant les indices de sommations $ia$ et $jb$ des deuxième et quatrième termes que l'on passe de la première à la deuxième ligne, et où ce sont les propriétés de symétrie des intégrales bi-électroniques qui permettent de passer de la deuxième à la troisième ligne. Un effort peut être fait pour réécrire la dernière équation avec la structure des matrices $\mathbf{K}$, en se souvenant que l'indice $t$ court sur les orbitales occupées ($k$) et virtuelles ($c$) :

$$\left(X_{ia,jb}\overset{*}{K}_{jb,ia}\right)^{(1)} = V^{\dagger}_{jk}K_{kb,ia}\left(\mathbf{X}+\mathbf{X}^{\dagger}\right)_{ia,jb} + V^{\dagger}_{jc}\overline{K}_{cb,ia}\left(\mathbf{X}+\mathbf{X}^{\dagger}\right)_{ia,jb} + V^{\dagger}_{bk}\overline{K}_{jk,ia}\left(\mathbf{X}+\mathbf{X}^{\dagger}\right)_{ia,jb} + V^{\dagger}_{bc}K_{jc,ia}\left(\mathbf{X}+\mathbf{X}^{\dagger}\right)_{ia,jb}$$
$$= V^{\dagger}_{jk}\left\{\mathbf{K},\mathbf{X}+\mathbf{X}^{\dagger}\right\}_{kj} + V^{\dagger}_{jc}\left\{\overline{\mathbf{K}},\mathbf{X}+\mathbf{X}^{\dagger}\right\}_{cj} + V^{\dagger}_{bk}\left\{\overline{\mathbf{K}},\mathbf{X}+\mathbf{X}^{\dagger}\right\}_{kb} + V^{\dagger}_{bc}\left\{\mathbf{K},\mathbf{X}+\mathbf{X}^{\dagger}\right\}_{cb}, \tag{F.1.3}$$

où j'utilise la notation $\{\sqcup,\sqcup\}$, vue équation (6.3.6), d'une sorte de « trace incomplète » qui dépend encore de deux indices parmi ceux qui composent les super-indices de deux matrices $\mathbf{X}$ et $\mathbf{Y}$ :

$$\begin{aligned}\{\mathbf{X},\mathbf{Y}\}_{ij} &= X_{\mathbf{i}a,kc}Y_{kc,\mathbf{j}a}\\ \{\mathbf{X},\mathbf{Y}\}_{ab} &= X_{i\mathbf{a},kc}Y_{kc,i\mathbf{b}}\end{aligned} \tag{F.1.4}$$

Remarquer que la notation $\overline{\mathbf{K}}$ est utilisée pour signifier un terme qui provient de l'expansion de $\mathbf{K}$, mais ne respecte plus sa structure de type *ia*, *jb* (on a, aussi : $\overline{\mathbf{K}'}$ et $\overline{\mathbf{J}}$).

En supposant que $\mathbf{X}$ est hermitien, on obtient l'expression finale du terme de premier ordre de l'expansion de $\left\langle\mathbf{X}\overset{*}{\mathbf{K}}\right\rangle$ :

$$\left\langle\mathbf{X}\overset{*}{\mathbf{K}}\right\rangle^{(1)} = 2V^{\dagger}_{jk}\{\mathbf{K},\mathbf{X}\}_{kj} + 2V^{\dagger}_{jc}\left\{\overline{\mathbf{K}},\mathbf{X}\right\}_{cj} + 2V^{\dagger}_{bk}\left\{\overline{\mathbf{K}},\mathbf{X}\right\}_{kb} + 2V^{\dagger}_{bc}\{\mathbf{K},\mathbf{X}\}_{cb} \tag{F.1.5}$$

## F.2 Fockienne longue-portée (et double comptage)

On veut dériver l'expression de

$$\mathrm{LR}^{(1)} = \left(\left\langle\left(\mathbf{d}^{(0)}+\mathbf{d}^{(2)}+\mathbf{z}\right)\overset{*}{\mathbf{f}}^{\mathrm{LR}}\right\rangle - \tfrac{1}{2}\left\langle\mathbf{d}^{(0)}\overset{*}{\mathbf{g}}^{\mathrm{LR}}\left[\mathbf{d}^{(0)}\right]\right\rangle\right)^{(1)} \tag{F.2.1}$$

On a donc besoin des expansions de $\overset{*}{\mathbf{h}}$ et $\overset{*}{\mathbf{g}}^{\mathrm{LR}}\left[\mathbf{d}^{(0)}\right]$ dans le contexte d'une trace avec un objet $\mathbf{X}$. Le terme de premier ordre d'une expansion de $\left\langle\mathbf{X}\overset{*}{\mathbf{h}}\right\rangle$ est simplement :

$$\begin{aligned}\left(X_{qp}\overset{*}{h}_{pq}\right)^{(1)} &= \left(X_{qp}\overset{*}{C}_{\mu p}h_{\mu\nu}\overset{*}{C}^{\dagger}{}_{q\nu}\right)^{(1)}\\ &= X_{qp}C_{\mu t}V_{tp}h_{\mu\nu}C^{\dagger}_{q\nu} + X_{qp}C_{\mu p}h_{\mu\nu}V^{\dagger}_{qt}C^{\dagger}_{t\nu}\\ &= V^{\dagger}_{pt}(\mathbf{hX})_{tp} + V^{\dagger}_{qt}\left(\mathbf{hX}^{\dagger}\right)_{tq},\end{aligned} \tag{F.2.2}$$

et une expansion du terme à deux électrons $\left\langle \mathbf{X}\overset{*}{\mathbf{g}}^{\mathrm{LR}}\left[\mathbf{d}^{(0)}\right]\right\rangle$ donne :

$$\begin{aligned}
X_{qp}\overset{*}{g}^{\mathrm{LR}}\left[\mathbf{d}^{(0)}\right]_{pq} &= X_{qp}d^{(0)}_{rs}\left[\langle p\overset{*}{r}|qs\rangle - \frac{1}{2}\langle p\overset{*}{s}|rq\rangle\right] \\
&= X_{qp}d^{(0)}_{rs}\overset{*}{C}_{\mu p}\overset{*}{C}_{\nu q}\left[\langle\mu\rho|\nu\sigma\rangle - \frac{1}{2}\langle\mu\sigma|\rho\nu\rangle\right]\overset{*}{C}^{\dagger}_{r\rho}\overset{*}{C}^{\dagger}_{s\sigma} \\
&= X_{qp}g^{\mathrm{LR}}\left[\mathbf{d}^{(0)}\right]_{pq} \\
&\quad + X_{qp}d^{(0)}_{rs}V_{tp}\left[\langle tr|qs\rangle - \frac{1}{2}\langle ts|rq\rangle\right] \\
&\quad + X_{qp}d^{(0)}_{rs}V_{tq}\left[\langle pr|ts\rangle - \frac{1}{2}\langle ps|rt\rangle\right] \\
&\quad + X_{qp}d^{(0)}_{rs}V_{tr}\left[\langle pt|qs\rangle - \frac{1}{2}\langle ps|tq\rangle\right] \\
&\quad + X_{qp}d^{(0)}_{rs}V_{ts}\left[\langle pr|qt\rangle - \frac{1}{2}\langle pt|rq\rangle\right] + O(\mathbf{V}^2)
\end{aligned} \tag{F.2.3}$$

Les deux premiers éléments du terme de premier ordre peuvent aisément être exprimés avec $\mathbf{g}^{\mathrm{LR}}\left[\mathbf{d}^{(0)}\right]$. Les deux derniers, en revanche, nécessitent une manipulation des intégrales faisant usage de leurs symétries pour faire émerger $\mathbf{g}^{\mathrm{LR}}\left[\mathbf{X}\right]$ :

$$\begin{aligned}
X_{qp}V_{tp}d^{(0)}_{rs}\left[\langle tr|qs\rangle - \frac{1}{2}\langle ts|rq\rangle\right] &= X_{qp}V_{tp}g^{\mathrm{LR}}\left[\mathbf{d}^{(0)}\right]_{tq} = V^{\dagger}_{pt}(\mathbf{g}^{\mathrm{LR}}\left[\mathbf{d}^{(0)}\right]\mathbf{X})_{tp} \\
X_{qp}V_{tq}d^{(0)}_{rs}\left[\langle pr|ts\rangle - \frac{1}{2}\langle ps|rt\rangle\right] &= X_{qp}V_{tq}g^{\mathrm{LR}}\left[\mathbf{d}^{(0)}\right]_{pt} = V^{\dagger}_{qt}(\mathbf{g}^{\mathrm{LR}}\left[\mathbf{d}^{(0)}\right]\mathbf{X}^{\dagger})_{tq} \\
d^{(0)}_{rs}V_{tr}X^{\dagger}_{pq}\left[\langle tp|sq\rangle - \frac{1}{2}\langle tq|ps\rangle\right] &= d^{(0)}_{rs}V_{tr}g^{\mathrm{LR}}\left[\mathbf{X}^{\dagger}\right]_{ts} = V^{\dagger}_{rt}(\mathbf{g}^{\mathrm{LR}}\left[\mathbf{X}^{\dagger}\right]\mathbf{d}^{(0)})_{tr} \\
d^{(0)}_{rs}V_{ts}X_{qp}\left[\langle tq|rp\rangle - \frac{1}{2}\langle tp|qr\rangle\right] &= d^{(0)}_{rs}V_{ts}g^{\mathrm{LR}}\left[\mathbf{X}\right]_{tr} = V^{\dagger}_{st}(\mathbf{g}^{\mathrm{LR}}\left[\mathbf{X}\right]\mathbf{d}^{(0)})_{ts}
\end{aligned} \tag{F.2.4}$$

On a tous les éléments à présent pour écrire le terme de premier ordre de l'expansion de la somme de la fockienne longue-portée et du double comptage longue-portée :

$$\left(\left\langle \mathbf{X}\overset{*}{\mathbf{f}}^{\mathrm{LR}}\right\rangle - \frac{1}{2}\left\langle \mathbf{d}^{(0)}\overset{*}{\mathbf{g}}^{\mathrm{LR}}\left[\mathbf{d}^{(0)}\right]\right\rangle\right)^{(1)} = V^{\dagger}_{pt}(\mathbf{hX})_{tp} + V^{\dagger}_{pt}(\mathbf{g}^{\mathrm{LR}}\left[\mathbf{d}^{(0)}\right]\mathbf{X})_{tp} + V^{\dagger}_{st}(\mathbf{g}^{\mathrm{LR}}\left[\mathbf{X}\right]\mathbf{d}^{(0)})_{ts}$$
$$- \tfrac{1}{2}V^{\dagger}_{pt}(\mathbf{g}^{\mathrm{LR}}\left[\mathbf{d}^{(0)}\right]\mathbf{d}^{(0)})_{tp} - \tfrac{1}{2}V^{\dagger}_{st}(\mathbf{g}^{\mathrm{LR}}\left[\mathbf{d}^{(0)}\right]\mathbf{d}^{(0)})_{ts}$$
$$+ V^{\dagger}_{qt}(\mathbf{hX}^{\dagger})_{tq} + V^{\dagger}_{qt}(\mathbf{g}^{\mathrm{LR}}\left[\mathbf{d}^{(0)}\right]\mathbf{X}^{\dagger})_{tq} + V^{\dagger}_{rt}(\mathbf{g}^{\mathrm{LR}}\left[\mathbf{X}^{\dagger}\right]\mathbf{d}^{(0)})_{tr}$$
$$- \tfrac{1}{2}V^{\dagger}_{qt}(\mathbf{g}^{\mathrm{LR}}\left[\mathbf{d}^{(0)}\right]\mathbf{d}^{(0)})_{tq} - \tfrac{1}{2}V^{\dagger}_{rt}(\mathbf{g}^{\mathrm{LR}}\left[\mathbf{d}^{(0)}\right]\mathbf{d}^{(0)})_{tr}$$
$$= V^{\dagger}_{pt}(\mathbf{f}^{\mathrm{LR}}\mathbf{X})_{tp} + V^{\dagger}_{st}(\mathbf{g}^{\mathrm{LR}}\left[\mathbf{X} - \mathbf{d}^{(0)}\right]\mathbf{d}^{(0)})_{ts}$$
$$+ V^{\dagger}_{qt}(\mathbf{f}^{\mathrm{LR}}\mathbf{X}^{\dagger})_{tq} + V^{\dagger}_{rt}(\mathbf{g}^{\mathrm{LR}}\left[\mathbf{X}^{\dagger} - \mathbf{d}^{(0)}\right]\mathbf{d}^{(0)})_{tr}, \quad \text{(F.2.5)}$$

qui s'écrit, dans le cas d'une trace avec la matrice hermitienne $\left(\mathbf{d}^{(0)} + \mathbf{d}^{(2)} + \mathbf{z}\right)$ :

$$\mathrm{LR}^{(1)} = 2V^{\dagger}_{pt}\left(\mathbf{f}^{\mathrm{LR}}\left(\mathbf{d}^{(0)} + \mathbf{d}^{(2)} + \mathbf{z}\right)\right)_{tp} + 2V^{\dagger}_{rt}\left(\mathbf{g}^{\mathrm{LR}}\left[\mathbf{d}^{(2)} + \mathbf{z}\right]\mathbf{d}^{(0)}\right)_{tr} \quad \text{(F.2.6)}$$

## F.3 Fockienne courte-portée (et double comptage)

On cherche à expliciter :

$$\left.\frac{\partial \mathrm{SR}}{\partial \mathbf{V}}\right|_{\mathbf{V=0}} = \frac{\partial}{\partial \mathbf{V}}\left(\left\langle\left(\mathbf{d}^{(0)} + \mathbf{d}^{(2)} + \mathbf{z}\right)\mathbf{f}^{\mathrm{SR}}\right\rangle + \Delta^{\mathrm{SR}}_{\mathrm{DC}}\right)_{\mathbf{V=0}} \quad \text{(F.3.1)}$$

La dérivée de la trace de la fockienne s'écrit comme suit :

$$X_{qp}\left.\frac{\partial f^{\mathrm{SR}}_{pq}}{\partial V_{ab}}\right|_{\mathbf{V=0}} = \sum_{A}\int dr\, \frac{\partial}{\partial V_{ab}}\left.\left(\frac{\partial F}{\partial \xi_A}\frac{\partial \xi_A}{\partial d^{(0)}_{pq}}\right)\right|_{V_{ab}=0} X_{qp}$$
$$= \underbrace{\sum_{AB}\int dr\, \frac{\partial^2 F}{\partial \xi_B \partial \xi_A}\left.\frac{\partial \xi_B}{\partial V_{ab}}\right|_{V_{ab}=0}\left(\frac{\partial \xi_A}{\partial d^{(0)}_{pq}}X_{qp}\right)}_{(a)} + \underbrace{\sum_{A}\int dr\, \frac{\partial F}{\partial \xi_A}\frac{\partial}{\partial V_{ab}}\left.\left(\frac{\partial \xi_A}{\partial d^{(0)}_{pq}}X_{qp}\right)\right|_{V_{ab}=0}}_{(b)}, \quad \text{(F.3.2)}$$

où les notations $\xi_A$ et $\xi_B$ dénotent deux dépendances (*a priori*) différentes, c'est-à-dire où notamment $\frac{\partial^2}{\partial \xi_A \partial \xi_B}$ représente une dérivée de la forme, par exemple $\frac{\partial^2}{\partial \rho_\alpha \partial \rho_\beta}$, $\frac{\partial^2}{\partial \rho_\alpha \partial \nabla \rho_\alpha \nabla \rho_\beta}$, *etc.*..

On va montrer que cette dérivation fait apparaître, de la même manière que dans le cas longue-portée, les termes suivants :

$$X_{qp}\left.\frac{\partial f^{\mathrm{SR}}_{pq}}{\partial V_{ab}}\right|_{\mathbf{V=0}} = 2\left(\mathbf{f}^{\mathrm{SR}}\mathbf{X}\right)_{ab} + 2\left(\mathbf{W}^{\mathrm{SR}}\left[\mathbf{X}\right]\mathbf{d}^{(0)}\right)_{ab}, \quad \text{(F.3.3)}$$

où le terme $\mathbf{W}^{\mathrm{SR}}[\mathbf{X}]\,\mathbf{d}^{(0)}$ est le miroir courte-portée du terme $\mathbf{g}^{\mathrm{LR}}[\mathbf{X}]\,\mathbf{d}^{(0)}$ trouvé dans l'équation (F.2.6). Pour s'en convaincre il faut dériver attentivement tous les éléments des termes (a) et (b). De manière complètement générale, j'écris que toutes les dépendances sont des produits de fonctions de la densité :

$$\xi_A = \prod_n \mathcal{X}_n^{\xi_A}\left(\rho^{(0)}\right) = \prod_n \mathcal{X}_n^{\xi_A}\left(d_{pq}^{(0)}\phi_p^\dagger\phi_q\right), \tag{F.3.4}$$

où $\mathcal{X}^{\xi_A}$ sont des fonctions différentes pour chaque $\xi_A$. Seules les fonctionnelles incluant explicitement des orbitales virtuelles ne peuvent s'écrire de cette manière. Avec ces notations, on peut écrire les éléments du terme $(a)$ :

$$(1):\quad \frac{\partial \xi_A}{\partial d_{pq}^{(0)}} = \sum_i \mathcal{X}_i^{\xi_A}\left(\phi_p^\dagger\phi_q\right)\prod_{n\neq i}\mathcal{X}_n^{\xi_A}\left(\rho^{(0)}\right) \tag{F.3.5}$$

$$\text{et}:\quad \frac{\partial \xi_A}{\partial d_{pq}^{(0)}}X_{qp} = \sum_i \mathcal{X}_i^{\xi_A}\left(\rho^{\mathbf{X}}\right)\prod_{n\neq i}\mathcal{X}_n^{\xi_A}\left(\rho^{(0)}\right) = \xi_A^{\mathbf{X}} \tag{F.3.6}$$

$$(2):\quad \left.\frac{\partial \xi_B}{\partial V_{ab}}\right|_{V_{ab}=0} = 2\sum_i \mathcal{X}_i^{\xi_B}\left(\phi_a^\dagger\phi_t\right)d_{tb}^{(0)}\prod_{n\neq i}\mathcal{X}_n^{\xi_B}\left(\rho^{(0)}\right), \tag{F.3.7}$$

où l'on utilise la dérivation suivante :

$$\sum_{pq}\left.\frac{\partial d_{pq}^{(0)}\phi_p^\dagger\phi_q}{\partial V_{ab}}\right|_{V_{ab}=0} = 2\phi_a^\dagger\phi_t d_{tb}^{(0)} \tag{F.3.8}$$

Dans l'élément (1) (équation (F.3.5)), une des occurrences de $\mathcal{X}^{\xi_A}$ est dérivée, les autres sont laissées inchangées, c'est-à-dire qu'*une* occurrence de $d_{pq}^{(0)}$ est « éliminée » dans $\xi_A$. On voit ensuite dans (F.3.6) que des termes $\xi_A^{\mathbf{X}}$ émergent. Ce sont des versions modifiées de $\xi_A$, c'est-à-dire des versions où *une* occurrence de $d_{pq}^{(0)}$ dans $\xi_A$ est remplacée par un autre élément de matrice densité, $X_{qp}$. Dans (F.3.7), on voit poindre une structure de type $\square_{at}d_{tb}^{(0)}$ qui mènera *in fine* à $W^{\mathrm{SR}}[\mathbf{X}]_{at}\,d_{tb}^{(0)}$ ; l'expression très générale du terme $(a)$ est d'ailleurs :

$$(a) = \int dr\,\frac{\partial^2 F}{\partial\xi_B\partial\xi_A}\left.\frac{\partial\xi_B}{\partial V_{ab}}\right|_{V_{ab}=0}\left(\frac{\partial\xi_A}{\partial d_{pq}^{(0)}}X_{qp}\right) = 2\left(\int dr\,\frac{\partial^2 F}{\partial\xi_B\partial\xi_A}\xi_A^{\mathbf{X}}\sum_i\mathcal{X}_i^{\xi_B}\left(\phi_a^\dagger\phi_t\right)\prod_{n\neq i}\mathcal{X}_n^{\xi_B}\left(\rho^{(0)}\right)\right)d_{tb}^{(0)}, \tag{F.3.9}$$

c'est-à-dire contribue uniquement au terme $W^{\mathrm{SR}}[\mathbf{X}]_{at}\,d_{tb}^{(0)}$.

La dérivation du terme $(b)$ est plus compliquée : dû à des effets de « dérivation d'un produit », il fournit à la fois des éléments pour reconstituer le terme $\mathbf{f}^{\mathrm{SR}}\mathbf{X}$ de l'équation (F.3.3) et des éléments supplémentaires de $\mathbf{W}^{\mathrm{SR}}[\mathbf{X}]\,\mathbf{d}^{(0)}$. En effet, la dérivation par $V_{ab}$ se fait soit sur un $\mathcal{X}^{\xi_A}$ « contaminé » par $\mathbf{X}$, soit sur un $\mathcal{X}^{\xi_A}$ « d'origine », c'est-à-dire que la dérivation par $V_{ab}$ va faire sortir soit un $X_{tb}$, soit un $d_{tb}^{(0)}$ :

$$
\begin{aligned}
(b) = \int dr\, \frac{\partial F}{\partial \xi_A} \frac{\partial}{\partial V_{ab}} \left( \frac{\partial \xi_A}{\partial d^{(0)}_{pq}} X_{qp} \right)_{V_{ab}=0} &= \int dr\, \frac{\partial F}{\partial \xi_A} \sum_i \left( \frac{\partial}{\partial V_{ab}} \mathcal{X}^{\xi_A}_i \left( \rho^{\mathbf{X}} \right) \right)_{V_{ab}=0} \prod_{n \neq i} \mathcal{X}^{\xi_A}_n \left( \rho^{(0)} \right) \\
&+ \int dr\, \frac{\partial F}{\partial \xi_A} \sum_{ij} \mathcal{X}^{\xi_A}_i \left( \rho^{\mathbf{X}} \right) \left( \frac{\partial}{\partial V_{ab}} \mathcal{X}^{\xi_A}_j (\rho) \right)_{V_{ab}=0} \prod_{\substack{n \neq i \\ n \neq j}} \mathcal{X}^{\xi_A}_n \left( \rho^{(0)} \right) \\
&= 2 \left( \int dr\, \frac{\partial F}{\partial \xi_A} \sum_i \mathcal{X}^{\xi_A}_i \left( \phi^\dagger_a \phi_t \right) \prod_{n \neq i} \mathcal{X}^{\xi_A}_n \left( \rho^{(0)} \right) \right) X_{tb} \\
&+ 2 \left( \int dr\, \frac{\partial F}{\partial \xi_A} \sum_{ij} \mathcal{X}^{\xi_A}_i \left( \rho^{\mathbf{X}} \right) \mathcal{X}^{\xi_A}_j \left( \phi^\dagger_a \phi_t \right) \prod_{\substack{n \neq i \\ n \neq j}} \mathcal{X}^{\xi_A}_n \left( \rho^{(0)} \right) \right) d^{(0)}_{tb},
\end{aligned}
\tag{F.3.10}
$$

où on ne fait qu'utiliser le résultat (F.3.8). Dans le premier terme, on voit l'expression de la fockienne (voir équation (F.3.5)) et le deuxième terme fournit de nouveaux éléments à $\mathbf{W}^{\mathrm{SR}}\left[\mathbf{X}\right]$, dont l'expression finale est :

$$
\begin{aligned}
W^{\mathrm{SR}}\left[\mathbf{X}\right]_{at} = \int dr \sum_{AB} \sum_{ij} \Big( & \frac{\partial^2 F}{\partial \xi_A \xi_B} \xi^{\mathbf{X}}_B \mathcal{X}^{\xi_A}_j \left( \rho^{(0)} \right) \\
& + \frac{\partial F}{\partial \xi_A} \mathcal{X}^{\xi_A}_j \left( \rho^{\mathbf{X}} \right) \Big) \mathcal{X}^{\xi_A}_i \left( \phi^\dagger_a \phi_t \right) \prod_{\substack{n \neq i \\ n \neq j}} \mathcal{X}^{\xi_A}_n \left( \rho^{(0)} \right)
\end{aligned}
\tag{F.3.11}
$$

Considérons à présent la dérivée du terme de double comptage $\Delta^{\mathrm{SR}}_{\mathrm{DC}} = E^{\mathrm{SR}}_{\mathrm{Hxc}} - \left\langle \mathbf{d}^{(0)} \mathbf{f}^{\mathrm{SR}} \right\rangle$. La dérivée de la fonctionnelle Hartree-échange-corrélation est :

$$
\left. \frac{\partial E^{\mathrm{SR}}_{\mathrm{Hxc}}}{\partial \mathbf{V}} \right|_{\mathbf{V}=\mathbf{0}} = \sum_A \int dr\, \frac{\partial F}{\partial \xi_A} \left. \frac{\partial \xi_A}{\partial \mathbf{V}} \right|_{\mathbf{V}=\mathbf{0}}
\tag{F.3.12}
$$

En se rappelant que l'expression de la fockienne est $\mathbf{f}^{\mathrm{SR}} = \sum_A \int dr\, \frac{\partial F}{\partial \xi_A} \frac{\partial \xi_A}{\partial \mathbf{d}^{(0)}}$, une comparaison des équations (F.3.5) et (F.3.7) permet d'écrire :

$$
\left. \frac{\partial E^{\mathrm{SR}}_{\mathrm{Hxc}}}{\partial \mathbf{V}} \right|_{\mathbf{V}=\mathbf{0}} = 2 \mathbf{f}^{\mathrm{SR}} \mathbf{d}^{(0)}
\tag{F.3.13}
$$

Ainsi, en utilisant (F.3.3) et (F.3.13), on montre que la dérivée du terme de double comptage s'écrit :

$$
\left. \frac{\partial \Delta^{\mathrm{SR}}_{\mathrm{DC}}}{\partial V_{ab}} \right|_{\mathbf{V}=\mathbf{0}} = 2 \mathbf{f}^{\mathrm{SR}} \mathbf{d}^{(0)} - 2 \left( \mathbf{f}^{\mathrm{SR}} \mathbf{d}^{(0)} \right)_{ab} - 2 \left( \mathbf{W}^{\mathrm{SR}} \left[ \mathbf{d}^{(0)} \right] \mathbf{d}^{(0)} \right)_{ab},
\tag{F.3.14}
$$

c'est-à-dire annule le terme $\mathbf{W}^{\text{SR}}\left[\mathbf{d}^{(0)}\right]\mathbf{d}^{(0)}$ qui émerge lors de la dérivation de la trace de la fockienne. On obtient ainsi l'expression finale pour la dérivée de SR par rapport à $\mathbf{V}$ :

$$\left.\frac{\partial \text{SR}}{\partial \mathbf{V}}\right|_{\mathbf{V}=\mathbf{0}} = 2\left(\mathbf{f}^{\text{SR}}\left(\mathbf{d}^{(0)}+\mathbf{d}^{(2)}+\mathbf{z}\right)\right)+2\mathbf{W}^{\text{SR}}\left[\mathbf{d}^{(2)}+\mathbf{z}\right]\mathbf{d}^{(0)} \tag{F.3.15}$$

## F.4 Un parallèle remarquable

Cette section reprend les résultats dérivés dans les sections F.2 et F.3 pour mieux mettre à jour les similarités dans la dérivation des contributions des termes « fockienne plus double comptage » de courte- et longue-portée aux équations de stationnarité du Lagrangien. Pour plus de clarté, on définit $\mathbf{d}=\mathbf{d}^{(0)}+\mathbf{d}^{(2)}+\mathbf{z}$. Dans la partie longue-portée, la dérivation du terme $\left\langle \mathbf{d}\mathbf{f}^{\text{LR}}\right\rangle$ fait émerger $\mathbf{f}^{\text{LR}}\mathbf{d}$ ainsi que, par un phénomène d'« interversion », $\mathbf{g}^{\text{LR}}\left[\mathbf{d}\right]\mathbf{d}^{(0)}$. L'effet de la dérivation du terme de double comptage est d'annuler la partie $\mathbf{g}^{\text{LR}}\left[\mathbf{d}^{(0)}\right]\mathbf{d}^{(0)}$ du terme d'interversion. On a ainsi :

$$\frac{1}{2}\frac{\partial}{\partial \mathbf{V}}\left.\left(\left\langle \mathbf{d}\mathbf{f}^{\text{LR}}\right\rangle+\text{DC}^{\text{LR}}\right)\right|_{\mathbf{V}=\mathbf{0}} = \mathbf{f}^{\text{LR}}\mathbf{d}+\mathbf{g}^{\text{LR}}\left[\mathbf{d}^{(2)}+\mathbf{z}\right]\mathbf{d}^{(0)} \tag{F.4.1}$$

De la même manière, pour la partie courte-portée, le terme $\left\langle \mathbf{d}\mathbf{f}^{\text{SR}}\right\rangle$ fait émerger $\mathbf{f}^{\text{SR}}\mathbf{d}$ ainsi que, par un phénomène d'« interversion » finalement comparable à précédemment, le nouvel objet que j'ai appelé $\mathbf{W}^{\text{SR}}\left[\mathbf{d}\right]\mathbf{d}^{(0)}$. Le terme de double comptage se comporte de la même manière que le double comptage longue-portée : il annule la partie $\mathbf{W}^{\text{SR}}\left[\mathbf{d}^{(0)}\right]\mathbf{d}^{(0)}$ dans le terme interverti. On a finalement :

$$\frac{1}{2}\frac{\partial}{\partial \mathbf{V}}\left.\left(\left\langle \mathbf{d}\mathbf{f}^{\text{SR}}\right\rangle+\text{DC}^{\text{SR}}\right)\right|_{\mathbf{V}=\mathbf{0}} = \mathbf{f}^{\text{SR}}\mathbf{d}+\mathbf{W}^{\text{SR}}\left[\mathbf{d}^{(2)}+\mathbf{z}\right]\mathbf{d}^{(0)} \tag{F.4.2}$$

**Les équations** (F.4.1) et (F.4.2) sont le miroir l'une de l'autre ; les termes $\mathbf{g}^{\text{LR}}\left[\mathbf{X}\right]\mathbf{d}^{(0)}$ et $\mathbf{W}^{\text{SR}}\left[\mathbf{X}\right]\mathbf{d}^{(0)}$ sont très semblables et émergent de l'exacte même manière au cours de la dérivation. Cela montre que je tiens ici une dérivation solide du gradient des énergies de méthodes mélangeant courte- et longue-portée.

## F.5 Lagrangien total

En observant les équations (F.2.5) et (F.3.15), on peut écrire :

$$\left.\frac{\partial \text{SR}+\text{LR}}{\partial \mathbf{V}}\right|_{\mathbf{V}=\mathbf{0}} = 2\mathbf{f}\left(\mathbf{d}^{(0)}+\mathbf{d}^{(2)}+\mathbf{z}\right)+2\mathbf{g}^{\text{LR}}\left[\mathbf{d}^{(2)}+\mathbf{z}\right]\mathbf{d}^{(0)}+2\mathbf{W}^{\text{SR}}\left[\mathbf{d}^{(2)}+\mathbf{z}\right]\mathbf{d}^{(0)} \tag{F.5.1}$$

On dispose à présent des équations (F.1.5) et (F.5.1) pour écrire les conditions stationnaires du Lagrangien que l'on rappelle ici :

$$\mathcal{L} = \Big\langle \underbrace{\Big(\mathbf{d}^{(0)} + \mathbf{z} + \mathbf{d}^{(2)}\Big)\mathbf{f}\Big\rangle + \Delta_{\mathrm{DC}}}_{\mathrm{SR}+\mathrm{LR}} + \langle \mathbf{KM} \rangle + \langle \mathbf{K}'\mathbf{N} \rangle + \langle \mathbf{JO} \rangle + \Big\langle \mathbf{x}(\mathbf{C}^\dagger \mathbf{SC} - \mathbf{1})\Big\rangle, \tag{6.3.7}$$

c'est-à-dire :

$$\begin{aligned} \frac{1}{2}\frac{\partial \mathcal{L}}{\partial \mathbf{V}} = {} & \mathbf{f}\Big(\mathbf{d}^{(0)} + \mathbf{d}^{(2)} + \mathbf{z}\Big) + \mathbf{g}^{\mathrm{LR}}\Big[\mathbf{d}^{(2)} + \mathbf{z}\Big]\mathbf{d}^{(0)} + \mathbf{W}^{\mathrm{SR}}\Big[\mathbf{d}^{(2)} + \mathbf{z}\Big]\mathbf{d}^{(0)} \\ & + \{\mathbf{K}, \mathbf{M}\} + \{\mathbf{K}', \mathbf{N}\} + \{\mathbf{J}, \mathbf{O}\} + \mathbf{x} \end{aligned} \tag{F.5.2}$$

Les expressions des éléments de matrice $\frac{\partial \mathcal{L}}{\partial V_{pq}}$ ne sont pas les mêmes selon la nature de $p$ et $q$. Les éléments $f_{pq}$, $d^{(0)}_{pq}$ et $d^{(2)}_{pq}$ sont non nuls si et seulement si $p$ et $q$ appartiennent à la même catégorie d'orbitale (orbitales occupées ou orbitales virtuelles), au contraire des éléments $z_{pq}$, qui sont non nuls si $p$ et $q$ appartiennent à des catégories d'orbitale différentes. Ainsi $(\mathbf{fd}^{(0)})_{pq}$ et $(\mathbf{fd}^{(2)})_{pq}$ sont non nuls si $p$ et $q$ appartiennent à la même catégorie d'orbitale et $(\mathbf{fz})_{pq}$ est non nul si $p$ et $q$ appartiennent à des catégories d'orbitale différentes. De plus les éléments de matrices clôturés par $\mathbf{d}^{(0)}$ ne sont non nuls que dans les domaines $ij$ et $aj$. Étant donnés ces remarques structurelles, les éléments de matrice $\frac{\partial \mathcal{L}}{\partial V_{pq}}$ s'écrivent :

$$\begin{aligned} \frac{\partial \mathcal{L}}{\partial V_{ij}} &= 2\Big(\mathbf{fd}^{(2)} + \mathbf{fd}^{(0)} + \mathbf{g}^{\mathrm{LR}}\Big[\mathbf{z} + \mathbf{d}^{(2)}\Big]\mathbf{d}^{(0)} + \mathbf{W}^{\mathrm{SR}}\Big[\mathbf{z} + \mathbf{d}^{(2)}\Big]\mathbf{d}^{(0)} + \{\mathbf{K}, \mathbf{M}\} + \{\mathbf{K}', \mathbf{N}\} + \{\mathbf{J}, \mathbf{O}\} + x\Big)_{ij} \\ \frac{\partial \mathcal{L}}{\partial V_{aj}} &= 2\Big(\mathbf{fz} + \mathbf{g}^{\mathrm{LR}}\Big[\mathbf{z} + \mathbf{d}^{(2)}\Big]\mathbf{d}^{(0)} + \mathbf{W}^{\mathrm{SR}}\Big[\mathbf{z} + \mathbf{d}^{(2)}\Big]\mathbf{d}^{(0)} + \Big\{\overline{\mathbf{K}}, \mathbf{M}\Big\} + \Big\{\overline{\mathbf{K}}', \mathbf{N}\Big\} + \Big\{\overline{\mathbf{J}}, \mathbf{O}\Big\} + x\Big)_{aj} \\ \frac{\partial \mathcal{L}}{\partial V_{ib}} &= 2\Big(\mathbf{fz} + \Big\{\overline{\mathbf{K}}, \mathbf{M}\Big\} + \Big\{\overline{\mathbf{K}}', \mathbf{N}\Big\} + \Big\{\overline{\mathbf{J}}, \mathbf{O}\Big\} + x\Big)_{ib} \\ \frac{\partial \mathcal{L}}{\partial V_{ab}} &= 2\Big(\mathbf{fd}^{(2)} + \{\mathbf{K}, \mathbf{M}\} + \{\mathbf{K}', \mathbf{N}\} + \{\mathbf{J}, \mathbf{O}\} + x\Big)_{ab} \end{aligned} \tag{F.5.3}$$

## F.6 Dérivée du Lagrangien

La dérivé du Lagrangien tel qu'écrit équation (6.3.7) est :

$$\mathcal{L}^{(x)} = \Big\langle\Big(\mathbf{d}^{(0)} + \mathbf{z} + \mathbf{d}^{(2)}\Big)\mathbf{f}^{(x)}\Big\rangle + \Delta^{(x)}_{\mathrm{DC}} + \Big\langle \mathbf{MK}^{(x)}\Big\rangle + \Big\langle \mathbf{NK}'^{(x)}\Big\rangle + \Big\langle \mathbf{OJ}^{(x)}\Big\rangle + \Big\langle \mathbf{xC}^\dagger\mathbf{S}^{(x)}\mathbf{C}\Big\rangle \tag{F.6.1}$$

Seule la dérivation des termes fockienne plus double comptage demandent une attention particulière. On obtient, pour les parties courte- et longue-portée :

$$\mathrm{SR}^{(x)} = E_{Hxc}^{\mathrm{SR}(x)} + \left\langle \left(\mathbf{d}^{(2)} + \mathbf{z}\right) \mathbf{f}^{\mathrm{SR}(x)} \right\rangle \tag{F.6.2}$$

$$\begin{aligned} \mathrm{LR}^{(x)} &= \left\langle \left(\mathbf{d}^{(0)} + \mathbf{d}^{(2)} + \mathbf{z}\right) \mathbf{h}^{(x)} \right\rangle \\ &+ \left( \left(\tfrac{1}{2}\mathbf{d}^{(0)} + \mathbf{z} + \mathbf{d}^{(2)}\right)_{pq} d_{rs}^{(0)} \right) (pq|rs)^{\mathrm{LR}(x)} \\ &- \frac{1}{2} \left( \left(\tfrac{1}{2}\mathbf{d}^{(0)} + \mathbf{z} + \mathbf{d}^{(2)}\right)_{pr} d_{ps}^{(0)} \right) (pq|rs)^{\mathrm{LR}(x)} \end{aligned} \tag{F.6.3}$$

Les dérivées apparaissant dans les termes de longue-portée ($\mathbf{h}^{(x)}$ et $(pq|rs)^{\mathrm{LR}(x)}$) sont connues, il nous faut expliciter les dérivées de la partie courte-portée, où les intégrales sont évaluées sur une grille de point $\{\omega_\lambda\}$. Deux termes vont donc émerger de la dérivation : des termes contenant les dérivées de l'intégrande *et* des termes traduisant la variation de la grille avec la modification des coordonnées atomiques :

$$\begin{aligned} \mathrm{SR}^{(x)} &= \sum_A \left( \int F(\xi_A) + \frac{\partial F}{\partial \xi_A} \frac{\partial \xi_A}{\partial d_{pq}^{(0)}} \left(\mathbf{d}^{(2)} + \mathbf{z}\right)_{qp} \right)^{(x)} \\ &= \sum_A \left\{ \omega_\lambda \left( F(\xi_A) + \frac{\partial F}{\partial \xi_A} \left( \xi_A^{\mathbf{d}^{(2)}} + \xi_A^{\mathbf{z}} \right) \right) \right\}^{(x)} \\ &= \sum_A \omega_\lambda^{(x)} \left( F(\xi_A) + \frac{\partial F}{\partial \xi_A} \left( \xi_A^{\mathbf{d}^{(2)}} + \xi_A^{\mathbf{z}} \right) \right) \\ &+ \sum_{AB} \omega_\lambda \left( \frac{\partial F}{\partial \xi_B} \xi_B^{(x)} + \frac{\partial^2 F}{\partial \xi_B \partial \xi_A} \left( \xi_A^{\mathbf{d}^{(2)}} + \xi_A^{\mathbf{z}} \right) \xi_B^{(x)} + \frac{\partial F}{\partial \xi_A} \left( \xi_A^{\mathbf{d}^{(2)}(x)} + \xi_A^{\mathbf{z}(x)} \right) \right) \end{aligned} \tag{F.6.4}$$

Soit, finalement :

$$\begin{aligned} \mathrm{SR}^{(x)} &= \sum_A \omega_\lambda^{(x)} \left( F(\xi_A) + \frac{\partial F}{\partial \xi_A} \left( \xi_A^{\mathbf{d}^{(2)}} + \xi_A^{\mathbf{z}} \right) \right) \\ &+ \sum_A \omega_\lambda \frac{\partial F}{\partial \xi_A} \left( \xi_A^{\mathbf{d}^{(0)}(x)} + \xi_A^{\mathbf{d}^{(2)}(x)} + \xi_A^{\mathbf{z}(x)} \right) + \sum_{AB} \omega_\lambda \frac{\partial^2 F}{\partial \xi_B \partial \xi_A} \left( \xi_A^{\mathbf{d}^{(2)}} + \xi_A^{\mathbf{z}} \right) \xi_B^{(x)} \end{aligned} \tag{F.6.5}$$

On peut écrire, en AO :

$$\mathcal{L}^{(x)} = \left\langle \mathbf{D}^1 \mathbf{H}^{(x)} \right\rangle + \left( \mathbf{D}^2 + \boldsymbol{\Gamma}^2 \right)_{\mu\nu,\rho\sigma} (\mu\nu|\rho\sigma)^{\mathrm{LR}(x)} + \left\langle \mathbf{X}\mathbf{S}^{(x)} \right\rangle + \mathrm{SR}^{(x)}, \tag{F.6.6}$$

où :

$$\begin{aligned} (\mathbf{D}^1)_{\mu\nu} &= C_{\mu p} \left( \mathbf{d}^{(0)} + \mathbf{d}^{(2)} + \mathbf{z} \right)_{pq} C_{q\nu}^\dagger = \left( \mathbf{D}^{(0)} + \mathbf{D}^{(2)} + \mathbf{Z} \right)_{\mu\nu} \\ (\mathbf{D}^2)_{\mu\nu,\sigma\rho} &= \left( \tfrac{1}{2}\mathbf{D}^{(0)} + \mathbf{D}^{(2)} + \mathbf{Z} \right)_{\mu\nu} D_{\rho\sigma}^{(0)} - \frac{1}{2} \left( \tfrac{1}{2}\mathbf{D}^{(0)} + \mathbf{D}^{(2)} + \mathbf{Z} \right)_{\mu\rho} D_{\nu\sigma}^{(0)} \\ (\boldsymbol{\Gamma}^2)_{\mu\nu,\sigma\rho} &= C_{\mu k} C_{\nu j} C_{c\rho}^\dagger C_{b\sigma}^\dagger (\mathbf{M})_{ia,kc} + C_{\mu k} C_{\nu j} C_{b\rho}^\dagger C_{c\sigma}^\dagger (\mathbf{N})_{ia,kc} + C_{\mu k} C_{\nu b} C_{j\rho}^\dagger C_{c\sigma}^\dagger (\mathbf{O})_{ia,kc} \\ (\mathbf{X})_{\mu\nu} &= C_{\mu p} (\mathbf{x})_{pq} C_{q\nu}^\dagger \end{aligned} \tag{F.6.7}$$

# Bibliographie

*Premières approches d'un problème à N-corps*

*RPA : Approximation de la Phase Aléatoire*

*Orbitales localisées*

*Visualisations dans l'espace réel*

*EED : Approximation du dénominateur effectif*

*Gradients analytiques des énergies RSH-RPA*

*Un contexte pour la RPA, des discussions autour des fonctions de réponse*

*Éléments d'intégration complexe*

*Détails de l'adaptation de spin*

*Dérivations autour de l'approximation EED*